\documentclass[12pt,letterpaper]{JHEP3}
\usepackage{amscd,amsmath,amssymb,amsfonts,xspace,mathrsfs,amsthm}
\usepackage{xcolor}

\usepackage{bbold}
\usepackage{latexsym}
\usepackage{graphicx}
\usepackage{dsfont}
\usepackage{color}
\usepackage{longtable}
\usepackage{tikz-cd}
\usetikzlibrary{bending}

\numberwithin{equation}{section}
 
\newcommand{\nn}{\nonumber}

\newcommand{\eps}{\epsilon}

\newcommand{\IR}{\mathds{R}}
\newcommand{\IC}{\mathds{C}}
\newcommand{\IF}{\mathds{F}}
\newcommand{\IZ}{\mathds{Z}}

\newcommand{\IQ}{\mathds{Q}}

\newcommand{\IP}{\mathds{P}}

\newcommand{\Tr}{\mbox{Tr}}

\newcommand{\sgn}{\mbox{sgn}}

\newcommand{\cA}{\mathcal{A}}
\newcommand{\cB}{\mathcal{B}}
\newcommand{\cC}{\mathcal{C}}
\newcommand{\cD}{\mathcal{D}}
\newcommand{\cE}{\mathcal{E}}

\newcommand{\cI}{\mathcal{I}}

\newcommand{\cL}{\mathcal{L}}
\newcommand{\cM}{\mathcal{M}}
\newcommand{\cN}{\mathcal{N}}
\newcommand{\cO}{\mathcal{O}}
\newcommand{\cP}{\mathcal{P}}
\newcommand{\cQ}{\mathcal{Q}}

\renewcommand{\Im}{{\rm Im}}
\newcommand{\q}{\mbox{q}}

\newcommand{\I}{\mathrm{i}}

\def\bOm{\bar\Omega}

\newcommand{\rk}{\mathrm{rk}}
\newcommand{\ch}{\mathrm{ch}}
\newcommand{\Td}{\mathrm{Td}}
\newcommand{\Ext}{\mathrm{Ext}}
\newcommand{\Hom}{\mathrm{Hom}}
\newcommand{\kahler}{K\"ahler }

\newcommand{\gtr}{g_{\rm tr}}
\newcommand{\Ftr}{F_{{\rm tr}}}

\def\bea{\begin{eqnarray}}
\def\eea{\end{eqnarray}}
\def\be{\begin{equation}}
\def\ee{\end{equation}}
\def\ba{\begin{align}}
\def\ea{\end{align}}
\def\bse{\begin{subequations}}
\def\ese{\end{subequations}}
\newcommand{\non}{\nonumber}
\def\OmS{\Omega_{\rm S}}
\def\Omstar{\Omega_{\star}}
\def\zetastar{\zeta^{\star}}
\def\dstar{d^{\star}}

\def\hVW{h^{S}}
\def\hrVW{h^{S,{\rm ref}}}
\def\hiVW{\mathfrak{h}}

\def\({\left(}
\def\){\right)}
\def\[{\left[}
\def\]{\right]}
\def\<{\left\langle}
\def\>{\right\rangle}
\def\hf{{1\over 2}}

\def\bp{}

\title{Vafa-Witten Invariants from Exceptional Collections}

\preprint{arXiv:2004.14466v5}

\author{Guillaume Beaujard$^{1}$, Jan Manschot$^{2,3}$, Boris Pioline$^{1}$
\\
$^1$ {\it Laboratoire de Physique Th\'eorique et Hautes
Energies (LPTHE), UMR 7589 CNRS-Sorbonne Universit\'e,
Campus Pierre et Marie Curie,
4 place Jussieu, F-75005 Paris, France}\\

$^2$ {\it School of Mathematics, Trinity College, Dublin 2, Ireland}\\

$^3$ {\it Hamilton Mathematical Institute, Trinity College, Dublin 2, Ireland}\\

\vspace*{2mm} {\tt e-mail:
\email{beaujard,pioline@lpthe.jussieu.fr,}
\email{manschot@maths.tcd.ie}
}
 
\vspace*{-3mm}

}

\abstract{Supersymmetric D-branes supported on the complex two-dimensional 
base $S$ of the local Calabi-Yau  threefold $K_S$
are described by semi-stable coherent sheaves on $S$.
 Under suitable conditions,
the BPS indices counting these objects (known as generalized Donaldson-Thomas invariants) 
coincide with the Vafa-Witten invariants of $S$  (which encode the Betti numbers of the moduli space of semi-stable sheaves). 
For surfaces which admit a strong collection of exceptional sheaves, we develop a general method for computing these invariants by exploiting the isomorphism between the derived category of coherent sheaves and the derived
category of representations of a suitable quiver with potential $(Q,W)$ constructed from the  exceptional collection. We spell out the dictionary between the Chern class $\gamma$ and polarization $J$ on $S$ vs. the dimension vector $\vec N$ and stability parameters $\vec\zeta$ on the quiver side. For all examples that we consider, which include all del Pezzo and Hirzebruch surfaces,
we find that the BPS indices $\Omega_\star(\gamma)$ at the attractor point (or self-stability condition)  vanish, except for dimension vectors corresponding to simple representations and  pure D0-branes. 
This opens up the possibility to compute the
BPS indices in any chamber using either the flow tree or the Coulomb branch formula. In all cases we find precise agreement with independent computations of Vafa-Witten invariants based on wall-crossing and blow-up formulae. This agreement suggests that i) generating functions of DT invariants for a large class of quivers coming from strong
exceptional collections  are mock modular functions of higher depth and ii)  non-trivial single-centered black holes and scaling solutions do not exist quantum mechanically in such local 
Calabi-Yau geometries. 
}

\begin{document}

\section{Introduction}

Unlike in situations with higher supersymmetry,  precision counting of BPS black hole microstates in string vacua with $\cN=2$ supersymmetry remains a challenge. In type IIA string theory compactified on a Calabi-Yau  three-fold $X$, a large class of BPS black holes can be constructed by wrapping a 
D4-brane wrapped on a complex codimension-one cycle $\cD\subset X$ divisor, or equivalently an M5-brane on $\cD\times S^1$ \cite{Maldacena:1996gb}. In this set-up, the generating function of BPS indices (defined mathematically as generalized Donaldson-Thomas invariants of the derived category of coherent sheaves) is identified with the elliptic genus of the $(0,4)$ superconformal field theory obtained by reducing the M5-brane along $\cD$, and is therefore expected to be modular \cite{deBoer:2006vg,Denef:2007vg}. This allows to determine it exactly in some simple cases \cite{Gaiotto:2006wm,Gaiotto:2007cd}. In general however, when the divisor 
$\cD$ is reducible, BPS indices have a complicated chamber structure as a function of \kahler moduli \cite{ks,Joyce:2008pc}, and the elliptic genus is only expected to be mock modular \cite{Alexandrov:2016tnf,Alexandrov:2017qhn}. While the modular anomaly has been fully characterized \cite{Alexandrov:2018lgp,Alexandrov:2019rth},  an explicit determination 
of the BPS indices remains difficult.
 
\medskip 

For a non-compact Calabi-Yau threefold, such as the total space $X={\rm Tot}(K_S)$ of the canonical bundle $K_S$
over a complex surface $S$,
the situation becomes more tractable: D4-branes wrapped on $S$ are described by a topological version of $\cN=4$ super Yang-Mills (SYM) theory with gauge group $U(N)$ \cite{Vafa:1994tf}, 
and the BPS indices are expected to coincide with the Vafa-Witten (VW) invariants \cite{Minahan:1998vr,Alim:2010cf,gholampour2017localized,Gholampour:2013jfa}. When $S$ is a Fano or almost Fano surface, vanishing theorems ensure that the gauge
theory localizes on solutions of Hermitian Yang-Mills equations, and Vafa-Witten invariants (in their
refined version) are given by the Euler number (more generally, the Poincar\'e polynomial) of the moduli space of semi-stable coherent sheaves on $S$, with the Chern vector  $\gamma=[N;\mu;n]$ determined by the D4-, D2- and D0-brane charges. This description arises by reducing the M5-brane world-volume theory along $S^1$ times the Euclidean time circle, rather than along the divisor $\cD$. The modular properties of the generating function of Vafa-Witten invariants $c_\gamma$ 
at fixed $(N,\mu)$ then follow from the invariance of $\cN=4$ SYM theory under $S$-duality. 
When $b_2^+(S)=1$, which happens whenever  $S$ is a rigid divisor inside $X$,  the VW invariants $c_{\gamma,J}$ start to depend on the polarization $J$ of $S$, which determines the \kahler moduli of $X$, and the
modular properties become anomalous due to boundary contributions
from reducible connections \cite{Vafa:1994tf} (see \cite{Dabholkar:2020fde,Manschot:2021qqe} for very recent
progress on this issue).

\medskip

When $S$ is a rational surface, meaning that it can be obtained by successive blow-ups from the
complete projective plane $\IP^2$ or from $\IF_0=\IP^1\times \IP^1$, VW invariants can be computed  in principle for any rank $N$, by  combining known results for special 
polarization \cite{Manschot:2011ym,Mozgovoy:2013zqx} with the blow-up \cite{Yoshioka:1996,Gottsche:1999ix} and wall-crossing formulae \cite{ks,Joyce:2008pc}. Explicit results have been obtained for $S=\IP^2$, any rank in \cite{Manschot:2014cca}, and up to rank 3 for any ruled surface in \cite{Manschot:2016gsx}. Alternative localization techniques are available for  toric surfaces \cite{weist2011torus,Nishinaka:2013mba,Kool2015,Bershtein:2015xfa}. 
While the structure of the blow-up and wall-crossing formulae basically guarantees that the generating functions of VW invariants of rational surfaces will be mock modular \cite{Toda:2014}, the precise determination of the modular anomaly has required some efforts \cite{Alexandrov:2016enp, Manschot:2017xcr}. The anomaly is now 
fully characterized for any rank $N$ and Fano surface $S$, using the connection with DT invariants \cite{Alexandrov:2019rth},  and has been exploited to conjecture an explicit form for VW invariants on Hirzebruch and del Pezzo surfaces for any rank \cite{Alexandrov:2020bwg,Alexandrov:2020dyy}.

\medskip

In this work, we develop yet another technique for computing Vafa-Witten invariants of complex surfaces, which relies on the isomorphism between the (bounded)\footnote{For brevity   
we drop the adjective `bounded' throughout this paper.} derived category
 of coherent sheaves on $S$ with the  derived category of representations of a suitable quiver with potential \cite{baer1988tilting,bondal1990representation,rickard1989morita}. The quiver $Q$ can be constructed from any full,  strongly exceptional collection of sheaves on $S$. Such collections are known explicitly for all del Pezzo and weak del Pezzo surfaces, and can in principle be constructed for any blow-up thereof. In the physics literature, the description of D-branes on the local Calabi-Yau $\mathrm{Tot}(K_S)$ in terms of the quiver $Q$ is well-known since the mid-90s \cite{Douglas:1996sw,Douglas:2000qw,Herzog:2003dj,Herzog:2003zc,Aspinwall:2004vm}.
The basic idea is that BPS states in any chamber of \kahler moduli space can be constructed as bound states of a few absolutely stable states, the dynamics of which is governed by a $0+1$ dimensional gauge theory with matter content encoded in the  quiver $Q$ \cite{Denef:2002ru}. This provides an efficient tool for deriving the BPS spectrum of {\bp four- and five-dimensional} gauge theories with $\cN=2$ supersymmetry \cite{Fiol:2000wx,Alim:2011kw,Cecotti:2012se,Chuang:2013wt,Cordova:2013bza,Cirafici:2017iju,Closset:2019juk}.

\medskip

Using this isomorphism, the computation of the VW invariant $c_{\gamma,J}$  is reduced to the computation of the Euler characteristics $\Omega(\vec N,\vec \zeta)$ (or the Poincar\'e polynomial $\Omega(\vec N,\vec \zeta,y)$ for refined invariants) of the moduli space of stable representations of $Q$, for a certain dimension vector $\vec N$ and stability parameter $\vec \zeta$ determined from $\gamma,J$.
If the quiver $Q$ had no closed loops and therefore no relations, one could then apply the general result of \cite{1043.17010} to obtain the indices in a straightforward fashion.   
Unfortunately, the quivers relevant for del Pezzo surfaces all involve closed loops and relations,
and therefore the result of {\it loc. cit.} applies only for special dimension vectors supported on a subquiver without loops. Since the superpotential is non-generic, the general localization result of 
\cite{Hori:2014tda} is also not directly applicable. 

\medskip

Nonetheless, in this paper we shall demonstrate that the quiver description can be turned into an effective computational tool, by exploiting the following key observation: 
for any dimension vector $\vec N$ corresponding to 
a torsion-free sheaf on $S$, there exists a value $\vec \zetastar$ of the stability parameters such that 
$\Omega(\vec N,\vec\zetastar)=0$. The parameter $\vec\zetastar$ depends on $\vec N$ through
$\vec\zetastar= -\kappa\cdot \vec N$, where $\kappa$ is the antisymmetric
adjacency matrix of the quiver. This particular choice of stability condition, 
known as the `self-stability condition' in the mathematics literature
\cite[Def. 11.3]{bridgeland2016scattering}, is the analogue 
 of the `attractor moduli' in supergravity \cite{Ferrara:1995ih} (more precisely, the large volume
 attractor point \cite{Denef:2007vg,deBoer:2008fk,Manschot:2009ia}); 
 indeed, just as in the supergravity case \cite{Denef:2000nb} one
can show that two-particle bound states are always unstable in the vicinity of the point 
$\vec\zeta=\vec\zetastar$ \cite{MPSunpublished,Alexandrov:2018iao}. 
Accordingly, 
the index $\Omstar(\vec N):=\Omega(\vec N,\vec\zetastar)$ is called the `attractor index'. 

\medskip

By analyzing the expected dimension of the moduli space of stable quiver representations 
for $\vec\zeta=\vec\zetastar$,   we shall prove that the attractor index 
$\Omstar(\vec N)$ vanishes unless\footnote{A possible exception is the case where $\vec N$ corresponds to the skyscraper sheaf $\cO_{pt}$ on $S$, or in physical parlance to a pure D0-brane.
 In this case, $\vec\zetastar$ vanishes identically and   $\Omega(\vec N,\vec\zeta)$ does not depend on $\vec\zeta$. The index  $\Omstar(\vec N)$ is still well-defined but need not
 vanish.
} 
 $\vec N$ is the  dimension vector  of a simple representation 
 (with dimension 1 on one node and zero elsewhere, in which case $\Omstar(\vec N)=1$). 
  The vanishing of attractor indices for generic $\vec N$  appears to be a remarkable property\footnote{This property goes under the name of `genteelness'
 in the maths literature \cite{bridgeland2016scattering}, and also occurs in 
 the study of framed BPS states \cite{Cordova:2013bza, Cirafici:2017iju}.} 
 of the quivers relevant for del Pezzo surfaces, and possibly of all rational surfaces. 
 It will transpire from our proof that this property critically depends  on the low rank
 of the antisymmetrized Euler form (which has rank 2 for all complex surfaces),
  on the low number of arrows between nodes as well as on the detailed form of superpotential. Moreover, we shall find evidence that the same vanishing property holds
for the single-centered indices $\OmS(\vec N)$, which are closely related to attractor indices 
but in general differ (see \S\ref{sec_flowtree} and \S\ref{sec_discuss} below).

\medskip
 
Given that $\Omega(\vec N,\vec\zeta)=0$ vanishes for $\vec\zeta=\vec\zetastar$, its value in other chambers can in principle be computed by 
following its value across all walls of marginal stability. In practice, it is much easier to apply the 
flow tree formula proposed in \cite{Alexandrov:2018iao} (extending earlier works \cite{Denef:2000nb,Denef:2001xn,Denef:2007vg, Manschot:2010nc,Manschot:2010xp}), which expresses $\Omega(\vec N,\vec\zeta)$
in terms as a sum of product of attractor indices $\Omstar(\vec N_j)$ for all possible
decompositions $\vec N=\sum_j \vec N_j$.  Although this formula is cumbersome to apply 
by hand, it is easily implemented on a computer, and is part of the Mathematica package 
{\tt CoulombHiggs.m}  originally released along with \cite{Manschot:2013sya}\footnote{The latest version is publicly available at {\tt https://github.com/bpioline/CoulombHiggs}}. 
Alternatively, one may use the Coulomb branch formula  (see \cite{Manschot:2014fua} for a concise review), implemented in the same package, to compute 
$c_{\gamma,J}$ from the single-centered indices. 
This gives an algorithmic way of computing 
the VW invariants $c_{\gamma,J}$ for any Chern class $\gamma=[N;\mu;n]$ and any
polarization $J$.

\medskip

In this work we shall demonstrate on a large set of examples
that  this  algorithm indeed gives a practical way of computing 
Vafa-Witten invariants for Fano and weak Fano surfaces, at least for low rank $N$ and instanton number $n$. Our examples include the projective plane $\IP^2$, Hirzebruch surfaces $\IF_m$
with $m\leq 2$, del Pezzo surfaces $dP_k$ with $k\leq 8$ as well as toric almost Fano surfaces. 
For each of these examples,
we consider one (or several) strong exceptional collection, spell out the dictionary between the Chern
vector $\gamma$ and the dimension vector $\vec N$, identify the stability condition $\vec\zeta^c$ relevant for the canonical polarization $J\propto c_1(S)$ (and for other polarizations in selected cases), and compute the quiver index  $\Omega_c(\vec N):=\Omega(\vec N,\vec \zeta^c)$
for dimension vectors
of moderate\footnote{Results for $h\leq 7$ are computable using  
{\tt CoulombHiggs.m} on a garden variety laptop in less than 
a minute; $h=9$ typically takes a few hours.}  height $h(\vec N):=\sum_i N_i$. 
In all cases, we find 
agreement with results  for VW indices
obtained  by combining blow-up and wall-crossing formulae.
This provides a striking validation of our method, and opens up several  directions
for future research (see \S\ref{sec_discuss}).

\medskip

The outline of this work is as follows. In Section \ref{sec_gen} we review basic facts about Vafa-Witten invariants of complex surfaces, exceptional collections of coherent sheaves and invariants of moduli spaces of quiver representations. 
In Section \ref{sec_p2} we consider the simplest rational surface, namely the complex projective plane $\IP^2$, and illustrate our procedure for computing Vafa-Witten invariants 
from the quiver description. In Section \ref{sec_Hirz},
we consider the Hirzebruch surfaces $\IF_m$ (with $\IF_0= \IP^1\times \IP^1$ and $\IF_1=dP_1$, the first del Pezzo surface), and check that  the wall-crossing phenomena on the quiver side take place precisely when sheaves become unstable, as long as  the slope
stays within the window
where the quiver description is valid. 
In Section \ref{sec_dP}, we consider a variety of strong exceptional collections on the del Pezzo surfaces $dP_2$ to $dP_8$, focussing on the canonical and blow-up chambers. 
For the non-toric $dP_4$ to $dP_8$ cases, we restrict to the special
three-block strong exceptional collections constructed in \cite{karpov1998three}. 
In \S\ref{sec_discuss}, we summarize our method and discuss some open directions.
In Appendix \ref{sec_genVW}, we collect known results on Vafa-Witten invariants
for rational surfaces, extend them to higher rank when possible and tabulate the first
few terms in the generating functions. This provides a reference point for comparison with
the quiver indices computed throughout the main body of the paper. 
 In Appendix \ref{sec_threeblockatt}, 
we prove  the vanishing of attractor indices for three-block collections.
Finally, in Appendix \ref{sec_weakFano}, we consider the case of toric weak Fano surfaces,
also known as pseudo del Pezzo surfaces, where the quiver typically involves bidirectional arrows.

\medskip
\medskip

\noindent {\it Note added:}  Further evidence of the vanishing of attractor indices for toric CY threefolds, was presented in \cite{Mozgovoy:2020has}, along with a general prediction for the 
 refined index for $p$ D0-branes.

\section{Generalities \label{sec_gen}}

In this section, we review some definitions and basic facts about the three main topics which underlie this work, namely Vafa-Witten invariants of complex surfaces, exceptional collections of coherent sheaves and invariants of moduli spaces of quiver representations. Most of the material
in this section is well known, albeit scattered over many different sources.
Throughout, $S$ is a simply  connected smooth complex surface with $b_1(S)=0$,  $\Lambda_S:= H^2(S,\IZ)$ is the second cohomology lattice, a unimodular lattice of signature $(b_2^+(S),b_2^-(S))$,  $c_1(K_S)=-c_1(S)\in\Lambda_S$ is the canonical class, and $J\in \Lambda_S \otimes \IR$ is the polarization (or \kahler form).

\subsection{Vafa-Witten invariants \label{sec_VW}}
Vafa-Witten theory is a topological field theory defined on any 4-manifold $S$, obtained as one
of the possible topologically twists of $\cN=4$ SYM theory \cite{Vafa:1994tf}. 
We restrict to the case
where the gauge group is $U(N)$ with $N\geq 1$. When $S$ is a polarized complex surface
such that  $J\cdot c_1(S)>0$, vanishing theorems ensure that 
the functional integral localizes on solutions of hermitian Yang-Mills equations \cite{Vafa:1994tf,Dijkgraaf:1997ce}\footnote{In general, Vafa-Witten invariants also involve contributions from the monopole branch. In this paper we restrict to cases where vanishing
theorems apply and such contributions are absent. This is the case for Fano and weak Fano surfaces \cite{Vafa:1994tf}. 
See \cite{Tanaka:2017jom,Tanaka:2017bcw,Gottsche:2017vxs,Laarakker:2018isn,Thomas:2018lvm} for recent progress in the general case.
}.
Solutions are classified by the rank $N,$ the first Chern class $\mu=c_1(F) \in \Lambda_S:= H^2(S,\IZ)$ and the second Chern class (or instanton number)
$n=\int_S c_2(F) \in \IZ$, where $F$ is the
field strength. We denote by $\gamma:=[N;\mu;n] = [N,\mu,\ch_2]$ the Chern character, where $\ch_2=\frac12\mu^2-n$ and $\Lambda= \IZ\oplus \Lambda_S \oplus \IQ$. 
The solutions  span a moduli space $\cM^S_{\gamma,J}$, which is invariant under positive rescaling
of $J$ but may depend on its direction. This moduli space has 
expected complex dimension \cite{huybrechts2010geometry}
\be
\label{dimM}
d_\IC (\cM^S_{\gamma,J}) = \mu^2 - 2 N \ch_2  - (N^2-1) \chi(\cO_S) 
=  2N^2\Delta(F) - (N^2-1) \chi(\cO_S)  
\ee
where $\Delta(F)$ is the Bogomolov discriminant,
\be
\label{defBog}
\Delta(F):= \frac{1}{N} \left( n - \frac{N-1}{2N} \mu^2 \right) 
\ee
and $\chi(\cO_S)=1-h_{0,1}(S)+h_{0,2}(S)$ is the holomorphic Euler characteristic, equal to 1 for all cases of interest in this paper. 
By the Donaldson-Uhlenbeck-Yau theorem, the moduli space $\cM^S_{\gamma,J}$ has a natural
compactification given by the 
moduli space of semi-stable coherent
sheaves $E$ in the sense of Gieseker-Maruyama \cite{huybrechts2010geometry}. 
Recall that Gieseker-stable sheaves are those
which do not admit any proper subsheaf $E' \subset E$ with larger slope $\nu_J(E')>\nu_J(E)$,
or with identical slope but  $\ch_2(E')/N'>\ch_2(E)/N$; here the slope is defined by 
\be
\label{defnuE}
\nu_J(E') = \frac{c_1(E') \cdot J}{N'}
\ee
for torsion-free sheaves (i.e. $N'\neq 0$), or $\nu(E')=+\infty$ for torsion sheaves ($N'=0$). 
Semi-stability is similarly defined
by replacing $>$ by $\geq$. Slope stability is defined by ruling out subsheaves with identical
slope, irrespective of $\ch_2$. 
Note that the moduli space thus defined a priori depends on $J$, and 
is invariant upon tensoring $E$ with any line bundle $\cL$; 
under this operation, the first Chern class shifts as 
\be
\label{specflow}
\mu\to \mu+N c_1(\cL)\ ,\quad \ch_2 \to \ch_2 - N c_1(E)\cdot c_1(\cL) + \frac12 N \, c_1(\cL)^2
\ee
while  $\Delta(E):=\Delta(F)$ and $N$ stay invariant.
The parameter $\mu$ (known as the 't Hooft flux) can therefore be restricted to
$\Lambda_S/ N \Lambda_S \simeq \IZ_N^{b_2(S)}$.  The moduli space is furthermore invariant under reflexion $\mu\to-\mu$, which leaves $\Delta$ invariant. 

\medskip

When the vector $\gamma=[N;\mu;n]$ is primitive, the
moduli space is either empty or a smooth projective variety of dimension \eqref{dimM}
with vanishing odd degree cohomology.
The Vafa-Witten invariant $c_{\gamma,J}$ and its refined version $c_{\gamma,J}(y)$
are then  defined as the Euler number $\chi(\cM)$ and Poincar\'e (Laurent) polynomial 
$P(\cM,y):=\sum_{p=0}^{d_\IC(\cM)} y^{2p-d_\IC(\cM)} b_p(\cM)$ of $\cM=\cM^S_{\gamma,J}$,
such that $P(\cM,1)=\chi(M)$.
More generally, if $\gamma$ is not primitive, then the Vafa-Witten invariants are defined by
\bea
c_{\gamma,J} &=&
\sum_{m|\gamma} (-1)^{\dim_\IC(\cM^S_{\gamma/m,J})-\dim_\IC(\cM_{\gamma,J})}\frac{\chi(\cM^S_{\gamma/m,J})}{m^2}\, ,
\label{defCNnJ}
\\
c_{\gamma,J}^{\rm ref}(y) &=&  \sum_{m|\gamma} \frac{(-1)^{m-1}(y-1/y) }{m(y^m-y^{-m})}\, P(\cM^S_{\gamma/m,J}, - (-y)^{m})\, ,
\label{defcref}
\eea
where  the sum runs over all positive integers $m$ such that $\gamma/m\in \Lambda$,
and the Euler or Betti numbers are defined using
intersection cohomology \cite{Yoshioka:1995,Manschot:2016gsx}. 
In either case, the VW invariants are
locally independent of the polarization of $J$, since they are quantized, but they could jump 
on real-codimension one loci; this turns out to happen only for $N>1$ and $b_2^+(S)=1$
(except when  $b_2(S)=1$, since $J$ is then uniquely fixed up to scale), which is the case of interest in this paper.

With these definitions, one expects that 
the generating function of refined VW invariants\footnote{The sum
runs over integers $n\geq \frac{N-1}{2N} \mu^2$ such that the dimension
\eqref{dimM} is positive, and converges in the upper half-plane $\Im\tau>0$.}
\bea
\label{defhVWref}
\hVW_{N,\mu,J}(\tau,w) &=&
\sum_{n}
\frac{ c_{[N;\mu;n],J}^{\rm ref}(y)}{y-y^{-1}}\,
\q^{n -\tfrac{N-1}{2N} \mu^2 - \tfrac{N \chi(S)}{24}},
\eea
where $\q=e^{2\pi\I\tau}, y=e^{2\pi \I w}$, will transform a a vector-valued holomorphic  Jacobi form 
 of weight $-\frac12 b_2(S)$,  index $-\frac16 K_S^2 (N^3-N)-2N$ with $K_S^2=c_1(K_S)^2$, under $\tau\to\frac{a\tau+b}{c\tau+d}, w\to\frac{w}{c\tau+d}$
with $\scriptsize{\begin{pmatrix} a & b \\c & d\end{pmatrix}} \in SL(2,\IZ)$  \cite{Alexandrov:2019rth}. This expectation follows from the invariance of $\cN=4$ SYM theory under S-duality, which acts
on the complexified gauge coupling  $\tau=\frac{\theta}{2\pi} + \frac{4\pi\I}{g^2_{\rm YM}}$ precisely in that manner. It is vindicated in the rank 1 case, where the generating function \eqref{defhVWref} is independent of $J,\mu$ and given by  \cite{Gottsche:1990},
\be
\label{h10anyS}
\hVW_{1}(\tau,w) = \frac{\I}{\theta_1(\tau,2w)\, \eta(\tau)^{b_2(S)-1}},
\ee
where $\eta$ and $\theta_1$ are the Dedekind eta function $\eta(\tau)=\q^{1/24}\prod_{n>0}(1-\q^n)$
and Jacobi theta functions $\theta_1(\tau,w)=\I \sum_{r\in\IZ+\frac12}(-1)^{r-\frac12} \q^{r^2/2} y^r$, which are both modular.  However, this reasoning
 ignores the possibility of non-holomorphic contributions from reducible connections. As a result, one expects that the generating function 
 \eqref{defhVWref} will transform as  a  
 {\it mock} Jacobi form of  depth $N-1$ \cite{Alexandrov:2019rth}.
 For $S=\IP^2$, $N=2$ it was observed in \cite{Klyachko:1991,Vafa:1994tf}  that the VW invariants reduce to Hurwitz class numbers, the generating function of which is a well-known mock modular form of
 depth one. This analysis was extended to other surfaces and higher $N$ in \cite{Yoshioka:1995,
 Manschot:2010nc, Manschot:2014cca, Manschot:2017xcr, gottsche2018refined}.
 For any $N\geq 2$ and weak Fano surface $S$, an explicit formula for the modular completion of  $\hrVW_{N,\mu,J}$ was proposed in \cite{Alexandrov:2019rth}. It is worth noting that  mock modularity in this context is not  tied to  wall-crossing, since it occurs already for $S=\IP^2$,
where the polarization is unique up to scale. 

\subsection{D-branes and coherent sheaves}

In the large volume limit, D-branes on a Calabi-Yau threefold $X$ can be viewed as objects in the derived category of coherent sheaves $\cD(X)$ \cite{MR1403918,Douglas:2000gi,Douglas:2000ah,Aspinwall:2001pu}. Recall that objects in this category are infinite complexes of coherent sheaves $\dots \rightarrow E_{-1} \rightarrow E_0 \rightarrow E_1 \rightarrow \dots$
where the subscript denotes the cohomological degree. 
Given a coherent sheaf $E$ on $X$, one constructs an infinite family of objects $E[k]$ of the derived category, where $E_m=0$ for $m\neq k$ and $E_k=E$. The Chern character 
$\ch(E) = \sum_m (-1)^m \ch(E_m)\in H^{\rm even}(X,\IQ)$ (or more precisely the K-theory class in $K(X)$) determines the D-brane charges, so the objects $E[k]$ with $k$ odd may be viewed as anti-branes.

\medskip

The global extension groups $\Ext^k(E,E')$ with $k\geq 0$
determine the spectrum of open strings between the D-branes associated to $E,E'$ \cite{Douglas:2000gi,Aspinwall:2004vm}:
\begin{itemize}
\item  $\Hom(E,E'):=\Ext^0_X(E,E')$ counts tachyonic strings stretched  between the two D-branes  (unless $E=E'$, in which case the tachyon is removed by the GSO projection);
\item $\Ext^1_X(E,E')$ counts light strings which may become massless at certain points in moduli space;
\item $\Ext^k_X(E,E')$ with $k>1$ corresponds to very massive strings which do not play any role at low energy 
\end{itemize}
When $X$ is the total space $\mathrm{Tot}(K_S)$ of the canonical bundle $K_S$ over a smooth complex surface $S$,
any object $E$ in the category $\cD(S)$ of coherent sheaves on $S$ can be lifted to an
object $i_*E$  in $\cD(X)$, corresponding to a D4-brane wrapped on $S$.
The $\Ext$ groups on $\cD(X)$ are related to those on 
$\cD(S)$ by 
\be
\label{Extq}
\Ext^k_X( i_* E, i_* E') = \Ext_S^k(E,E') \oplus \Ext_S^{3-k}(E',E) 
\ee
where the second factor is understood through Serre duality on $X$.  Thus,
light open strings stretched between D-branes $E$ to $E'$ originate both from $\Ext^1_S(E,E')$ and $\Ext^2_S(E,E')$, while $\Ext^0_S(E,E')$ and $\Ext^3_S(E,E')$ lead to tachyons. 

\medskip

With the help of vanishing theorems, the dimension of 
extension groups can often be computed from the 
Euler form $\chi(E,E')$, which is in turn given by  the Riemann-Roch formula
\be
\chi(E,E') := \sum_{k\geq 0} (-1)^k \dim \Ext_S^k(E,E') 
= \int_S \ch(E^*)\, \ch(E')\, \Td(S)\ .
\ee
Here  $E^*$ is the dual sheaf to $E$, whose Chern character $\ch(E^*)$ coincides with $\ch(E)$ up to a sign $(-1)^{k}$ 
on terms of degree $2k$,   $\Td(S)=1+\frac12 c_1(S)+\frac{1}{12}(c_1(S)^2+c_2(S))$ is the Todd class of $S$. The r.h.s can be evaluated in terms of the components $\ch=[\rk,c_1,\ch_2]$ of the Chern character, and in particular the degree  $\deg(E):=c_1(S) \cdot c_1(E)$,
\be
\begin{split}
\label{chigen}
\chi(E,E') =& \rk(E) \rk(E') + \rk(E)\, \ch_2(E')
+\rk(E')\, \ch_2 (E) - c_1(E)\cdot c_1(E') \\ & + \frac12 \left[ \rk (E)\, \deg (E')-\rk (E')\, \deg (E) \right]
\end{split} 
\ee
The first line is symmetric under exchange of $E,E'$ and determines the dimension 
\eqref{dimM} of the moduli space of stable sheaves via 
\be
\label{dimM2}
d_\IC(\cM^S_{\gamma,J}) = 1 - \chi(E,E)\ ,
\ee
while the second line in \eqref{chigen} 
is antisymmetric.  We shall denote the antisymmetrized Euler form by 
\be
\label{chiAS}
\langle E,E'\rangle = \chi(E,E')-\chi(E',E) = \rk (E)\, \deg (E')-\rk (E')\,\deg (E)
\ee
which depends only on the rank and degree.  An important fact is that 
\eqref{chiAS} has rank 2, with kernel spanned by Chern vectors such that $\rk(E)=\deg(E)=0$,
sometimes known as exceptional branes.

\medskip

A special class of coherent sheaves are invertible sheaves
$E=\cO_S(D)$, where $D$ is a divisor on $S$. They correspond to a single D4-brane wrapped 
on $S$, carrying an Abelian electromagnetic flux $F=c_1(E)$ but no D0-brane (since $c_2(E)=0$). 
For $D=0$, $\cO=\cO_S(0)$ is the structure sheaf.
In this case, the extension groups 
are just the sheaf cohomology groups 
\be
\Ext_S^k ( \cO(D), \cO(D') ) = H^k(S, \cO(D'-D)) \ ,
\ee
while the Euler form \eqref{chigen} 
reduces to  
\be
\chi(\cO(D), \cO(D')) = 1+\frac12 [D'-D] \cdot [D'-D + c_1(S)]\ .
\ee
As is standard, we denote $\chi(D)=\chi(\cO(0),\cO(D))$.
Other special  cases include the skyscraper sheaf $\cO_{pt}$, corresponding to a single D0-brane on $S$, and torsion sheaves supported on a divisor $D$, corresponding to D2-D0-brane bound states.

\subsection{Exceptional collections}

A standard way of describing the derived category of coherent sheaves on a complex surface $S$ relies on a choice of strong exceptional collection. Recall that a coherent sheaf $E$ on $S$ (or more generally an
object in the derived category $\cD(S)$) is called {\it exceptional} if
\be
\Ext_S^0(E,E)\simeq \IC, \quad \Ext_S^k(E,E)=0\quad \forall k>0\ .
\ee
In particular, $\chi(E,E)=1$ so the space of deformations has vanishing dimension and the 
sheaf is rigid.

\medskip
 An {\it exceptional collection} on $S$ is an ordered set of exceptional objects $\cC=(E^1,\dots E^r)$ such that 
 \be
 \Ext_S^k(E^i,E^j)=0 \quad \forall k\geq 0, \ 1\leq j < i \leq  r\ .
 \ee 
For del Pezzo surfaces (more generally, for smooth projective surfaces whose anticanonical class is generated by global sections), one can show that when $i<j$, $\Ext_S^k(E^i,E^j)=0$ unless $k\in\{0,1\}$ \cite[Lemma 2.4]{kuleshov1995exceptional}. We shall only be interested in {\it full} exceptional collections, such that the Chern characters $\ch(E^i)$ give a basis of the lattice $K(S)$; for a simply connected surface $S$, this implies in particular that $r=\chi(S)=b_2(S)+2$.  A {\it strong} exceptional collection is an exceptional collection such that for all $i,j$, $\Ext_S^k(E^i,E^j)=0$
unless $k=0$.  The matrix of Euler products (which we refer to as the Euler matrix)
\be
S^{ij}=\chi(E^i,E^j)
\ee
is then upper triangular with ones on the diagonal and positive integers above it. 
If the ranks $\rk(E^i)$ are all positive, then it follows that the slopes (defined in \eqref{defnuE} with $J=c_1(S)$) are increasing, namely
$\nu(E^i)\leq\nu(E^j)$ if $i\leq j$. In the special case where the first object $E^1$ in $\cC$ 
is the structure sheaf $\cO_S$, with vanishing slope, then the slopes of the other objects can be read off from the first row of the matrix $S$,
$\nu(E^i) =\chi(\cO_S,E^i)/\rk(E^i)=S^{1i}/\rk(E^i)$.
An {\it exceptional block} is an exceptional collection such that $S^{ij}=\delta^{ij}$. In such case 
the ranks $\rk(E^i)$ and slopes $\nu(E^i)$ are independent of $i$ \cite[Proposition 1.6]{karpov1998three}.

\medskip

In order to construct a quiver for the derived category $\cD(S)$,  one requires a collection $\cC^\vee=(E_1^\vee, \dots, E_r^\vee)$ of objects in $\cD(S)$ which are dual to $(E^1,\dots, E^r)$ in the sense that 
\be
\label{chidual}
\chi(E^i, E_j^\vee) = \delta^i_{\;j}
\ee
This implies that $\ch(E_i^\vee)=\sum_j S^\vee_{ji} \, \ch(E^j)$ where $S^\vee$ is the inverse of $S$, with entries (note the transposition) 
\be
S^\vee_{ij} = \chi(E_j^\vee, E_i^\vee) 
\ee
The matrix  $S^\vee$ is then upper triangular with ones along the diagonal, but its off-diagonal elements are not necessarily 
positive. For a suitable choice of   $E_i^\vee$ (see \S\ref{sec_mut} below),
the dual collection is exceptional,  but not strongly so. The D-branes associated to 
$i_*(E_i^\vee)$ provide a basis of fractional branes on $X$ \cite{Aspinwall:2004bs}. 

\medskip 

Using these collections, the Chern character of any object $E$ in $\cD(S)$ 
can be expressed as a linear combination of the Chern characters of either $E^i$ or $E_i^\vee$, 
\be
\label{chexp}
\ch(E) = \sum_i  \chi(E^i,E)\, \ch(E_i^\vee) =  \sum_i  \chi(E,E_i^\vee)\, \ch(E^i) 
\ee
so that the Euler form may be written in either of the two forms
\be
\chi(E,E') = \sum_i  \chi(E,E_i^\vee)  \,  \chi(E^i,E')
=\sum_i  \chi(E_i^\vee,E')  \,  \chi(E^i,E)\ .
\ee
In particular, for the skyscraper sheaf $\cO_{pt}$ corresponding to D0-branes, with Chern vector 
$[0,0,-1]$,
\be
\ch(\cO_{pt}) = - \sum_i  \rk(E^i)\, \ch(E_i^\vee) =  \sum_i  \rk(E_i^\vee)\, \ch(E^i) 
\ee
Since $\chi(\cO_{pt},\cO_{pt})=0$, it follows that 
\be
\label{sumrkrk}
 \sum_i  \rk(E^i)\,  \rk(E_i^\vee) = 0\ .
\ee
In particular, if the ranks $\rk(E^i)$ of the original strong collection are all positive, then some of the ranks $\rk(E_i^\vee)$ of the dual collection must be negative, so the $E_i^\vee$'s necessarily
live in the derived category.
Similarly, for the structure sheaf $\cO_S$ corresponding to D4-branes, 
with Chern vector $[1,0,0]$,
\bea
\ch(\cO_{S}) &=& \sum_i  \left[ \rk(E^i) + \ch_2(E^i) -\frac12 \deg(E^i)\right] \, \ch(E_i^\vee)  \nn\\
&=&  \sum_i  \left[ \rk(E_i^\vee) + \ch_2(E^\vee_i) +\frac12\deg(E_i^\vee)\right] \, \ch(E^i) 
\eea
Since $\chi(\cO_{pt},\cO_{S})=\chi(\cO_{S},\cO_{pt})=-1$, it follows that 
\be
 \sum_i  \rk(E^i)\,  \left[ \ch_2(E_i^\vee) +\frac12 \deg (E^\vee_i)\right]= 
 \sum_i  \rk(E_i^\vee)\,  \left[ \ch_2(E^i) +\frac12 \deg (E^i)\right]=1
 \ .
\ee
Using the fact that 
$\chi(\cO_{pt},\cO_{S}\otimes K_S)=\chi(\cO_{S}\otimes K_S,\cO_{pt})=-1$,
we find the stronger conditions 
\bea
\label{sumrkch2}
 \sum_i  \rk(E^i)\,  \ch_2(E_i^\vee) &=& \sum_i  \rk(E_i^\vee)\,  \ch_2(E^i)=1\ ,\nn\\
 \label{sumrkdeg}
 \sum_i  \rk(E^i) \, \deg (E^\vee_i) &=& \sum_i  \rk (E_i^\vee) \,  \deg (E^i)=  0\ .
\eea
Using these properties, it is straightforward to show that 
\be
\label{trsst}
\Tr S^{-1} S^t=\chi(S)\ ,
\ee
a property
which plays an important role in \cite{Herzog:2003dj,Herzog:2003zc}. 

\subsection{Mutations and dual collections \label{sec_mut}}

Mutations provide a useful way of constructing a collection $\cC^\vee$ dual
to $\cC$.  
 Recall that for an exceptional pair of coherent sheaves $(E,F)$,
the left mutation $L_E F$ of $F$ with respect to $E$ is defined by one of the  short exact sequences (see e.g. \cite{karpov1998three})\footnote{When $E,F$ are objects in the derived
category of coherent sheaves, one should instead write distinguished triangles in the derived category.}
\bea
0 & \to& L_E F \to \Hom(E,F)\otimes E \to F \to 0 \quad (\mbox{division}) \nn\\
0 &\to&  \Hom(E,F)\otimes E \to F \to  L_E F \to 0  \quad (\mbox{recoil})  \\
0 &\to&  F \to  L_E F \to \Ext^1(E,F) \otimes E \to 0  \quad (\mbox{extension}) \nn
\eea
The first two possibilities occur when $\Hom(E,F)\neq 0$, while the last one occurs when 
$\Ext^1(E,F)\neq 0$. Similarly, the right mutation $R_F^E$ is defined by 
\bea
0 & \to& E  \to \Hom(E,F)^* \otimes F \to R_F E \to 0 \quad (\mbox{division}) \nn\\
0 &\to& R_F E \to  E \to   \Hom(E,F)^* \otimes F \to 0  \quad (\mbox{recoil})  \\
0 &\to& \Ext^1(E,F)^* \otimes F   \to  R_F E \to E \to 0  \quad (\mbox{extension}) \nn
\eea
The pairs 
$(L_E F,E)$ and $(F,R_F E)$ are then exceptional. 
Note that $L_E(F)$ and $R_F(E)$ really belong to the derived category. 
Using the fact that $\ch E = \ch E' + \ch E''$ for short exact sequences $0\to E' \to E \to E'' \to 0$,
one finds that the Chern character of $L_E(F)$ is given by \cite{Wijnholt:2002qz}
\be
\ch (L_E F)  = \pm \left[ \chi(E,F)\, \ch(E) -\ch(F) \right]\ ,\quad 
\ch (R_F E)  = \pm \left[ \chi(E,F)\, \ch(F) -\ch(E) \right]\
\ee
where the $+$ sign holds for division and recoil, and the $-$ sign for extension. 

\medskip

Given a strong exceptional collection $\cC=(E^1,\dots E^r)$, one can show that  the  collection 
$\cC^\vee=(E_1^\vee, \dots, E_r^\vee)$ defined by 
\be
\label{Edualmut}
E_i^\vee =    L_{E^1}\ \dots L_{E^{i-1}} E^i 
\ee
satisfies the duality condition \eqref{chidual} and  is
exceptional \cite{bondal1990representation,Herzog:2006bu}. Clearly, this cannot 
be a strong exceptional collection, since the entries of $S_{ij}^\vee$ are not positive. 

\medskip

As explained below \eqref{Extq}, 
in order to ensure that the fractional D-branes
associated to the dual collection $\cC^\vee$  do not have tachyons, one 
should require that  
\be
\label{notachyon}
\Ext_S^0(E_i^\vee, E_j^\vee)= \Ext_S^3(E_i^\vee, E_j^\vee)=0 \quad \forall i\neq j 
\ee
While the vanishing for $i<j$ is automatic, it is not for $i>j$.  The condition \eqref{notachyon}  
can be ensured
by restricting to full strong exceptional collections $\cC$ which form the foundation
of a helix  \cite{bondal1990representation,Tomasiello:2000ym,Herzog:2006bu}. Unfortunately,  
this notion turns out to be too constraining as it rules out all rational surfaces except $\IP^2$ \cite{bondal1993homological}. 
A weaker notion is to require that $\cC$ 
is a cyclic full  strongly exceptional collection, defined as follows \cite{hille2011exceptional}: extend 
$\cC=(E^1,\dots, E^r)$ to a  bi-infinite sequence $E^{i}, i\in \IZ$ by requiring 
$E^{i+r} = E^i \otimes K_S^{-1}$. Then $\cC$  is a cyclic (full, strong) exceptional collection
if any subsequence $E^i, E^{i+1}, \dots, E^{i+r}$ of length $r$ is a (full, strong)  exceptional collection. 
Similar to the case of strong helixes \cite{Herzog:2006bu}, one can show that when $\cC$ is a cyclic full strong exceptional collection, the dual collection $\cC^\vee$ defined by \eqref{Edualmut} is also exceptional 
and satisfies the no-tachyon condition \eqref{notachyon} \cite{Perling-private}.

\subsection{Toric surfaces \label{sec_toric}}
Exceptional collections are readily available for smooth toric surfaces. Recall that those are described by a fan $\Delta$ inside a two-dimensional lattice, which can be identified as $\IZ^2$ up to the action of $GL(2,\IZ)$. The two-dimensional cones $\sigma_1, \dots, \sigma_r$ correspond to affine patches, glued along effective divisors $D_1,\dots, D_r$ 
associated to the one-dimensional
cones (or rays) $\rho_1,\dots, \rho_r$. Here the index is understood modulo $r=\chi(S)=b_2(S)+2$. We denote by $v_i$ the primitive vectors generating the ray $\rho_i$.
The toric divisors $D_i$ satisfy the linear relations 
\be
\label{eqRS}
\sum_i (u,v_i) \, D_i=0
\ee
 for any lattice vector $u$ and form an over-complete basis of $H^2(S,\IZ)$. The intersection product $D_i \cdot D_j$ vanishes unless $i-j \in \{-1,0,1\} \ ({\rm modulo}\  r)$, and satisfies $D_i \cdot D_{i+1}=1, D_i\cdot D_i=a_i$ where the integers $a_i$ are determined by the relations  
\be
\label{eqDij}
v_{i-1}+ v_{i+1} +a_i\, v_i = 0 \ .
\ee
Clearly, the sequence of integers $a_i, i=1\dots  r$ determine the vectors $v_i$ and hence the fan, up to the action of $GL(2,\IZ)$, but allowed sequences are restricted by the periodicity condition $v_i=v_{i+r}$.  
The Chern class is given by
\be
c(S) = \prod_i (1+D_i)
\ee
in particular the first Chern class satisfies 
\be
c_1(S) = \sum_i D_i\ ,\quad D_i \cdot c_1(S) = a_i+2\ .
\ee
Under an equivariant blow-up at the point $D_i \cap D_{i+1}$, the cone $\sigma_i$ bounded by $(v_i,v_{i+1})$
splits into two cones bounded by $(v_i, \tilde v)$ and $(\tilde v, v_{i+1})$, such that the divisor $\tilde D$ associated to $\tilde v$ has self-intersection $-1$.  The sequence of self-intersection numbers is
therefore extended to $a_1,\dots, a_{i-1}, a_i-1, -1, a_{i+1}-1, a_{i+2}, \dots a_n$. Conversely, given 
a toric fan $\sigma_1,\dots \sigma_r$ with a divisor $D_i$ such that $a_i=-1$,  the toric fan
of the blown-down surface along $D_i$ is obtained by merging the cones $\sigma_i$, $\sigma_{i+1}$
into a single cone, and the sequence of self-intersection numbers is reduced to 
$a_1,\dots, a_{i-2},a_{i-1}+1,a_{i+1}+1,a_{i+2}, \dots a_r$. Note that the sum $\sum_i a_i$ changes 
by $\pm 3$ under this process, {\bp indeed $\frac13\sum_i a_i$ is equal to the signature $\sigma(S)$.}

\medskip

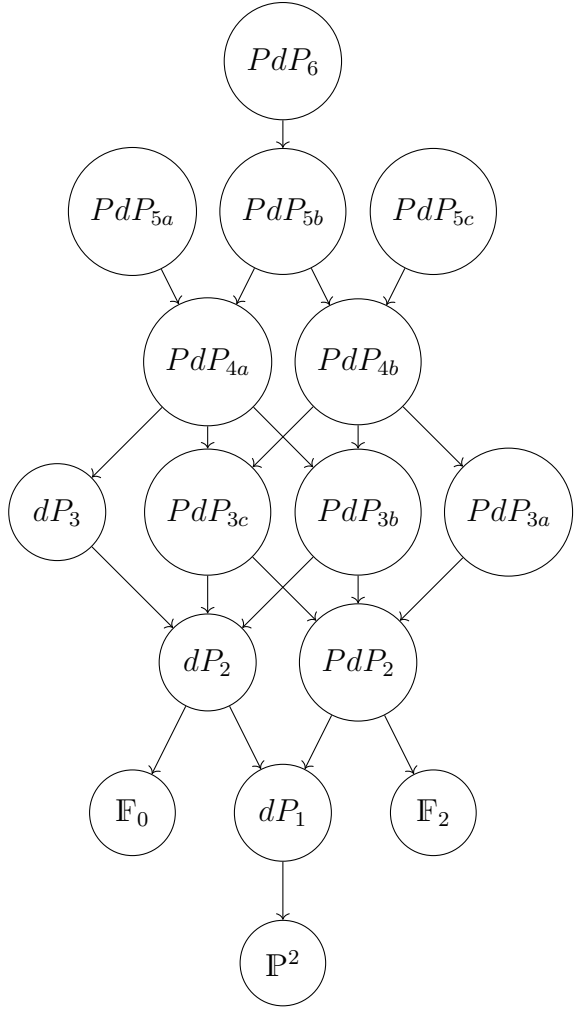
\begin{figure}[h]
 \begin{center}
\begin{tikzpicture}[inner sep=2mm,scale=2]
 \node (6) at (0,5) [circle,draw] {$PdP_6$};
 \node (5a) at (-1,4) [circle,draw] {$PdP_{5a}$};
 \node (5b) at (0,4) [circle,draw] {$PdP_{5b}$};
 \node (5c) at (1,4) [circle,draw] {$PdP_{5c}$};
  \node (4a) at (-.5,3) [circle,draw] {$PdP_{4a}$};
 \node (4b) at (.5,3) [circle,draw] {$PdP_{4b}$};
  \node (3a) at (-1.5,2) [circle,draw] {$dP_3$};
 \node (3b) at (-.5,2) [circle,draw] {$PdP_{3c}$};
 \node (3c) at (.5,2) [circle,draw] {$PdP_{3b}$};
  \node (3d) at (1.5,2) [circle,draw] {$PdP_{3a}$};
 \node (2) at (-.5,1) [circle,draw] {$dP_2$};
 \node (2a) at (.5,1) [circle,draw] {$PdP_2$}; 
 \node (f0) at (-1,0) [circle,draw] {$\IF_0$};
 \node (1) at (0,0) [circle,draw] {$dP_1$};
 \node (f2) at (1,0) [circle,draw] {$\IF_2$};
  \node (0) at (0,-1) [circle,draw] {$\IP^2$};
 \draw [->] (6) to node[auto] {$ $} (5b);
 \draw [->] (5a) to node[auto] {$ $} (4a);
 \draw [->] (5b) to node[auto] {$ $} (4a);
 \draw [->] (5b) to node[auto] {$ $} (4b);
\draw [->] (5c) to node[auto] {$ $} (4b);
 \draw [->] (4a) to node[auto] {$ $} (3a);
  \draw [->] (4a) to node[auto] {$ $} (3b);
  \draw [->] (4a) to node[auto] {$ $} (3c);
  \draw [->] (4b) to node[auto] {$ $} (3b);
  \draw [->] (4b) to node[auto] {$ $} (3c);
  \draw [->] (4b) to node[auto] {$ $} (3d);
  \draw [->] (3a) to node[auto] {$ $} (2); 
  \draw [->] (3b) to node[auto] {$ $} (2); 
  \draw [->] (3b) to node[auto] {$ $} (2a); 
    \draw [->] (3c) to node[auto] {$ $} (2); 
  \draw [->] (3c) to node[auto] {$ $} (2a); 
\draw [->] (3d) to node[auto] {$ $} (2a); 
 \draw [->] (2) to node[auto] {$ $} (f0); 
\draw [->] (2) to node[auto] {$ $} (1); 
 \draw [->] (2a) to node[auto] {$ $} (1); 
 \draw [->] (2a) to node[auto] {$ $} (f2); 
 \draw [->] (1) to node[auto] {$ $} (0); 
\end{tikzpicture}
\end{center}
\caption{The 16 toric  weak Fano surfaces, including 
the 5 toric Fano surfaces, and their relations via blow-downs. \label{figtoricblow}}
\end{figure}

For a smooth Fano surface, the condition $D_i \cdot c_1(S)>0$ implies $a_i\geq -1$ for all $i$. 
The vectors $v_i$ form a convex reflexive polygon, with a single lattice point in the interior and whose boundary contains no lattice point besides the vertices.
There exist 5 such smooth, toric, compact Fano surfaces, namely the projective plane $\IP^2$,
the Hirzebruch surfaces $\IF_0=\IP^1\times\IP^1$ and $\IF_1=dP_1$, and the del Pezzo surfaces 
$dP_2$ and $dP_3$. For a weak Fano surface (which means that $c_1(S)$ is nef, i.e.
 $D_i \cdot c_1(S)\geq 0$) the condition $a_i\geq -1$ is relaxed to $a_i\geq -2$. There are 
 11 smooth, toric, compact weak Fano surfaces, sometimes known as pseudo-del Pezzo surfaces. The toric diagrams are displayed e.g. in \cite[Fig. 1]{Hanany:2012hi}, and have lattice points on the edges, corresponding to the divisors $D_i$ such that 
$D_i \cdot c_1(S)=0$. These toric surfaces are related by sequences of toric blow-ups as shown in Figure \ref{figtoricblow}. 

%\FIGURE{

\medskip

Given a smooth toric surface with a given numbering of the toric divisors $D_1,\dots D_r$, let 
us consider the collection of invertible sheaves  \cite{Herzog:2005sy,hille2011exceptional}
\be
\label{toricol}
\cO(0), \cO(D_1), \cO(D_1+D_2), \dots, \cO(D_1+\dots +D_{r-1})
\ee
The matrix of Euler numbers $\chi(E_i,E_j)$ is then upper triangular, given by 
\be
S= \begin{pmatrix} 
1 & 2+a_1 & 4 + a_1+a_2 & 6+a_1+a_2+a_3 &  8+a_1+a_2+a_3+a_4 &  \dots \\
0 & 1 & 2+a_2 & 4 + a_2+a_3 &   6+a_2+a_3+a_4 & \dots \\
0 & 0 &  1 & 2+a_3 & 4 + a_3+a_4 & \dots  \\
\vdots & \vdots & 0 & \ddots & \ddots & \ddots \\
\vdots & \vdots & \ddots & \ddots &1 & 2+ a_{n-1} \\
0 & \dots & \dots & \dots & 0 & 1
\end{pmatrix} 
\ee
More generally, one may take any length-$r$ subset
of the cyclic sequence
\be
\label{torhelix}
\dots, \cO(-D_r), \cO(0), \cO(D_1), \cO(D_1+D_2), \dots, \cO(D_1+\dots +D_{r-1}), \cO(c_1(S)),  \cO(c_1(S)+D_1), \dots\ ,
\ee
or reverse the labelling of the toric divisors $D_i$. One obtains in this way  different strong exceptional collections, but the quiver defined below does not depend on the choice of length-$r$ subset. 

\medskip 

When $S$ is weak Fano, the collection \eqref{toricol} is {\it not} strongly exceptional,
despite the fact that all entries above the diagonal in the matrix $S$ are positive. 
The reason is that\footnote{We are grateful to M. Perling for explaining this 
to us, and further discussions about weak Fano surfaces.}, for a  toric divisor $D_i$ with $c_1(S) \cdot D_i=0$,
one has $\chi(D_i)=D_i^2+2=0$ hence
$\dim H^1(D_i) = \dim H^0(D_i) + \dim H^2(D_i)$. However, $\dim H^0(D_i)>0$ 
since $D_i$ is effective
and therefore $\dim H^1(D_i)>0$.
In \cite{hille2011exceptional}, for any smooth toric surface 
the authors construct an  alternative sequence 
$\tilde D_1,\dots,\tilde D_r$ with $\sum_i \tilde D_i = c_1(S)$, called  `toric system', 
such that 
\be
\label{toricolt}
\cO(0), \cO(\tilde D_1), \cO(\tilde D_1+\tilde D_2), \dots, \cO(\tilde D_1+\dots +\tilde D_{r-1})
\ee
forms a cyclic full strong exceptional collection. 
The toric system is constructed iteratively via an augmentation process
at each step of the blow-up process, and is tabulated in \cite[Table 2]{hille2011exceptional}
for each of the 11 weak Fano cases (see Appendix \ref{sec_weakFano} below for more details).

\subsection{Quiver quantum mechanics and quiver moduli \label{sec_quiv}}

As shown in \cite{baer1988tilting,bondal1990representation,rickard1989morita}, given a cyclic 
full strong exceptional collection $\cC$ on $S$,  the derived category of coherent sheaves $\cD(S)$
on a complex surface is isomorphic to the derived category of representations $\cD(Q)$ of the quiver $Q$ with set of vertices $Q_0$ and of arrows $Q_1$ constructed as follows: 
\begin{itemize}
\item each object $E_i^\vee$, $i=1,\dots, r$ in the dual collection $\cC^\vee$ corresponds to a vertex
 $v_i\in Q_0$;
\item for each pair $i\neq j$ with $S^\vee_{ij}<0$, $Q_1$ includes $-S^\vee_{ij}$ arrows $v_i\to v_j$;
\item for each pair $i\neq j$ with $S^\vee_{ij}>0$, $Q_1$ includes $S^\vee_{ij}$ arrows $v_j\to v_i$;
\end{itemize}
Note that our convention for the orientation of the arrows is opposite to the usual convention in  mathematics, since $\Ext^1(E_j^\vee,E_i^\vee)$ contributes positively to the number of 
arrows $v_i\to v_j$,
while $\Ext^2(E_j^\vee,E_i^\vee)$ contributes negatively.
The  signed number of arrows from $v_i$ to $v_j$ (such that 
arrows from  $v_j$ to $v_i$ are counted negatively), defines the adjacency matrix 
\be
\label{defkappa}
\kappa_{ij} = S^\vee_{ji} - S^\vee_{ij} =  \langle E_i^\vee, E_j^\vee \rangle\ .
\ee
Now, for a given object $E\in \cD(S)$, one introduces the integer vector $\vec n=(n_1,\dots, n_r)$, whose components are the coefficients $n_i=\chi(E^i,E)$  in the expansion \eqref{chexp}  of 
$\gamma=\ch(E)$  
on the basis of vectors $\gamma_i=\ch E_i^\vee$. 
Since $\chi(E^i,E)=\rk E^i \times \ch_2(E)+\dots$ and $\rk E^i>0$, the 
vector $\vec n$ has negative entries for large instanton number $c_2(E)$, and we pick instead\footnote{It would be useful to have a conceptual understanding of this sign flip, which could help to resolve the sign issue mentioned in \cite{Alexandrov:2019rth}, footnote 24. }
$\vec N=-\vec n$ as the relevant dimension vector, such that  $\gamma=-\sum_i N_i \gamma_i$.

\medskip

Physically, 
the quiver $Q$ and dimension vector $\vec N$ encodes the matter content of the supersymmetric quantum mechanics  describing the interactions of the exceptional branes associated to  
$E_i^\vee$, with each species appearing in $N_i$ copies \cite{Douglas:1996sw,Denef:2002ru}. 
Each node corresponds to a vector multiplet transforming in the adjoint representation in the $i$-th 
factor of  the gauge group $G=\prod_i U(N_i)$, and each arrow $v_i\to v_j$ to a chiral multiplet $\Phi_{ij}^\alpha$ (with $\alpha$ running from 1 to $|S_{ij}^\vee|$)
transforming in the bifundamental representation $(N_i, \overline{N_j})$ of $U(N_i)\times U(N_j)$.
Edge loops $v_i\to v_i$ correspond to additional chiral multiplets in the adjoint. The entries $\kappa_{ij}$ in the matrix \eqref{defkappa} are (up to sign) the Dirac-Schwinger-Zwanziger products between the
electromagnetic charges $\gamma_i=\ch(E_i^\vee)$, which determine the net number of light open strings between the two exceptional branes.   
The supersymmetric Lagrangian for this system also depends on a choice of real vector $\vec\zeta=(\zeta_1,\dots, \zeta_r)$ such that $\sum_i N_i \zeta_i=0$,  corresponding to the Fayet-Iliopoulos parameters, and 
when the quiver admits oriented loops, a superpotential $W(\Phi)$, which is a  sum of traces of products of $\Phi_{ij}^\alpha$ associated to closed 
loops. For a given exceptional collection, a general prescription for computing the superpotential was discussed in \cite{Aspinwall:2004bs,Aspinwall:2005ur}.  For toric surfaces, the superpotential
can also be read off from the associated brane tiling \cite{Franco:2005rj}.
There are also additional parameters such as  gauge and kinetic couplings which do not play a r\^ole in the existence of supersymmetric bound states.

\medskip

A quiver representation $R$, corresponding physically to a vacuum configuration to the chiral fields in the supersymmetric quantum mechanics,  is a  collection of vector spaces  $\IC^{N_i}$ for each
vertex $v_i\in Q_0$,  and  maps $\Phi_{ij}^{\alpha}$ for each arrow $(v_i\to v_j) \in Q_1$, subject to relations $\partial W/\partial \Phi_{ij}^\alpha=0$. Two notable classes of representations are the {\it simple} representations $R_i^\vee$, with dimension vector $N_j=\delta_{ij}$, and the {\it projective}  representations $R^i$, with
dimension vector $N_j=S^{ij}$. As the notation suggests, these representations are associated to the
sheaves $E_i^\vee$ and $E^i$ respectively. The representations $R_i^\vee$ are rigid, while
the representations $R^i$ form a complete set of projective objects, such that any representation $R$
has a projective resolution in terms of sums of $R^i$ (see e.g. \cite{Aspinwall:2004vm}). The  derived category of quiver representation $\cD(Q)$ is defined by considering infinite complexes of representations $\dots  \rightarrow R_{-1} \rightarrow R_0 \rightarrow R_1 \rightarrow \dots$. Given a 
representation $R$ with dimension vector $\vec N$, the object $R[k]$ has $R_m=0$ for $m\neq k$, $R_k=R$ and dimension
vector $(-1)^k \vec N$. 

\medskip

The space of representations admits a natural action of the complexified
gauge group $G_\IC=\prod_v GL(N_v,\IC)$ by
changing basis in each of the vector spaces  $\IC^{N_v}$. The moduli space of quiver representations,
given by the quotient of the space of arbitrary quiver representations by $G$, is in general singular.
In order to obtain smooth projective moduli spaces, at least when the
dimension vector is coprime, one restricts to semi-stable representations, where semi-stability
is defined as follows: let $\vec\zeta=(\zeta_1,\dots,\zeta_r)$ be a real vector  such that $\sum_i N_i \zeta_i=0$.  A representation $R$ of dimension vector $\vec N'$ is stable (resp., semi-stable) if it admits no proper subrepresentation $R'$ such that $\nu_{\vec\zeta}(R')>\nu_{\vec\zeta}(R)$, where the slope  is defined by 
\be
\label{defnuR}
\nu_{\vec\zeta} (R') = \sum_i N'_i \zeta_i\ .
\ee
As shown in \cite{zbMATH00720513}, the $G_\IC$-orbits of semi-stable representations are
 in one to one correspondence with the $G$-orbits of representations satisfying the conditions
\be
\label{Dterm}
\sum_{j:S_{ij}^\vee<0 }  \sum_{\alpha=1}^{-S_{ij}^\vee} (\Phi_{ij}^\alpha) ^\dagger  \Phi_{ij}^\alpha
- \sum_{j: S_{ij}^\vee>0}  \sum_{\alpha=1}^{S_{ij}^\vee}  
(\Phi_{ji}^\alpha)^\dagger   \Phi_{ji}^\alpha =\zeta_i
\ee
 at each node $i$, which are known as the
D-term conditions in the supersymmetric quantum mechanics \cite{Denef:2002ru}.

\medskip

The moduli space $\cM^Q_{\vec N,\vec \zeta}$ is then the set of isomorphism classes of 
semi-stable quiver representations,  modulo the action of the the complexified gauge group 
$G_\IC$. If the dimension vector is coprime and the superpotential is generic,
$\cM^Q_{\vec N,\vec \zeta}$ is a smooth projective variety with vanishing odd cohomology, and the (refined) index is defined as the Poincar\'e polynomial. More generally, we define the rational index 
\be
\label{defbOm}
\bOm_{Q}(\vec N,\vec \zeta,y) = \sum_{m|\vec N} \frac{y-1/y}{m(y^m-1/y^m)}\, 
\Omega_{Q}(\vec N/m, \vec \zeta,y^m) ,
\ee
where $\Omega_{Q}(\vec N,\vec \zeta,y)$ is the Poincar\'e polynomial 
defined using intersection homology,
\be
\label{defOm}
\Omega_{Q}(\vec N,\vec \zeta,y) = \sum_{p=0}^{d_\IC(\cM)} (-y)^{2p-d_\IC(\cM)} 
b_p(\cM)\ ,\quad \cM:=\cM^Q_{\vec N,\vec \zeta}
\ee
Note that this differs from \eqref{defCNnJ}, \eqref{defcref} by a change of variable $y\to -y$,
needed to match conflicting conventions in the literature on Vafa-Witten invariants and on quivers. When the dimension vector has negative entries, we define 
$\Omega_{Q}(\vec N,\vec \zeta,y)=\Omega_{Q}(-\vec N,\vec \zeta,y)$ (in particular the sign flip 
mentioned below \eqref{defkappa} is harmless). 
If the entries do not have identical signs, then  $\Omega_{Q}(\vec N,\vec \zeta,y)$ is defined 
to vanish. Furthermore, transposing the maps $\Phi_{ij}^\alpha$ leads to the equality 
$\Omega_{Q}(\vec N,\vec \zeta,y)=\Omega_{Q^t}(\vec N,-\vec \zeta,y)$
where $Q^t$ denotes the transposed 
quiver with vertex set $Q_0$ and with reversed arrows from $Q_1$.  In addition, 
BPS indices are invariant under Seiberg dualities, which correspond to specific sequences of mutations on the exceptional collection \cite{Herzog:2004qw}. We refer to \cite{Manschot:2013dua} for details on the action of Seiberg dualities on the dimension and stability vectors. Henceforth we shall denote the index \eqref{defOm} as $\Omega(\vec N,\vec \zeta,y)$, omitting the subscript $Q$.

\subsection{Beilinson quiver and stability conditions\label{sec_stab}}

While the Baer-Bondal-Rickard theorem guarantees the isomorphism 
$\cD(S) \sim \cD(Q)$ between the derived categories of coherent sheaves and quiver representations, one also expects that the moduli spaces of semi-stable objects 
$\cM^S_{\gamma,J}$ and $\cM^Q_{\vec N,\vec\zeta}$
will be in bijective correspondence
for a suitable choice 
of stability parameter $\vec\zeta(J)$, generalizing the well-known 
result for $S=\IP^2$ \cite{drezet1985fibres}.
 A basic requirement on the assignment $J\mapsto \vec\zeta$ is that the expected 
 dimensions of the moduli spaces $\cM^S_{\gamma,J}$ and $\cM^Q_{\vec N,\vec \zeta}$ should match. 
 
\medskip

On the quiver side, recall that semi-stable representations typically have\footnote{From the viewpoint of the moduli space of sheaves,
 the arrows in $Q_1$ correspond to the Higgs fields arising in the dimensional reduction of HYM equations on $X=\mathrm{Tot}(K_S)$ down to $S$; 
 the vanishing of  the arrows in $Q'_1$ is tantamount to the fact that all semi-stable sheaves on $X$
 are of the form $i_*E$ where $E$ is a semi-stable sheaf on $S$. This vanishing requires that 
 the superpotential be linear with respect to a subset of arrows, and is closely related to the
phenomenon known as dimensional reduction in the mathematics literature, see \cite[\S4.8]{Kontsevich:2010px}, \cite[App. A]{Davison:2013nza}. We are grateful to  Pierrick Bousseau and Olivier Schiffmann for discussions on this issue.\label{foodimred}
 } 
{\bp $\Phi_{i'j'}^{\alpha}=0$ for 
$(i',j',\alpha) $  in a cut $Q'_1$ of the superpotential $W$, depending
on the stability parameters $\vec\zeta$; a cut being  a 
subset $Q'_1\subset Q_1$ of the set of arrows $Q_1$  such that 
$W$ is linear in the $\Phi_{i'j'}^{\alpha}$'s. }
The remaining edges in $\hat Q_1 = Q_1\backslash Q'_1$
then form a spanning tree on the quiver $Q$. The  maps $\Phi_{ij}^\alpha$ with
$(i,j)\in \hat Q_1$
are then subject to the relations $\partial W/\partial \Phi_{i'j'}^\alpha=0$ for each $(i',j')\in Q'_1$. 
The subquiver $\hat Q$ with vertices $Q_0$ and edges in $\hat Q_1$ has no oriented loops, but instead relations $\{\partial W/\partial \Phi_{i'j'}^\alpha=0, (i',j')\in Q'_1\}$. For toric surfaces,
the relevant {\bp cuts} 
%subquivers (which we refer to as `branches') 
are in one-to-one correspondence with internal perfect matchings on the 
associated brane tiling \cite{Hanany:2006nm}.

\medskip

For the purpose of describing coherent sheaves on $S$, the appropriate choice of  
subquiver is the 
Beilinson quiver, defined as the subquiver $\hat Q$ with vertices $\hat Q_0=Q_0$, arrows $v_i\to v_j$ for each $(i,j)$ such that $S^\vee_{ij}<0$ (coming from $\Ext^1(E_j^\vee, E_i^\vee))$, and relations  $\partial W/\partial \Phi_{i'j'}^\alpha=0$ for each $(i',j')$ such that 
 $S^\vee_{i'j'}>0$ (coming from $\Ext^2(E_{j'}^\vee, E_{i'}^\vee)$). For this  subquiver, 
the expected dimension of the moduli space of semi-stable representations is given by 
\be
d_\IC(\cM^Q_{\vec N,\vec \zeta}) = \sum_{i\neq j, S^\vee_{ij}<0} |S^\vee_{ij}| N_i N_j - \sum_{i'\neq j', S^\vee_{i'j'}>0} S^\vee_{i'j'} \, N_{i'} N_{j'} - \( \sum_i N_i^2 - 1\) =
1- \sum_{i,j} N_i \, S^\vee_{ij}\, N_j
\ee
where the  term in round brackets 
comes from modding out by the action of the complexified gauge group 
$\prod_i GL(N_i,\IC)$ modulo its center. Remarkably, this agrees with the expected dimension
\eqref{dimM2} on the sheaf side. As we shall see later, the Beilinson quiver $
\hat Q$ can be viewed as a generalization of Beilinson's monad construction of sheaves on the projective plane $\IP^2$. 

\medskip

It remains to find the precise map $J\to\vec\zeta$, and the conditions on the Chern vector $\gamma$
such that semi-stable representations are supported on the  Beilison quiver $\hat Q$. 
For the first issue,
recall that stability parameters 
can be read off from the central charges of the corresponding D4-D2-D0 branes on the local Calabi-Yau $X=\mathrm{Tot}(K_S)$, via \cite{Manschot:2010qz}
\be
\label{zetafromZ}
\zeta_i = \lambda\, \Im( Z_{-\gamma_i} \, \overline{Z_\gamma} )\ ,\quad \lambda\in \IR^+
\ee
where $\lambda$ is an irrelevant scale factor and $\Im$ denotes the imaginary part. Here $\gamma$ is the Chern character of the sheaf $E$,
 while $-\gamma_i$ are the Chern characters of the objects $E_i^\vee[1]$ in the dual exceptional collection, corresponding to the vertices of the quiver. 
 Note that the stability parameters \eqref{zetafromZ} automatically satisfy $\sum_i N_i \zeta_i=0$ 
 where $N_i$ are the coefficients in the decomposition $\gamma=-\sum_i N_i \gamma_i$. Moreover,
 for any vector $\gamma'=-\sum_i N'_i \gamma_i$ one has 
 \be
 \sum_i N'_i \zeta_ i=  \lambda\, \Im( Z_{\gamma'} \, \overline{Z_\gamma} )\ .
\ee
The corresponding stability condition, based on the full quantum corrected central charge, is known as $\Pi$-stability \cite{Douglas:2000ah}. 
 Now, at large volume
the central charge $Z_\gamma$ for a D4-D2-D0-brane bound state simplifies to \cite{Diaconescu:2007bf}
\be
\label{ZGamma}
Z_\gamma= - \int_X e^{-J} \Gamma\ ,\quad \Gamma= \ch(i_\star(E))\, \sqrt{\Td(X)}
\ee
where $E$ is a coherent sheaf on $S$, and $i_*(E)$ its push-forward on $X$. 
The class $\Gamma\in H_{\rm even}(X)$,  sometimes known as the Mukai vector, evaluates to 
\be
\label{Mukai}
\Gamma = N\, [S] + \I_\star \left[c_1(E) + \frac{N}{2} c_1(S) \right] + q_0\, \omega_S
\ee
where $\omega_S$ is the unit volume form on $S$, and $q_0$ is the D0-brane charge,
\be
q_0 = - \left( c_2(E) - \frac{N-1}{2N} c_1(E)^2 \right) + N \frac{\chi(S)}{24} + \frac12
\int_S \left( c_1(E) + \frac{N}{2} c_1(S) \right)^2
\ee
Inserting \eqref{Mukai} in \eqref{ZGamma} we get 
\bea
\label{Zgen}
Z_\gamma &= & -\frac{N}{2} \int_S J^2 + \int_S J \wedge  \left( c_1(E) + \frac{N}{2} c_1(S) \right) - q_0
\nn\\
&=& -N \left[ \frac{1}{2} \int_S J^2 
-\frac12 \int_S J\, c_1(S)  + \frac18 \int_S c_1(S)^2  +  \frac{\chi(S)}{24}  \right]
+ \int_S \left[J - \frac12 c_1(S)\right]   c_1(E) - \ch_2(E)
\nn\\
&=&  -N \left[ \frac{1}{2} \int_S J'^2 +  \frac{\chi(S)}{24} \right] 
+ \int_S J' \, c_1(E) - \ch_2(E)
\eea
where $J'=J-\frac12 c_1(S)$. Rescaling  $J':=z J_0$ by a complex number $z=s+\I t$
with large imaginary part $t\gg 1$, this becomes  
\bea
\label{ZZbgen}
\Im [Z_{\gamma'}  \overline{Z_{\gamma}} ] &= & 
\frac12\left( (t^3-s^2 t) \, J_0^2- \frac{\chi(S)}{24}\right)  \, 
\left[ N \int_S J_0 \cdot c_1(E') - N' \int_S J_0 \cdot c_1(E) \right] 
\\
&+& t \left[ \ch_2(E') \int_S J_0 \cdot c_1(E) - \ch_2(E)  \int_S J_0 \cdot c_1(E') \right] 
+ s\, t\, J_0^2\, \left[ N'\, \ch_2(E) - N \ch_2 (E') \right] \nn
\eea
where $J_0^2:=\int_S J_0^2$. We may therefore choose the stability parameters $\vec\zeta(J_0)$  such that\footnote{This identification generalizes the prescription in \cite{drezet1985fibres,king1995chow,ohkawa2010moduli} 
for $S=\IP^2$, and was derived independently for any $S$ 
in \cite{Maiorana:2017wdq}. We thank Pierrick Bousseau for drawing
our attention to this reference.} 
\be
 \label{stabsum}
 \sum_i N'_i \zeta_ i=  \rho \left[ N\, \int_S J_0\cdot c_1(E') - N'\, \int_S J_0\cdot c_1(E)  \right] 
 + N'\, \ch_2(E) - N \ch_2(E')
\ee
for all $\gamma'=\[N',c_1(E'),\ch_2(E')\]$, for suitably large $\rho\sim t^2/2$. Indeed, for this choice
 the slope \eqref{defnuR} of a subrepresentation agrees with 
the slope of a subsheaf \eqref{defnuE}  at leading order in $\rho$. The subleading term in \eqref{stabsum} is only
relevant when the leading term vanishes, and is chosen such that  the stability condition
for quiver representations agree with the Gieseker stability condition for subsheaves.
The same is true of \eqref{ZZbgen} when $t\gg 1$ and $s<0$ \cite{Diaconescu:2007bf},
although there is no value of $s,t$ for which\eqref{ZZbgen} and \eqref{stabsum} would become
identical. 

\medskip 
Having identified the relevant stability parameters for a given polarization, it remains to 
check that these stability parameters are consistent with the truncation to the Beilinson quiver $\hat Q$, which is required in order that the dimensions of the moduli spaces $\cM^S_{\gamma,J}$ and $\cM^Q_{\vec N,\vec \zeta}$ agree. In particular, to be consistent with the D-term conditions \eqref{Dterm},  one should have $\zeta_i\geq 0$ and $\zeta_j\leq 0$ for any source $i$ or sink $j$ in the quiver $\hat Q$, respectively.  As we shall show on examples, this restricts the possible range of slopes of the sheaf $E$ to a certain window
\be
\nu_- \leq \nu(E) \leq \nu_+
\ee
which is set by the slopes of the sheaves $E_i^\vee$  (when all the slopes are finite,
then $\nu_i\nu_-$ and $\nu_+$ are the minimal and maximal slopes).
Depending on its width, this window may  
or  may not be reachable by applying a  spectral flow 
\eqref{specflow}.

\subsection{Canonical and attractor indices \label{sec_omstar}}

For the particular case of the canonical polarization $J_0=c_1(S)$,  the leading term in 
 \eqref{stabsum} is proportional to the antisymmetrized Euler form $ \langle E', E \rangle$
 defined in  \eqref{chiAS}. Thus, we get 
\be
\label{zetacan}
 \zeta_i^c = \rho\, \kappa_{ij} N^j + \eta_i
\ee
where $\kappa_{ij}$ is the antisymmetrized Euler matrix \eqref{defkappa}, $\vec \eta$
is a fixed vector such that $\sum_i N'_i \eta_i =  N'\, \ch_2(E) - N \ch_2(E')$ and $\rho\gg 1$. 
Up to a large
{\it negative} coefficient, the leading term in \eqref{zetacan} is recognized as the 
attractor stability parameter 
\be
\label{zetaatt}
 \zetastar_i = - \kappa_{ij} N^j
 \ee
 which plays a central r\^ole in the attractor flow tree conjecture \cite{Alexandrov:2018iao}.\footnote{
In the mathematics literature, \eqref{zetacan} is known as the `self-stability condition' \cite[Def. 11.3]{bridgeland2016scattering}.}
For the value $\vec\zeta=\vec\zetastar$, one may easily show that two-particle bound states
with charge $(\gamma_L,\gamma_R)$ such that $\gamma=\gamma_L+\gamma_R$ and 
$\langle \gamma_L,\gamma_R \rangle \neq 0$ are never stable. Marginal bound states
with $\langle \gamma_L,\gamma_R \rangle=0$ may occur, but they are not expected
to contribute to the index.  By contrast, two-particle bound states are typically allowed in the
chamber around \eqref{zetacan}; the perturbation $\vec\eta$ ensures that $\vec\zeta^c$
does not sit on a wall of marginal stability. We shall denote by $\Omstar(\vec N,y)$ and
$\Omega_c(\vec N,y)$ the value of the index $\Omega(\vec N,\vec \zeta,y)$ at 
$\vec\zeta=\vec\zetastar$ and $\vec\zeta=\vec\zeta^c$ and refer to these values 
as attractor and canonical indices, respectively. Importantly, using  \cite[Eq. (1.7)]{Manschot:2013dua} one can check that the condition $\zeta_i=\pm \kappa_{ij} N^j$ is invariant under Seiberg dualities, therefore the same holds for the attractor and canonical indices.

\medskip

Note that both \eqref{zetaatt} and the leading term in 
\eqref{zetacan} vanish when $\vec N$ lies in the kernel of the adjacency matrix $\kappa$.
This happens  for torsion sheaves with $N=\deg(E)=0$, sometimes
known as exceptional branes, and includes the case of pure D0-branes.  
In that case, it follows that $\langle \gamma_L,\gamma_R \rangle =\langle \gamma_L,\gamma\rangle=0$
for all splittings, therefore the index  cannot jump and $\Omega(\vec N,\vec\zeta,y)$ becomes 
independent of $\vec\zeta$.

\medskip

Remarkably, we shall find that with the exception of dimension vectors 
associated to simple representation (in which case $\Omstar(\vec N,y)=1$)
and pure D0-branes (in which case $\Omstar(\vec N,y)$ is undetermined), the
attractor index $\Omstar(\vec N,y)$ always vanishes
for quivers associated to del Pezzo surfaces (and plausibly for any rational surface). 
To establish this claim, we shall produce a positive quadratic form $\cQ(\vec N)$ such that the expected dimension of the moduli space $\cM^Q_{\vec N,\vec\zetastar}$ 
in the attractor chamber can be written as
 \be
 \label{dstareps}
 \dstar_\IC= 1 - \cQ(\vec N) - \sum_i \lambda_i N_i \zetastar_i
 \ee
for suitable choices of $\lambda_i\in \IQ$ such that $\lambda_i=0$ or $\sgn(\lambda_i)=\sgn(\zetastar_i)$ in 
 the branch where \eqref{dstareps} applies. By construction, the quadratic form $\cQ(\vec N)$ is integer valued and positive but degenerate along the dimension vector corresponding to pure D0-branes, since $\dstar_\IC=1$, $\vec\zetastar=0$ in that case. Moreover, it takes the value $\cQ(\vec N)=1$ whenever $\vec N$ is the dimension vector of a simple representation, since $N_i \zetastar_i=0$ for any $i$ in that case. Since these are the only possibilities which allow $\dstar_\IC \geq 0$, we conclude that $\Omstar(\vec N,y)=0$ when neither of these conditions hold. Unfortunately. we do not have a geometric understanding of the quadratic form $\cQ(\vec N)$, and can only find it currently  
 by brute force computer search. 
 We emphasize that 
the vanishing of $\Omstar(\vec N,y)$ depends on the special structure of the adjacency matrix 
$\kappa$ and superpotential $W$, which restricts the allowed branches in the attractor chamber.

\subsection{The flow tree and Coulomb branch formulae \label{sec_flowtree}} 

Having identified the relevant dimension vector and stability parameters on the quiver side, we can now apply the arsenal of techniques for computing the quiver indices $\bOm(\vec N,\vec \zeta,y)$, and compare with expected results for the Vafa-Witten invariants  $c_{\gamma,J}^{\rm ref}(y)$. When the dimension vector $\vec N$ has support on a subquiver $Q$ without loop, and if $\vec N$ is a primitive vector, then Reineke's formula \cite{1043.17010} can be used to compute $\bOm(\vec N,\vec \zeta,y)$. When $\vec N$ is not primitive, 
the same formula can be used to compute the  stack invariant $\cI(\vec N,\vec\zeta,y)$, which can be converted into the rational invariant  $\bOm(\vec N,\vec \zeta,y)$  and in turn into the integer invariant  $\bOm(\vec N,\vec \zeta,y)$
as explained in Appendix \ref{sec_genVW} and \cite{Manschot:2013sya}.  When the support of the dimension vector has loops,
one can try to apply Seiberg dualities as in \cite{Manschot:2013dua}: if the new dimension vector has  support on a subquiver without loop, the previous method can be applied again;  if the entries
of the new dimension vector do not all have the same sign, then the index $\bOm(\vec N,\vec \zeta,y)$ must vanish.

\medskip

For general Chern classes, the dimension vector is supported on the full quiver $Q$, which necessarily has loops and these methods fail. However, we can use the fact that the index vanishes in the attractor
chamber  to compute the index in the chamber of interest  \eqref{zetacan} by 
applying the wall-crossing formula across each of the walls of marginal stability which separate the two chambers. In fact, there is a more efficient way to proceed, which is to use the flow tree formula developped in \cite{Alexandrov:2018iao}, which directly expresses the index 
$\Omega(\gamma,\zeta,y)$ for any stability condition,
in terms of attractor indices $\Omstar(\alpha_j,y)$ for all possible decompositions $\gamma=\sum_j\alpha_j$ of the dimension vector $\gamma$ (for simplicity, we identify $\gamma$ 
with the dimension vector $\vec N$). 
More precisely, the rational index is given by
\bea
\label{FTform}
\bOm(\gamma,\zeta,y) =
\sum_{\gamma=\sum_{j=1}^n \alpha_j}
\frac{\gtr(\{\alpha_j,c_j\},y)}{|{\rm Aut}\{\alpha_j\}|}\,
\prod_{j=1}^n \bOm_\star(\alpha_j,y).
\eea
where the sum runs over all distinct unordered splittings of $\gamma$ into sums of vectors
$\alpha_j$ with non-negative entries, $|{\rm Aut}\{\alpha_j\}|$ is the
order of the subgroup of the permutation group $S_n$ preserving the ordered set $\{\alpha_j\}$, and
$\gtr(\{\alpha_j,c_j\},y)$ is the `tree index' for $n$ dyons with charges $\alpha_j$ and 
stability parameters $c_j=\sum_v n_{j,v} \zeta_v$, where $n_{j,v}$ are the coefficients of the
vector  $\alpha_j$ on the basis $\gamma_v$; note that these parameters add up to zero,
$\sum_{j=1}^n c_j=0$. 
The tree index  is in turn defined by
\be
\gtr(\{\alpha_j,c_j\}, y)
= \frac{(-1)^{n-1+\sum_{i<j} \langle\alpha_i,\alpha_j \rangle}}{(y-y^{-1})^{n-1}} \,
\sum_{\sigma\in S_n}\,
\Ftr(\{\alpha_{\sigma(i)},c_{\sigma(i)}\})\,
y^{\sum_{i<j} \langle\alpha_{\sigma(i)},\alpha_{\sigma(j)} \rangle}
\label{gtF}
\ee
where $\sigma$ runs over all permutations of $\{1,\dots, n\}$ and 
$\Ftr(\{\alpha_j,c_j\})$ is  
the `partial tree index'; the latter is defined as 
a sum over all  planar flow trees with $n$
leaves carrying ordered charges $\alpha_1,\dots, \alpha_n$.
It is most conveniently evaluated using the recursive formula \cite[(2.59)]{Alexandrov:2018iao},
\be
\Ftr(\{\alpha_j,c_j\})=\hf\sum_{\ell=1}^{n-1} \bigl( \sgn(S_\ell)
-\sgn (\Gamma_{n\ell})\bigr)\,
\Ftr(\{\alpha_j,c_j^{(\ell)}\}_{j=1}^\ell)\,
\Ftr(\{\alpha_j,c_j^{(\ell)}\}_{j=\ell+1}^n),
\label{inductFn}
\ee
where 
\be
S_\ell=\sum_{j=1}^\ell c_j,
\quad
\Gamma_{k\ell}=\sum_{i=1}^k\sum_{j=1}^\ell 
\langle\alpha_i,\alpha_j\rangle,
\quad
\beta_{k\ell}=\sum_{i=1}^k \langle\alpha_i,\alpha_\ell\rangle,
\quad 
c_i^{(\ell)}= c_i -\frac{\beta_{ni}}{\Gamma_{n\ell}}\, S_{\ell}.
\ee
The parameters $c_i^{(\ell)}$ correspond to the stability parameters at the point where
 the attractor flow crosses the wall for the decay 
$\gamma\to(\alpha_1+\dots+\alpha_\ell,\alpha_{\ell+1}+\cdots + \alpha_n)$, and they satisfy
the condition $\sum_{j=1}^\ell c_j^{(\ell)}=\sum_{j=\ell+1}^n c_j^{(\ell)}=0$. 
The recursion is initiated by the value $\Ftr(\{\alpha,c\})=1$ for a single dyon of charge $\alpha$
and stability parameter $c=0$. Using \eqref{FTform} and the vanishing of $\Omstar(\alpha,y)$ for
non-simple dimension vectors\footnote{While  the attractor index may be non-zero
for pure D0-branes, it turns out that it does not contribute to the flow tree formula \eqref{FTform} for dimension vectors 
$\vec N$ associated to torsion-free sheaves (see the discussion Eq. (4.4) in \cite{Manschot:2009ia}).}, 
one may compute the index $\Omega(\vec N,\vec \zeta)$
in principle for any dimension vector and any (generic) stability condition.

\medskip

Alternatively, one may apply the Coulomb branch formula developped in 
\cite{Manschot:2011xc,Manschot:2012rx,Manschot:2013sya} (see \cite{Manschot:2014fua} for a concise review), to compute the index in terms of `single-centered indices'  $\OmS(\alpha_i,y)$. 
In a nutshell, the Coulomb branch formula reads
\bea
\label{CBform}
\Omega(\gamma,\zeta,y) &=&  \sum_{\gamma=\sum_{j=1}^n m_j \alpha_j}
\frac{\widehat{g}_C(\{\alpha_j,m_j, c_j\},y)}{|{\rm Aut}\{\alpha_j,m_j\}|}
\prod_{j=1}^n \OmS(\alpha_j,y^{m_j})
\eea
where the sum runs over unordered decompositions $\alpha   =\sum_{j=1}^n m_j\alpha_j$ with $m_j\geq 1$,  ${\rm Aut}\{\alpha_j,m_j\}$ denotes the subgroup of $S_n$ which preserves the pairs $(\alpha_j,m_j)$, and $\widehat{g}_C(\{\alpha_j,m_j, c_j\},y)$ is the modified Coulomb  index,
obtained from a sum over collinear configurations by applying the minimal modification hypothesis \cite{Manschot:2011xc}. Unlike the attractor index $\Omstar(\alpha,y)$, the single-centered index
 $\OmS(\alpha,y)$ is not known to coincide  with the index $\Omega(\gamma,\vec\zeta^S,y)$ in a
 putative chamber $\vec\zeta^S$, indeed it currently does not have an intrinsic mathematical definition.
Instead, it is defined recursively by inverting the Coulomb branch formula \eqref{CBform} for any (generic) choice of $\vec\zeta$; by virtue of the compatibility of the Coulomb branch formula with the wall-crossing formula, the resulting value of   $\OmS(\gamma,y)$  is independent of the choice
of $\vec\zeta$. Choosing $\vec\zeta=\vec\zetastar$ allows to express $\Omstar(\gamma,y)$ as a
polynomial in the $\OmS(\alpha,y)$'s, and vice-versa. Physically, $\Omstar(\vec N)$ includes both contributions from single-centered black holes counted by $\OmS(\vec N)$, as well as from multi-centered solutions with `scaling behavior' \cite{Bena:2006kb,Alexandrov:2018iao}.

\medskip

By evaluating \eqref{CBform} in the attractor chamber, we shall find evidence for the conjecture that for quivers associated to Fano surfaces, the single-centered indices
$\OmS(\alpha,y)$ are zero unless $\alpha$ is the dimension vector of a simple representation
or corresponds to a pure D0-brane,  just like the attractor indices $\Omstar(\alpha,y)$. If this conjecture is true, one may  use indifferently the flow tree or Coulomb branch formulae for evaluating the index  $\Omega(\vec N, \vec \zeta,y)$ for arbitrary values of the stability parameters.

\section{Complex projective plane \label{sec_p2}}

In this section, we discuss the simplest case in which the general considerations of the previous section apply, namely  the complex projective plane $S=\IP^2$. Vector bundles and coherent sheaves on $\IP^2$ have been discussed extensively in the mathematics literature, starting with the work of 
Beilinson, Dr\'ezet and Le Potier \cite{beilinson1978coherent,drezet1985fibres,le1994propos}.
In the physics literature, it was revisited in \cite{Diaconescu:1999dt,Douglas:2000qw}
in the context of D-branes on the orbifold $\IC^3/\IZ_3$, or equivalently the local Calabi-Yau 
$K_{\IP^2}$.

\subsection{From exceptional collection to BPS quiver}

The projective complex plane is the simplest example of a 
toric Fano surface, with toric fan generated by the 3 vectors,
\be
v_i = \begin{pmatrix} 
1 & 0 & -1  \\
0 & 1 & -1 
\end{pmatrix}
\ee
The divisors  $D_1=D_2=D_3=H$ are identified with the hyperplane class  $H$, and satisfy $H^2=1$. The canonical class is $c_1(S)=3H$ so the degree is $c_1(S)^2=9$. The toric collection \eqref{toricol}
is then the standard strong exceptional collection  
\be
\cC = (\cO(0),\cO(1),\cO(2) )
\ee
(considered for example in
Example 4.4 in \cite{Herzog:2004qw}), which is known to be strongly cyclic.
The Chern characters of the sheaves $E^i$ are
\be
\gamma^1=[1,0,0]\ , \quad \gamma^2=[1,1,\frac12]\ ,\quad \gamma^3=[1,2,2] \ ,
\ee
with increasing slopes $0,3,6$.
Here and henceforth we denote $\gamma=[N,c_1,\ch_2]$ or, when we prefer to display the integer-valued second Chern class, $\gamma=[N;c_1;c_2]$.
The Euler matrix and its inverse evaluate to 
\be
S=
\begin{pmatrix} 1 & 3 & 6 \\ 0 & 1 & 3 \\
0 & 0 & 1 \end{pmatrix} \ ,\quad 
S^\vee=
\begin{pmatrix} 1 & -3 & 3 \\ 0 & 1 & -3 \\
0 & 0 & 1 \end{pmatrix}\
\ee
The dual Chern characters 
\be
 \gamma _1 =[1,0,0]\ ,\quad 
 \gamma _2 =[-2,1,\frac{1}{2}]\ ,\quad 
 \gamma _3 =[1,-1,\frac{1}{2}]\ ,
\ee
with decreasing slopes $0,-3/2,-3$,
are recognized as the characters of  the exceptional collection
\be
\cC^\vee = \left( \cO, \Omega(1)[1], \cO(-1)[2] \right) 
\ee
where $\Omega(1)$ 
is the twisted cotangent bundle defined by the short exact sequence 
$0\to \Omega(1) \to \IC^3\otimes \cO \to \cO(1) \to 0$. 

Following the procedure outlined in \S\ref{sec_quiv}, the corresponding quiver has 
adjacency matrix 
\be
\kappa = (S^\vee)^t - S^\vee = \begin{pmatrix} 0 & 3 & -3 \\ -3 & 0 & 3 \\
3 & -3 & 0 \end{pmatrix}\
\ee
This is recognized as the cyclic 3-node quiver with $3$ arrows between each subsequent node,
familiar from the discussion of D-branes on the orbifold $\IC^3/\IZ_3$ \cite{Douglas:1996sw},
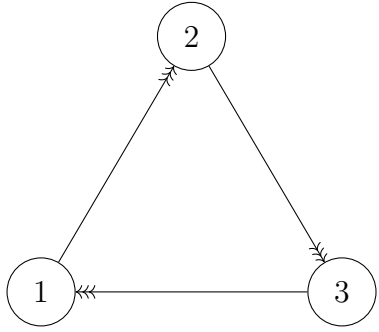
 \begin{center}
\begin{tikzpicture}[inner sep=2mm,scale=2]
  \node (a) at ( -1,0) [circle,draw] {$1$};
  \node (b) at ( 0,1.7) [circle,draw] {$2$};
  \node (c)  at ( 1,0) [circle,draw] {$3$};
 \draw [->>>] (a) to node[auto] {$ $} (b);
 \draw [->>>] (b) to node[auto] {$ $} (c);
 \draw [->>>] (c) to node[auto] {$ $} (a);
\end{tikzpicture}
\end{center}
The superpotential is obtained by evaluating the cubic superpotential 
of $\cN=4$ SYM theory on $\IZ^3$-invariant D-brane configurations \cite{Douglas:1996sw,Douglas:1997de},\footnote{{\bp Here and elsewhere, we omit the trace and write the product of chiral operators along a closed path from left to right, for easier reading.}}
\be
W= \sum_{(\alpha\beta\gamma)\in S_3}  \sgn(\alpha\beta\gamma)\, 
\Phi_{12}^\alpha \Phi_{23}^\beta \Phi_{31}^\gamma
\ee
where $\sgn(\alpha\beta\gamma)$ is the signature of the permutation. For rank $(1,1,1)$ and vanishing Fayet-Iliopoulos terms, the space of supersymmetric vacua reproduces 
the orbifold $\IC^3/\IZ^3$ probed by D0-branes.

\medskip

Given a coherent sheaf $E$ on $S$, its Chern character  decomposes as $\gamma=\sum_i n_i \gamma_i$ where
\be
 n_1 =\frac{3}{2} c_1 +  \ch_2+ N \ ,\quad 
 n_2 = \frac{1}{2} c_1 + \ch_2 \ ,\quad 
 n_3 = -\frac{1}{2}c_1+ \ch_2 
\ee
or conversely
\be
N = n_1-2 n_2+n_3 \ ,\quad 
c_1= n_2-n_3 \,\quad 
\ch_2 = \frac{1}{2} \left(n_2+n_3\right)
\ee
Note that the quiver is symmetric under cyclic permutations of the nodes, and under exchange of two nodes provided the arrows are reversed. In particular, the exchange of $n_1 \leftrightarrow n_3$ corresponds to the symmetry 
\be
[N, c_1, \ch_2 ] \to [N, -N-c_1, \ch_2+ c_1 + \frac{N}{2} ]\ ,
\ee
which is a combination of the reflexion $c_1\to -c_1$ and a spectral flow \eqref{specflow}.  
In order that the all entries in the dimension vector be positive for large  positive $c_2$, 
we choose 
\be
\vec N:=(N_1,N_2,N_3)=-(n_1,n_2,n_3) \ .
\ee 
Identifying  the
D-brane charges 
$(Q_4,Q_2,Q_0)$ in \cite{Douglas:2000qw} with $[N,c_1,\ch_2]$, we see that
the dimension  vector $(n_1,n_2,n_3)$ from \cite[(2.9)]{Douglas:2000qw} is $(N_3,N_2,N_1)$
in our notations. In either conventions, the height is $h(\vec N)=N_1+N_2+N_3=-N-\frac32 c_1(c_1+1)+3 c_2$
 and the dimension vector for a single D0-brane is $
\vec N=(1,1,1)$.

\subsection{Beilinson quiver and stability conditions}

\medskip
For dimension vector $\vec N\neq (p,p,p)$ and generic stability parameters $\vec \zeta$,
one of the three set of arrows typically vanishes\footnote{See footnote \ref{foodimred} on page \pageref{foodimred}. Note that the reasoning in \cite[\S 5.2.3]{Denef:2007vg} 
{\bp cannot be used to justify this assertion, }
since the superpotential is not generic. For $\vec N=(p,p,p)$ corresponding to $p$ D0-branes,
{\bp the assertion fails, since}
the moduli space is expected to be the Hilbert scheme of $p$ points on $X$, with dimension
$3p$, while the vanishing of $\Phi_{31}^\alpha$ would result in a moduli space of dimension 2.}, 
and the cyclic quiver reduces to a linear 3-node
quiver with relations. Without loss of generality, we focus on the chamber where the maps $\Phi_{31}^\alpha$ vanish, corresponding to the Beilinson quiver 
 
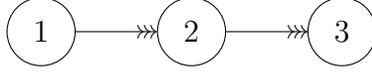
\begin{center}
\begin{tikzpicture}[inner sep=2mm,scale=2]
  \node (a) at ( -1,0) [circle,draw] {$1$};
  \node (b) at ( 0,0) [circle,draw] {$2$};
  \node (c)  at ( 1,0) [circle,draw] {$3$};
 \draw [->>>] (a) to node[auto] {$ $} (b);
 \draw [->>>] (b) to node[auto] {$ $} (c);
% \draw [->>>] (c) to node[auto] {$ $} (a);
\end{tikzpicture}
\end{center}
with relations $\sum_{\beta,\gamma} \sgn(\alpha\beta\gamma) \Phi_{12}^\beta \Phi_{23}^\gamma=0$ for $\alpha=1,2,3$. 
As noted in \cite{Douglas:2000qw}, 
representations of this reduced quiver reproduce the monad construction of bundles on 
$\IP^2$ \cite{beilinson1978coherent}.
The expected dimension of the quiver moduli space in this chamber is 
\be
\label{dimP2}
d_\IC =  3(N_1 N_2+N_2 N_3-N_3 N_1)-N_1^2-N_2^2-N_3^2+1 = c_1^2 -2 N \ch_2 - N^2 + 1  
\ee
in agreement with the expected dimension \eqref{dimM} of the moduli space of stable coherent sheaves. 

\medskip

In order that this chamber be consistent with the D-term conditions \eqref{Dterm}, we need that the
stability parameters $\vec\zeta$ satisfy $\zeta_1\geq 0, \zeta_3\leq 0$. According to 
our general prescription \eqref{zetacan}, they are given by 
$\vec\zeta^c \propto \kappa\cdot \vec N$, up to a small correction necessary to agree with Gieseker stability,
\bea
 \zeta _1 &=&3\rho(N_2-N_3) - \frac12(N_2+N_3) =  -3 \rho\,  c_1 + \ch_2 \nn\\
 \zeta _2 &=&3\rho(N_3-N_1) +\frac12(N_1+3N_3) =  3 \rho  (2 c_1+N)   -2 \ch_2-\frac{N}{2} \nn\\
 \zeta _3 &=&3\rho(N_1-N_2) +\frac12(N_1-3N_2) =  -3 \rho (c_1+N)   + \ch_2-\frac{N}{2} 
\label{FIP2}
\eea
These stability parameters can also be obtained from the central charge \eqref{Zgen}
\be
\label{Zgamma}
\begin{split}
Z_\gamma = & - \left( \frac{z^2}{2} + \frac18\right)\, N + z\, c_1  - \ch_2 \\
= & N_1 \left( \frac12 z^2 + \frac18 \right)   - N_2 \left(z^2+z-\frac14\right) 
+ N_3 \left( \frac12 z^2+z+\frac58\right)  
\end{split} 
\ee
where $J'=z \, c_1(S)/3$, $z=s+\I t$.  In the large volume limit $t\to\infty$, we get 
\be
\begin{split}
\zeta_1 =&\Im(Z_{-\gamma_1} \bar Z_\gamma)/t =  \frac12 t^2 (N_2-N_3) 
- \frac{1}{8} \left(N_3 \left(4 s^2+4 s-1\right)+N_2 \left(-4 s^2+4 s+1\right)\right) + \dots ,\\
\zeta_2 =&\Im(Z_{-\gamma_2} \bar Z_\gamma)/t =  \frac12 t^2 ( N_3-N_1)- \frac{1}{8} \left(N_1 \left(4 s^2-4 s-1\right)-N_3 \left(4 s^2+12 s+7\right)\right) + \dots \\
\zeta_3 =&\Im(Z_{-\gamma_3} \bar Z_\gamma)/t =  \frac12 t^2 \, (N_1-N_2) - \frac{1}{8} \left(N_2 \left(4 s^2+12 s+7\right)+N_1 \left(-4 s^2-4 s+1\right)\right) + \dots  \ ,
\end{split}
\ee
where the dots are terms of order $1/t$.
The leading terms agree with \eqref{FIP2} upon setting $\rho=t^2/6$. There is no choice of $s$ such that subleading terms would agree, however it can be checked that they lead to usual Gieseker  stability conditions for $s<0$.  Moreover, it can be checked that the stability condition
following from \eqref{FIP2} is equivalent to the prescription in \cite{le1994propos,king1995chow,ohkawa2010moduli}
despite the fact that subleading terms
are different.

\medskip

The conditions $\zeta_1\geq 0, \zeta_3\leq 0$ then require that the first Chern class should lie in the window
\be
\label{slopeP2}
-N \leq  c_1 \leq  0\ ,
\ee
or equivalently $-3 \leq \nu(E) \leq 0$, where $\nu(E)$ is the slope \eqref{defnuE} for $J=c_1(S)$.  Note that the window \eqref{slopeP2}
is set by the range of slopes of the dual sheaves $E_i^\vee$. 
For $N\geq 1$, the condition \eqref{slopeP2} can always be satisfied by applying the spectral flow \eqref{specflow}, which in the present case reduces to 
\be
\label{flow}
 c_1 \mapsto c_1 + \epsilon N\ ,\quad 
 {\rm ch}_2 \mapsto {\rm ch}_2 + \epsilon c_1 + \frac12 N c_1^2\ ,\quad
 c_2 \mapsto c_2 + \frac12  \epsilon (N-1) ( 2 c_1 + N\epsilon )
\ee
with $\epsilon\in\IZ$. 

\subsection{Quiver moduli \label{sec_P2moduli}} 

Under the condition \eqref{slopeP2}, one expects that the moduli space of 
semi-stable coherent sheaves  will be isomorphic to the moduli space
of semi-stable quiver representations,
\be
\cM^S_{[N;c_1;c_2],J}\simeq  \cM^Q_{\vec N,\vec \zeta}
\ee
This isomorphism, if true, has several practical consequences. 

\medskip

Firstly, the moduli space must be empty unless 
$N_1,N_2,N_3$ have the same sign, a  condition which is stronger than $d_\IC\geq 0$ in \eqref{dimP2}.
For example, for $N=p+q$, $c_1=0$ and $c_2=p\geq 0$, 
the expected dimension $1+p^2-q^2$ is positive when $0< q\leq p$, 
however the dimension vector $(p,p,-q)$ is not positive, so the moduli space must in fact be empty.
This implies that for $c_1=0$, the moduli space is empty unless $c_2\geq N$. 
{\bp One can easily generalize this
argument to cases
with $c_1\neq 0$ and conclude that the Bogomolov discriminant $\Delta$ and slope $\nu\in[-1,0]$ must satisfy $\Delta\ge \frac12 \nu^2 + {\rm max}(\frac32\nu+1,-\frac12\nu)\geq \frac38$, which is consistent with the
domain of existence of stable vector bundles  
analysed in  \cite{drezet1985fibres}. }

\medskip

Second, in  the case where one of the $N_i$'s vanish, the superpotential relations
become trivial and the quiver reduces to a generalized Kronecker quiver\footnote{We denote by $K_m(N_1,N_2)$ the two-node quiver with $m$ arrows $1\to 2$ and dimension vector 
$(N_1,N_2)$.}  $K_3(N_1,N_2)$
or $K_3(N_2,N_3)$. The index for the 
latter  can be computed using Reineke's formula and compared with
the modular prediction. This simplification occurs when either 
\begin{itemize}
\item $c_2=\frac12 c_1(c_1-1)$ and $-N<c_1\leq -N/2$  so that $N_3=0$;
\item  $c_2=\frac12 c_1(c_1+1)$  implies $N_2=0$ and $n_1=c_1<0$, so the moduli space is empty
\item $c_2=N + \frac12 c_1(c_1+3)$ and $-N/2\leq c_1 \leq 0$, so that $N_1=0$;
\end{itemize}
In particular, we get 
\be
\begin{array}{|l|c|l|}
\hline
\[N;c_1;c_2\] & \vec N &\Omega_c(\vec N)  \\
\hline
\[1;0;1\] & (0,1,1) & y^2+1+1/y^2\\
\[2;-1;1\] & (0,1,0) &   1 \\
\[2;0;2\] & (0,2,2) &  -y^5-y^3-y -\dots \\
\[3;-1;2\] & (0,2,1) &  y^2+1+1/y^2\\
\[3;-2;3\] & (1,2,0) &   y^2+1+1/y^2\\
\[3;0;3\]& (0,3,3) &  y^{10}+y^8 + 2y^6 + 2 y^4 +2 y^2 + 2 + \dots \\
\[4;-2;3\] & (0,2,0) & 0  \\
\[4;-1;3\]& (0,3,2) & y^6+y^4+3 y^2 + 3+\dots   \\
\[4;-3;6\] & (2,3,0)  &y^6+y^4+3 y^2 + 3 +\dots \\
\[4;0;4\]& (0,4,4) &  -y^{17} -y^{15}-3y^{13} -4y^{11}-6 y^{9} -6 y^7 -7 y^5- 7 y^3 - 7 y -\dots\\
\[5;-2;4\] &  (0,3,1) & 1 \\
\[5;-1;4\] & (0,4,3)  & y^{12}+y^{10}+3 y^8 + 5 y^6 + 8 y^4 + 10 y^2 + 12 + \dots \\
\hline
\end{array}
\nn
\ee
in perfect agreement with the analysis in Appendix \ref{sec_VWP2}. 

\medskip

Third, in the case where Reineke's formula is not applicable, we can use the flow tree  formula \eqref{FTform} to compute the index in the chamber \eqref{FIP2} of interest, provided the attractor indices $\Omstar(\vec N):=\Omega(\vec N,\vec\zetastar)$ are known. Fortunately, we can show that $\Omstar(\vec N)=0$
unless $\vec N$ is equal to $(1,0,0), (0,1,0), (0,0,1)$ or $(p,p,p)$ for some positive integer $p$, up to an overall sign. In fact,
we can show a more general statement:

\medskip

\noindent {\bp Proposition:} Consider a cyclic 3-node quiver with adjacency matrix $\kappa_{12}=a$, $\kappa_{23}=b$ $\kappa_{31}=c$. If $a,b,c$ satisfy\footnote{Up to permutations, the solutions to \eqref{dtotP2gencond} are $(1,1,c)$ with $1\leq c\leq 5$, $(1,2,c)$ with $2\leq c\leq 5$, $(1,3,c)$ with $3\leq c\leq 4$, $(2,2,c)$ with $2\leq c\leq 4$, $(2,3,3)$
and $(3,3,3)$.}
\be
\label{dtotP2gencond}
1\leq a,b,c \leq 6\ ,\quad 
a^2+b^2+c^2 + \frac13 abc \leq 36
\ee
and none of the following conditions are satisfied, 
\begin{itemize}
\item  $\vec N \in  \{ (\pm 1,0,0), (0,\pm 1,0), (0,0,\pm 1)\}$
\item $(a,b,c)=(2,2,2)$ and $\vec N=\pm (1,1,1)$
\item $(a,b,c)=(3,3,3)$ and $\vec N=\pm (p,p,p)$ for any $p\geq 1$.
\end{itemize}
then $\Omstar(\vec N)=0$.

\medskip

\noindent {\it Proof:}  If one of the $N_i$'s vanish then the 3-node quiver reduces to a Kronecker quiver, and the attractor indices are known to vanish except for basis vectors. Thus we assume that all of the $N_i$'s are strictly positive. The attractor stability parameters are
\be
\label{ze123abc}
\zetastar_1=c N_3-aN_2\ ,\quad \zetastar_2=a N_1-b N_3\ ,\quad \zetastar_3=bN_2- cN_1
\ee
Depending on the signs of the $\zetastar_i$'s, in the attractor chamber we have either  
\begin{itemize}
\item $\Phi_{31}^\alpha=0$ when $\zetastar_1\geq 0, \zetastar_3\leq 0$, with expected dimension
\bea
\label{dimPabc}
d_\IC &=& a N_1 N_2+ b N_2 N_3- c N_3 N_1 -N_1^2-N_2^2-N_3^2+1  
\nn\\&=& 
1 - \cQ(\vec N) - \frac23 N_1 \zetastar_1 +  \frac23 N_3 \zetastar_3 
\eea
\item  $\Phi_{12}^\alpha=0$ when $\zetastar_2\geq 0, \zetastar_1\leq 0$,  with expected dimension
\bea
d'_\IC &=& - a N_1 N_2+ b N_2 N_3+ c N_3 N_1 -N_1^2-N_2^2-N_3^2+1 \nn\\
&=&1  - \cQ(\vec N)- \frac23 N_2 \zetastar_2 +  \frac23 N_1 \zetastar_1 
\eea
\item  $\Phi_{23}^\alpha=0$ when $\zetastar_3\geq 0, \zetastar_2\leq 0$,  with expected dimension
\bea
d''_\IC &=& a N_1 N_2- b N_2 N_3+ c N_3 N_1 -N_1^2-N_2^2-N_3^2+1   \nn\\
&=&1 - \cQ(\vec N)- \frac23 N_3 \zetastar_3 +  \frac23 N_2 \zetastar_2 
\eea
\end{itemize}
where $\cQ$ is the quadratic form 
\be
\cQ(\vec N)= \frac12 \begin{pmatrix} N_1 & N_2 & N_3 \end{pmatrix} 
\begin{pmatrix} 2 & -a/3 & -c/3  \\ -a/3 & 2 & -b/3 \\ -c/3 & -b/3 & 2 \end{pmatrix}
\begin{pmatrix} N_1 \\ N_2 \\ N_3 \end{pmatrix} 
\ee 
When the conditions \eqref{dtotP2gencond} are satisfied, the quadratic form $\cQ$ is positive,
and it follows that the relevant dimension 
$\dstar_\IC\in \{d_\IC,d'_\IC,d''_\IC\}$ is less or equal to 1. 
It is equal to 1 only when  $\cQ$ is degenerate (which occurs only when $(a,b,c)=(3,3,3)$) and
when $\vec N$ lies in the kernel of $Q$, i.e.  $\vec N=(p,p,p)$. The dimension $\dstar_\IC$ vanishes 
for $(a,b,c)=(2,2,2)$ and $\vec N=(1,1,1)$, and is strictly negative in all other cases. Hence
the attractor index $\Omstar(\vec N)$ vanishes, except in the cases listed above. QED.

\medskip

For the case $(a,b,c)=(3,3,3)$ relevant for $\IP^2$, we conclude that the attractor indices vanish unless $(N_1,N_2,N_3)=(p,p,p)$
for some integer $p$. Applying the flow tree formula,
we obtain a prediction for the indices in the canonical chamber. For low values of $N$ and $-N\leq c_1\leq 0$,  we find 
\be
\begin{array}{|c|c|l|}
\hline
\[N;c_1;c_2\] & \vec N  &\Omega_c(\vec N)  \\
\hline
\[ 1;0;2\] &(1,2,2) & y^4+2y^2+3+\dots  \nn\\
\[ 1;0; 3\] &(2,3,3) & y^6+2y^4+5y^2+6+\dots  \nn\\
\[ 2;0; 3\] &(1,3,3) &-y^9-2 y^7-4 y^5-6 y^3-6 y-\dots \nn\\
\[ 2;-1; 2\] &(1,2,1) &y^4 + 2y^2 + 3+\dots \nn\\
\[ 2;-1; 3\] &(2,3,2) &y^8+2y^6+6y^4+9y^2+12+\dots \nn\\
\[ 3;-1; 3\] & (1,3,2) &y^8+2y^6+5y^4+8y^2+10+\dots \nn\\
\[ 4;-1;4\] &(1,4,3) &y^{14}+2 y^{12}+5 y^{10}+10 y^8+18 y^6+28 y^4+38 y^2+42+\dots \nn\\
\[ 4;-2; 4\] &(1,3,1) &y^5+y^3+y+\dots \nn\\
\[ 4;-2;5\] & (2,4,2) & -y^{13}-2y^{11}-6y^9-10y^7-17y^5-21y^3-24y- \dots 
\\
\hline 
\end{array}
\ee
Remarkably, this agrees with 
the results of the analysis in Section \ref{sec_VWP2}. 

\medskip

Finally, applying the Coulomb branch formula \eqref{CBform}
in the attractor chamber we find circumstancial 
evidence that  the single-centered invariants satisfy the following conjecture:

\medskip

\noindent {\bp Conjecture :} the single-centered invariants 
$\OmS(\vec N)$ vanish unless 
\be (N_1,N_2,N_3)\in 
\{ (\pm 1,0,0), (0,\pm 1,0), (0,0,\pm 1), (p,p,p), p\geq 0\}
\ee

\medskip

Specifically, using the mathematica package {\tt CoulombHiggs} we find 
$\Omstar(\vec N)=\OmS(\vec N)$ for 
\be
\vec N \in \{ (2,2,1), (1,2,1), (3,3,1), (1,3,3), (2,3,1), (1,3,2), (1,3,1) \}\ ,
\ee
as a result of large number of cancellations when $(a,b,c)$ are set to $(3,3,3)$. 
For $\vec N=(2,3,2)$, we find a more complicated combination
\be
\Omstar(2,3,2) = \OmS(2,3,2)  - (y+1/y) \, \OmS(1,3,1) + \OmS(1,2,1)
\ee
but in all these cases, the vanishing of $\Omstar(\vec N)$ implies that $\OmS(\vec N)$ also vanishes. 
In contrast, for a pure D0-brane, we get (upon perturbing around the attractor stability vector 
$\vec\zetastar=0$)
\be
\label{OmD0}
\Omstar(1,1,1) = -y - y^{-1} + \OmS(1,1,1) 
\ee 
so that $\Omstar(1,1,1)$ and $\OmS(1,1,1)$ cannot both vanish. The relation \eqref{OmD0} is independent on the perturbation used to evaluate the left-hand side, and it appears 
 to extend to bound states of $p>1$ D0-branes with dimension vector $(p,p,p)$, taking into account the vanishing of $\OmS(\vec N)$ for generic $\vec N$.

\section{Hirzebruch surfaces \label{sec_Hirz}}

We now turn to the next simplest toric surfaces, namely the Hirzebruch surfaces $S=\mathbb{F}_m$. Since $b_2(S)=2$, the \kahler cone is non-trivial and the Vafa-Witten invariants exhibit wall-crossing phenomena. We shall check that the wall-crossing phenomena on the quiver side take place precisely when sheaves become unstable, as long as the degree stays within the window
where the quiver description is valid.

\medskip

 The Hirzebruch surface $S=\mathbb{F}_m$ is defined as the projectivization of the rank 2 bundle $\cO\oplus \cO(m)$ over $\IP^1$. The toric fan is generated by 4 vectors,
\be
v_i = 
\begin{pmatrix}
1 & 0 & -1 & 0\\
0 & 1 & m & -1
\end{pmatrix}
\ee
which is convex for $|m|\leq 2$. By symmetry we may assume $m\geq 0$.
The linear relations $D_1=D_3$, $D_4=D_2+m D_3$ from \eqref{eqRS} are consistent with the identifications
$D_1=F, D_2=C$ where $F$ and $C$ are the fiber and basis of the fibration $\IF_m\to \IP^1$.
Using \eqref{eqDij} one finds $C^2=-m, F^2=0, C\cdot F=1$, so the intersection form 
$C^{\alpha\beta}$  in the basis
$(C,F)$  and its inverse $C_{\alpha\beta}$ are
\be
C^{\alpha\beta} = \begin{pmatrix} -m & 1 \\ 1 & 0\end{pmatrix}\ ,\quad 
C_{\alpha\beta} = \begin{pmatrix} 0 & 1 \\ 1 & m\end{pmatrix}
\ee
The anticanonical class $c_1(S)=2 C + (m+2) F$ satisfies $c_1(S)^2=8$. Since $c_1(S)\cdot C=2-m$, $c_1(S)\cdot F=2$, the Hirzebruch surface 
$\IF_m$ is Fano for $m=0,1$ and weak Fano for $m=2$. For these values of $m$, $\IF_m$ has the following alternative descriptions:
\begin{itemize}
\item $\IF_0$ is simply the product 
$\IP^1\times \IP^1$, and the total space of the canonical bundle over $\IF_0$ is a $\IZ_2$ orbifold of the conifold. 
\item  $\IF_1$ is the blow-up of $\IP^2$ at one point, so also known as the first del Pezzo surface $dP_1$. 
\item  $\IF_2$ is also known
as $Y^{2,2}$, and the total space of the canonical bundle over $\IF_2$ is the orbifold $\IC^3/\IZ_4$ with
 action $(1,1,2)$.
 \end{itemize} 
  It is worth noting that the surfaces  $\mathbb{F}_m$ and
 $\mathbb{F}_{m+2}$ are diffeomorphic as smooth manifolds, but not complex
 diffeomorphic \cite{hirzebruch1978}. Still, their homology lattice and canonical class match under shifting $C\to C-F$ (see e.g. \cite[A.3]{Berglund:2016yqo}). 
 
 \medskip
 
 The K\"ahler form
 $J_{m_1,m_2}=m_1(C+m F) + m_2 F$ lies in the K\"ahler cone for $2m_2+(m+2)m_1>0$.
 The  canonical chamber $J\propto c_1(S)$ corresponds to $\eta=m_1/m_2=2/(2-m)$, or $\eta=+\infty$ when $m=2$. The central 
 charge is given by
 \be
 Z _\gamma= - N \left[ \frac{m}{2} m_1^2 + m_1 m_2 +\frac16 \right] + m_1 \tilde c_{1,F} + m_2  \tilde c_{1,C} -\ch_2
 \ee
 where $( \tilde c_{1,C},\tilde c_{1,F})$ are the coefficients of the first Chern class $c_1(E)$ on the basis $( C,F)$. The degree is 
 \be
 \deg E = c_1(S)\cdot c_1(E) = (2-m) \tilde c_{1,C}+ 2 \tilde c_{1,F}\ .
 \ee

 \medskip
 
Exceptional  collections of invertible sheaves on $\IF_m$ are obtained from toric systems via \eqref{toricolt}. Toric systems leading to cyclic strongly exceptional collections were classified in \cite[Prop. 5.2]{hille2011exceptional} (with $(P,Q)$ in {\it loc. cit}. identified with $(F,C+mF)$). 

\subsection{$\mathbb{F}_0$}
In view of the isomorphism  $\mathbb{F}_0=\IP^1\times\IP^1$, we denote$(c_{1,1},c_{1,2})$=$( \tilde c_{1,C},\tilde c_{1,F})$, and  by $\cO(p,q)$ the 
line bundle with $(c_{1,1},c_{1,2})=(p,q)$.

\subsubsection{Phase I \label{sec_F01}}

We consider the cyclic strong 
exceptional collection \cite{kuleshov1997moduli,Maiorana:2017wdq,perling2003some}
\be
\cC=\left( \cO(0,0),\cO(1,0),\cO(0,1),\cO(1,1) \right)
\ee
associated to the toric system $\tilde D_i=(C,F-C,C,F+C)$.
 The Chern vectors of the objects $E^i$ and dual objects $E_i$ are 
\be
\begin{array}{ccl}
 \gamma^1 &=& \[1,(0,0),0\] \\
 \gamma^2 &=& \[1,(1,0),0\] \\
 \gamma^3 &=& \[1,(0,1),0\] \\
 \gamma^4 &=& \[1, (1,1),1\] \\
\end{array}
\qquad
\begin{array}{ccl}
 \gamma_1 &=& \[1,(0,0),0\] \\
 \gamma_2 &=& \[-1,(1,0),0\] \\
 \gamma_3 &=& \[-1,(0,1),0\] \\
 \gamma_4 &=& \[1,(-1,-1),1\] \\
\end{array}
\ee
with slopes $0,2,2,4$ and $0,-2,-2,-4$, respectively.
The Euler matrix, its inverse and the adjacency matrix evaluate to 
\be
\label{SF01}
S=\left(
\begin{array}{cccc}
 1 & 2 & 2 & 4 \\
 0 & 1 & 0 & 2 \\
 0 & 0 & 1 & 2 \\
 0 & 0 & 0 & 1 \\
\end{array}
\right)
\ ,\quad
S^\vee=\left(
\begin{array}{cccc}
 1 & -2 & -2 & 4 \\
 0 & 1 & 0 & -2 \\
 0 & 0 & 1 & -2 \\
 0 & 0 & 0 & 1 \\
\end{array}
\right)
\ ,\quad
\kappa
=\left(
\begin{array}{cccc}
 0 & 2 & 2 & -4 \\
 -2 & 0 & 0 & 2 \\
 -2 & 0 & 0 & 2 \\
 4 & -2 & -2 & 0 \\
\end{array}
\right)
\ee 
corresponding to the four-node quiver 
 
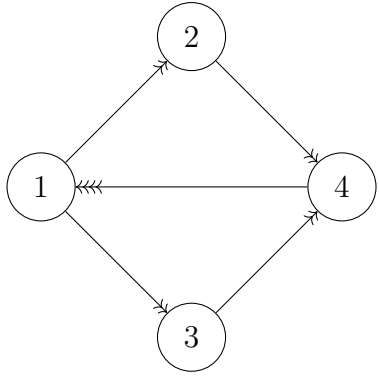
\begin{center}
\begin{tikzpicture}[inner sep=2mm,scale=2]
  \node (a) at ( -1,0) [circle,draw] {$1$};
  \node (b) at ( 0,1) [circle,draw] {$2$};
  \node (c)  at ( 0,-1) [circle,draw] {$3$};
  \node (d)  at ( 1,0) [circle,draw] {$4$};
 \draw [->>] (a) to node[auto] {$ $} (b);
 \draw [->>] (a) to node[auto] {$ $} (c);
 \draw [->>] (b) to node[auto] {$ $} (d);
 \draw [->>] (c) to node[auto] {$ $} (d);
  \draw [->>>>] (d) to node[auto] {$ $} (a);
\end{tikzpicture}
\end{center}
with cubic superpotential  \cite{Feng:2000mi}
\bea
W&=&\Phi_{12}^{1} \Phi_{24}^{1} \Phi_{41}^{4}-\Phi_{12}^{1} \Phi_{24}^{2} \Phi_{41}^{3}-\Phi_{12}^{2} \Phi_{24}^{1}
   \Phi_{41}^{2}+\Phi_{12}^{2} \Phi_{24}^{2} \Phi_{41}^{1} \nn\\
   &&-\Phi_{13}^{1} \Phi_{34}^{1} \Phi_{41}^{4}+\Phi_{13}^{1} \Phi_{34}^{2}
   \Phi_{41}^{2}+\Phi_{13}^{2} \Phi_{34}^{1} \Phi_{41}^{3}-\Phi_{13}^{2} \Phi_{34}^{2} \Phi_{41}^{1}
\eea
A general Chern character $\gamma=[N,c_{1,1},c_{1,2},\ch_2]$ decomposes as $\gamma=
\sum_i n_i \gamma_i$ with 
\bea
n_1 &=& c_{1,1} + c_{1,2} + \ch_2 + N \nn\\
n_2 &=& c_{1,1} + \ch_2\nn\\
n_3 &=& c_{1,2}+ \ch_2 \nn\\
n_4 &=& \ch_2
\eea
or conversely
\bea
N &=& n_1 - n_2 - n_3 + n_4 \ ,\quad \ch_2 = n_4\ ,\quad 
c_{1,1} = n_2 - n_4\ ,\quad 
c_{1,2} = n_3 - n_4 
\eea
Note that the exchange of $c_{1,1}$ and $c_{1,2}$ amounts to  exchanging of $n_2$ and $n_3$,
which is a symmetry of the quiver. 
For large positive $c_2$, the entries in $\vec n$ are all negative, so we 
consider the opposite dimension vector $\vec N=-\vec n$.

\medskip

In the canonical chamber $J\propto c_1(S)$, 
the  stability parameters  $\vec\zeta=\vec\zeta^c$ in \eqref{zetacan} are
\bea
\label{F0FI}
\zeta_1&=& \rho(2N_2 + 2N_3- 4 N_4) -N_4 = -  \rho\, \deg +\ch_2   \nn\\
\zeta_{2,3} &=& \rho(2N_4 -2N_1) +N_4  =  \rho  (\deg+2N)- \ch_2 \nn\\
\zeta_4 &=& \rho(4 N_1 -2 N_2 -2N_3) + N_1-N_2-N_3 =-\rho (\deg+4N) + \ch_2 -N  
\eea
Note that the leading term vanishes for $\vec N$ in the span of $(1,0,1,0)$ and $(0,1,0,1)$, corresponding to D0-branes and D2-branes wrapped on the exceptional curve $F-C$.

\medskip

For the Beilinson subquiver with $\Phi_{41}^\alpha=0$, the dimension of the moduli space of quiver
representations
\be
\label{dimBeiF0I}
d_\IC = 2 N_1 N_2 + 2 N_1 N_3 + 2 N_2 N_4 + 2 N_3 N_4 - 4 N_1 N_4 - N_1^2 - N_2^2 
- N_3^2 -N_4^2+1
\ee
matches the expected dimension \eqref{dimM} of the moduli space of sheaves. 
This is consistent with  the D-term conditions \eqref{Dterm} provided  $\zeta_1\geq 0, \zeta_4\leq 0$ hence
\be
\label{winF01}
-2N \leq   c_{1,1} + c_{1,2}  \leq   0
\ee
or equivalently $-4\leq \nu(E) \leq 0$, where $\nu(E)$ is the slope \eqref{defnuE} with $J=c_1(S)$.
This window is set by the range of slopes of the objects in the dual collection $\cC^\vee$.

\medskip

In the attractor chamber $\vec\zeta=\vec\zetastar$, given by the opposite of the leading term in \eqref{F0FI}, one has either\footnote{For generic $\vec\zeta$, in additional to the three chambers listed
below there are two other chambers,
namely $\Phi_{12}^\alpha=\Phi_{34}^\alpha=0$ or $\Phi_{13}^\alpha=\Phi_{24}^\alpha=0$. These last two do not arise when $\zeta_2$ and $\zeta_3$ have the same sign. The five chambers correspond
to the five perfect matchings $s_3,s_1,s_2,s_4,s_5$ listed in \cite[\S 17.2]{Hanany:2012hi}. }
\begin{itemize}
\item  $\Phi_{41}^\alpha=0$ when $\zetastar_1\geq 0, \zetastar_4\leq 0$: the expected dimension 
\eqref{dimBeiF0I} can be written as 
\be
d_\IC = 1 -\cQ(\vec N) 
+\frac12 ( N_4 \zetastar_4 - N_1 \zetastar_1) 
\ee
\item $\Phi_{12}^\alpha=\Phi_{13}^\alpha=0$ when $\zetastar_1\leq 0, \zetastar_{2,3}\geq 0$, 
with expected dimension
\be
d'_\IC = 1- \cQ(\vec N) + N_1 \zetastar_1 - \frac12 N_2 \zetastar_2 - \frac12 N_3 \zetastar_3 
\ee
\item $\Phi_{24}^\alpha=\Phi_{34}^\alpha=0$ when $\zetastar_{2,3}\leq 0, \zetastar_4\geq 0$,
with expected dimension
\be
d''_\IC =  1- \cQ(\vec N) + \frac12 N_2 \zetastar_2+ \frac12 N_3 \zetastar_3  - N_4 \zetastar_4 
\ee
\end{itemize}
where $\cQ(\vec N)$ is the quadratic form
\be
\cQ(\vec N) =  \frac12\left[ (N_1-N_2)^2 +(N_1-N_3)^2+(N_2-N_4)^2+(N_3-N_4)^2 \right]
\ee
Since $\cQ$ is positive, and degenerate along the direction $ (1,1,1,1)$, the expected dimension is negative unless $\vec N$ corresponds to a simple representation or to a D0-brane. 
Therefore the attractor index
 $\Omstar(\vec N)$ vanishes except in those cases.

\medskip

Applying the flow tree formula \eqref{FTform} in the canonical chamber, we get  
\be
\begin{array}{|c|c|l|}
\hline
\[N; c_{1,1},c_{1,2};c_2\] & \vec N &\Omega_c(\vec N)  \\ \hline
\[1;0,0;1\] &(0,1,1,1)&y^2+2+1/y^2\nn\\
\[1;0,0;2\] &(1,2,2,2)& y^4+3 y^2 + 6 + 3/y^2 + 1/y^4\nn\\
\[2;0,0;1\]&(-1,1,1,1)&0 \nn\\
\[2;-1,0;1\] &(0,2,1,1) & -y -1/y \nn\\
 \[2;-1,-1;1\] &(0,1,1,0) & 0 \nn\\
 \[2;0,0;2\]&(0,2,2,2) &-y^5 - 2 y^3 -3 y -\dots\nn\\
 \[2;-1,0;2\] &(1,3,2,2) &  -y^5 - 3 y^3 -7 y -\dots \nn\\
 \[2;-1,-1;2\] &(1,2,2,1) &  -y^3- y -1/ y -1/y^3 \nn\\
   \[3;0,0;3\] &(0,3,3,3) & y^{10}+2 y^8+5   y^6 +8 y^4 + 9 y^2 + 10 +\dots\nn\\
  \[3;-1,0;2\] &(0,3,2,2) &  y^4+2y^2 +4+\dots\nn\\
  \[3;-1,-1;2\] &(0,2,2,1) &  1\nn\\
   \[3;-1,-2;3\] &(1,2,3,1) &  0\nn\\
 \hline
\end{array} 
\ee
in agreement with the analysis in \S\ref{sec_genVW}, see 
\eqref{VW1}, \eqref{F0mod2} and \eqref{F0mod3}. 
 Applying the Coulomb branch formula \eqref{CBform}
in the attractor chamber,
we find circumstancial evidence that the single centered indices $\OmS(\vec N)$ vanish, just
like the attractor indices $\Omstar(\vec N)$.

\medskip

For general \kahler parameters $z_i= \I \, m_i$ with $m_i\gg 1$, 
the stability  parameters are instead given by\footnote{See also \cite{kuleshov1997moduli} and 
\cite[\S 5]{Maiorana:2017wdq} for an independent mathematical 
derivation of the stability parameters in this model.} 
\bea
\zeta_1 &=& m_{1} (N_{3}-N_{4})+m_{2}(N_{2}-N_{4}) - N_4, \nn\\
\zeta_2 &=& m_{1} (N_{4}-N_3) +m_{2}  (N_{3}-N_{1}) + N_4, \nn\\
\zeta_3&=&  m_{1} ( N_{2}-N_1) + m_{2} (N_{4}-N_2) + N_4,\nn\\
\zeta_4&=& m_{1}  ( N_{1}- N_{2})+m_{2} (N_{1}- N_{3}) + N_1 - N_2-N_3
   \eea
   This satisfies $\sum_i N_i \zeta_i=0$ and 
\be
 \sum_i N'_i \zeta_i = N \, (m_1 c_{1,2}' + m_2 c_{1,1}') - N' \, (m_1 c_{1,2} + m_2 c_{1,1}) +  N' \ch_2   - N \ch'_2
\ee
in agreement with \eqref{stabsum}. 
Note that under exchange of $(n_2,n_3)$ and $(m_1,m_2)$, $(\zeta_2,\zeta_3)$ swaps as well.
The chamber $\Phi_{41}=0$ is consistent with the D-term conditions \eqref{Dterm} provided 
\be
- (m_1+m_2) N \leq m_1\, c_{1,2}+m_2\, c_{1,1} \leq 0
\ee
For $N=2$, by examining the contributions to the flow tree formula we find the following wall-crossing phenomena:
\begin{itemize}
\item for $(c_{1,1},c_{1,2})=(0,0)$, $1\leq c_2\leq 3$ there is no wall-crossing at $\eta>0$;
\item for $(c_{1,1},c_{1,2})=(-1,0)$, $c_2=1$ we find that $\Omega(0,2,1,1)$ jumps at $\eta=m_1/m_2=1/2$, due to a bound state of $\gamma_L=(0,0,1,0)=[1;0,-1;0]$ 
and $\gamma_R=(0,2,0,1)=[1;-1,1;0]$  with $\langle\gamma_L,\gamma_R\rangle =2$.   
\item for $(c_{1,1},c_{1,2})=(-1,0)$,  $c_2=2$, the index  $\Omega(1,3,2,2)$ jumps at
\begin{itemize}
\item  $\eta=3/2$, due 
to bound states of 
$\gamma_L=(0,0,2,1)=[1;1,-1;0]$ and $\gamma_R=(1,3,0,1)=[1;-2,1;0]$ 
with $\langle\gamma_L,\gamma_R\rangle =-2$;
\item  $\eta=1/2$, due to bound states of 
$\gamma_L=(0,2,0,1)=[1;-1,1;0]$  and $\gamma_R=(1,1,2,1)=[1;0,-1;1]$ 
with  $\langle\gamma_L,\gamma_R\rangle =-2$;
and of  
$2\gamma_L=2(0,0 ,1,0)=2[1;0,-1;0]$  and $\gamma_R=(1,3,0,2)=[0;-1,2;0]$
with $\langle\gamma_L,\gamma_R\rangle =2$;
\item $\eta=1/4$, due to bound states of 
$\gamma_L=(0,3,0,2)=[1;-1,2;0]$ and $\gamma_R=(1,0,2,0)=[1;0,-2;0]$ 
with $\langle\gamma_L,\gamma_R\rangle =-6$. 
\end{itemize}
\item For $(c_{1,1},c_{1,2})=(-1,-1)$, $c_2=2$
the index $\Omega(1,2,2,1)$ jumps at
\begin{itemize}
\item  $\eta=3$,
 due to bound states of 
$\gamma_L=(0,0,2,1)=[1;1,-1;0]$ and $\gamma_R=(1,2,0,0)=[1;-2,0;0]$ 
with  $\langle\gamma_L,\gamma_R\rangle =-4$; 
\item $\eta=1/3$, 
 due to bound states of 
$\gamma_L=(0,2,0,1)=[1;-1,1;0]$ and $\gamma_R=(1,0,2,0)=[1;0,-2;0]$.
\end{itemize}
\end{itemize}
These results are consistent with the analysis in \S\ref{VWF0}, upon identifying $(\alpha,\beta)=(-c_{1,2},c_{1,1})$.

\medskip

It is interesting to consider the boundary chamber  $\eta\to 0^+$, where the generating functions of VW invariants are given by \eqref{defHN}--\eqref{htoH}, and controlled by the 
Hall algebra of $\IP^1$  \cite{Mozgovoy:2013zqx}.
In this chamber, the quiver description is valid for $-N\leq c_{1,1}\leq 0$. From the flow tree formula
we get 
\be
\begin{array}{|c|c|l|}
\hline
\[N; c_{1,1},c_{1,2};c_2\] & \vec N &\Omega(\vec N, \eta\to 0^+)  \\ \hline
\[2;0,-1;1\] &(0,1,2,1) & -y -1/y\ \nn\\
\[2;-1,0;1\] &(0,2,1,1) & 0\ \nn\\
\[2;-1,-1;1\] &(0,1,1,0) & 0 \nn\\
\[2;0,0;2\] &(0,2,2,2) & -y^5-2y^3- 3y-\dots \nn\\
\[3;0,-1;2\] &(0,2,3,2) &y^4+2y^2+4+\dots \nn\\
\[3;-1,0;2\] &(0,3,2,2) &0\nn\\
\[3;-1,-1;2\] &(0,2,2,1) &0 \nn\\
\[3;0,0;3\] &(0,3,3,3) &y^{10}+2y^8+5y^6+8y^4+9y^2+10+\dots \nn\\
\[4;0,-1;3\] &(0,3,4,3) &-y^{9}-2y^7-6y^5-11y^3-15 y +\dots \nn\\
\[4;-1,0-;3\] &(0,4,3,3) &0\nn\\
\[4;-1,-1;3\] &(0,3,3,2) &0\nn\\
\[4;0,-2;2\] &(0,2,4,2) &0\\
\hline
\end{array}
\ee
in precise agreement with the $q$-expansions in \eqref{hFmboundary2}--\eqref{hFmboundary4}.  In particular, the index vanishes whenever $\mu\cdot F = c_{1,1}\neq 0\mod N$.

\subsubsection{Phase II}
We now consider the strong exceptional collection 
\be
\cC = \left( \cO(0), \cO(1,0),\cO(1,1)\ ,\cO(2,1) \right) 
\ee
associated via  \eqref{toricol} to the toric system $D_i=(C,F,C,F)$.
The Chern vectors of the objects $E^i$ and dual objects $E_i$ are 
\be
\begin{array}{ccl}
 \gamma^1 &=& \[1,(0,0),0\] \\
 \gamma^2 &=& \[1,(1,0),0\] \\
 \gamma^3 &=& \[1,(1,1),1\] \\
 \gamma^4 &=& \[1,(2,1),2\] \\
\end{array}
\qquad 
\begin{array}{ccl}
 \gamma_1 &=& \[1,(0,0),0\] \\
 \gamma_2 &=& \[-1,(1,0),0\] \\
 \gamma_3 &=& \[-1,(-1,1),1\] \\
 \gamma_4 &=& \[1,(0,-1),0\] \\
\end{array}
\ee
with slope $0,2,4,6$ and $0,-2,0,-2$, respectively. 
The Euler matrix, its inverse and the adjacency matrix evaluate to 
\be
\label{SF02}
S= \left(
\begin{array}{cccc}
 1 & 2 & 4 & 6 \\
 0 & 1 & 2 & 4 \\
 0 & 0 & 1 & 2 \\
 0 & 0 & 0 & 1 \\
\end{array}
\right),\quad 
S^\vee=\left(
\begin{array}{cccc}
 1 & -2 & 0 & 2 \\
 0 & 1 & -2 & 0 \\
 0 & 0 & 1 & -2 \\
 0 & 0 & 0 & 1 \\
\end{array}
\right)
\ ,\quad 
\kappa =\left(
\begin{array}{cccc}
 0 & 2 & 0 &- 2 \\
 -2 & 0 & 2 & 0 \\
 0 & -2 & 0 & 2 \\
 2 & 0 & -2 & 0 \\
\end{array}
\right)
\ee
corresponding to the cyclic quiver 
 \begin{center}
\begin{tikzpicture}[inner sep=2mm,scale=2]
  \node (a) at ( -1,1) [circle,draw] {$1$};
  \node (b) at ( 1,1) [circle,draw] {$2$};
  \node (c)  at ( 1,-1) [circle,draw] {$3$};
  \node (d)  at ( -1,-1) [circle,draw] {$4$};
 \draw [->>] (a) to node[auto] {$ $} (b);
 \draw [->>] (b) to node[auto] {$ $} (c);
 \draw [->>] (c) to node[auto] {$ $} (d);
 \draw [->>] (d) to node[auto] {$ $} (a);
\end{tikzpicture}
\end{center}
with quartic superpotential \cite{Feng:2000mi},\cite[(2.2)]{Feng:2001xr}
\be
W = \sum_{(\alpha\beta)\in S_2} \sum_{(\gamma\delta)\in S_2} 
\sgn(\alpha,\beta) \,  \sgn(\gamma,\delta) \,  \Phi_{12}^\alpha \, \Phi_{23}^\gamma \, \Phi_{34}^\beta\, \Phi_{41}^\delta\ .
\ee
{\bp Note that one recovers the Phase I quiver \eqref{SF01} upon applying Seiberg duality on any node. }
A general Chern character $\gamma=[N,c_{1,1},c_{1,2},\ch_2]$ decomposes as $\gamma=\sum_i n_i \gamma_i$ with 
\bea
n_1 &=& c_{1,1} + c_{1,2} + \ch_2 + N \nn\\
n_2 &=& c_{1,1} + \ch_2\nn\\
n_3 &=&\ch_2 \nn\\
n_4 &=& -c_{1,2} + \ch_2
\eea
or conversely
\bea
N &=& n_1 - n_2 - n_3 + n_4 \ ,\quad \ch_2 = n_3 \nn\\
c_{1,1} &=& n_2 - n_3\ ,\quad 
c_{1,2} = n_3 - n_4 
\eea
Note that the symmetry exchanging $c_{1,1}$ and $c_{1,2}$ is no longer manifest. 
In order to have a positive dimension vector for large positive $c_2$, we set $\vec N=-\vec n$. 

\medskip

In the canonical chamber $J\propto c_1(S)$, 
the  stability parameters  $\vec\zeta=\vec\zeta^c$ in \eqref{zetacan} are then
\be
\label{FIKingF0}
\begin{split}
\zeta_1 =&  \rho \, (N_2-N_4) -N_3  = -\rho\,\deg+\ch_2\\
\zeta_2 =&  -\zeta_4 = \rho ( N_3-N_1)+ N_3 = \rho (2N+\deg) -\ch_2 \\
\zeta_3 =& \rho (N_4-N_2)- N_2+N_1+N_4 = \rho\deg -\ch_2-N
\end{split}
\ee
Note that the leading term vanishes for $\vec N$ in the span of $(1,0,1,0)$ and $(0,1,0,1)$, which again corresponds to D0-branes and D2-branes wrapped on $F-C$.

\medskip

For the Beilinson subquiver  with $\Phi_{41}=0$, the dimension of the moduli space of quiver representations
\be
\label{dimF0II}
d_{\IC} = 
2 N_{1} N_{2}+2 N_{2}   N_{3}+2 N_{3} N_{4}-2 N_{1} N_{4}-N_{1}^2-N_{2}^2-N_{3}^2-N_{4}^2+1
\ee 
matches the expected dimension \eqref{dimM} of the moduli space of sheaves. In this chamber, the 
D-term conditions \eqref{Dterm} require  $\zeta_1\geq 0, \zeta_4\leq 0$
hence 
\be
\label{wF0II}
-N \leq c_{1,1}+c_{1,2} \leq 0
\ee 
or equivalently $-2\leq \nu(E) \leq 0$.
This window is again set by the range of slopes of the objects in the dual collection $\cC^\vee$.

\medskip

In the attractor chamber, the stability parameters are given by
\be
\zetastar_1 = 2 (N_4-N_2) = -\zetastar_3\ ,\quad \zetastar_2 = 2 (N_1-N_3) = -\zetastar_4\ .
\ee
Using symmetry under cyclic permutations we can assume without loss of generality that $\zetastar_{1,2}\geq 0,
\zetastar_{3,4}\leq 0$, so that $\Phi_{41}$ vanishes. The dimension \eqref{dimF0II} can be written as 
\be
d_\IC  = 1- \left[ (N_1-N_2)^2 + (N_3-N_4)^2  \right] - N_1 \zetastar_1 - N_2 \zetastar_2
\ee
This is manifestly negative, unless $N_1=N_2$ and $N_3=N_4$. These equalities however imply 
$\zetastar_1=-\zetastar_2$, so that both have to vanish, hence all $N_i$'s are equal. 
It follows that  the attractor index vanishes unless $\vec N$ corresponds to a simple representation
or to a pure D0-brane. 

\medskip

In the canonical chamber, the flow tree formula \eqref{FTform} leads to 
\be
\begin{array}{|c|c|l|}
\hline
\[N; c_{1,1},c_{1,2};c_2\] & \vec N  &\Omega_c(\vec N)  \\ \hline
\[1;0,0;1\] &(0,1,1,1)&y^2+2+1/y^2\ \nn\\
\[1;0,0;2\] &(1,2,2,2)&y^4+3 y^2 + 6 + 3/y^2 + 1/y^4 \nn\\
 \[2;0,0;1\] &(-1,1,1,1) &0 \nn\\
 \[2;-1,0;1\] &(0,2,1,1) & -y -1/y \nn\\
  \[2;0,-1;1\] &(0,1,1,0) & -y -1/y \nn\\
   \[2;-1,-1;1\] &(0,1,0,-1) & 0  \nn\\
  \[2;0,0;2\]&(0,2,2,2) &-y^5 - 2 y^3 -3 y -\dots\nn\\
 \[2;-1,0;2\] &(1,3,2,2) &  -y^5 - 3 y^3 -7 y -\dots \nn\\
 \[2;0,-1;2\] &(1,2,2,1) &  -y^5 - 3 y^3 -7 y -\dots\nn\\
 \[2;-1,-1;2\] &(1,2,1,0) &  -y^3- y -1/ y -1/y^3 \\
  \[3;0,0;3\] &(0,3,3,3) & y^{10}+2 y^8+5   y^6 +8 y^4 + 9 y^2 + 10 +\dots\nn\\
  \[3;-1,0;2\] &(0,2,2,1) &  y^4+2y^2 +4+\dots\nn\\
  \[3;-1,-1;2\] &(0,2,1,0) &  1\nn\\
   \[3;-1,-2;3\] &(1,2,1,-1) & 0 \nn\\
   \[3;-1,-2;4\] &(2,3,2,0) &  y^8+2y^6+4y^4+4y^4+5+\dots\nn\\
 \hline
 \end{array}
\ee
 in agreement  with  \eqref{VW1},  \eqref{F0mod2},  \eqref{F0mod3}. Applying the Coulomb branch formula \eqref{CBform} in the attractor chamber, we find that the single centered indices $\OmS(\vec N)$  vanish, just like the attractor indices $\Omstar(\vec N)$.

\medskip

More generally, for polarisation $J'=\I m_1(C+F)+ \I m_2 F$ with $m_1,m_2\gg 1$, 
the stability parameters are given by 
\bea
\zeta_1 &=& - m_1\, c_{1,2} - m_2 \, c_{1,1} + \ch_2  \nn\\
\zeta_2 &=& m_1\, c_{1,2} + m_2 \, (N+c_{1,1})  - \ch_2 \nn\\
\zeta_3 &=& m_1\, (N+c_{1,2})+ m_2 \,(-N+ c_{1,1}) -\ch_2 - N \nn\\
\zeta_4 &=& - m_1\, (N+ c_{1,2}) - m_2 \, c_{1,1} + \ch_2
\eea
so the quiver description is valid when $\eta c_{1,2} + c_{1,1} <0 , \eta(N+c_{1,2}) + c_{1,1} >0$.
For Chern classes $[2;-1,0;1]$, $[2;0,-1;1]$, $[2,-1,-1;2]$, we find  walls at 
$\eta=1/2$, $\eta=2$, $\eta=3$ (but the wall at $\eta=1/3$ is not accessible) 
while $[2;0,0,2]$ has no walls at $\eta>0$, in agreement with the analysis in  \S\ref{VWF0}.
The chamber $J_{\epsilon,1}$ corresponds to $\eta\to 0^+$, and the quiver description is valid provided $c_{1,1}=0$. In this chamber, we get 
\be
\begin{array}{|c|c|l|}
\hline
\[N; c_{1,1},c_{1,2};c_2\] & (\vec N)  &\Omega(\vec N, \eta\to 0^+)  \\ \hline
\[2;0,-1;1\] &(0,1,1,0) & -y -1/y\ \nn\\
\[2;0,-1;2\] &(1,2,2,1) & -y^5-3y^3-7y-\dots \ \nn\\
\[2;0,0;2\] &(0,2,2,2) & -y^5-2y^3- 3y-\dots \nn\\
\[3;0,-1;2\] &(0,2,2,1) &y^4+2y^2+4+\dots \nn\\
\[3;0,0;3\] &(0,3,3,3) &y^{10}+2y^8+5y^6+8y^4+9y^2+10+\dots \nn\\
\[4;0,-1;3\] &(0,3,3,2) &-y^{9}-2y^7-6y^5-11y^3-15 y +\dots \nn\\
\[4;0,-2;3\] &(1,3,3,1) &-y^{9}-3y^7-7y^5-11y^3-16 y +\dots \nn\\
\[4;0,-3;3\] &(0,3,3,2) &-y^{9}-2y^7-6y^5-11y^3-15 y +\dots \nn\\
\hline
\end{array}
\ee
in precise agreement with \eqref{hFmboundary2}--\eqref{hFmboundary4}.

\subsection{$\mathbb{F}_1$} 
Since $\IF_1$ is a blow-up of $\IP^2$ at one point, we abuse notation and denote by $H=C+F$ the pull-back of 
the hyperplane class of $\IP^2$, while  $C$ is identified as the exceptional divisor.
In the basis $(H,C)$, the intersection form is then 
\be
C^{\alpha\beta}=\begin{pmatrix} 1 & 0 \\ 0 & -1\end{pmatrix}\ ,\quad 
C_{\alpha\beta}=\begin{pmatrix} 1 & 0 \\ 0 & -1 \end{pmatrix}
\ee
We denote the components of the \kahler class $J'$ and first Chern class $c_1(E)$ on this basis by $(z_{,H},z_{C})$ and $(c_{1,H},c_{1,C})$, respectively, such that $z_C=-z_2, z_H=z_1+z_2$, $\tilde c_{1,C}=c_{1,H}+c_{1,C}, \tilde c_{1,F}=c_{1,H}$.  The central charge in the large volume limit is 
then
\be
\begin{split}
Z_\gamma = & 
-N \left( \frac12 z_H^2 - \frac12 z_C^2 +\frac16 \right)  + c_{1,H} z_H + c_{1,C} z_C -  {\rm ch}_2 \\
= & 
-N \left( \frac12 z_1^2 + z_1 z_2 +\frac16 \right)  + \tilde c_{1,F} \, z_1  
+ \tilde c_{1,C} \,  z_2  -  {\rm ch}_2 \
\end{split}
\ee
and the degree is 
\be
\deg E = 3 c_{1,H} + c_{1,C}\ .
\ee

Following  \cite[\S5.1]{Herzog:2005sy} we consider the strong exceptional collection 
\footnote{This coincides with the second 
exceptional collection $\cO,\cO(0,1),\cO(1,1),\cO(2,2)$ 
in  \cite{perling2003some}; the first collection $\cO,\cO(1,0),\cO(1,1),\cO(2,1)=\cO, \cO(H-C), \cO(H), \cO(2H-C)$ in 
{\it loc. cit.}, also studied in \cite[(59)]{Herzog:2003dj}, leads to the same quiver, since it corresponds to the same toric system $(H,C,C,H-C,H)$ up to cyclic
permutation.}  
\be
\cC = \left(  \cO, \cO(C), \cO(H), \cO(2H) \right)
\ee
associated via \eqref{toricol} to the toric system $D_i=(C,H-C,H,H-C)$. 
The Chern characters $[N,(c_{1,H},c_{1,C}),\ch_2]$ of the projective and simple representations
are 
\be
\begin{array}{ccl}
 \gamma^1 &=& \[1,(0,0),0\] \\
 \gamma^2 &=& \[1,(0,1),-\frac{1}{2}\] \\
 \gamma^3 &=& \[1,(1,0),\frac{1}{2}\] \\
 \gamma^4 &=& \[1,(2,0),2\] \\
\end{array}
\qquad
\begin{array}{ccl}
 \gamma_1 &=& \[1,(0,0),0\] \\
 \gamma_2 &=& \[0,(0,1),-\frac{1}{2}\] \\
 \gamma_3 &=&  \[-2,(1,-2),\frac{3}{2}\] \\
 \gamma_4 &=& \[1,(-1,1),0\] \\
\end{array}
\ee
with slopes $0,1,3,6$ and  $0,\infty,-1/2,-2$, respectively. 
The Euler matrix, its inverse and the adjacency matrix are given by 
\be
\label{SF11}
S=
 \left(
\begin{array}{cccc}
 1 & 1 & 3 & 6 \\
 0 & 1 & 2 & 5 \\
 0 & 0 & 1 & 3 \\
 0 & 0 & 0 & 1 \\
\end{array}
\right)
\ ,\quad
S^\vee =
\left(
\begin{array}{cccc}
 1 & -1 & -1 & 2 \\
 0 & 1 & -2 & 1 \\
 0 & 0 & 1 & -3 \\
 0 & 0 & 0 & 1 \\
\end{array}
\right)\ ,\quad 
\kappa =
\left(
\begin{array}{cccc}
 0 & 1 & 1 & -2 \\
 -1 & 0 & 2 & -1 \\
 -1& -2 & 0 & 3 \\
2 & 1 & -3 & 0 \\
\end{array}
\right)\ ,\
\ee
corresponding to the 4-node quiver \cite[Fig. 7]{Feng:2001xr}, \cite{Aspinwall:2005ur}
 \begin{center}
\begin{tikzpicture}[inner sep=2mm,scale=2]
  \node (a) at ( -1,1) [circle,draw] {$1$};
  \node (b) at ( 1,1) [circle,draw] {$2$};
  \node (c)  at ( 1,-1) [circle,draw] {$3$};
  \node (d)  at ( -1,-1) [circle,draw] {$4$};
 \draw [->] (a) to node[auto] {$ $} (b);
 \draw [->>] (b) to node[auto] {$ $} (c);
 \draw [->>>] (c) to node[auto] {$ $} (d);
 \draw [->>] (d) to node[auto] {$ $} (a);
  \draw [->] (a) to node[near end,below] {$ $} (c);
    \draw [->] (d) to node[near end,above] {$ $} (b);
\end{tikzpicture}
\end{center}
The  superpotential  is given by  
 \cite{Feng:2001xr,Aspinwall:2005ur}
\be
W=-\Phi_{12} \Phi_{23}^{1} \Phi_{34}^{2} \Phi_{41}^{2}+\Phi_{12} \Phi_{23}^{2} \Phi_{34}^{2} \Phi_{41}^{1}+\Phi_{13} \Phi_{34}^{1} \Phi_{41}^{2}-\Phi_{13}
   \Phi_{34}^{3} \Phi_{41}^{1}+\Phi_{23}^{1} \Phi_{34}^{3} \Phi_{42}-\Phi_{23}^{2} \Phi_{34}^{1} \Phi_{42}
\ee
A general Chern character $\gamma=[N,c_{1,H},c_{1,C},\ch_2]$ decomposes as $\gamma=\sum_i n_i \gamma_i$ with 
\bea
n_1 &=&\frac{1}{2} (c_{1,C}+3 c_{1,H}) +\ch_2 + N ,\nn\\
n_2 &=& \frac{1}{2} (3 c_{1,C}+3 c_{1,H}) + \ch_2 \nn\\
n_3 &=&   \frac{1}{2} (c_{1,C}+c_{1,H}) + \ch_2 \nn\\
   n_4 &=& \frac{1}{2} (c_{1,C}-c_{1,H}) + \ch_2 
\eea
or conversely
\bea
N &=& n_{1}-2 n_{3}+n_{4}, \quad \ch_2=   \frac{1}{2} (3 n_{3}-n_{2}) \nn\\
c_{1,H} &=& n_{3}-n_{4} \ ,\quad  c_{1,C} =n_{2}-2  n_{3}+n_{4}
\eea
In order that the dimension vector be positive for large positive $c_2$, 
we set  $\vec N=-\vec n$ as usual. The stability parameters  $\vec\zeta=\vec\zeta^c$ in \eqref{zetacan} for $J\propto c_1(S)$ are given by
\bea
\zeta_1&=& \rho(N_2 + N_3 - 2 N_4)+\frac12(N_2-3N_3)= -2 \deg \, \rho  + \ch_2 \nn\\
\zeta_2 &=& \rho(- N_1 + 2 N_3-N_4) +\frac12(2N_3-N_1-N_4)= N \, \rho + \frac12 N \nn \\
\zeta_3 &=& \rho(- N_1 - 2 N_2 +3 N_4)+\frac12(3N_1+3N_4-2N_2) = (2 \deg + N)\rho
-2\ch_2-\frac32 N, \nn\\
\zeta_4 &=& \rho(2 N_1 + N_2 -3 N_3) +\frac12(N_2-3N_3) = -( \deg+ 2 N)\rho+\ch_2
\eea
The leading order term vanishes for $\vec N$ in the span of $(1,1,1,1)$ and $(0,3,1,2)$,
which corresponds to D0-branes and D2-branes wrapped on the exceptional curve $3C-H$.

\medskip

For the Beilinson subquiver with $\Phi_{41}=\Phi_{42}=0$, 
the dimension of the moduli space of quiver representations
\be
\begin{split}
d_\IC=& N_{1} N_{2}+N_{1} N_{3} +2
   N_{2} N_{3}+3 N_{3} N_{4} -N_{2} N_{4}
  -2 N_{1} N_{4}-N_{1}^2-N_{2}^2-  N_{3}^2 -  N_{4}^2+1 
\end{split}
\ee
agrees with the expected dimension of the moduli space of sheaves \eqref{dimM}. This requires 
 $\zeta_1\geq 0, \zeta_4\leq 0$ hence
\be
\label{windowF1can}
-2N \leq  \deg \leq 0
\ee
Unlike the cases of $\IP^2$ and $\IF_0$, this window is no longer fixed by the range of slopes
 in the dual collection $\cC^\vee$, which is unbounded due to the vanishing rank of the object $E_2^\vee$.

\medskip

In the attractor chamber, we have either (corresponding to the perfect matchings $s_3, s_1,s_2,s_4$ in \cite[\S 16]{Hanany:2012hi})
\begin{itemize}
\item $\Phi_{41}=\Phi_{42}=0$ when $\zetastar_1\geq 0, \zetastar_4\leq 0$ (and therefore $\zetastar_2\geq 0$), with expected dimension 
\bea
d_\IC &=&
1 - \cQ(\vec N) + \frac23 (-N_1 \zetastar_1- N_2 \zetastar_2  + N_4 \zetastar_4 )
\eea
\item $\Phi_{34}=0$ when $\zetastar_3\leq 0, \zetastar_4\geq 0$ with expected 
dimension
\bea
d'_\IC 
&=& 1 - \cQ(\vec N) + \frac23  (N_3 \zetastar_3 - N_4 \zetastar_4)
\eea
\item $\Phi_{23}=\Phi_{13}=0$, when $\zetastar_2\leq 0, \zetastar_3\geq 0$ (and therefore $\zetastar_1\leq 0$), with expected dimension
\bea
d''_\IC 
&=&1 - \cQ(\vec N) + \frac23  (N_1 \zetastar_1 + N_2 \zetastar_2 -N_3 \zetastar_3 )
\eea
\item $\Phi_{12}=\Phi_{13}=\Phi_{42}=0$ when $\zetastar_1\leq 0, \zetastar_2\geq 0$ 
(and therefore $\zetastar_4\leq 0$) with expected dimension
\be
d'''_\IC
=1 - \cQ(\vec N) + \frac23  ( 2 N_1 \zetastar_1 - N_2 \zetastar_2  +N_4 \zetastar_4)
\ee
\end{itemize}
where $\cQ$ is the positive quadratic form
\bea
\cQ(\vec N) &=&\frac12 \left[(N_1 - N_2)^2 + (N_3 - N_4)^2\right]) + \frac13 \left[(N_2 - N_3)^2 + (N_1 - N_4)^2\right]  \nn\\&&+ 
 \frac16 \left[(N_2 - N_4)^2 + (N_1 - N_3)^2 \right]
\eea
In all cases the expected dimension is negative unless $\vec N$ corresponds to a simple representation or to a pure D0-brane, therefore  the attractor index 
$\Omstar(\vec N)$ vanishes.

\medskip

For the canonical polarization $J\propto c_1(S)$, the flow tree formula leads to 
\be
\begin{array}{|c|c|l|}
\hline
\[N; c_{1,H},c_{1,C};c_2\] & \vec N  &\Omega_c(\vec N)  \\ \hline
\[1;0,0;1\] &(0,1,1,1)&y^2+2+1/y^2\nn\\
\[1;0,0;2\] &(1,2,2,2)&y^4+3 y^2 + 6 + 3/y^2 + 1/y^4\nn\\
\[2;0,0;1\]&(-1,1,1,1)&0 \nn\\
 \[2;0,0;2\]&(0,2,2,2) &-y^5 - 2 y^3 -3 y -\dots\nn\\
 \[2;-1,0;2\]&(1,3,2,1)&y^4+3y^2+5+\dots \nn\\
   \[3;-1,0;2\]&(0,3,2,1)&y^2+1+1/y^2 \nn\\
     \[3;0,0;3\]&(0,3,3,3)&y^{10} + 2 y^8 + 5 y^6 + 8 y^4 + 10 y^2 + 11 +\dots \nn\\
  \[3;-2,0;3\]&(1,4,2,0)&y^2+1+1/y^2 \nn\\
    \[4;-1,0;3\]&(0,4,3,2)&y^6 + 2 y^4 + 5 y^2 + 6 +\dots% \nn\\
    \\ \hline
    \end{array}
\ee
\medskip
 in agreement with the analysis in \S\ref{sec_genVW}, see in particular \eqref{VW1}, \eqref{F1mod2}, \eqref{F1mod300}. Applying the Coulomb branch formula in the attractor chamber,
we find evidence that the single centered indices $\OmS(\vec N)$ vanish, just
like the attractor indices $\Omstar(\vec N)$.

\medskip

For general  polarization $J'=\I m_1(C+F) +\I m_2 F$, the stability parameters
are instead given by 
\bea
\zeta_1&=&m_2 ( N_2-N_3)+m_1(N_3-N_4) + \frac12 (N_2-3N_3)  \nn\\
  \zeta_2 &=& m_2 ( 2N_3 - N_1 -N_4) + \frac12 (2N_3-N_1-N_4) \nn\\
\zeta_3 &=&  m_1 ( N_4-N_1) + m_2 (N_1-2N_2+N_4) + \frac12 (3N_1-N_2+3N_4) \nn\\
   \zeta_4 &=& m_1 (N_1-N_3) + m_2 (N_2-N_3)  + \frac12 (N_2-3N_3)  
   \eea
 where we recall that $m_H=m_1+m_2, m_C=-m_2$.  
The chamber $\Phi_{41}=\Phi_{42}=0$ is consistent with the D-term conditions \eqref{Dterm} provided
\be
\label{windowF1gen}
- (m_C+ m_H) N \leq  m_H\, c_{1,H} - m_C\,c_{1,C} \leq  0
\ee
For $(m_H,m_C)=(3,-1)$, this reproduces \eqref{windowF1can}. In the boundary chamber  $J_{\epsilon,1}$, where the generating function of VW invariants \eqref{defHN}  is controlled by the Hall algebra of $\IP^1$,
corresponding to $(m_H,m_C)=(1+\epsilon,-1)$, this window shrinks to zero size. For fixed Chern vector
with $N>0$ and $-N<c_{1,H}<0$, the condition \eqref{windowF1gen} holds provided the ratio 
$\eta=m_1/m_2$ satisfies
\be
\label{windoweta}
\eta\geq  - \frac{c_{1,H}+c_{1,C}}{c_{1,H}+N}
\ee
 
 By examining the  trees contributing to  the flow tree formula, 
we find that the  indices in the first three lines of the previous table, namely 
$\Omega(0,1,1,1)$,  $\Omega(1,2,2,2)$, $\Omega(0,2,2,2)$ have no chamber dependence, 
consistently with the analysis in Section \ref{VWF1}.
However for $\gamma= \[2,-1,0,2\]$, we find $\Omega(1,3,2,1)=y^4+3y^2+5+\dots$
for $\eta>1$, $y^2+2+1/y^2$ for $\eta<1$, with a jump $\Delta\Omega=(y^2+1+1/y^2)^2$
due to two-particle bound states with charges $(\gamma_L,\gamma_R)
=([1,0,-1,-1/2],[1,-1,1,-1])$ and $([3,-2,1,-3/2],[-1,1,-1,0])$ (with $\langle \gamma_L,\gamma_R\rangle=-1$ in both cases). 
In contrast, the analysis of VW invariants in Section \ref{VWF1} gives the same value for $\eta>1$ but 
$y^4+y^2+1+\dots$ for $1/3<\eta<1$, $0$ for $\eta<1/3$. This is not in contradiction with the quiver
index since the condition \eqref{windoweta} requires $\eta>1$.

\subsection{Higher $\IF_m$}

We consider the strong exceptional  collection \cite[p100]{rudakov1990helices} (as quoted in \cite{Herzog:2006bu})
\be
\cC= \left( \cO, \cO(F),\cO(C+m F), \cO(C+(m+1) F) \right)
\ee
corresponding to the toric system $\tilde D_i=(F,C+(m-1) F, F, C+F)$.
The Chern vectors of the projective and simple representations
are 
\be
\begin{array}{ccl}
\gamma^1&=&[1,(0,0),0] \\
\gamma^2&=&[1,(0,1),0] \\ 
\gamma^3&=&[1,(1,m),\frac{m}{2}]\\
 \gamma^4&=&[1,(1,m+1),\frac{m+2}{2}]
 \end{array}
 \qquad 
\begin{array}{ccl}
\gamma_1&=& [1,(0,0),0]\\
\gamma_2&=& [-1,(0,1),0]\\
\gamma_3&=& [-1,(1,0),\frac{m}{2}]\\
\gamma_4&=&[1,(-1,-1),\frac{2-m}{2}] 
\end{array}
\ee
with slopes $0,2,m+2,m+4$ and  $0,-2,m-2,m-4$.
The Euler matrix, its inverse and the adjacency matrix are given by  
\bea
\label{SFm}
S&=&\left(
\begin{array}{cccc}
 1 & 2 & m+2 & m+4 \\
 0 & 1 & m & m+2 \\
 0 & 0 & 1 & 2 \\
 0 & 0 & 0 & 1 \\
\end{array}
\right)\ ,\quad 
S^\vee = \left(
\begin{array}{cccc}
 1 & -2 & m-2 & 4-m \\
 0 & 1 & -m & m-2 \\
 0 & 0 & 1 & -2 \\
 0 & 0 & 0 & 1 \\
\end{array}
\right)
\nn\\
\kappa &=& 
 \left(
\begin{array}{cccc}
 0 & 2 & 2-m & m-4 \\
 -2 & 0 & m & 2-m \\
 m-2 & -m & 0 & 2 \\
 4-m & m-2 & -2 & 0 \\
\end{array}
\right)
\eea
This leads to the quiver $Q_m$ 
 (possibly up to bidirectional arrows and edge loops)
 
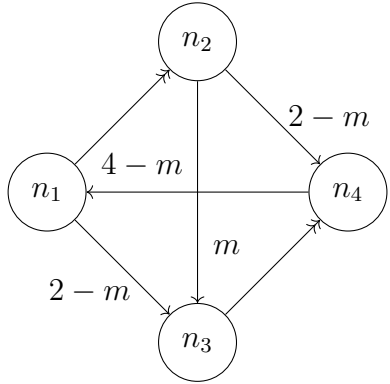
\begin{center}
\begin{tikzpicture}[inner sep=2mm,scale=2]
  \node (a) at ( -1,0) [circle,draw] {$n_1$};
  \node (b) at ( 0,1) [circle,draw] {$n_2$};
  \node (c)  at ( 0,-1) [circle,draw] {$n_3$};
  \node (d)  at ( 1,0) [circle,draw] {$n_4$};
 \draw [->>] (a) to node[auto] {$ $} (b);
 \draw [->] (a) to node[near end, left] {$2-m$} (c);
 \draw [->] (b) to node[near end, right] {$m$} (c);
 \draw [->] (b) to node[right] {$2-m$} (d);
 \draw [->>] (c) to node[auto] {$ $} (d);
  \draw [->] (d) to node[above, near end] {$4-m $} (a);
\end{tikzpicture}
\end{center}
A general Chern character $\gamma=[N,c_1,\ch_2]$ decomposes as $\gamma=
\sum_i n_i \gamma_i$ with 
\bea
n_1 &=& \frac{2-m}{2} c_{1,1} +c_{1,2}+\ch_2+N, \nn\\
n_2 &=&
   -\frac{ m}{2}c_{1,1}+c_{1,2}+\ch_2,\nn\\
n_3 &=& -\frac{  m}{2}c_{1,1} +c_{1,1}+\ch_2 \nn\\
n_4 &=& \ch_2-\frac{ m}{2}c_{1,1}
   \eea
   or conversely,
\bea
N &=& n_1-n_2-n_3+n_4,\ \quad \ch_2 =  n_4+ \frac{1}{2} m (n_3-n_4) \nn \\ 
c_{1,1}&=&n_3-n_4, \ ,\quad
c_{1,2} =  n_2-n_4 
\eea   
As usual, we take $\vec N=-\vec n$ so that $N_i$ is positive for large $c_2$. 
For $J\propto c_1(S)$, the  stability parameters \eqref{zetacan} are 
\bea
\zeta_1 &=& % \rho (-(m-2) N_{3}+(m-4) N_{4}+2 N_{2}) + \dots  =
 -\rho \deg + \ch_2 \nn \\
\zeta_2 & =&% &\rho( m \, N_{3}-(m-2) N_{4}-2  N_{1}) +  \dots 
= (\deg +2N) \rho -\ch_2 \nn \\
\zeta_3 &=& %\rho (m \,N_{1}-m N_{2}-2 N_{1}+2 N_{4}) +\dots =
 (\deg - (m-2)N) \rho-\ch_2-\frac{m}{2}N
\nn  \\
\zeta_4 &=&%=\rho (-(m-4) N_{1}+m\, N_{2}-2 (N_{2}+N_{3})) + \dots =
 ((m-4)N-\deg)\rho+\ch_2+\frac{m-2}{2} N 
\eea
Depending on $m$, we find the following results:
\begin{itemize}
\item For $m=0$, $Q_0$ coincides with the quiver \eqref{SF01} for $\IF_0$, phase I.
\item For $m=1$, $Q_1$ coincides with the quiver  \eqref{SF11} for $\IF_1$, 
up to a permutation of the nodes.
\item {\bp For $m=2$, $Q_2$ reduces to the same quiver \eqref{SF02} as for $\IF_0$, phase II,  up to a  bidirectional arrows between nodes $(1,3)$ and $(2,4)$ which are not visible from the adjacency matrix $\kappa$ but follow from the brane tiling description  } \cite{Feng:2004uq}\footnote{We are grateful to Sebastian Franco and Yang-Hui He for discussions on quivers for $\IF_m$ with $m\geq 2$.}. Since for each of the internal perfect matchings, one arrow in each pair vanishes, the dimension of the quiver moduli space is unaffected, and the
vanishing of attractor indices {\bp should} follow by the same arguments as for  $\IF_0$, phase II.  
Moreover, the dimension vectors are identified provided the components of Chern class are shifted,  
$(c_{1,1},c_{1,2})\to (c_{1,1},c_{1,2}-c_{1,1})$. 
\item 
For $m=3$, $Q_3$ is identical to the  quiver for $\IF_1$ in \eqref{SF11},  possibly up to bidirectional arrows; the Beilinson quiver
$\hat Q_3$ with  $\Phi_{31}=\Phi_{42}=\Phi_{41}=0$ requires $\zeta_1>0, \zeta_4<0$, so 
is valid in the range $-N\leq \deg\leq 0$.
\item 
For $m=4$,  $Q_4$ is mapped by Seiberg duality with respect to node 1 or 4 (assuming that Seiberg duality is not spoilt by potential bidirectional arrows or edge loops) to the quiver \eqref{SF01} as  for $\IF_0$, model I; the Beilinson quiver
$\hat Q_4$ with  $\Phi_{31}=\Phi_{42}=0$ 
is valid only for sheaves with $\deg =0$.
\item
For $m\geq 5$, the Beilinson quiver $\hat Q_5$ 
with $\Phi_{31}=\Phi_{42}=0$ requires $\zeta_1>0, \zeta_4<0$, hence $(m-4) N \leq  \deg \leq 0$ hence $N=0$. Moreover, $Q_5$ appears to belong to a different mutation class than $Q_1$ (assuming that Seiberg duality applies).
\end{itemize} 
 The equivalences $Q_2\sim Q_0$, $Q_3\sim Q_1$, $Q_4\sim Q_2$ noted above 
are consistent with the analysis in Section \ref{sec_VWFm}, which shows that at least at rank 2, the Vafa-Witten invariants for $\IF_m$ are related to those of $\IF_{m+2}$ by shifting the first Chern class and the polarization. We shall leave a more detailed study of quivers for $\IF_m$ with $m\geq 3$ for future work.

\section{del Pezzo surfaces \label{sec_dP}}

We now turn to the del Pezzo surfaces $S=dP_k$, defined as the blow-up of $\IP^2$ at $k$ points in generic position (such that no three points should be collinear and no six points should lie on a conic).
For $k\geq 2$, $dP_k$ is isomorphic to the blow-up of $\IF_0=\IP^1\times\IP^1$ at $k-1$ points,
while $dP_1=\mathbb{F}_1$ as mentioned earlier. The del Pezzo surfaces surfaces
are toric for $k\leq 3$, Fano for $k\leq 8$ and weak Fano for $k=9$. Note that del Pezzo surfaces 
with $3\leq k\leq 8$ admit a positive curvature K\"ahler-Einstein metric, hence can serve as the 
base of a Sasaki-Einstein 5-dimensional space. 
\medskip

Viewing $dP_k$ as the $k$-point blow-up of $\IP^2$, the second homology $H_2(S,\IZ)$ has dimension $b_2(S)=k+1$ and 
is spanned
by the hyperplane class $H$ of $\IP^2$ and by the classes of the exceptional divisors $C_1,\dots, C_k$, with intersection numbers
\be
H\cdot H=1\ ,\quad C_i \cdot C_j = -\delta_{ij} \ ,\quad H \cdot C_i = 0
\ee
The anticanonical class is 
\be
\label{c1dP}
c_1(S) = 3 H - \sum_{i=1}^k C_i
\ee
hence the degree is $c_1(S)^2=9-k$.
As explained in \cite{Iqbal:2001ye}, $H_2(S,\IZ)$ admits an action of the 
Weyl group $W_k$ of the simple Lie group  $E_k$ 
(with $E_{2}=O(1,1) \times SL(2)$, $E_{3}=SL(2)\times SL(3)$, $E_4=SL(5)$, $E_5=O(5,5)$
and $E_{6,7,8}$ the exceptional series) -- the same Lie group which arises as 
the U-duality group in M-theory compactified on $T^k$.  The finite group $W_k$ is generated by Weyl reflections with respect to $C_i-C_j$ (corresponding to permutations of the exceptional divisors $C_i$) and with respect to $H-C_i-C_j-C_k$ for any triplet $i<j<k$, which maps 
\be
(H,C_i,C_j,C_k) \to (2H-C_i-C_j-C_k, H-C_j-C_k, H-C_i-C_k, H-C_i-C_j)
\ee
leaving  \eqref{c1dP} invariant. 

\medskip

For any $k\leq 8$, the collection of invertible sheaves
\be
\label{excepdP}
\left( \cO, \ \cO(C_1), \dots ,\cO(C_k), \ \cO(H), \ \cO(2H)  \right)
\ee
is known to be strongly exceptional, and all other strongly exceptional collections on $dP_k$ are related to 
\eqref{excepdP} by mutations \cite{kuleshov1995exceptional}. However, \eqref{excepdP} is not in general a cyclic strongly exceptional collection. For $k\leq 3$, $S$ is toric and suitable collections
can be constructed as in \S\ref{sec_toric}.  For any $k\leq 8$, exceptional collections were proposed  using brane web techniques in \cite{Hanany:2001py}, but their status for $k\geq 4$ is not fully understood. We shall instead rely on the three-block exceptional collections constructed
for any $2\leq k\leq 8$ in  \cite{karpov1998three}, which are automatically tachyon-free. 
 For the 
purpose of comparing with the predictions in Appendix \S\ref{sec_VWdP} based on the blow-up formula, we shall mostly 
focus on the Vafa-Witten invariants in the canonical chamber  and in the 
blow-up chamber $J'\propto H_\epsilon$ where $H_\epsilon=H+\epsilon c_1(S)$ with $0<\epsilon\ll 1$ is  a small perturbation of the pull-back of the hyperplane class of $\IP^2$; this perturbation is
necessary in order to avoid walls of marginal stability. 

\subsection{Three-block collections}

In \cite{karpov1998three}, by a sequence of blow-ups and mutations, 
the authors construct  three-block-shaped strong exceptional collections for any del Pezzo surface $dP_k$ with $k\geq 3$.  These collections are classified by solutions of the 
Markov-type equation\footnote{This  equation
was interpreted in terms of NSVZ beta-functions in \cite{Herzog:2003dj}. }
\be
\label{Markov}
\alpha x^2 + \beta y^2 + \gamma z^2 = xyz \, \sqrt{K_S^2\, \alpha\beta\gamma} 
\ee
where  $\alpha,\beta,\gamma$ correspond to the size of the three blocks,
$x,y,z$ to the rank of the sheaves $E_i$ in 
each block and $K_S^2=c_1(K_S)^2=9-k$. Each of these collections is full so $\alpha+\beta+\gamma=\chi(S)=k+3$.
 We shall restrict to the solutions of \eqref{Markov} with the smallest value of $x+y+z$
with $x,y,z\geq 1$, since all other solutions can be obtained by a sequence of
transformations 
\be
(x,y,z) \mapsto 
(b\gamma z-x   , y,z ) \ \mbox{or} \   (x, c\alpha x - y ,z)  
 \ \mbox{or} \  (x ,y,a\beta y - z) 
\ee
corresponding to a product of Seiberg dualities on either of the three blocks. The minimal solutions to the diophantine equation \eqref{Markov} are tabulated in Table \ref{tab}, along with additional data described below. For given degree $K_S^2=9-k$, the possible values of $(\alpha,\beta,\gamma)$
are  in one-to-one correspondence with maximal subgroups 
of type $SL(\alpha) \times SL(\beta) \times SL(\gamma) \subset E_{k}$ \cite{Herzog:2003dj}, and the corresponding three-block collections are manifestly invariant under a subgroup $S_\alpha\times S_\beta \times S_\gamma$ of the Weyl group $W_k$.

\begin{table}
$$
\begin{array}{|c|c|c|c|c|c|c|}
\hline
S & K_S^2 & \# & (\alpha,\beta,\gamma) & (x,y,z,-y') & (a,b,c,b') & (\cA,\cB,\cC) 
%& \Delta_a & \Delta_b & \Delta_c  
\\ \hline
\IP_2 &9		& (1) & (1,1,1) & (1,1,1,2) & (3,3,3,6) & (9,9,9) \\ \hline %& 0 & 9 &9  \\
\IP_1\times \IP_1 &8  &(2) &  (1,2,1) & (1,1,1,1) & (2,4,2,4) & (8,16,8) \\ \hline %  & 5 & 32 & 32 \\
dP_3  &6		& (3) & (1,2,3) & (1,1,1,2) & (1,2,3,4) &  (6,12,18) \\ \hline %& 9 & 96 & 0 \\
dP_4	  &5		& (4) & (1,1,5) & (1,2,1,3) & (1,2,5,3) &  (5,20,25) \\ \hline %& 23 &320 & 0\\
dP_5	  &4		& (5) & (2,2,4) & (1,1,1,3) & (1,1,2,3) &  (8,8,16) \\ \hline % & 5 &32 & 0  \\
dP_6	  &3		& (6.1) & (3,3,3) & (1,1,1,2) & (1,1,1,2) &  (9,9,9) \\ %& 0 & 9 & 9\\
 	 &		& (6.2) & (1,2,6) & (2,1,1,5) & (1,1,3,5) & (12,6,18)\\ \hline % & 9 & 12 & 0 \\
dP_7	 &2 		& (7.1) & (1,1,8) & (2,2,1,6) & (1,1,4,3) & (8,8,16) \\% & 5 & 32 & 0 \\
	&		& (7.2) & (2,4,4) & (2,1,1,3) & (1,1,1,3) & (16,8,8)\\%  & 5 & 0 & 32 \\
	&		& (7.3) & (1,3,6) & (3,1,1,5) & (1,1,2,5) & (18,6,12)  \\\hline % & 9 & 0 & 12 \\
dP_8	 &1		& (8.1) & (1,1,9) &(3,3,1,6) & (1,1,3,2)  & (9,9,9)  \\%& 0 & 9 & 9 \\
        & 		& (8.2) & (1,2,8) &(4,2,1,6) & (1,1,2,3)  & (16,8,8) \\%& 5 &0 & 32 \\
	&		& (8.3) & (2,3,6) &(3,2,1,4) & (1,1,1,2)  & (6,12,18) \\%& 9 & 96 & 0 \\
	&		& (8.4) & (1,5,5) &(5,2,1,3) & (1,2,1,3)  & (25,20,5) \\\hline %& 23 & 0 & 320  \\
ShdP_5 & 4 & (9.1) & (2,2,1) & (1,1,2,3) & (2,2,2,4) & (8,8,16)  \\\hline % &5 & 32 &0\\
ShdP_7 & 2 & (9.2) & (2,1,4) & (2,2,1,4) & (2,1,2,3) & (16,8,8) %& 5 & 0& 32
 \\ \hline
\end{array}
$$
\caption{List of minimal solutions to Markov's equation \eqref{Markov}, from \cite{karpov1998three,Benvenuti:2004dw}.
The last two lines
correspond to the two additional solutions found in \cite{Benvenuti:2004dw}, whose geometric interpretation remains unclear. \label{tab}}
\end{table}

\medskip

For each three-block exceptional collections of the form
\be
\cC = (E^1,\dots, E^{\alpha};  E^{\alpha+1},\dots, E^{\alpha+\beta}; E^{\alpha+\beta+1},\dots 
E^{\alpha+\beta+\gamma})\ ,
\ee
the Euler form, its inverse and the adjacency matrix have a $3\times 3$ block structure
\be
S= 
\begin{pmatrix}
\mathbb{1}_\alpha & C & B' \\ & \mathbb{1}_\beta & A \\ 
& & \mathbb{1}_\gamma
\end{pmatrix}
\ ,\quad
S^\vee =\begin{pmatrix}
\mathbb{1}_\alpha & - C & B \\ & \mathbb{1}_\beta & -A \\ 
& & \mathbb{1}_\gamma
\end{pmatrix} \ ,\quad 
\kappa=
\begin{pmatrix}
\mathbb{0}_\alpha & C & -B \\
 -C^t & \mathbb{0}_\beta & A \\ 
B^t & -A^t & \mathbb{0}_\gamma
\end{pmatrix} \ ,\quad 
\ee
where $A,B,B',C$ are $\beta\times\gamma, \alpha\times\gamma,\alpha\times\gamma,\alpha\times\beta$ matrices with 
all entries equal to $a,b,b',c$, respectively. The latter are given by 
\be
a=\alpha x \, K', \quad 
b=\beta y \, K'\ ,\quad
c=\gamma z \, K'\ ,\quad 
b'=\beta a c - b 
\ee
where $K'=\sqrt{K_S^2/(\alpha\beta\gamma)}$. It will be useful to define 
\be
\label{defABC}
\mathcal{A} =a^2 \beta \gamma\ ,\quad 
\mathcal{B} =b^2 \alpha \gamma\ ,\quad 
 \mathcal{C}=c^2 \alpha \beta , 
\ee
such that the Markov equation \eqref{Markov} becomes 
\be
\label{Markov2}
\mathcal{A}+\mathcal{B}+\mathcal{C}=\sqrt{\mathcal{A}\mathcal{B}\mathcal{C}}
\ee
The values of $(\alpha,\beta,\gamma)$, $(x,y,z)$,  $(a,b,c)$, $(\cA,\cB,\cC)$ for each collection are displayed in Table \ref{tab}.
Using \eqref{Edualmut}, the dual collection is given by 
\be
\cC^\vee = (E^\vee_1,\dots, E^\vee_{\alpha};  E^\vee_{\alpha+1},\dots, E^\vee_{\alpha+\beta}; E^\vee_{\alpha+\beta+1},\dots 
E^\vee_{\alpha+\beta+\gamma})
\ee
where 
\be
\label{Evee3}
E^\vee_i = \begin{cases} E_i & \mbox{for}\quad  i=1\dots \alpha 
\\
L_{\cE} E_i   & \mbox{for} \quad i=\alpha+1,\dots,  \alpha+\beta
\\ 
E_i \otimes K_S & \mbox{for} \quad i=\alpha+\beta+1,\dots,  \alpha+\beta+\gamma
\end{cases}
\ee
where $L_{\cE}=L_{E_1}\cdot \dots \cdot L_{E_\alpha}$.
In particular,  the rank of the sheaves in the dual collection 
are $(x',y',z')=(x,y-c\alpha x,z)$, such that $b'=-\beta y' K'$. 
It is easy to check that these data satisfy the constraints \eqref{sumrkrk} and \eqref{trsst},
along with 
\be
a x - b' y + c z = a x' + b y' + c z' = 0
\ee
It is straightforward to show that  the slopes  $\nu^i$ and $\nu_i^\vee$ of the sheaves are uniquely 
from \eqref{Evee3} and \eqref{sumrkdeg}, up to overall translation, 
\bea
\nu_1&=& \nu_1^\vee, \quad \nu_3^\vee=\nu_3-K_S^2 \nn\\
\nu_1-\nu_2 &=& -\frac{c}{xy}\ ,\quad \nu_2-\nu_3=-\frac{a}{yz}\ , \quad \nu_3-\nu_1 = \frac{b'}{xz} \nn \\
\nu^\vee_1-\nu^\vee_2 &=& \frac{c}{x'y'}\ ,\quad 
\nu^\vee_2-\nu^\vee_3=\frac{a}{y'z'}\ ,\quad 
\nu^\vee_3-\nu^\vee_1 = -\frac{b}{x'z'} 
\eea
in such a way that $\nu_1<\nu_2<\nu_3$ while $\nu_1^\vee>\nu_2^\vee>\nu_3^\vee$. The superpotential is a sum of cubic terms, and is known explicitly only for a few cases \cite{Beasley:2001zp,Feng:2002zw,Wijnholt:2002qz}.

\medskip

In the next subsections, we shall discuss each three-block exceptional collection in detail (along with some additional non-three-block collections for $dP_2$ and $dP_3$). It is useful however to 
highlight some general properties. 
First, in all cases the adjacency matrix $\kappa$ has rank 2, with a $k+1$-dimensional kernel spanned by the dimension vector $\vec N=-(x,\dots;y,\dots;z,\dots)$ for pure D0-branes, and the 
dimension vector for the  $k$ remaining exceptional D2-branes with $\deg E=0$. 
The stability parameters
$\vec\zeta=\rho(\varsigma_1,\dots;\varsigma_2,\dots;\varsigma_3,\dots)+\cO(1)$ in the canonical chamber are independent of $i$ in each block (up to subleading corrections as $\rho\to\infty$), and given by 
\be
\label{stabdPc}
\varsigma_1 = c\, \cN_2 - b\, \cN_3\ ,\quad 
\varsigma_2 = a\, \cN_3 - c\, \cN_1\ ,\quad 
\varsigma_3 = b\, \cN_1 - a\, \cN_2\ ,
\ee
where
\be
\label{defcN123}
\cN_1=\sum_{i=1}^{\alpha} N_i\ ,\quad 
\cN_2=\sum_{i=\alpha+1}^{\alpha+\beta} N_i\ ,\quad
\cN_3=\sum_{i=\alpha+\beta+1}^{\alpha+\beta+\gamma} N_i\ 
\ee
In terms of the rank and degree
\bea
\rk(E)&= &- x'\, \cN_1- y'\, \cN_2- z'\, \cN_3:= N\ \quad  \nn\\
\deg(E)&=& -x'\nu_1^\vee \cN_1- y'\nu_2^\vee \cN_2- z'\nu_3^\vee \cN_3:= N\, \nu
\eea
the stability parameters in the canonical chamber \eqref{zetacan} can be re-expressed as 
\be
\label{stabdP}
\varsigma_1 = x' N (\nu_1^\vee-\nu)\ ,\quad 
\varsigma_2 = y' N (\nu_2^\vee- \nu)\ ,\quad 
\varsigma_3 = z' N (\nu_3^\vee-\nu)\ .
\ee
By construction, the  dimension in the chamber $\Phi_{ij}=0$ where $i=\alpha+\beta+1,\dots, r$ and $j=1\dots\alpha$, given by
\be
d_\IC=c\, \cN_1\cN_2+a \cN_2\cN_3-b\cN_1\cN_3-\sum_i N_i^2 + 1, 
\ee
 agrees with the expected dimension \eqref{dimM} of the moduli space of coherent sheaves. 
 This is consistent with the stability parameters \eqref{stabdP} provided the slope $\nu$ lies in the interval
\be
\nu_3^\vee \leq \nu \leq  \nu_1^\vee
\ee
In Appendix \ref{sec_threeblockatt}, we show that the attractor index $\Omstar(\vec N)$ vanishes
unless
$\vec N$ is the dimension vector of a simple representation, or is proportional to the 
 dimension vector $(x,\dots;y,\dots; z,\dots)$ associated to pure D0-branes. 
Using this result, we evaluate the BPS indices in the blow-up chamber $J'\propto H_\epsilon$ for a variety of dimension vectors, and find perfect agreement with 
the generating functions listed in Appendix \S\ref{sec_VWdP}.

\subsection{$dP_2$}
The toric fan is generated by 5 vectors,
\be
v_i = \begin{pmatrix} 
1 & 0 & -1 & -1 & 0 \\
0 & 1 & 0 & -1 & -1
\end{pmatrix}
\ee
The corresponding divisors satisfy the linear relations
\be
D_1=D_3+D_4\ ,\quad D_2=D_4+D_5
\ee
and form an overcomplete basis of $H_2(S,\IZ)$, with intersection numbers
\be
D_i^2 = ( 0,0,-1,-1,-1)\ .
\ee
One may identify 
\be
D_1=H-C_2\ ,\quad D_2=H-C_1\ ,\quad D_3=C_1\ ,\quad D_4=H-C_1-C_2\ ,\quad D_5=C_2
\ee
where $H,C_1,C_2$ are the hyperplane class of $\IP^2$ and the two exceptional divisors, respectively,
such that  the intersection matrix in the basis $H,C_1,C_2$ becomes diagonal,
\be
C^{\alpha\beta} = 
 \left(
\begin{array}{ccc} 1 & \\ & -1 \\ & & -1  \end{array}
\right)
\ee
The square of the canonical class $c_1(S)=\sum_i D_i=3H-C_1-C_2$ 
evaluates to $7$ as expected.

\subsubsection{Model I}

We start with the standard exceptional collection  $(\cO,\cO(C_1),\cO(C_2),\cO(H),\cO(2H))$ from \eqref{excepdP}. In the language of \S\ref{sec_toric}, this is obtained via \eqref{toricolt} from the 
toric system
\be
\label{tDdP2}
\tilde D_i = (C_1, C_2-C_1, H-C_2, H, H-C_1- C_2)\ .
\ee 
 The Chern vectors of the projective and primitive objects are 
\be
\begin{array}{ccl}
 \gamma^1 &=& \[1,\(0,0,0\),0\] \\
 \gamma^2 &=& \[1,\(0,1,0\),-\frac{1}{2}\] \\
 \gamma^3 &=& \[1,\(0,0,1\),-\frac{1}{2}\] \\
 \gamma^4 &=& \[1,\(1,0,0\),\frac{1}{2}\] \\
 \gamma^5 &=& \[1,\(2,0,0\),2\] \\
\end{array}
\qquad
\begin{array}{ccl}
 \gamma_1 &=& \[1,\(0,0,0\),0\] \\
 \gamma_2 &=& \[0,\(0,1,0\),-\frac{1}{2}\] \\
 \gamma_3 &=& \[0,\(0,0,1\),-\frac{1}{2}\] \\
 \gamma_4 &=& \[-2,\(1,-2,-2\),\frac{5}{2}\] \\
 \gamma_5 &=& \[1,\(-1,1,1\),-\frac{1}{2}\] \\
\end{array}
\ee
with slopes $0,1,1,3,6$ and $0,\infty,\infty,1/2,-1$, respectively. 
The Euler matrix has a four-block structure, 
\be
\label{SdP2I}
S=
\left(
\begin{array}{ccccc}
 1 & 1 & 1 & 3 & 6 \\
 0 & 1 & 0 & 2 & 5 \\
 0 & 0 & 1 & 2 & 5 \\
 0 & 0 & 0 & 1 & 3 \\
 0 & 0 & 0 & 0 & 1 \\
\end{array}
\right)\ ,\quad 
S^\vee=
\left(
\begin{array}{ccccc}
 1 & -1 & -1 & 1 & 1 \\
 0 & 1 & 0 & -2 & 1 \\
 0 & 0 & 1 & -2 & 1 \\
 0 & 0 & 0 & 1 & -3 \\
 0 & 0 & 0 & 0 & 1 \\
\end{array}
\right)\ ,\quad
\kappa=\left(
\begin{array}{ccccc}
 0 & 1 & 1 & -1 & -1 \\
 -1 & 0 & 0 & 2 & -1 \\
 -1 & 0 & 0 & 2 & -1 \\
 1 & -2 & -2 & 0 & 3 \\
 1 & 1 & 1 & -3 & 0 \\
\end{array}
\right)
\ee
corresponding to the quiver
 
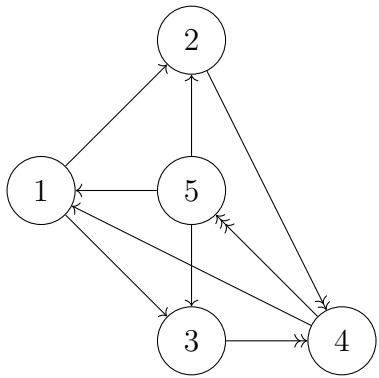
\begin{center}
\begin{tikzpicture}[inner sep=2mm,scale=2]
  \node (a) at ( -1,0) [circle,draw] {$1$};
  \node (b) at ( 0,1) [circle,draw] {$2$};
  \node (c)  at ( 0,-1) [circle,draw] {$3$};
    \node (d)  at ( 1,-1) [circle,draw] {$4$};
      \node (e)  at ( 0,0) [circle,draw] {$5$};
 \draw [->] (a) to node[auto] {$ $} (b);
 \draw [->] (a) to node[auto] {$ $} (c);
  \draw [->>] (b) to node[auto] {$ $} (d);
 \draw [->>] (c) to node[auto] {$ $} (d);
 \draw [->>>] (d) to node[auto] {$ $} (e);
\draw [->] (e) to node[auto] {$ $} (a);
\draw [->] (e) to node[auto] {$ $} (b);
\draw [->] (e) to node[auto] {$ $} (c);
\draw [->] (d) to node[bend right=45] {$ $} (a);
\end{tikzpicture}
\end{center}
The superpotential was found in \cite[p20]{Feng:2002zw} (see also model 12b in \cite{Hanany:2012hi})
\be
\begin{split}
W=&
-\Phi_{24}^2  \Phi_{45}^3 \Phi_{52}
+\Phi_{34}^2 \Phi_{45}^3 \Phi_{53} 
-\Phi_{34}^1\Phi_{45}^2  \Phi_{53} 
+\Phi_{24}^1\Phi_{45}^1  \Phi_{52} 
-\Phi_{12} \Phi  _{24}^1 \Phi_{41} \\ & 
+\Phi_{13} \Phi_{34}^1 \Phi_{41}
 +\Phi_{12}   \Phi_{24}^2  \Phi_{45}^2 \Phi_{51}
-\Phi_{13}   \Phi_{34}^2  \Phi_{45}^1 \Phi_{51}
\end{split}
\ee
A general Chern character $\gamma=[N,c_1,\ch_2]$ decomposes as $\gamma=
\sum_i n_i \gamma_i$ with 
\bea
   n_1&=&  \frac{1}{2} (c_{1,1}+c_{1,2}+3 c_{1,H})+ \ch_2+ N\nn\\
   n_2&=&  \frac{1}{2} (3 c_{1,1}+c_{1,2}+3 c_{1,H})+  \ch_2 \nn\\
   n_3&=& 
   \frac{1}{2} (c_{1,1}+3 c_{1,2}+3 c_{1,H})+ \ch_2\nn\\
   n_4&=&  \frac{1}{2}
   (c_{1,1}+c_{1,2}+c_{1,H})+  \ch_2\nn\\
   n_5&=&  \frac{1}{2}
   (c_{1,1}+c_{1,2}-c_{1,H})+ \ch_2
   \eea
or conversely,
\be
\begin{split}
N=  n_1-2 n_4+n_5, \quad \ch_2=  \frac{1}{2} (-n_2-n_3+5   n_4-n_5)
\\
c_{1,H}=  n_4-n_5,\quad 
c_{1,1}=  n_2-2 n_4+n_5, \quad 
c_{1,2}=  n_3-2 n_4+n_5,   
\end{split}
\ee   
Note that under  exchanging $c_{1,1} \leftrightarrow c_{1,2}$, $n_1, n_4, n_5$ stay invariant while $n_2,n_3$ get exchanged. For large $c_2$, the entries in $\vec n$ are all negative, so we should
consider the dimension vector $\vec N=-\vec n$.
The relevant stability parameters   in \eqref{zetacan}
 in the canonical chamber $J\propto c_1(S)$ are 
\bea
\zeta_1&=&\rho(N_2 + N_3 -N_4 - N_5  ) + \frac{1}{2} (N_{2}+N_{3}+N_{5}-5 N_{4})
=-\rho \, \deg +\ch_2 \nn\\
\zeta_{2,3}&=&  \rho(2 N_4  - N_1 -  N_5)-
\frac{1}{2} (N_{1}-2 N_{4}+N_{5}) 
= \rho\, N + \frac{N}{2} \nn\\
\zeta_4&=& \rho( N_1 + 3 N_5 -2 N_2 - 2 N_3)   
+\frac{5  N_{1}}{2}-N_{2}-N_{3}+\frac{3 N_{5}}{2} \nn\\
&=&\rho( 2\deg - N)-2\ch_2-\frac52 N, \nn\\
\zeta_5&=&  \rho( N_1 + N_2 + N_3 -3 N_4 ) -  \frac{1}{2}
   (N_{1}-N_{2}-N_{3}+3 N_{4})  \nn\\ &&
   = - \rho ( \deg + N )+\ch_2+\frac{N}{2}
\eea
where $\deg=3 c_{1,H}+ c_{1,2} + c_{1,1}$. In the chamber where 
$\Phi_{41}=\Phi_{51}=\Phi_{52}=\Phi_{53}=0$,  the dimension 
\be
d_\IC=N_{1} N_{2}+N_{1} N_{3}+2 N_{2} N_{4}+2 N_{3}
   N_{4}-N_{1} N_{4}-N_{1}
   N_{5}-N_{2} N_{5}-N_{3} N_{5}+3 N_{4} N_{5}-\sum_i N_i^2+1
\ee
coincides with the expected dimension \eqref{dimM}. 
This is consistent with 
$\zeta_1\geq 0, \zeta_5 \leq 0$ provided 
\be
- N \leq \deg \leq 0\ .
\ee
Note that this differs from the range of slopes in the dual collection $\cC^\vee$, which is unbounded due to the vanishing rank of the objects $E_2^\vee$ and $E_3^\vee$.

\medskip

In the attractor chamber, we have either (corresponding to the perfect matchings $s_6,s_1,s_2,s_5$ in  \cite{Hanany:2012hi}, $s_3, s_4$ being incompatible with $\zetastar_2=\zetastar_3$)
 \begin{itemize}
 \item $\Phi_{41}=\Phi_{51}=\Phi_{52}=\Phi_{53}=0$ when  $\zetastar_1\geq 0, \zetastar_5\leq 0$
(which implies $\zetastar_{2,3}\geq 0$), with dimension
\bea
d_\IC 
&=&  1-\cQ(\vec N)+   N_5 \zetastar_5 - N_1 \zetastar_1 - \frac12 N_2   \zetastar_2 
- \frac12 N_3 \zetastar_3  
\eea
\item $\Phi_{45}=\Phi_{41}=0$ when $\zetastar_4\leq 0, \zetastar_5 \geq 0$ (which implies $ \zetastar_1>0$), with dimension
\bea
d'_\IC
&=& 1-\cQ(\vec N)+  \frac12 ( N_4 \zetastar_4 -  N_1 \zetastar_1 - N_5  \zetastar_5)
\eea
\item   $\Phi_{24}=\Phi_{34}=0$ when $\zetastar_{2,3}\leq 0, \zetastar_4\geq 0$, with dimension
\bea
d''_\IC 
&=& 1-\cQ(\vec N)- N_4 \zetastar_4 + \frac12 N_2 \zetastar_2 + \frac12 N_3   \zetastar_3
\eea
\item $\Phi_{12}=\Phi_{13}=\Phi_{52}=\Phi_{53}=0$ when  $\zetastar_1\leq 0, \zetastar_{2,3}\geq 0$
(which implies $\zetastar_5<0$), with dimension
\bea
d'''_\IC
&=&  1-\cQ(\vec N)+  N_1 \zetastar_1 + N_5 \zetastar_5 - \frac12 N_2   \zetastar_2 - \frac12 N_3 \zetastar_3  
\eea
\end{itemize}
where
\bea
\cQ(\vec N) &=&  
\frac14\left[ (N_1-N_2)^2+(N_1-N_3)^2+(N_2-N_5)^2+(N_3-N_5)^2 \right] \nn\\&&
+ \frac12 \left[ (N_4-N_2)^2+  (N_4-N_3)^2+ (N_1-N_5)^2\right]
\eea
Since $\cQ$ is a positive quadratic form, degenerate along the direction $(1,1,1,1,1)$.
 the expected dimension is manifestly negative unless $\vec N$ corresponds to a simple representation or a pure D0-brane. 
Applying the  Coulomb branch formula in the attractor chamber, we find evidence that 
single-centered invariants also vanish under the same condition on $\vec N$.

\medskip

For $\gamma=[1;0,0,0;1],[1;0,0,0;2]$ in the canonical chamber, using the flow tree formula one deduces 
\bea
\Omega(0,1,1,1,1,\vec\zeta^c)&=&y^2+3+1/y^2 \nn\\
\Omega(1,2,2,2,2,\vec\zeta^c) &=& y^4+4y^2+10+4/y^2+1/y^4  
\eea
 in agreement with \eqref{VW1}. The same result is expected to hold for any $J'$ with $J'\cdot c_1(S)>0$.  
 
 \medskip
 
In the blow-up chamber $J'\propto H+\epsilon c_1(S)$, one should use 
\be
\vec\zeta^H=(-c_{1,H},0,0,2c_{1,H}+N, -N- c_{1,H}) +\epsilon \vec\zeta^c
\ee
so that the stability condition \eqref{stabsum} reduces to 
$\sum_i N'_i \zeta_i = N c'_{1,H}-N' c_{1,H} + \cO(\epsilon)$.
 This is consistent with the D-term conditions 
$\zeta_1\geq 0, \zeta_5\leq 0$ provided $-N\leq c_{1,H}\leq 0$. 
We get
\be
\begin{array}{|c|c|l|}
\hline
\[N; c_{1,H},c_{1,1},c_{1,2};c_2\] & \vec N  &\Omega(\vec N, \vec\zeta^H)  \\ \hline
\[2;-1,1,0;1\] &(0,1,2,1,0)&-y-1/y \nn \\
 \[2;-1,0,1;1\] &(0,2,1,1,0)&-y-1/y \nn \\
 \[2;-1,1,1;1\] &(0,1,1,1,0)&y^2+2+1/y^2 \nn\\
 \[2;-1,2,0;1\]  &(1,1,3,2,1)&y^4+4y^2+7+\dots \nn\\
 \[2;0,0,0;2\]  &(0,2,2,2,2)&-y^5-3y^3-6y-\dots \nn\\
\[ 3;-1,3,0;-1\] & (0,0,3,2,1) &  y^2+1+1/y^2 \nn\\
\[ 3;-1,2,1;1\] & (0,1,2,2,1) &  y^6+3y^4+6y^2+7+\dots 
    \\ \hline
    \end{array}
\ee
in agreement with the analysis in  \S\ref{sec_VWdP}.

\subsubsection{Model II}

We now consider  the exceptional collection  
\be
\left( \cO, \cO(H-C_2),  \cO(2H-C_1-C_2),  \cO(2H-C_2),  \cO(3H-C_1-2C_2) \right)
\ee
following from \eqref{toricol} and discussed in  \cite{Herzog:2005sy}. The Chern vectors of the objects $E^i$ and dual objects $E_i$ are 
\be
\begin{array}{ccl}
 \gamma^1 &=& \[1,\(0,0,0\),0\] \\
 \gamma^2 &=& \[1,\(1,0,-1\),0\] \\
 \gamma^3 &=& \[1,\(2,-1,-1\),1\] \\
 \gamma^4 &=& \[1,\(2,0,-1\),\frac{3}{2}\] \\
 \gamma^5 &=& \[1,\(3,-1,-2\),2\] \\
\end{array}
\qquad
\begin{array}{ccl}
 \gamma_1 &=& \[1,\(0,0,0\),0\] \\
 \gamma_2 &=& \[-1,\(1,0,-1\),0\] \\
 \gamma_3 &=& \[-1,\(0,-1,1\),1\] \\
 \gamma_4 &=& \[0,\(-1,1,1\),\frac{1}{2} \] \\
 \gamma_5 &=& \[1,\(0,0,-1\),-\frac{1}{2} \] \\
\end{array}
\ee
with slopes $0,2,4,5,6$ and $0,-2,0,\infty,-1$, respectively. 
The Euler matrix, its inverse and the adjacency matrix are 
\be
\label{SdP2II}
S = 
\left(
\begin{array}{ccccc}
 1 & 2 & 4 & 5 & 6 \\
 0 & 1 & 2 & 3 & 4 \\
 0 & 0 & 1 & 1 & 2 \\
 0 & 0 & 0 & 1 & 1 \\
 0 & 0 & 0 & 0 & 1 \\
\end{array}
\right)\ ,\quad
S^\vee= 
\left(
\begin{array}{ccccc}
 1 & -2 & 0 & 1 & 1 \\
 0 & 1 & -2 & -1 & 1 \\
 0 & 0 & 1 & -1 & -1 \\
 0 & 0 & 0 & 1 & -1 \\
 0 & 0 & 0 & 0 & 1 \\
\end{array}
\right)\ ,\quad
\kappa=\left(
\begin{array}{ccccc}
 0 & 2 & 0 & -1 & -1 \\
 -2 & 0 & 2 & 1 & -1 \\
 0 & -2 & 0 & 1 & 1 \\
 1 & -1 & -1 & 0 & 1 \\
 1 & 1 & -1 & -1 & 0 \\
\end{array}
\right)
\ee
corresponding to the quiver 
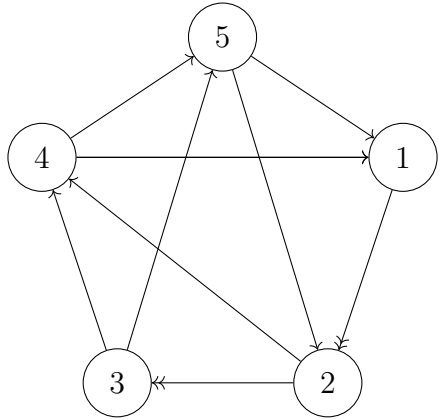
\begin{center}
\begin{tikzpicture}[inner sep=2mm,scale=2]
  \node (a) at ( 1.2,.5) [circle,draw] {$1$};
  \node (b)  at ( .7,-1) [circle,draw] {$2$};
  \node (c)  at ( -.7,-1) [circle,draw] {$3$};
    \node (d) at ( -1.2,.5) [circle,draw] {$4$};
    \node (e) at ( 0,1.3) [circle,draw] {$5$};
 \draw [->>] (a) to node[auto] {$ $} (b);
 \draw [->>] (b) to node[auto] {$ $} (c);
 \draw [->] (c) to node[auto] {$ $} (d);
 \draw [->] (d) to node[auto] {$ $} (e);
 \draw [->] (e) to node[auto] {$ $} (a);
 \draw [->] (d) to node[near end,below] {$ $} (a);
 \draw [->] (b) to node[near end,below] {$ $} (d);
 \draw [->] (c) to node[near end,above] {$ $} (e);
 \draw [->] (e) to node[near end,above] {$ $} (b);
 \draw [->] (d) to node[near end,above] {$ $} (a);
\end{tikzpicture}
\end{center}
with superpotential \cite[(4.10)]{Feng:2002zw} (see also model 12a in \cite{Hanany:2012hi})
\be
\begin{split}
W=&-  \Phi_{12}^2 \Phi_{24} \Phi_{41}-\Phi_{23}^2 \Phi_{35} \Phi_{52}
+\Phi_{24} \Phi_{45} \Phi_{52}
\\&
+\Phi_{12}^2 \Phi_{23}^1 \Phi_{35} \Phi_{51} 
+ \Phi_{12}^1 \Phi  _{23}^2 \Phi_{34} \Phi_{41}
-\Phi_{12}^1 \Phi_{23}^1 \Phi_{34} \Phi_{45} \Phi_{51}  
\end{split}
\ee
Note that under Seiberg duality with respect to node 3, one recovers the quiver \eqref{SdP2I}
of model I in the previous subsection.
A general Chern character $\gamma=[N,c_1,\ch_2]$ decomposes as $\gamma=
\sum_i n_i \gamma_i$ with 
\bea
 n_1 &=& \frac{1}{2} (c_{1,1}+c_{1,2}+3 c_{1,H})+ \ch_2+N \nn\\
 n_2 &=& \frac{1}{2} (c_{1,1}-c_{1,2}+c_{1,H})+ \ch_2 \nn\\
 n_3 &=& \frac{1}{2} (-c_{1,1}-c_{1,2}-c_{1,H})+ \ch_2  \nn\\
 n_4 &=& \frac{1}{2} (c_{1,1}-c_{1,2}-c_{1,H})+ \ch_2 \nn\\
  n_5 &=& \frac{1}{2} (-c_{1,1}-3 c_{1,2}-3 c_{1,H})+2 \ch_2 
  \eea
or conversely
\bea
 N &=& n_1-n_2-n_3+n_5 \ ,\quad \ch_2 = n_3 +  \frac{1}{2} (n_4-n_5) \nn\\
 c_{1,H} &=& n_2-n_4\ ,\quad  c_{1,1} = n_4-n_3 \ ,\quad 
 c_{1,2} = -n_2+n_3+n_4-n_5 
\eea
For large $c_2$, the entries in $\vec n$ are all negative, so we should
consider the dimension vector $\vec N=-\vec n$.
The relevant stability parameters  \eqref{zetacan}
 in the canonical chamber $J\propto c_1(S)$ are   
\bea
 \zeta_1 &=& -\deg  \rho  + \ch_2 \nn\\
 \zeta_2 &=& (\deg +2 N) \rho  -\ch_2 \nn\\
 \zeta_3 &=& \deg  \rho   -\ch_2-N \nn\\
 \zeta_4 &=& -N \rho  -\frac{N}{2} \nn\\
 \zeta_5 &=&- (\deg +N) \rho  + \ch_2+\frac{N}{2} 
\eea
The dimension agrees with \eqref{dimM} in the chamber where $\Phi_{41}=\Phi_{51}=\Phi_{52}=0$.
This requires $\zeta_1\geq 0,\zeta_5\leq 0$ hence
\be
\label{wdP2II}
-N \leq  \deg \leq 0
\ee
This differs from the range of slopes in the dual collection $\cC^\vee$, which is unbounded due to 
the vanishing rank of the object $E_4^\vee$.

\medskip

In the attractor chamber \eqref{zetaatt}, we have either (corresponding to the perfect matchings 
$s_1,s_2,s_3,s_4,s_5$ in  \cite{Hanany:2012hi})
\begin{itemize}
\item $\Phi_{41}=\Phi_{51}=\Phi_{52}=0$ when $\zetastar_1\geq 0, \zetastar_5\leq 0$ (which implies $\zetastar_{3,4}\leq 0$), with dimension
\bea
d_\IC
&=&1 - \cQ(\vec N) - \frac12 N_1 \zetastar_1  + \frac12 N_3 \zetastar_3 + N_4 \zetastar_4 + N_5 \zetastar_5 
\eea
\item  $\Phi_{23}=\Phi_{24}=0$ when  $\zetastar_2\leq 0$,  $\zetastar_3\geq 0$ (which implies $\zetastar_{1}\leq 0$),
with dimension
\bea
d'_\IC
&=&1 - \cQ(\vec N) + \frac12 N_1 \zetastar_1 + N_2 \zetastar_2 - \frac12 N_3 \zetastar_3 
\eea
\item  $\Phi_{12}=\Phi_{52}=0$ when  $\zetastar_1\leq 0$,  $\zetastar_2\geq 0$ (which implies $\zetastar_{3}\geq 0$), with dimension
\bea
d''_\IC
&=& 1 - \cQ(\vec N) + \frac12 N_1 \zetastar_1 - N_2 \zetastar_2 - \frac12 N_3 \zetastar_3
\eea
\item $\Phi_{24}=\Phi_{34}=\Phi_{35}=0$ when $\zetastar_3\leq 0, \zetastar_4\geq 0$ (and therefore
$ \zetastar_{1,5}\geq 0$),
with dimension
\bea
d'''_\IC
&=&1 - \cQ(\vec N)  - \frac12 N_1 \zetastar_1 + \frac12 N_3 \zetastar_3
- N_4 \zetastar_4  - N_5 \zetastar_5 
 \eea
\item $\Phi_{41}=\Phi_{35}=\Phi_{45}=0$ when $\zetastar_4\leq 0, \zetastar_5\geq 0$ (and therefore
$ \zetastar_1\geq 0,  \zetastar_3\leq 0$),
with dimension
\bea
d''''_\IC
&=&1 - \cQ(\vec N) - \frac12 N_1 \zetastar_1 + \frac12 N_3 \zetastar_3 + N_4 \zetastar_4 - N_5 \zetastar_5
\eea
\end{itemize}
where $\cQ$ is the positive quadratic form
\bea
\cQ(\vec N)&=& \frac12 \left[ (N_2 - N_3)^2 + (N_2 - N_1)^2 + (N_5 - N_4)^2\right] \nn\\&& + 
 \frac14 \left[(N_4 - N_1)^2 + (N_4 - N_3)^2 + (N_5 - N_1)^2 + (N_5 - N_3)^2\right]
\eea
In all cases, the expected dimension is manifestly negative unless $\vec N$ corresponds to a 
simple representation or a pure D0-brane.  Applying the Coulomb branch formula in the attractor chamber, we find evidence that the 
single-centered invariants $\OmS(\vec N)$ 
also vanish, under the same conditions on $\vec N$.

 \medskip
 
 For  $\gamma=[1;0,0,0;1]$ in the canonical chamber, we get from the flow tree formula 
\be
\Omega(0,1,1,1,1,\vec\zeta^c)=y^2+3+1/y^2
\ee
in agreement with \eqref{VW1}. The same result is expected to hold for any $J'$ with $J'\cdot c_1(S)>0$.  

\medskip

In the blow-up chamber $J'\propto H_\epsilon$, one should instead use 
\be
\vec\zeta^H=(-c_{1,H},c_{1,H}+N,c_{1,H},-N,- c_{1,H})+\epsilon \vec\zeta^c
\ee
 so that the stability condition \eqref{stabsum} reduces to 
$\sum_i N'_i \zeta_i = N c'_{1,H}-N' c_{1,H} + \cO(\epsilon)$.
This is consistent with the conditions
 $\zeta_1\geq 0, \zeta_5\leq 0$ only when $c_{1,H}=0$.  From the flow tree formula we get
 \be
\begin{array}{|c|c|l|}
\hline
\[N; c_{1,H},c_{1,1},c_{1,2};c_2\] & \vec N &\Omega(\vec N, \vec\zeta^H)  \\ \hline
  \[2;0,-1,0;1\] & (0,2,1,2,1) &y^2+2+1/y^2 \nn\\
\[2; 0,-1,-1;1\] &  (1,2,1,2,0) & -y^3-2y-\dots \nn\\
  \[2; 0,0,0;2\] & (0,2,2,2,2) &-y^5-3y^3-6y-\dots \nn\\
 \[ 3;0,1,-2;0\]  &(0,1,2,1,0)&y^2+1+1/y^2 \nn\\
  \[ 3;0,-1,-1;1\]  &(0,2,1,2,0)&y^2+1+1/y^2 \nn\\
\[ 3;0,0,-1;2\]  &(0,2,2,2,1)&y^6+3y^4+8y^2+10+\dots 
    \\ \hline
    \end{array}
\ee
in agreement with the analysis in  \S\ref{sec_VWdP}.

\subsection{$dP_3$}
The toric fan is generated by 6 vectors,
\be
v_i = \begin{pmatrix} 
1 & 1 & 0 & -1 &-1 & 0 \\
0 & 1 & 1 & 0 & -1 & -1
\end{pmatrix}
\ee
The corresponding divisors satisfy the linear relations
\be
D_1+D_2=D_4+D_5\ ,\quad D_2+D_3=D_5+D_6
\ee
and form an overcomplete basis of $H_2(S,\IZ)$, with intersection numbers 
$D_i^2 = -1$ for all $i$. 
According to \cite[\S 2.6]{Iqbal:2001ye}, the divisors $D_i$ can be expressed in terms of $H, C_1, C_2, C_3$ associated to the hyperplane class of $\IP^2$ and the three exceptional divisors via
\be
D_1=C_1,\ D_2 =H-C_1-C_2, \ D_3=C_2, \  
D_4=H-C_2-C_3, \ D_5 =C_3, \ D_6=H-C_1-C_3
\ee
In the basis $H, C_1, C_2, C_3$, the intersection matrix becomes diagonal,
\be
C^{\alpha\beta} = 
\left(
\begin{array}{cccc} 1 & \\ & -1 \\ & & -1 \\ & & & -1  \end{array}
\right)
\ee
The square of the canonical class $c_1(S)=\sum_i D_i=3H-C_1-C_2-C_3$ evaluates to $6$ as expected. We shall consider four different collections, corresponding to models I to IV in \cite{Beasley:2001zp,Feng:2002zw}.

\subsubsection{Model I}
We first consider  the exceptional collection obtained from \eqref{toricol}
\be
\cC = \left( \cO(0),\cO(C_1),\cO(H-C_2),\cO(H),\cO(2H-C_2-C_3), \cO(2H-C_2) \right)\ ,
\ee
which is equivalent to the one used in \cite[(3.10)]{Herzog:2005sy}, upon exchanging $C_1$ and $C_2$. The Chern vectors of the objects $E^i$ and dual objects $E_i$ are 
\be
\begin{array}{ccl}
 \gamma^1 &=& \[1,\(0,0,0,0\),0\] \\
 \gamma^2 &=& \[1,\(0,0,1,0\),-\frac{1}{2}\] \\
 \gamma^3 &=& \[1,\(1,-1,0,0\),0\] \\
 \gamma^4 &=& \[1,\(1,0,0,0\),\frac{1}{2}\] \\
 \gamma^5 &=& \[1,\(2,-1,0,-1\),1\] \\
 \gamma^6 &=& \[1,\(2,-1,0,0\),\frac{3}{2}\] \\
\end{array}
\qquad
\begin{array}{ccl}
 \gamma_1 &=& \[1,\(0,0,0,0\),0\] \\
 \gamma_2 &=& \[0,\(0,0,1,0\),-\frac{1}{2}\] \\
 \gamma_3 &=& \[-1,\(1,-1,-1,0\),\frac{1}{2}\] \\
 \gamma_4 &=& \[-1,\(0,1,-1,0\),1\] \\
 \gamma_5 &=& \[0,\(0,0,0,-1\),\frac{1}{2}\] \\
 \gamma_6 &=& \[1,\(-1,0,1,1\),-\frac{1}{2}\] \\
\end{array}
\ee
with slopes $0,1,2,3,4,5$ and  $0,\infty,-1,0,\infty,-1$.
The Euler matrix, its inverse and the adjacency matrix are given by 
\be
\label{EulerdP3HP}
S =\left(
\begin{array}{cccccc}
 1 & 1 & 2 & 3 & 4 & 5 \\
 0 & 1 & 1 & 2 & 3 & 4 \\
 0 & 0 & 1 & 1 & 2 & 3 \\
 0 & 0 & 0 & 1 & 1 & 2 \\
 0 & 0 & 0 & 0 & 1 & 1 \\
 0 & 0 & 0 & 0 & 0 & 1 \\
\end{array}
\right)\ ,\quad
S^\vee=
\left(
\begin{array}{cccccc}
 1 & -1 & -1 & 0 & 1 & 1 \\
 0 & 1 & -1 & -1 & 0 & 1 \\
 0 & 0 & 1 & -1 & -1 & 0 \\
 0 & 0 & 0 & 1 & -1 & -1 \\
 0 & 0 & 0 & 0 & 1 & -1 \\
 0 & 0 & 0 & 0 & 0 & 1 \\
\end{array}
\right)\ ,\quad 
\kappa=
\left(
\begin{array}{cccccc}
 0 & 1 & 1 & 0 & -1 & -1 \\
 -1 & 0 & 1 & 1 & 0 & -1 \\
 -1 & -1 & 0 & 1 & 1 & 0 \\
 0 & -1 & -1 & 0 & 1 & 1 \\
 1 & 0 & -1 & -1 & 0 & 1 \\
 1 & 1 & 0 & -1 & -1 & 0 \\
\end{array}
\right)
\ee
This reproduces the quiver in \cite{Hanany:2001py}, known as model I, invariant
under the dihedral group $D_6$,
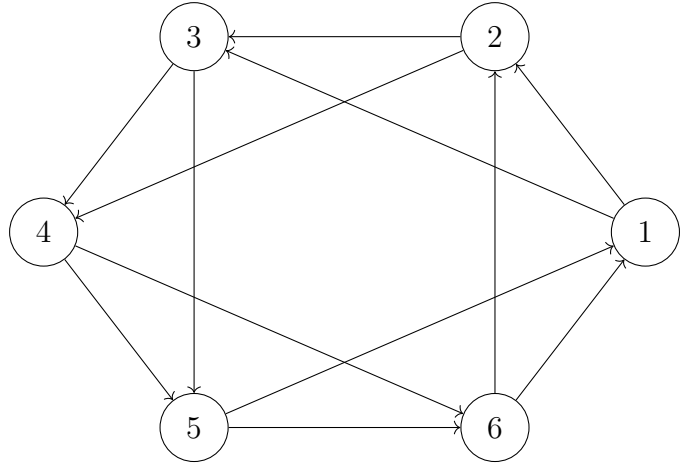
\begin{center}
\begin{tikzpicture}[inner sep=2mm,scale=2]
  \node (a) at ( 2,0) [circle,draw] {$1$};
    \node (f)  at ( 1,-1.3) [circle,draw] {$6$};
  \node (b) at ( 1,1.3 ) [circle,draw] {$2$};
  \node (c) at ( -1, 1.3) [circle,draw] {$3$};
  \node (d)  at ( -2,0 ) [circle,draw] {$4$};
  \node (e)  at ( -1,-1.3) [circle,draw] {$5$};
 \draw [->] (a) to node[auto] {$ $} (b);
 \draw [->] (a) to node[auto] {$ $} (c);
 \draw [->] (b) to node[auto] {$ $} (d);
 \draw [->] (b) to node[auto] {$ $} (c);
 \draw [->] (c) to node[auto] {$ $} (d);
 \draw [->] (c) to node[auto] {$ $} (e);
 \draw [->] (d) to node[auto] {$ $} (e);
 \draw [->] (d) to node[auto] {$ $} (f);
 \draw [->] (e) to node[auto] {$ $} (a);
 \draw [->] (e) to node[auto] {$ $} (f);
 \draw [->] (f) to node[auto] {$ $} (b);
  \draw [->] (f) to node[auto] {$ $} (a);
\end{tikzpicture}
\end{center}
The superpotential is given by \cite{Beasley:2001zp,Feng:2002zw} (see also model 10a in \cite{Hanany:2012hi})
\be
\begin{split}
W=& \Phi_{12}\Phi_{23}\Phi_{34}\Phi_{45}\Phi_{56}\Phi_{61} - 
\left( \Phi_{23} \Phi_{35}\Phi_{56}\Phi_{62} +
\Phi_{13}  \Phi_{34} \Phi_{46}\Phi_{61}+
\Phi_{12}\Phi_{24}   \Phi_{45} \Phi_{51} \right) \\ & 
+ \Phi_{13}\Phi_{35}\Phi_{51} +  \Phi_{24}\Phi_{46}\Phi_{62} 
\end{split}
\ee
A general Chern character $\gamma=[N,c_1,\ch_2]$ decomposes as $\gamma=
\sum_i n_i \gamma_i$ with 
\bea
n_1 &=&  \frac{1}{2} (c_{1,1}+c_{1,2}+c_{1,3}+3 c_{1,H})+ \ch_2+ N\nn\\
   n_2 &=&  \frac{1}{2} (c_{1,1}+3c_{1,2}+c_{1,3}+3 c_{1,H})+   \ch_2\nn\\
   n_3 &=&  \frac{1}{2} (-c_{1,1}+c_{1,2}+c_{1,3}+c_{1,H}) +\ch_2\nn\\
   n_4 &=&  \frac{1}{2} (c_{1,1}+c_{1,2}+c_{1,3}+c_{1,H})+  \ch_2\nn\\
   n_5 &=&  \frac{1}{2} (-c_{1,1}+c_{1,2}-c_{1,3}-c_{1,H})+ \ch_2\nn\\
   n_6 &=&  \frac{1}{2} (-c_{1,1}+c_{1,2}+c_{1,3}-c_{1,H})+\ch_2
\eea
or conversely
\be
\begin{split}
N =  n_1-n_3-n_4+n_6\ ,\quad   \ch_2 =  \frac{1}{2} (-n_2+n_3+2 n_4+n_5-n_6)\\ 
c_{1,H} =  n_3-n_6 ,\quad 
   c_{1,1} =  n_4-n_3 ,\quad 
c_{1,2} =    n_2-n_3-n_4+n_6 ,\quad 
   c_{1,3} =  n_6-n_5 ,\quad 
\end{split}
\ee
Setting $\vec N=-\vec n$, the stability parameters in the canonical chamber $J\propto c_1(S)$
 are 
\bea
\zeta_1&=&  \rho( N_2+N_3 - N_5 - N_6) + \frac{1}{2} (N_{2}-N_{3}-2 N_{4}-N_{5}+N_{6})
\nn\\
&=& -\rho \,\deg + \ch_2,  
\nn\\
\zeta_2&=&-\zeta_5 =  \rho(N_3 + N_4 - N_1-N_6) + \frac{1}{2} (N_{3}+N_{4}-N_{1}-N_{6})
= \rho\, N + \frac{N}{2} \nn\\
\zeta_3&=& -\zeta_6 = \rho(N_4 + N_5 - N_1 - N_2) + \frac{1}{2}  (N_{1}-N_{2}+N_{4}+N_{5})
\nn\\
&=& \rho( \deg + N) -\ch_2 -\frac{N}{2}\nn\\
\zeta_4&=&  \rho(N_5 + N_6 - N_2 - N_3) + \frac{1}{2} (2 N_{1}-N_{2}-N_{3}+N_{5}+N_{6})
\nn\\
&=&  \rho\, \deg -\ch_2 -N 
   \eea
The dimension $d_\IC$  in the chamber where $\Phi_{51}=\Phi_{61}=\Phi_{62}=0$ 
coincides with the expected
dimension \eqref{dimM}. This is consistent with $\zeta_1\geq 0, \zeta_6\leq 0$ provided
\be
\label{wdP3I}
-N \leq \deg  \leq  0
\ee
This differs from the range of slopes in the dual collection $\cC^\vee$, which is unbounded due to 
the vanishing rank of $E_2^\vee$ and $E_5^\vee$.
%Note that a cyclic permutation of $c_{1,1}, c_{1,2}, c_{1,3}$ leaves $n_1,n_6$ invariant and maps
%\be
%n_2\to 2n_4-n_6,\quad
%n_3\to n_4+n_5-n_6\ ,\quad
%n_5\to -n_2+2n_4+n_5-n_6\ ,\quad
%n_6\to n_4+n_5-n_3
%\ee
%The action of the dihedral group $D_6$ does not seem to lead to a simple action on the Chern character. The fractional branes, in the kernel of the adjacency matrix, have dimension vector $(1,0,0,1,0,0)$, $(0,1,0,0,1,0)$ 
%and $(1,0,1,0,1,0)$.

\medskip
In the attractor chamber, by cyclic symmetry there is no loss of generality in assuming that 
$\zetastar_1\leq 0, \zetastar_6\geq 0$ hence $\zetastar_2\leq 0, \zetastar_3\leq 0,
\zetastar_4\geq 0$, consistently with $\Phi_{24}=\Phi_{34}=\Phi_{35}=0$. The expected dimension
in this chamber can be written as 
\bea
\dstar_\IC&=&  - \frac12\left[ (N_1-N_2)^2 + (N_2-N_3)^2+(N_3-N_1)^2 + 
 (N_4-N_5)^2 + (N_5-N_6)^2+(N_6-N_4)^2 \right] \nn\\&&
 + N_1 \zetastar_1 +N_2 \zetastar_2+N_3 \zetastar_3 + 1
\eea
which is manifestly negative, unless $N_1=N_2=N_3$ and $N_4=N_5=N_6$. For those values however,
one has $\zetastar_1=2(N_4-N_1)=-\zetastar_3$, while 
$\zetastar_1$ and $\zetastar_3$ must be both negative, so all $N_i$'s must be equal, corresponding to a pure D0-brane. We conclude
that the attractor index $\Omstar(\vec N)$ vanishes except for simple representations or D0-branes.

\medskip

For $\gamma=[1;0,0,0,0;1]$  in the canonical chamber,
 we find using the flow tree formula
 \be
\Omega(0,1,1,1,1,1,\vec\zeta^c) = y^2 + 4 + 1/y^2 \ .
\ee
in agreement with \eqref{VW1}. 
The same result is expected to hold for any $J'$ with $J'\cdot c_1(S)>0$.  Applying the  flow tree formula in the attractor chamber, we find evidence that 
single-centered invariants vanish, just like attractor invariants.

\medskip

In the blow-up chamber $J'\propto H_\epsilon$, one should instead use 
\be
\vec\zeta^H=(-c_{1,H},0,c_{1,H}+N,c_{1,H},0,- c_{1,H}-N)
+\epsilon \vec\zeta^c
\ee
This is consistent with
 $\zeta_1\geq 0,\zeta_6\leq 0$ provided $-N\leq c_{1,H} \leq 0$.
From the flow tree formula we get
\be
\begin{array}{|c|c|l|}
\hline
\[N;c_1;c_2\] & \vec N   &\Omega(\vec N, \vec\zeta^H)  \\
\hline
\[2;-1,0,0,0;1\] &(0,2,1,1,0,0)&1 \nn \\
\[2;-1,1,0,0;1\] &(2,2,1,1,1,0)&-y-1/y \nn \\
\[2;-1,1,1,0;1\] &(0,1,2,1,1,1)&y^2+ 2+ 1/y^2 \nn\\
\[2;0,-1,0,0;1\] &(0,2,1,2,1,1)&y^2+ 3+ 1/y^2 \nn\\
\[2;0,-1,1,0;0\] &(0,1,1,2,1,1)&-y^3-3y-\dots \nn\\
\[2;0,-1,1,-1;0\]&(0,1,1,2,0,1)&1\nn\\
\[3;-1,0,3,0;-1\]&(0,0,2,2,1,1)&y^2+1+1/y^2\nn\\
\[3;-1,0,2,0;1\]&(0,1,2,2,1,1)&y^4+2y^2+3+\dots\nn\\
\[3;0,-2,2,0;-1\]&(0,1,1,3,1,1)&0 \nn\\
\hline
\end{array}
\nn
\ee
in agreement with the analysis in   \S\ref{sec_VWdP}.

\medskip

After  left-mutation with respect to node 2 and applying the  permutation $142356$ on the nodes, we obtain the model II below.

\subsubsection{Model II}

We now consider the exceptional collection obtained by Seiberg duality on any node of
the previous model. Specifically, we consider the Chern vectors
\be
\label{gdP3II}
\begin{array}{ccl}
 \gamma^1 &=& \[1,\(0,0,0,0\),0\] \\
 \gamma^2 &=& \[1,\(1,-1,0,0\),0\] \\
 \gamma^3 &=& \[1,\(1,0,0,0\),\frac{1}{2}\] \\
 \gamma^4 &=& \[1,\(2,-1,-1,0\),1\] \\
 \gamma^5 &=& \[1,\(2,-1,0,-1\),1\] \\
 \gamma^6 &=& \[1,\(2,-1,0,0\),\frac{3}{2}\] \\
\end{array}
\qquad
\begin{array}{ccl}
 \gamma_1 &=& \[1,\(0,0,0,0\),0\] \\
 \gamma_2 &=& \[-1,\(1,-1,0,0\),0\] \\
 \gamma_3 &=& \[-1,\(0,1,0,0\),\frac{1}{2}\] \\
 \gamma_4 &=& \[0,\(0,0,-1,0\),\frac{1}{2}\] \\
 \gamma_5 &=& \[0,\(0,0,0,-1\),\frac{1}{2}\] \\
 \gamma_6 &=& \[1,\(-1,0,1,1\),-\frac{1}{2}\] \\
\end{array}
\ee
with slopes $0,2,3,4,4,5$ and $0,-2,-1,\infty,\infty,-1$, respectively.
The Euler matrix has a five-block structure, 
\be
\label{SdP3II}
S=\left(
\begin{array}{cccccc}
 1 & 2 & 3 & 4  & 5 \\
 0 & 1 & 1 & 2  & 3 \\
 0 & 0 & 1 & 1  & 2 \\
 0 & 0 & 0 & \mathbb{1}_2 & 1 \\
 0 & 0 & 0 & 0 & 1 \\
\end{array}
\right),\ 
S^\vee= 
\left(
\begin{array}{cccccc}
 1 & -2 & -1 & 1  & 1 \\
 0 & 1 & -1 & -1 & 1 \\
 0 & 0 & 1 & -1 & 0 \\
 0 & 0 & 0 & \mathbb{1}_2 & -1 \\
 0 & 0 & 0 & 0 & 1 \\
\end{array}
\right), \ 
\kappa=
\left(
\begin{array}{cccccc}
 0 & 2 & 1 & -1 & -1 \\
 -2 & 0 & 1 & 1 & -1 \\
 -1 & -1 & 0 & 1 & 0 \\
 1 & -1 & -1 & \mathbb{0}_2 & 1 \\
 1 & 1 & 0 & -1  & 0 \\
\end{array}
\right)
\ee
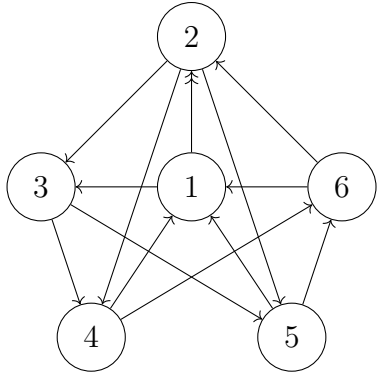
 \begin{center}
\begin{tikzpicture}[inner sep=2mm,scale=2]
  \node (a) at ( -1,0) [circle,draw] {$3$};
  \node (b) at ( 0,1) [circle,draw] {$2$};
  \node (c)  at ( 2/3,-1) [circle,draw] {$5$};
  \node (d)  at ( -2/3,-1) [circle,draw] {$4$};
    \node (e)  at ( 0,0) [circle,draw] {$1$};
      \node (f)  at ( 1,0) [circle,draw] {$6$};    
 \draw [->] (b) to node[auto] {$ $} (a);
 \draw [->] (a) to node[auto] {$ $} (c);
 \draw [->] (a) to node[auto] {$ $} (d);
  \draw [->>] (e) to node[auto] {$ $} (b);
 \draw [->] (c) to node[auto] {$ $} (e);
 \draw [->] (d) to node[auto] {$  $} (e);
 \draw [->] (f) to node[auto] {$ $} (e);
\draw [->] (f) to node[auto] {$  $} (b);
\draw [->] (c) to node[auto] {$  $} (f);
\draw [->] (d) to node[auto] {$  $} (f);
\draw [->] (e) to node[auto] {$ $} (a);
\draw [->] (b) to node[auto] {$ $} (c);
\draw [->] (b) to node[auto] {$ $} (d);
\end{tikzpicture}
\end{center}
with superpotential \cite{Beasley:2001zp,Feng:2002zw}  (see also model 10b 
in \cite{Hanany:2012hi})
\bea
\label{WdP3II}
W&=& \Phi_{12}^2 \Phi_{24} \Phi_{41}- \Phi_{12}^1 \Phi_{25} \Phi_{51}
-\Phi_{13} \Phi_{34} \Phi_{41}+\Phi_{13} \Phi_{35} \Phi_{51} \nn\\&&
-\Phi_{24} \Phi_{46} \Phi_{62}+\Phi_{25} \Phi_{56} \Phi_{62}
- \Phi_{12}^2 \Phi_{23} \Phi_{35} \Phi_{56} \Phi_{61}
+ \Phi_{12}^1 \Phi_{23} \Phi_{34} \Phi_{46} \Phi_{61}
   \eea
A general Chern character $\gamma=[N,c_1,\ch_2]$ decomposes as $\gamma=
\sum_i n_i \gamma_i$ with  
\bea
 n_1 &=& \frac{1}{2} (c_{1,1}+c_{1,2}+c_{1,3}+3 c_{1,H})+ \ch_2+N \nn\\
 n_2 &=& \frac{1}{2} (-c_{1,1}+c_{1,2}+c_{1,3}+c_{1,H})+ \ch_2 \nn\\
 n_3 &=& \frac{1}{2} (c_{1,1}+c_{1,2}+c_{1,3}+c_{1,H})+ \ch_2 \nn\\
 n_4 &=& \frac{1}{2} (-c_{1,1}-c_{1,2}+c_{1,3}-c_{1,H})+ \ch_2 \nn\\
 n_5 &=& \frac{1}{2} (-c_{1,1}+c_{1,2}-c_{1,3}-c_{1,H})+ \ch_2 \nn\\
 n_6 &=& \frac{1}{2} (-c_{1,1}+c_{1,2}+c_{1,3}-c_{1,H})+ \ch_2 
\eea
or conversely
\bea
 N &=& n_1-n_2-n_3+n_6 \ ,\quad
  \ch_2 = \frac{1}{2} \left(n_3+n_4+n_5-n_6\right)  \nn\\
 c_{1,H} &=& n_2-n_6 \ ,\quad 
 c_{1,1} = n_3-n_2\ ,\quad 
 c_{1,2} = n_6-n_4\ ,\quad 
 c_{1,3} = n_6-n_5
\eea
As usual, we take $\vec N=-\vec n$. 
The stability parameter  $\vec\zeta=\vec\zeta^c$ in \eqref{zetacan}
in the canonical chamber $J\propto c_1(S)$ is then 
\bea
 \zeta_1 &=& -\deg  \rho  + \ch_2 \nn\\
 \zeta_2 &=& (\deg +2 N) \rho  -\ch_2 \nn\\
 \zeta_3 &=& -\zeta_6= (\deg +N) \rho  -\ch_2-\frac{N}{2} \nn\\
 \zeta_{4,5} &=& -N \rho  -\frac{N}{2} %\nn\\
% \zeta_6 &=& -(\deg +N) \rho + \ch_2+\frac{N}{2} \nn\\
\eea
In the chamber where $\Phi_{41}=\Phi_{51}=\Phi_{61}=\Phi_{62}=0$,
the expected dimension 
\be
d_\IC=N_1(2N_2+N_3)+N_2(N_3+N_4+N_5)+(N_4+N_5)(N_3+N_6)
- N_1(N_4+N_5+N_6) - N_2 N_6 -\sum_i N_i^2 +1
\ee
agrees with \eqref{dimM} 
This requires $\zeta_{1}\geq 0$, $\zeta_{6}\leq 0$ hence
\be
-N \leq \deg \leq 0\ .
\ee
Again, this differs from the range of slopes in the dual collection $\cC^\vee$, which is unbounded due to the vanishing rank of $E_4^\vee$ and $E_5^\vee$.

\medskip

 In the attractor chamber,  depending on the signs of the $\zetastar_i$'s we have the following possibilities (corresponding to the perfect matchings $s_3,\dots, s_7$ in \cite{Hanany:2012hi}): 
\begin{itemize}
\item $\Phi_{41}=\Phi_{51}=\Phi_{46}=\Phi_{56}=0$ when $\zetastar_{4,5}\leq 0, \zetastar_6\geq 0$
hence $\zetastar_3\leq 0$, with  dimension
\be
d_\IC=1 - \cQ(\vec N) + N_3 \zetastar_3 + N_4 \zetastar_4 + N_5 \zetastar_5 - N_6 \zetastar_6
\ee 
\item $\Phi_{24}=\Phi_{34}=\Phi_{25}=\Phi_{35}=0$ when $\zetastar_3\leq 0, \zetastar_{4,5}\geq 0$,
hence $\zetastar_6\geq 0$, with dimension
\be
d'_\IC=1 - \cQ(\vec N) + N_3 \zetastar_3 - N_4 \zetastar_4 - N_5 \zetastar_5 - N_6 \zetastar_6
\ee 
\item $\Phi^\alpha_{12}=\Phi_{13}=\Phi_{62}=0$ when $\zetastar_1\leq 0, \zetastar_{2}\geq 0$, with dimension
\be
d''_\IC=1 - \cQ(\vec N) + N_1 \zetastar_1 - N_2 \zetastar_2
\ee 
\item $\Phi_{41}=\Phi_{51}=\Phi_{61}=\Phi_{62}=0$  when $\zetastar_1\geq 0, \zetastar_{6}\leq 0$ hence $\zetastar_2\geq 0$, with dimension
\be
d'''_\IC=1 - \cQ(\vec N) - N_1 \zetastar_1 - N_2 \zetastar_2
\ee 
\item $\Phi_{13}=\Phi_{23}=\Phi_{24}=\Phi_{25}=0$ when $\zetastar_2\leq 0, \zetastar_{3}\geq 0$ hence $\zetastar_1\leq 0$, with dimension
\be
d''''_\IC=1 - \cQ(\vec N) + N_1 \zetastar_1 + N_2 \zetastar_2
\ee 
\end{itemize}
where $\cQ$ is  the positive quadratic form 
\be
\cQ(\vec N)= (N_1 - N_2)^2 + \frac12 \left[ (N_3 - N_4)^2 + (N_3 - N_5)^2 + (N_4 - N_6)^2 + (N_5 - N_6)^2 \right]
\ee
Thus the expected dimension is strictly negative unless $N_1=N_2$ and $N_3=N_4=N_5=N_6$. In fact, one can check that it is positive only when all $N_i$'s are equal,
corresponding to a pure D0-brane, or for dimension vectors corresponding to simple representations.

\medskip

After right-mutation on the node 4 and applying the permutation $134256$, we get the model I above.
After right-mutation on the node 6 and exchanging the nodes 4 and 6, we get the model III described next.

\subsubsection{Model III}
We now consider the exceptional collection with Chern vectors
\be
\begin{array}{ccl}
 \gamma^1 &=& \[1,\(0,0,0,0\),0\] \\
 \gamma^2 &=& \[1,\(1,-1,0,0\),0\] \\
 \gamma^3 &=& \[1,\(1,0,0,0\),\frac{1}{2}\] \\
 \gamma^4 &=& \[1,\(2,-1,-1,-1\),\frac{1}{2}\] \\
 \gamma^5 &=& \[1,\(2,-1,0,-1\),1\] \\
 \gamma^6 &=& \[1,\(2,-1,-1,0\),1\] \\
\end{array}
\qquad
\begin{array}{ccl}
 \gamma_1 &=& \[1,\(0,0,0,0\),0\] \\
 \gamma_2 &=& \[-1,\(1,-1,0,0\),0\] \\
 \gamma_3 &=& \[-1,\(0,1,0,0\),\frac{1}{2}\] \\
 \gamma_4 &=& \[-1,\(1,0,-1,-1\),\frac{1}{2}\] \\
 \gamma_5 &=& \[1,\(-1,0,1,0\),0\] \\
 \gamma_6 &=& \[1,\(-1,0,0,1\),0\] \\
\end{array}
\ee
with slopes $0,2,3,4,4$ and $0,-2,-1,-1,-2,-2$, respectively. The Euler matrix has now a four-block structure, 
\be
S=\left(
\begin{array}{cccccc}
 1 & 2 & 3 & 4  \\
 0 & 1 & 1 & 2 \\
 0 & 0 & \mathbb{1}_2  & 1 \\
 0 & 0 & 0 &  \mathbb{1}_2  \\
\end{array}
\right),\ 
S^\vee=\left(
\begin{array}{cccccc}
 1 & -2 & -1 & 2 \\
 0 & 1 & -1 & 0 \\
 0 & 0 & \mathbb{1}_2 & -1  \\
 0 & 0 & 0 &  \mathbb{1}_2  \\
\end{array}
\right),\ 
\kappa=
\left(
\begin{array}{cccccc}
 0 & 2 & 1 & -2 \\
 -2 & 0 & 1 & 0 \\
 -1 & -1 & \mathbb{0}_2 & 1 \\
 2 & 0 & -1 &   \mathbb{0}_2   \\
\end{array}
\right)
\ee
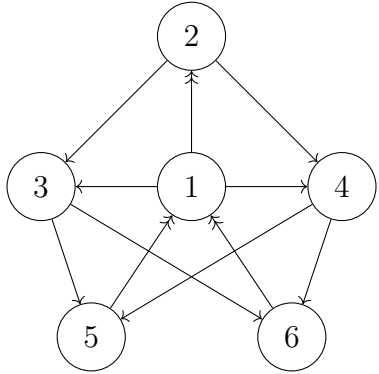
 \begin{center}
\begin{tikzpicture}[inner sep=2mm,scale=2]
  \node (a) at ( -1,0) [circle,draw] {$3$};
  \node (b) at ( 0,1) [circle,draw] {$2$};
  \node (c)  at ( 2/3,-1) [circle,draw] {$6$};
  \node (d)  at ( -2/3,-1) [circle,draw] {$5$};
    \node (e)  at ( 0,0) [circle,draw] {$1$};
      \node (f)  at ( 1,0) [circle,draw] {$4$};    
 \draw [->] (b) to node[auto] {$ $} (a);
 \draw [->] (a) to node[auto] {$ $} (c);
 \draw [->] (a) to node[auto] {$ $} (d);
  \draw [->>] (e) to node[auto] {$ $} (b);
 \draw [->>] (c) to node[auto] {$ $} (e);
 \draw [->>] (d) to node[auto] {$  $} (e);
 \draw [->] (e) to node[auto] {$ $} (f);
\draw [->] (b) to node[auto] {$  $} (f);
\draw [->] (f) to node[auto] {$  $} (c);
\draw [->] (f) to node[auto] {$  $} (d);
\draw [->] (e) to node[auto] {$ $} (a);
\end{tikzpicture}
\end{center}
with superpotential \cite{Beasley:2001zp,Feng:2002zw} (see also model 10c in \cite{Hanany:2012hi})
\bea
W&=&\Phi_{12}^2\ \Phi_{23} \Phi_{36} \Phi_{61}^2 
+\Phi_{12}^2\ \Phi_{24} \Phi_{45} \Phi_{51} 
-\Phi_{12}\Phi_{23} \Phi_{35} \Phi_{51}^2 
   - \Phi_{12} \ \Phi_{24} \Phi_{46} \Phi_{61}\nn\\&&
   -\Phi_{14} \Phi_{45} \Phi_{51}^2+\Phi_{14} \Phi_{46} 
   \Phi _{61}^2+\Phi_{13} \Phi_{35} \Phi_{51}-\Phi_{13} \Phi_{36} \Phi_{61}
      \eea
A general Chern character $\gamma=[N,c_1,\ch_2]$ decomposes as $\gamma=
\sum_i n_i \gamma_i$ with 
\bea
 n_1 &=& \frac{1}{2} (c_{1,1}+c_{1,2}+c_{1,3}+3 c_{1,H})+ \ch_2+ N \nn\\
 n_2 &=& \frac{1}{2} (-c_{1,1}+c_{1,2}+c_{1,3}+c_{1,H})+ \ch_2 \nn\\
 n_3 &=& \frac{1}{2} (c_{1,1}+c_{1,2}+c_{1,3}+c_{1,H})+ \ch_2 \nn\\
 n_4 &=& \frac{1}{2} (-c_{1,1}-c_{1,2}-c_{1,3}-c_{1,H})+ \ch_2 \nn\\
 n_5 &=& \frac{1}{2} (-c_{1,1}+c_{1,2}-c_{1,3}-c_{1,H})+\ch_2 \nn\\
 n_6 &=& \frac{1}{2} (-c_{1,1}-c_{1,2}+c_{1,3}-c_{1,H})+ \ch_2 
\eea
or conversely
\bea
 N &=& n_1-n_2-n_3-n_4+n_5+n_6 \ ,\quad   \ch_2 = \frac{1}{2} \left(n_3+n_4\right)  \nn\\
 c_{1,H} &=& n_2+n_4-n_5-n_6 \ ,\quad %\nn\\
 c_{1,1} = n_3-n_2 \ ,\quad 
 c_{1,2} = n_5-n_4 \ ,\quad 
 c_{1,3} = n_6-n_4 
 \eea
As usual we take $\vec n=-\vec N$. The stability parameter  $\vec\zeta=\vec\zeta^c$ in \eqref{zetacan} for $J\propto c_1(S)$  is
\bea
 \zeta_1 &=& -\deg  \rho  + \ch_2 \nn\\
 \zeta_2 &=& (\deg +2 N) \rho - \ch_2 \nn\\
 \zeta_{3,4} &=& (\deg +N) \rho -\ch_2-\frac{N}{2} \nn\\
 \zeta_{5,6} &=& -(\deg +2 N) \rho +  \ch_2 
\eea
In the chamber where $\Phi_{51}=\Phi_{61}=0$, the expected dimension
\begin{multline}
    d_\IC = 1-\sum_i N_i^2 +2N_1 N_2+(N_1+N_2)(N_3+N_4)+(N_3+N_4)(N_5+N_6) - 2 N_1(N_5+N_6)
\end{multline}
agrees with \eqref{dimM}.
This requires $\zeta_{1}\geq 0$, $\zeta_{6}\leq 0$ hence
\be
-2N \leq  \deg \leq 0
\ee
in agreement with the range of slopes in the collection $\cC^\vee$.

\medskip

In the attractor chamber,  depending on the signs of the $\zetastar_i$'s we have the following possibilities (corresponding to the perfect matchings $s_8,s_1,s_6,s_3$ in \cite{Hanany:2012hi}): 
\begin{itemize}
\item $\Phi^\alpha_{51}=\Phi_{61}^\alpha=0$ when $\zetastar_{1,2}\geq 0, \zetastar_{5,6}\leq 0$, with dimension 
\be
d_\IC=1-\cQ(\vec N) - N_1\zetastar_1- N_2\zetastar_2
\ee
\item $\Phi_{35}=\Phi_{36}=\Phi_{45}=\Phi_{46}=0$ when $\zetastar_{3,4}\leq 0, \zetastar_{5,6}\geq 0$,  with dimension 
\be
d'_\IC=1-\cQ(\vec N) + N_3\zetastar_3+ N_4\zetastar_4 - N_5\zetastar_5- N_6\zetastar_6
\ee
\item $\Phi_{13}=\Phi_{14}=\Phi_{23}=\Phi_{24}=0$  when $\zetastar_{1,2}\leq 0, \zetastar_{3,4}\geq 0$,  with dimension 
\be
d''_\IC=1-\cQ(\vec N) + N_1\zetastar_1 + N_2\zetastar_2
\ee
\item $\Phi_{12}^\alpha=\Phi_{13}=\Phi_{14}=0$ when $\zetastar_1\leq 0, \zetastar_{2}\geq 0$, with dimension 
\be
d'''_\IC=1-\cQ(\vec N) + N_1\zetastar_1- N_2\zetastar_2
\ee
\end{itemize}
where $\cQ$ is  the positive quadratic form 
\be 
\cQ(\vec N)=(N_1-N_2)^2 + 
\frac12 \big[ (N_5-N_3)^2 + (N_5-N_4)^2 \\+ (N_6-N_3)^2 + (N_6-N_4)^2 \big] 
\ee
Thus the expected dimension is strictly negative unless $N_1=N_2$ and $N_3=N_4=N_5=N_6$. In fact, as in the previous case one can check that it is positive only when all $N_i$'s are equal,
corresponding to a pure D0-brane, or for dimension vectors corresponding to simple representations.

\medskip

After left-mutation on the node 4 and exchanging the nodes $4$ and $6$,  we get the model II above. 
After left-mutation on the node 2 and applying the permutation $142356$ to the nodes, we get the model IV below.

\subsubsection{Model IV}

We finally consider the three-block exceptional collection from \cite{karpov1998three},
also studied in \cite[\S3.1]{Wijnholt:2002qz},
\be
\left( \cO, \, \cO(H), \, \cO(2H-C_1-C_2-C_3), \, \cO(2H-C_2-C_3)\ ,\cO(2H-C_1-C_3)\ , \cO(2H-C_1-C_2) \right)
\ee
 The Chern vectors of the projective and simple representations
are 
\be
\label{gdP3IV}
\begin{array}{ccl}
 \gamma^1 &=& \[1,\(0,0,0,0\),0\] \\
 \gamma^2 &=& \[1,\(1,0,0,0\),\frac{1}{2}\] \\
 \gamma^3 &=& \[1,\(2,-1,-1,-1\),\frac{1}{2}\] \\
 \gamma^4 &=& \[1,\(2,0,-1,-1\),1\] \\
 \gamma^5 &=& \[1,\(2,-1,0,-1\),1\] \\
 \gamma^6 &=& \[1,\(2,-1,-1,0\),1\] \\
\end{array}
\qquad
\begin{array}{ccl}
 \gamma_1 &=& \[1,(0,0,0,0),0\] \\
 \gamma_2 &=& \[-2,(1,0,0,0),\frac{1}{2}\] \\
 \gamma_3 &=& \[-2,(2,-1,-1,-1),\frac{1}{2}\] \\
 \gamma_4 &=& \[1,(-1,1,0,0),0\] \\
 \gamma_5 &=& \[1,(-1,0,1,0),0\] \\
 \gamma_6 &=& \[1,(-1,0,0,1),0\] \\
\end{array}
\ee
with slope $0,3,4$ and $0,-3/2,-2$. 
The Euler matrix has a three-block form
\be
\label{SdP3IV}
S=\left(
\begin{array}{ccc}
1 & 3 & 4 \\
 0 &  \mathbb{1}_2 & 1 \\
 0 & 0 &  \mathbb{1}_3 \\
\end{array}
\right),\quad 
S^\vee=\left(
\begin{array}{ccc}
 1 & -3 & 2 \\
 0 &  \mathbb{1}_2 & -1 \\
 0 & 0 &  \mathbb{1}_3 \\
\end{array}
\right), \quad 
\kappa=\left(
\begin{array}{ccc}
 0 & 3 & -2 \\
 -3 &  \mathbb{0}_2 & 1 \\
 2 & -1 & \mathbb{0}_3 \\
\end{array}
\right)
\ee
corresponding to the quiver
 
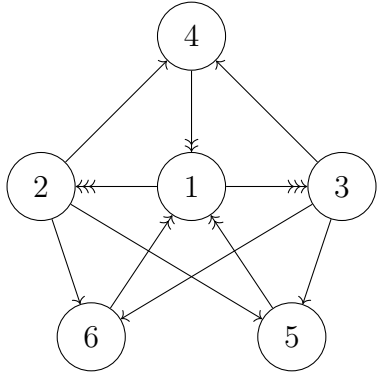
\begin{center}
\begin{tikzpicture}[inner sep=2mm,scale=2]
  \node (a) at ( -1,0) [circle,draw] {$2$};
  \node (b) at ( 0,1) [circle,draw] {$4$};
  \node (c)  at ( 2/3,-1) [circle,draw] {$5$};
  \node (d)  at ( -2/3,-1) [circle,draw] {$6$};
    \node (e)  at ( 0,0) [circle,draw] {$1$};
      \node (f)  at ( 1,0) [circle,draw] {$3$};    
 \draw [->] (a) to node[auto] {$ $} (b);
 \draw [->] (a) to node[auto] {$ $} (c);
 \draw [->] (a) to node[auto] {$ $} (d);
  \draw [->>] (b) to node[auto] {$ $} (e);
 \draw [->>] (c) to node[auto] {$  $} (e);
 \draw [->>] (d) to node[auto] {$ $} (e);
 \draw [->>>] (e) to node[auto] {$  $} (f);
\draw [->] (f) to node[auto] {$ $} (b);
\draw [->] (f) to node[auto] {$  $} (c);
\draw [->] (f) to node[auto] {$  $} (d);
\draw [->>>] (e) to node[auto] {$ $} (a);
\end{tikzpicture}
\end{center}
The  superpotential was obtained in \cite{Feng:2002zw} (see also model 10d in \cite{Hanany:2012hi}),
\be
\begin{split}
W=&
    \Phi _{12}^1 \Phi _{24} \Phi _{41}^1 
      - \Phi _{12}^2 \Phi _{24} \Phi _{41}^2
 + \Phi _{12}^3 \Phi _{25} \Phi _{51}^1
   - \Phi _{12}^1 \Phi _{25} \Phi _{51}^2
   -\Phi _{12}^3 \Phi _{26} \Phi _{61}^2 
   +\Phi _{12}^2 \Phi _{26} \Phi _{61}^1 
  \nn\\& 
   +\Phi _{13}^1 \Phi _{34} \Phi _{41}^2
      -\Phi _{13}^2\Phi _{34} \Phi _{41}^1
   +\Phi_{13}^3 \Phi _{35} \Phi _{51}^2
      -\Phi _{13}^1 \Phi _{35} \Phi _{51}^1
   +\Phi _{13}^2 \Phi _{36} \Phi _{61}^2
   -\Phi _{13}^3 \Phi _{36} \Phi_{61}^1
   \end{split}
\ee
A general Chern character $\gamma=[N,c_1,\ch_2]$ decomposes as $\gamma=
\sum_i n_i \gamma_i$ with  
\bea
n_1&=&
\frac{1}{2} (c_{1,1}+c_{1,2}+c_{1,3}+3 c_{1,H})+ \ch_2+ N, \nn\\
n_2&=&   \frac{1}{2} (c_{1,1}+c_{1,2}+c_{1,3}+c_{1,H})+  \ch_2,\nn\\
n_3&=&   \frac{1}{2} (-c_{1,1}-c_{1,2}-c_{1,3}-c_{1,H})+ 
   \ch_2,\nn\\
n_4&=&   \frac{1}{2} (c_{1,1}-c_{1,2}-c_{1,3}-c_{1,H})+
   \ch_2,\nn\\
n_5&=&   \frac{1}{2} (-c_{1,1}+c_{1,2}-c_{1,3}-c_{1,H})+
   \ch_2,\nn\\
n_6&=&   \frac{1}{2} (-c_{1,1}-c_{1,2}+c_{1,3}-c_{1,H})+ \ch_2
\eea
or conversely
\bea
N&=&n_{1}-2 n_{2}-2 n_{3}+n_{4}+n_{5}+n_{6}, \quad 
\ch_2=\frac{n_{2}+n_{3}}{2} \\
c_{1,H}&=&n_{2}+2 n_{3}-n_{4}-n_{5}-n_{6}, \quad 
c_{1,1}=n_{4}-n_{3},\quad
c_{1,2}=n_{5}-n_{3},\quad
c_{1,3}=n_{6}-n_{3} \nn
\eea
As usual we set $\vec N=-\vec n$. 
    The total dimensions for the three blocks
 are then 
 \bea
 \cN_1 &=& - \frac{1}{2}\deg- \ch_2-  N \ ,\quad 
  \cN_2 =-2 \ch_2 \ ,\quad  
   \cN_3 = \frac{1}{2} \deg - 3 \ch_2 \nn\\
 \cN &=&-6 \ch_2-N
  \eea
while the degree is
\be
\deg= 3c_{1,H}+\sum_i c_{1,i} = -3 N_2-3 N_3+2 N_4+2 N_5+2 N_6
\ee
 The stability parameters  $\vec\zeta=\vec\zeta^c$ in \eqref{zetacan} for $J\propto c_1(S)$ are 
\bea
\zeta_1&=& -\rho\, \deg +\ch_2 \nn\\
\zeta_{2,3}&=& \rho(2 \deg+3 N ) -2\ch_2 -\frac{N}{2}\nn\\
\zeta_{4,5,6} &=& -\rho( \deg +2 N) +\ch_2
\eea
The dimension agrees with \eqref{dimM} in the chamber where $\Phi_{41}=\Phi_{51}=\Phi_{61}=0$. This requires $\zeta_1\geq 0, \zeta_{4,5,6}\leq 0$ hence
\be
-2N \leq  \deg   \leq 0 \ ,
\ee
in agreement with the range of slopes of the stable objects. The vanishing of the attractor
indices follows from the arguments in \S\ref{sec_threeblockatt}.  
Alternatively, we can write the expected dimension in the three possible chambers as
\begin{itemize}
\item In the chamber $\Phi_{41}=\Phi_{51}=\Phi_{61}=0$, valid for $\zetastar_1\geq 0, \zetastar_{4,5,6}\leq 0$,
\bea
d_\IC &=& 1 - \cQ(\vec N)  - N_1 \zetastar_1 + \frac13(N_4+N_5+N_6) \zetastar_4 
\eea
\item In the chamber $\Phi_{12}=\Phi_{13}=0$, 
valid for $ \zetastar_{2,3}\geq 0, \zetastar_1\leq 0$,
\bea
 d'_\IC &=&  1 - \cQ(\vec N)  +\frac23 N_1\zetastar_1  - \frac13(N_2+N_3) \zetastar_2 
\eea
\item In the chamber $\Phi_{ij}=0$ with $i=2,3$, $j=4,5,6$, valid for $\zetastar_{4,5,6}\geq 0$, 
$\zetastar_{2,3}\leq 0$, 
\bea
 d''_\IC &=&  1 - \cQ(\vec N) + (N_2+N_3) \zetastar_2  
 - \frac23(N_4+N_5+N_6) \zetastar_4 
 \eea
 \end{itemize}
 where $\cQ$ is the positive quadratic form
 \bea
 \cQ(\vec N) &=& \frac13 \sum_{i=1}^3 \sum_{j=4}^6 (N_i-N_j)^2
  \eea
 The dimension is manifestly negative, except for dimension vectors corresponding to 
 simple representations or D0-branes.
 Computing the
attractor index using  the Coulomb branch formula  \eqref{CBform}, 
we find evidence that single-centered invariants also vanish.

\medskip

For $\gamma=[1;0,0,0,0;1]$ we get from the
Reineke formula (since the resulting quiver has no loops)  the expected result in the canonical chamber
\be
\Omega(0,1,1,1,1,1,\vec\zeta^c) = y^2 + 4 + 1/y^2 \ .
\ee
The same result is expected to hold for any $J'$ with $J'\cdot c_1(S)>0$.  

\medskip

In the blow-up chamber $J'\propto H_\epsilon$, one should instead use 
\be
\vec\zeta^H=(-c_{1,H},2c_{1,H}+N,2c_{1,H}+2N,- c_{1,H}-N,- c_{1,H}-N,- c_{1,H}-N)
+\epsilon \vec\zeta^c\ ,
\ee
which is consistent with $\zeta_1\geq 0,\zeta_{4,5,6}\leq 0$ provided $-N\leq c_{1,H}\leq 0$. 
From the flow tree formula we get 
\be
\begin{array}{|c|c|l|}
\hline
\[N; c_1 ;c_2\] & \vec N &\Omega(\vec N, \vec\zeta^H)  \\ \hline
\[2;-1,0,0,0;1\] &(0,1,0,0,0,0,0)&1 \nn \\
 \[2;-1,1,0,0;1\] &(0,1,1,0,1,1)&-y-1/y \nn \\
 \[2;-1,1,1,0;1\] &(0,1,2,1,1,2)&y^2+ 2+ 1/y^2  \nn\\
 \[2;0,-1,0,0;1\] &(0,2,1,2,1,1)&y^2+3+1/y^2 \nn\\
 \[2;0,-1,-1,-1;0\] &(1,3,0,1,1,1)&1  \nn\\
 \[2;-2,1,1,1;1\] &(0,0,1,0,0,0)&1  \nn\\
 \[2;-2,1,1,1;2\] &(1,1,2,1,1,1)&y^4+5y^2+12+\dots  \nn\\
 \[2;-2,1,1,0;0\] &(1,1,1,0,0,1)&-y^3-3y-\dots   \nn\\
 \[3;-3,1,1,1;4\] & (1,1,1,0,0,0) & y^4+2y^2+3+\dots\nn\\
 \[3;-3,2,1,1;3\] & (1,1,2,0,1,1) &  y^4+4y^2+8+\dots
    \\ \hline
    \end{array}
\ee
in agreement with the analysis in  \S\ref{sec_VWdP}.

\medskip

After right-mutation on the node 4 and applying the permutation 
$134256$, on the nodes, we get the model III above.

\medskip

\subsection{$dP_4$}

The fourth del Pezzo surface is no longer toric, but it admits 
a three-block strong exceptional collection constructed in \cite{karpov1998three}, and further
studied in \cite[\S3.2]{Wijnholt:2002qz}, 
\be
\begin{split} 
\cC= \left( \cO,\,F, \, \cO(H), \, \cO(2H-C_2-C_3-C_4), \, 
 \cO(2H-C_1-C_3-C_4), \, \right. \\
 \left.   \cO(2H-C_1-C_2-C_4), \,  \cO(2H-C_1-C_2-C_3) \right) \, 
 \end{split}
\ee
where $F$ is a rank 2 bundle  with Chern character $[2,(3,-1,-1,-1,-1),1/2]$, 
defined by the short exact sequence 
\be
0\to \cO(2H-C_1-C_2-C_3-C_4)  \to F \to \cO(H) \to 0
\ee 
The Chern vectors of the objects $E^i$ and dual objects $E_i^\vee$ are
\be
\begin{array}{ccl}
 \gamma^1 &=& \[1,\(0,0,0,0,0\),0\] \\
 \gamma^2 &=& \[2,\(3,-1,-1,-1,-1\),\frac{1}{2}\] \\
 \gamma^3 &=& \[1,\(1,0,0,0,0\),\frac{1}{2}\] \\
 \gamma^4 &=& \[1,\(2,0,-1,-1,-1\),\frac{1}{2}\] \\
 \gamma^5 &=& \[1,\(2,-1,0,-1,-1\),\frac{1}{2}\] \\
 \gamma^6 &=& \[1,\(2,-1,-1,0,-1\),\frac{1}{2}\] \\
 \gamma^7 &=& \[1,\(2,-1,-1,-1,0\),\frac{1}{2}\] \\
\end{array}
\qquad
\begin{array}{ccl}
 \gamma_1 &=& \[1,\(0,0,0,0,0\),0\] \\
 \gamma_2 &=& \[-3,\(3,-1,-1,-1,-1\),\frac{1}{2}\] \\
 \gamma_3 &=& \[1,\(-2,1,1,1,1\),0\] \\
 \gamma_4 &=& \[1,\(-1,1,0,0,0\),0\] \\
 \gamma_5 &=& \[1,\(-1,0,1,0,0\),0\] \\
 \gamma_6 &=& \[1,\(-1,0,0,1,0\),0\] \\
 \gamma_7 &=& \[1,\(-1,0,0,0,1\),0\] \\
\end{array}
\ee
with slope $0,5/2,3$ and $0,-5/3,-2$, respectively.
The    Euler form has the three-block structure
\be
S=\left(
\begin{array}{ccc}
 1 & 5 & 3 \\
 0 & 1 & 1 \\
 0 & 0 & \mathbb{1}_5 \\
\end{array}
\right),\ S^\vee=
\left(
\begin{array}{ccc}
 1 & -5 & 2 \\
 0 & 1 & -1 \\
 0 & 0 & \mathbb{1}_5  \\
\end{array}
\right),\ 
\kappa=
\left(
\begin{array}{ccc}
 0 & 5 & -2 \\
 -5 & 0 & 1 \\
 2 & -1 & \mathbb{0}_5  \\
\end{array}
\right)
\ee
A general Chern character $\gamma=[N,c_1,\ch_2]$ decomposes as $\gamma=
\sum_i n_i \gamma_i$ with  
\bea
n_1&=&
\frac{1}{2} (c_{1,1}+c_{1,2}+c_{1,3}+c_{1,4}+3 c_{1,H})+ \ch_2+N,\nn\\
n_2&=&   2 \ch_2,\nn\\
n_3&=&   \frac{1}{2}
   (c_{1,1}+c_{1,2}+c_{1,3}+c_{1,4}+c_{1,H})+ \ch_2\nn\\
n_4&=&   \frac{1}{2}
   (c_{1,1}-c_{1,2}-c_{1,3}-c_{1,4}-c_{1,H})+\ch_2\nn\\
n_5&=&   \frac{1}{2}
   (-c_{1,1}+c_{1,2}-c_{1,3}-c_{1,4}-c_{1,H})+ \ch_2\nn\\
n_6&=&   \frac{1}{2}
   (-c_{1,1}-c_{1,2}+c_{1,3}-c_{1,4}-c_{1,H})+ \ch_2\nn\\
n_7&=&   \frac{1}{2}
   (-c_{1,1}-c_{1,2}-c_{1,3}+c_{1,4}-c_{1,H})+ \ch_2 
\eea
or conversely
\bea
N&=&n_{1}-3 n_{2}+n_{3}+n_{4}+n_{5}+n_{6}+n_{7},\quad 
\ch_2=   \frac{n_{2}}{2}\nn\\
c_{1,H}&=&3 n_{2}-2  n_{3}-n_{4}-n_{5}-n_{6}-n_{7},\quad 
c_{1,1}=-n_{2}+n_{3}+n_{4}, \nn\\
c_{1,2}&=&-n_{2}+   n_{3}+n_{5}, \quad 
c_{1,3}=-n_{2}+n_{3}+n_{6}, \quad 
c_{1,4}=-n_{2}+n_{3}+n_{7},
\eea
 The relevant dimension vector is then $N_i=-n_i$.
   The total dimensions for the three blocks and the degree
 are then 
 \bea
 \cN_1 &=& -\frac{1}{2}\deg-\ch_2-   N\ ,\quad 
  \cN_2 =-2 \ch_2  \ ,\quad 
   \cN_3 =\frac{1}{2} \deg -5 \ch_2 \nn\\
   \deg&=& 3c_{1,H}+\sum_i c_{1,i} = -5 N_2+2 N_3+2 N_4+2 N_5+2 N_6+2 N_7
% \cN &=& -8 \ch_2-N\
  \eea
The stability parameters  $\vec\zeta=\vec\zeta^c$ in \eqref{zetacan} for $J\propto c_1(S)$ are 
\bea
\zeta_1&=&-\rho \,\deg  +\ch_2\ ,\quad % \nn\\
\zeta_2=\rho( 3\deg +5 N) -3\ch_2 -\frac{N}{2}\ ,\quad 
\zeta_{3,4,5,6,7}=   -\rho( \deg +2 N)+\ch_2 \nn\\
\eea
The dimension agrees with \eqref{dimM} in the chamber where $\Phi_{13}=\Phi_{14}=\Phi_{15}=\Phi_{16}=\Phi_{17}=0$. This requires $\zeta_1\geq  0, \zeta_{3,4,5,6,7}\leq  0$ hence
\be
-2N \leq  \deg   \leq 0 
\ee
in agreement with the range of slopes of the stable objects. The vanishing of the attractor
indices follows from the arguments in \S\ref{sec_threeblockatt}. Computing the
attractor index using  the Coulomb branch formula  \eqref{CBform}, 
we find evidence that single-centered invariants also vanish.

\medskip

 For $\gamma=[1;0,0,0,0,0;1]$ we get from the
Reineke formula (since the resulting quiver has no loops)  the expected result 
\be
\Omega(0,2,1,1,1,1,1) = y^2 + 5 + 1/y^2 
\ee
In the blow-up chamber $J'\propto H_\epsilon$, one should instead use 
\be
\vec\zeta^H=(-c_{1,H}, 3 c_{1,H} + 3 N, -c_{1,H} - 2 N, -c_{1,H} - N, -c_{1,H} - N, -c_{1,H} - 
  N, -c_{1,H} - N)+\epsilon \vec\zeta^c
  \ee
  so that $\sum_i n'_i \zeta_i = N c'_{1,H}-N' c_{1,H}$. This is consistent with 
$\zeta_1\geq  0,\zeta_{3,4,5,6,7}\leq  0$  provided $-N\leq c_{1,H}\leq  0$. 
From the flow tree formula we get
\be
\begin{array}{|c|c|l|}
\hline
\[N; c_1 ;c_2\] &\vec N &\Omega(\vec N, \vec\zeta^H)  \\ \hline
 \[2;-1,0,0,0,0;1\] &(0,1,1,0,0,0,0)&1 \nn \\
 \[2;-1,1,0,0,0;1\] &(0,2,1,0,1,1,1)&-y-1/y \nn \\
 \[2;-1,1,1,0,0;1\] &(0,3,1,1,1,2,2)&y^2+ 2+ 1/y^2  \nn\\
 \[2;-2,1,1,1,0;1\] &(0,1,0,0,0,0,1)&1  \nn\\
 \[2;-2,1,1,1,1;1\] &(0,2,0,1,1,1,1)&-y-1/y  \nn\\
 \[2;-2,1,1,0,0;2\] &(1,2,1,0,0,1,1)&-y^3-4y-\dots \nn\\
 \[3;-2,0,0,0,0;3\] & (1,2,2,0,0,0,0) & y^2+1+1/y^2 \nn\\
 \[3;-3,1,1,1,0;4\] & (1,2,1,0,0,0,1) & y^4+3y^2+4+\dots
    \\ \hline
    \end{array}
\ee
in agreement with the analysis in  \S\ref{sec_VWdP}.

\subsection{$dP_5$}
We consider the three-block strong exceptional collection from \cite{karpov1998three}, also studied in 
\cite[\S3.3]{Wijnholt:2002qz}
\be
\begin{split}
\cC= \left( \cO(C_4), \, \cO(C_5), \, \cO(H), \, \cO(2H-C_1-C_2-C_3), \  
\cO(3H-C_1-C_2-C_3-C_4),  \right. \\ 
\left. 
 \cO(2H-C_1-C_2), \,  \cO(2H-C_2-C_3), \,  \cO(2H-C_1-C_3) \right)\ . \, 
 \end{split}
\ee
The Chern vectors of the objects $E^i$ and dual objects $E_i^\vee$ are
\be
\label{gdP5}
\begin{array}{ccl}
 \gamma^1 &=& \[1,\(0,0,0,0,1,0\),-\frac{1}{2}\] \\
 \gamma^2 &=& \[1,\(0,0,0,0,0,1\),-\frac{1}{2}\] \\
 \gamma^3 &=& \[1,\(1,0,0,0,0,0\),\frac{1}{2}\] \\
 \gamma^4 &=& \[1,\(2,-1,-1,-1,0,0\),\frac{1}{2}\] \\
 \gamma^5 &=& \[1,\(3,-1,-1,-1,-1,-1\),2\] \\
 \gamma^6 &=& \[1,\(2,-1,-1,0,0,0\),1\] \\
 \gamma^7 &=& \[1,\(2,0,-1,-1,0,0\),1\] \\
 \gamma^8 &=& \[1,\(2,-1,0,-1,0,0\),1\] \\
\end{array}
\qquad
\begin{array}{ccl}
 \gamma_1 &=& \[1,\(0,0,0,0,1,0\),-\frac{1}{2}\] \\
 \gamma_2 &=& \[1,\(0,0,0,0,0,1\),-\frac{1}{2}\] \\
 \gamma_3 &=& \[-3,\(1,0,0,0,-2,-2\),\frac{5}{2}\] \\
 \gamma_4 &=& \[-3,\(2,-1,-1,-1,-2,-2\),\frac{5}{2}\] \\
 \gamma_5 &=& \[1,\(0,0,0,0,0,0\),0\] \\
 \gamma_6 &=& \[1,\(-1,0,0,1,1,1\),-1\] \\
 \gamma_7 &=& \[1,\(-1,1,0,0,1,1\),-1\] \\
 \gamma_8 &=& \[1,\(-1,0,1,0,1,1\),-1\] \\
\end{array}
\ee
with slope $1,3,4$ and $1, 1/3, 0$, respectively. The Euler matrix has a three-block structure
\be
\label{SdP5}
S=\left(
\begin{array}{ccc}
 \mathbb{1}_2  & 2 & 3 \\
 0 & \mathbb{1}_2  & 1 \\
 0 & 0 & \mathbb{1}_4  \\
\end{array}
\right),\ 
S^\vee=\left(
\begin{array}{ccc}
  \mathbb{1}_2 & -2 & 1 \\
 0 &  \mathbb{1}_2 & -1 \\
 0 & 0 &  \mathbb{1}_4 \\
\end{array}
\right),\ 
\kappa=\left(
\begin{array}{ccc}
  \mathbb{0}_2 & 2 & -1 \\
 -2 &  \mathbb{0}_2 & 1 \\
 1 & -1 &  \mathbb{0}_4 \\
\end{array}
\right)\
\ee
Note that the superpotential may in principle depend on the complex structure of $dP_5$. 
A general Chern character $\gamma=[N,c_1,\ch_2]$ decomposes as $\gamma=
\sum_i n_i \gamma_i$ with  
\bea
n_1&=&
\frac{1}{2} (c_{1,1}+c_{1,2}+c_{1,3}+3 c_{1,4}+c_{1,5}+3 c_{1,H})+  \ch_2 \nn\\
n_2&=&   \frac{1}{2} (c_{1,1}+c_{1,2}+c_{1,3}+c_{1,4}+3 c_{1,5}+3 c_{1,H})+ \ch_2\nn\\
n_3&=&   \frac{1}{2}  (c_{1,1}+c_{1,2}+c_{1,3}+c_{1,4}+c_{1,5}+c_{1,H})+ \ch_2\nn\\
n_4&=&   \frac{1}{2} (-c_{1,1}-c_{1,2}-c_{1,3}+c_{1,4}+c_{1,5}-c_{1,H})+  \ch_2\nn\\
n_5&=&   \frac{1}{2} (-c_{1,1}-c_{1,2}-c_{1,3}-c_{1,4}-c_{1,5}-3 c_{1,H})+ \ch_2+ N\nn\\
n_6&=&   \frac{1}{2} (-c_{1,1}-c_{1,2}+c_{1,3}+c_{1,4}+c_{1,5}-c_{1,H})+ \ch_2\nn\\
n_7&=&   \frac{1}{2} (c_{1,1}-c_{1,2}-c_{1,3}+c_{1,4}+c_{1,5}-c_{1,H})+   \ch_2\nn\\
n_8&=&   \frac{1}{2}  (-c_{1,1}+c_{1,2}-c_{1,3}+c_{1,4}+c_{1,5}-c_{1,H})+ \ch_2
\eea
or conversely
\bea
N&=&n_{1}+n_{2}-3 n_{3}-3 n_{4}+n_{5}+n_{6}+n_{7}+n_{8},\nn\\
c_{1,H} &-& n_{3}+2  n_{4}-n_{6}-n_{7}-n_{8}, \quad c_{1,1} =n_{7}-n_{4}, \quad
c_{1,2} =n_{8}-n_{4}, \quad
c_{1,3} =n_{6}-n_{4} \nn\\
c_{1,4} &=&n_{1}-2 n_{3}-2 n_{4}+n_{6}+n_{7}+n_{8},\quad 
c_{1,5} =n_{2}-2 n_{3}-2 n_{4}+n_{6}+n_{7}+n_{8} \nn\\
\ch_2 &=&\frac{1}{2} \left(5n_{3}+5 n_{4}-n_{1}-n_{2}\right)-n_{6}-n_{7}-n_{8}
\eea
 As usual we set $\vec N=-\vec n$. 
  The total dimensions for the three blocks and degree
 are then 
 \bea
 \cN_1 &=& -c_{1,1}-c_{1,2}-c_{1,3}-2 c_{1,4}-2 c_{1,5}-3 c_{1,H}-2
   \ch_2  \nn\\
  \cN_2 &=&-c_{1,4}-c_{1,5}-2  \ch_2 \nn \\
   \cN_3 &=&c_{1,1}+c_{1,2}+c_{1,3}-c_{1,4}-c_{1,5}+3 c_{1,H}-4
   \ch_2-N\nn \\
  % \\ \cN &=& -4 c_{1,4}-4 c_{1,5}-8 \ch_2-N
\deg&=& 3c_{1,H}+\sum_i c_{1,i} = -N_1-N_2+N_3+N_4
\eea
The stability parameters  $\vec\zeta=\vec\zeta^c$ in \eqref{zetacan} for $J\propto c_1(S)$ are 
\bea
\zeta_{1,2}&=&-\rho( \deg -N) +\ch_2 + \frac{N}{2} \nn\\
\zeta_{3,4}  &= & \rho( 3 \deg-N) 
-3\ch_2 - \frac{5N}{2} \nn\\
\zeta_{5} &=& 
   -\rho \, \deg  +\ch_2 \ , \quad
   \zeta_{6,7,8} = 
   -\rho\, \deg  +\ch_2+N 
\eea
The dimension agrees with \eqref{dimM} in the chamber where $\Phi_{ij}=0$ with 
$i=5,6,7,8$ and $j=1,2$. 
This requires $\zeta_{1,2}\geq  0, \zeta_{5,6,7,8}\leq  0$ hence
\be
0 \leq \deg  \leq  N
\ee
in agreement with the range of slopes of the stable objects. 
The vanishing of the attractor
indices follows from the arguments in \S\ref{sec_threeblockatt}. Computing the
attractor index using  the Coulomb branch formula  \eqref{CBform}, 
we find evidence that single-centered invariants also vanish.

\medskip

For $\gamma=[1;0,0,0,0,0,0;1]$  in the canonical chamber, we get from the
flow tree formula the expected result
\be
\Omega(1,1,1,1,0,1,1,1,\vec\zeta^c) = y^2 + 6 + 1/y^2 \ .
\ee
The same result is expected for any polarization such that $J'\cdot c_1(S)>0$. 

\medskip
In the blow-up chamber $J'\propto H_\epsilon$, one should instead use 
\be
\vec\zeta^H=
(-c_{1,H}, -c_{1,H}, 3 c_{1,H} + N,  3 c_{1,H} + 2 N, -c_{1,H}, -c_{1,H} - N, -c_{1,H} - N, -c_{1,H} - N)
 +\epsilon \vec\zeta^c
 \ee
This is consistent with $\zeta_{1,2}\geq  0,\zeta_{5,6,7,8}\leq  0$ only when $c_{1,H}=0$. From the flow tree formula we get 
\be
\begin{array}{|c|c|l|}
\hline
\[N; c_1 ;c_2\] & \vec N  &\Omega(\vec N, \vec\zeta^H)  \\ 
\hline
 \[2;0,0,0,0,0,1;1\] &(1,0,1,1,0,1,1,1) &y^2+5+1/y^2\nn\\ 
\[2;0,0,0,0,1,1;1\] &(0,0,1,1,1,1,1,1) & -y^3-5y-\dots \nn\\
\[2;0,-1,0,0,1,1;0\] & (0,0,1,0,0,0,1,0) & 1 \nn\\
\[2;0,-1,-1,0,1,1;0\] & (1,1,2,0,0,0,1,1) & -y-1/y \nn\\
\[3;0,0,0,0,1,1;1\] & (0,0,1,1,0,1,1,1) & y^2+4+1/y^2
    \\ \hline
    \end{array}
\ee
in agreement with the analysis in  \S\ref{sec_VWdP}.

\medskip

After mutating with respect to nodes 5 and 8, in that order, and applying the cyclic permutation $(2, 6, 5, 4, 3, 7, 8)$ (or one of its images under the dihedral group $D_8$, which is a symmetry of the resulting quiver) one obtains the same quiver 
as in \cite[(5.5)]{Hanany:2001py}.

\subsection{$dP_6$}

\subsubsection{Three-block collection (6.1) of type $(1,1,1)$}
We consider the three-block exceptional collection (6.1) from \cite{karpov1998three} (also studied in 
\cite[\S3.4]{Wijnholt:2002qz})
\be
\begin{split}
\cC=& \left( \cO(C_4), \, \cO(C_5), \, \cO(C_6),  \cO(H-C_1), \, \cO(H-C_2), \,   \cO(H-C_3), \right. \\
&\left. 
\cO(3H-C_1-C_2-C_3-C_4-C_5-C_6), \, 
 \cO(H), \,  \cO(2H-C_1-C_2- C_3) \right) 
 \end{split}
\ee
The Chern vectors of the objects $E^i$ and dual objects $E_i^\vee$ are 
\be
\label{gdP61}
\begin{array}{ccl}
 \gamma^1 &=& \[1,\(0,0,0,0,1,0,0\),-\frac{1}{2}\] \\
 \gamma^2 &=& \[1,\(0,0,0,0,0,1,0\),-\frac{1}{2}\] \\
 \gamma^3 &=& \[1,\(0,0,0,0,0,0,1\),-\frac{1}{2}\] \\
 \gamma^4 &=& \[1,\(1,-1,0,0,0,0,0\),0\] \\
 \gamma^5 &=& \[1,\(1,0,-1,0,0,0,0\),0\] \\
 \gamma^6 &=& \[1,\(1,0,0,-1,0,0,0\),0\] \\
 \gamma^7 &=& \[1,\(3,-1,-1,-1,-1,-1,-1\),\frac{3}{2}\] \\
 \gamma^8 &=& \[1,\(1,0,0,0,0,0,0\),\frac{1}{2}\] \\
 \gamma^9 &=& \[1,\(2,-1,-1,-1,0,0,0\),\frac{1}{2}\] \\
\end{array}
\qquad
\begin{array}{ccl}
 \gamma_1 &=& \[1,\(0,0,0,0,1,0,0\),-\frac{1}{2}\] \\
 \gamma_2 &=& \[1,\(0,0,0,0,0,1,0\),-\frac{1}{2}\] \\
 \gamma_3 &=& \[1,\(0,0,0,0,0,0,1\),-\frac{1}{2}\] \\
 \gamma_4 &=& \[-2,\(1,-1,0,0,-1,-1,-1\),\frac{3}{2}\] \\
 \gamma_5 &=& \[-2,\(1,0,-1,0,-1,-1,-1\),\frac{3}{2}\] \\
 \gamma_6 &=& \[-2,\(1,0,0,-1,-1,-1,-1\),\frac{3}{2}\] \\
 \gamma_7 &=& \[1,\(0,0,0,0,0,0,0\),0\] \\
 \gamma_8 &=& \[1,\(-2,1,1,1,1,1,1\),-1\] \\
 \gamma_9 &=& \[1,\(-1,0,0,0,1,1,1\),-1\] \\
\end{array}
\ee
with slope $1,2,3$ and $1,1/2,0$, The Euler matrix has a three-block form
\be
\label{SdP61}
S=\left(
\begin{array}{ccc}
  \mathbb{1}_3 & 1 & 2 \\
 0 &   \mathbb{1}_3  & 1 \\
 0 & 0 &   \mathbb{1}_3  \\
\end{array}
\right),\ 
S^\vee=\left(
\begin{array}{ccc}
   \mathbb{1}_3  & -1 & 1 \\
 0 &   \mathbb{1}_3  & -1 \\
 0 & 0 &   \mathbb{1}_3  \\
\end{array}
\right), 
\kappa=\left(
\begin{array}{ccc}
   \mathbb{0}_3  & 1 & -1 \\
 -1 &   \mathbb{0}_3  & 1 \\
 1 & -1 &   \mathbb{0}_3  \\
\end{array}
\right)
\ee
A general Chern character $\gamma=[N,c_1,\ch_2]$ decomposes as $\gamma=
\sum_i n_i \gamma_i$ with  
\bea
 n_1 &=& \frac{1}{2} (c_{1,1}+c_{1,2}+c_{1,3}+3 c_{1,4}+c_{1,5}+c_{1,6}+3
   c_{1,H})+ \ch_2 \nn\\
 n_2 &=& \frac{1}{2} (c_{1,1}+c_{1,2}+c_{1,3}+c_{1,4}+3 c_{1,5}+c_{1,6}+3
   c_{1,H})+ \ch_2 \nn\\
 n_3 &=& \frac{1}{2} (c_{1,1}+c_{1,2}+c_{1,3}+c_{1,4}+c_{1,5}+3 c_{1,6}+3
   c_{1,H})+ \ch_2 \nn\\
 n_4 &=& \frac{1}{2}
   (-c_{1,1}+c_{1,2}+c_{1,3}+c_{1,4}+c_{1,5}+c_{1,6}+c_{1,H})+ \ch_2
   \nn\\
 n_5 &=& \frac{1}{2}
   (c_{1,1}-c_{1,2}+c_{1,3}+c_{1,4}+c_{1,5}+c_{1,6}+c_{1,H})+ \ch_2
   \nn\\
 n_6 &=& \frac{1}{2}
   (c_{1,1}+c_{1,2}-c_{1,3}+c_{1,4}+c_{1,5}+c_{1,6}+c_{1,H})+ \ch_2
   \nn\\
 n_7 &=& \frac{1}{2} (-c_{1,1}-c_{1,2}-c_{1,3}-c_{1,4}-c_{1,5}-c_{1,6}-3
   c_{1,H})+ \ch_2+ N \nn\\
 n_8 &=& \frac{1}{2}
   (c_{1,1}+c_{1,2}+c_{1,3}+c_{1,4}+c_{1,5}+c_{1,6}+c_{1,H})+ \ch_2
   \nn\\
 n_9 &=& \frac{1}{2}
   (-c_{1,1}-c_{1,2}-c_{1,3}+c_{1,4}+c_{1,5}+c_{1,6}-c_{1,H})+ \ch_2
\eea
or conversely
\bea
 N &=& n_1+n_2+n_3-2 n_4-2 n_5-2 n_6+n_7+n_8+n_9 \nn\\
 c_{1,H} &=& n_4+n_5+n_6-2 n_8-n_9 \nn\\
 c_{1,1} &=& n_8-n_4 \ ,\quad 
 c_{1,2} = n_8-n_5 \ ,\quad
 c_{1,3} = n_8-n_6 \nn\\
 c_{1,4} &=& n_1-n_4-n_5-n_6+n_8+n_9 \nn\\
 c_{1,5} &=& n_2-n_4-n_5-n_6+n_8+n_9 \nn\\
 c_{1,6} &=& n_3-n_4-n_5-n_6+n_8+n_9 \nn\\
 \ch_2 &=& \frac{1}{2} \left(3 n_4+3 n_5+3 n_6-n_1-n_2-n_3\right) -n_8-n_9
\eea
As usual we set $\vec N=-\vec n$. 
  The total dimensions for the three blocks and degree
 are then 
 \bea
 \cN_1 &=&\frac{1}{2} (-3 c_{1,1}-3 c_{1,2}-3 c_{1,3}-5 c_{1,4}-5 c_{1,5}-5
   c_{1,6}-9 c_{1,H})-3 \ch_2
   \nn\\
  \cN_2 &=&\frac{1}{2} (-c_{1,1}-c_{1,2}-c_{1,3}-3
   c_{1,4}-3 c_{1,5}-3 c_{1,6}-3 c_{1,H})-3 \ch_2
   \nn\\
   \cN_3 &=&\frac{1}{2}
   (c_{1,1}+c_{1,2}+c_{1,3}-c_{1,4}-c_{1,5}-c_{1,6}+3 c_{1,H}) -3
   \ch_2- N \nn\\
%   \nn\\
% \cN &=& \frac{1}{2} (-3 c_{1,1}-3 c_{1,2}-3 c_{1,3}-9 c_{1,4}-9
%   c_{1,5}-9 c_{1,6}-9 c_{1,H})-9 \ch_2- N
\deg&=& 3c_{1,H}+\sum_i c_{1,i} = -N_1-N_2-N_3+N_4+N_5+N_6
\eea
The stability parameters  $\vec\zeta=\vec\zeta^c$ in \eqref{zetacan}  for $J\propto c_1(S)$ are 
\bea
 \zeta_{1,2,3} &=& -\rho\left(  \deg -N \right) + \ch_2+\frac12 N\ , \quad 
 \zeta_{4,5,6} = \rho\left (2 \deg -N \right) -2\ch_2-\frac32 N \nn\\
 \zeta_{7} &=& -\rho\,  \deg + \ch_2\,\quad   \zeta_{8,9} = -\rho\, \deg   + \ch_2 + N 
\eea
The dimension agrees with \eqref{dimM} in the chamber where $\Phi_{ij}=0$
where $i=1,2,3$ and $j=7,8,9$. 
This requires $\zeta_{1,2,3}\geq  0, \zeta_{7,8,9}\leq  0$ hence
\be
0\leq \deg \leq N
\ee
in agreement with the range of slopes of the stable objects. 
The vanishing of the attractor
indices follows from the arguments in \S\ref{sec_threeblockatt}.  Applying the Coulomb branch formula in the attractor chamber, we find evidence that 
single-centered invariants also vanish, under the same conditions on $\vec N$.

\medskip

 For $\gamma=[1;0,0,0,0,0,0,0;1]$ in the canonical 
 chamber we get from the flow tree formula
 the expected result 
\be
\Omega(1,1,1,1,1,1,0,1,1, \vec\zeta^c) = y^2 + 7 + 1/y^2 
\ee 
The same result is expected for any polarization such that $J'\cdot c_1(S)>0$. 

\medskip
In the blow-up chamber $J'\propto H_\epsilon$, one should instead use 
\be
\vec\zeta^H=(-c_{1,H}, -c_{1,H}, -c_{1,H}, 2 c_{1,H} + N, 2 c_{1,H} + N, 
 2 c_{1,H} + N, -c_{1,H}, -c_{1,H} - 2 N, -c_{1,H} - N)+\epsilon \vec\zeta^c
 \ee
This is  consistent with $\zeta_{1,2,3}\geq  0,\zeta_{7,8,9}\leq  0$ only for $c_{1,H}=0$. From the flow tree formula we get 
\be
\begin{array}{|c|c|l|}
\hline
\[N; c_1 ;c_2\] &\vec N &\Omega(\vec N, \vec\zeta^H)  \\ \hline
\[ 2;0, -1, 0, 0, 1, 1, 1;0\] &(0, 0, 0, 0, 1, 1, 1, 1, 0) & -y-1/y \nn\\
 \[2;0,0,0,0,0,1;1\] &(1, 1, 0, 1, 1, 1, 0, 1, 1) &y^2+6+1/y^2 \nn \\
\[2;0,0,0,1,0,1;1\] &(0, 1, 0, 1, 1, 1, 1, 1, 1) &-y^3-6y-\dots  \nn\\
\[ 3;0,0,0,0,1,1,0;1 \] & (0,0,1,1,1,1,0,1,1) & y^2+5+1/y^2  \nn\\
\[3;0,-1,0,1,1,1,1;0\] & (0,0,0,0,1,2,1,1,1) & y^2 +2 +1/y^2\nn\\
\[3;0,0,0,0,1,1,1;1\] & (0,0,0,1,1,1,1,1,1) & y^4+5y^2+6+\dots 
   \\ \hline
    \end{array}
\ee
in agreement with the analysis in \S\ref{sec_VWdP}.

\medskip

It is worth noting that the quiver is symmetric under independent permutations of the nodes (123), (456), (789), and (for example) under the circular permutation $(147258369)$. The resulting group has order $648$, and is a subgroup of index $80$ inside the Weyl group of $E_6$.

\subsubsection{Three-block collection (6.2) of type $(2,1,1)$}

The second strong exceptional collection (6.2) from \cite{karpov1998three} is no longer
made of invertible sheaves, but involves a rank two sheaf $E^1$. 
%%\be
%\begin{split} 
%\cC=& \left( F,\    \cO(H)\ ,  \cO(3H-C_1-C_2-C_3-C_4-C_5-C_6), \right. \\
%&\cO(3H-C_2-C_3-C_4-C_5-C_6)\ ,\quad 
%\cO(3H-C_1-C_3-C_4-C_5-C_6)\ ,\quad 
%\\
%&\cO(3H-C_1-C_2-C_4-C_5-C_6)\ ,\quad 
%\cO(3H-C_1-C_2-C_3-C_5-C_6)\ ,\quad 
%\\
%&\left. \cO(3H-C_1-C_2-C_3-C_4-C_6)\ ,\quad 
%\cO(3H-C_1-C_2-C_3-C_4-C_5)
%\right) 
% \end{split}
%\ee
%where $F$ is a rank 2 sheaf with Chern character $[2,(1,0,0,0,0,0),-1/2]$.
The Chern vectors of the objects $E^i$ and dual objects $E_i^\vee$ are 
\be
\begin{array}{ccl}
 \gamma^1 &=& \[2,\(1,0,0,0,0,0,0\),-\frac{1}{2}\] \\
 \gamma^2 &=& \[1,\(1,0,0,0,0,0,0\),\frac{1}{2}\] \\
 \gamma^3 &=& \[1,\(3,-1,-1,-1,-1,-1,-1\),\frac{3}{2}\] \\
 \gamma^4 &=& \[1,\(3,0,-1,-1,-1,-1,-1\),2\] \\
 \gamma^5 &=& \[1,\(3,-1,0,-1,-1,-1,-1\),2\] \\
 \gamma^6 &=& \[1,\(3,-1,-1,0,-1,-1,-1\),2\] \\
 \gamma^7 &=& \[1,\(3,-1,-1,-1,0,-1,-1\),2\] \\
 \gamma^8 &=& \[1,\(3,-1,-1,-1,-1,0,-1\),2\] \\
 \gamma^9 &=& \[1,\(3,-1,-1,-1,-1,-1,0\),2\] \\
\end{array}
\qquad
\begin{array}{ccl}
 \gamma_1 &=& \[2,\(1,0,0,0,0,0,0\),-\frac{1}{2}\] \\
 \gamma_2 &=& \[-5,\(-2,0,0,0,0,0,0\),2\] \\
 \gamma_3 &=& \[-5,\(0,-1,-1,-1,-1,-1,-1\),3\] \\
 \gamma_4 &=& \[1,\(0,1,0,0,0,0,0\),-\frac{1}{2}\] \\
 \gamma_5 &=& \[1,\(0,0,1,0,0,0,0\),-\frac{1}{2}\] \\
 \gamma_6 &=& \[1,\(0,0,0,1,0,0,0\),-\frac{1}{2}\] \\
 \gamma_7 &=& \[1,\(0,0,0,0,1,0,0\),-\frac{1}{2}\] \\
 \gamma_8 &=& \[1,\(0,0,0,0,0,1,0\),-\frac{1}{2}\] \\
 \gamma_9 &=& \[1,\(0,0,0,0,0,0,1\),-\frac{1}{2}\] \\
\end{array}
\ee
with slope $3/2,3,4$ and $3/2,6/5,1$, respectively. The Euler matrix has a three-block structure,
\be
S=\left(
\begin{array}{ccc}
 1 & 3 & 5 \\
 0 &   \mathbb{1}_2  & 1 \\
 0 & 0 &   \mathbb{1}_6  \\
\end{array}
\right), \ 
S^\vee=\left(
\begin{array}{ccc}
 1 & -3 & 1 \\
 0 &   \mathbb{1}_2  & -1 \\
 0 & 0 &   \mathbb{1}_6  \\
\end{array}
\right),\ 
\kappa=\left(
\begin{array}{ccc}
 0 & 3 & -1 \\
 -3 &   \mathbb{0}_2  & 1 \\
 1 & -1 &   \mathbb{0}_6  \\
\end{array}
\right)
\ee
A general Chern character $\gamma=[N,c_1,\ch_2]$ decomposes as $\gamma=
\sum_i n_i \gamma_i$ with  
\bea
 n_1 &=& c_{1,1}+c_{1,2}+c_{1,3}+c_{1,4}+c_{1,5}+c_{1,6}+2 c_{1,H}+2
   \ch_2 \nn\\
 n_2 &=& \frac{1}{2}
   (c_{1,1}+c_{1,2}+c_{1,3}+c_{1,4}+c_{1,5}+c_{1,6}+c_{1,H})+ \ch_2
   \nn\\
 n_3 &=& \frac{1}{2} (-c_{1,1}-c_{1,2}-c_{1,3}-c_{1,4}-c_{1,5}-c_{1,6}-3
   c_{1,H})+ \ch_2+N \nn\\
 n_4 &=& \frac{1}{2} (c_{1,1}-c_{1,2}-c_{1,3}-c_{1,4}-c_{1,5}-c_{1,6}-3
   c_{1,H})+ \ch_2+N \nn\\
 n_5 &=& \frac{1}{2} (-c_{1,1}+c_{1,2}-c_{1,3}-c_{1,4}-c_{1,5}-c_{1,6}-3
   c_{1,H})+\ch_2+N \nn\\
 n_6 &=& \frac{1}{2} (-c_{1,1}-c_{1,2}+c_{1,3}-c_{1,4}-c_{1,5}-c_{1,6}-3
   c_{1,H})+\ch_2+N \nn\\
 n_7 &=& \frac{1}{2} (-c_{1,1}-c_{1,2}-c_{1,3}+c_{1,4}-c_{1,5}-c_{1,6}-3
   c_{1,H})+ \ch_2+ N \nn\\
 n_8 &=& \frac{1}{2} (-c_{1,1}-c_{1,2}-c_{1,3}-c_{1,4}+c_{1,5}-c_{1,6}-3
   c_{1,H})+ \ch_2+ N \nn\\
 n_9 &=& \frac{1}{2} (-c_{1,1}-c_{1,2}-c_{1,3}-c_{1,4}-c_{1,5}+c_{1,6}-3
   c_{1,H})+ \ch_2+N 
\eea
or conversely
\bea
 N &=& 2 n_{1}-5 n_{2}-5 n_{3}+n_{4}+n_{5}+n_{6}+n_{7}+n_{8}+n_{9} \nn\\
 c_{1,H} &=& n_{1}-2 n_{2} \ ,\quad
 c_{1,1} = n_{4}-n_{3} \ ,\quad 
 c_{1,2} = n_{5}-n_{3} \ ,\quad
 c_{1,3} = n_{6}-n_{3} \nn\\
 c_{1,4} &=& n_{7}-n_{3} \ ,\quad
 c_{1,5} = n_{8}-n_{3} \ ,\quad
 c_{1,6} = n_{9}-n_{3} \nn\\
 \ch_2 &=& -\frac{1}{2} \left( n_{1}+n_{4}+n_{5}+n_{6}+n_{7}+n_{8}+n_{9}\right) +2 n_{2}+3 n_{3}
\eea
 The relevant dimension vector is then $N_i=-n_i$.
  The total dimensions for the three blocks and the degree
 are then 
 \bea
 \cN_1 &=& -c_{1,1}-c_{1,2}-c_{1,3}-c_{1,4}-c_{1,5}-c_{1,6}-2 c_{1,H}-2
   \ch_2\nn\\
  \cN_2 &=&c_{1,H}-2 \ch_2-N \nn\\
   \cN_3 &=&2 c_{1,1}+2 c_{1,2}+2 c_{1,3}+2
   c_{1,4}+2 c_{1,5}+2 c_{1,6}+9 c_{1,H}-6 \ch_2-6 N\nn\\
%   \\
% \cN &=& c_{1,1}+c_{1,2}+c_{1,3}+c_{1,4}+c_{1,5}+c_{1,6}+8 c_{1,H}-10
%   \ch_2-7 N
\deg&=& 3c_{1,H}+\sum_i c_{1,i} = -3 N_1+6 N_2+6 N_3-N_4-N_5-N_6-N_7-N_8-N_9
\eea
The stability parameters  $\vec\zeta=\vec\zeta^c$ in \eqref{zetacan}  for $J\propto c_1(S)$ are 
\bea
 \zeta_1 &=& \rho\left( -2 \deg +3 N \right) +2\ch_2+\frac12 N\ ,
 \quad
  \zeta_{4,5,6,7,8,9} =\rho \left(  -\deg +N  \right) + \ch_2 + \frac12 N \nn\\
 \zeta_{2} &=& \rho\left( 5 \deg -6 N \right) -5\ch_2 -2 N \ ,\quad 
    \zeta_{3} = \rho\left( 5 \deg -6 N \right) -5\ch_2 -3 N  
\eea
The dimension agrees with \eqref{dimM} in the chamber where $\Phi_{1j}=0$
with $j=4,5,6,7,8,9$. This requires $\zeta_{1}\geq  0, \zeta_{4,5,6,7,8,9}\leq  0$ hence
\be
N \leq  \deg   \leq \frac32N
\ee
in agreement with the range of slopes of stable objects. 

\medskip
The vanishing of the attractor
indices follows from the arguments in \S\ref{sec_threeblockatt}. Applying the Coulomb branch formula in the attractor chamber, we find evidence that 
single-centered invariants also vanish.

\medskip

For $\gamma=[1;0,1,0,0,0,0,0;1]$ in the canonical chamber 
we get from the flow tree formula
 the expected result 
\be
\Omega(1,1,1,1,1,1,0,1,1) = y^2 + 7 + 1/y^2 
\ee 
The same result is expected for any polarization such that $J'\cdot c_1(S)>0$. 

\medskip
In the blow-up chamber $J'\propto H_\epsilon$, one should instead use 
\be
\vec\zeta^H=(-2 c_{1,H} + N, 5 c_{1,H} - 2 N, 5 c_{1,H}, -c_{1,H}, -c_{1,H}, -c_{1,H}, -c_{1,H}, -c_{1,H}, -c_{1,H})
 +\epsilon \vec\zeta^c
 \ee
 This is  consistent with $\zeta_{1}\geq  0,\zeta_{4,5,6,7,8,9}\leq  0$ for $0\leq c_{1,H}\leq N/2$. 
 From the flow tree formula we get 
\be
\begin{array}{|c|c|l|}
\hline
\[N; c_1 ;c_2\] & \vec N &\Omega(\vec N, \vec\zeta^H)  \\ \hline
 \[2;1,-1,0,0,0,0,0;1\] &(1, 1, 0, 1, 0, 0, 0, 0, 0) & -y-1/y\nn \\
  \[3;1,0,0,0,0,0,0;2\] &(1, 1, 0, 0, 0, 0, 0, 0, 0) &y^2+1+1/y^2 \nn \\
 \[3;0,1,1,1,1,0,-1;0\] & (2,1,1,0,0,0,0,1,2) & y^2+1+1/y^2 
    \\ \hline
    \end{array}
\ee
in agreement with the analysis in \S\ref{sec_VWdP}.

\medskip

Note that the automorphism group of the quiver is $\IZ_2\times S_6$, corresponding to independent permutations of $(23)$ and $(456789)$. It is a subgroup of order 1440 and index 36 inside the Weyl group of $E_6$.

\subsection{$dP_7$}

The authors of  \cite{karpov1998three} provide three distinct three-block exceptional 
collections on $dP_7$. As noted in  \cite{Wijnholt:2002qz}, the quivers for the collections
$(7.2)$, $(7.3)$ can be obtained from the one for $(8.2)$ by applying a sequence of Seiberg dualities
and permutations of the nodes,\footnote{We denote by $SD[\cC,\cI,\sigma]$ the sequence of Seiberg dualities, starting from the quiver $Q$ associated to the collection $\cC$, dualizing successively with respect to each node in the list $\cI$, and applying the permutation $\sigma$ on the nodes of the final quiver.}
\begin{itemize}
\item  $(7.2) \simeq SD[ (7.1), \{3,4,5,6,2\},{\rm Id}]$  
\item  $(7.3) \simeq SD[ (7.2), \{3,4,5,1,2\},(1,3,4,5,2,6,7,8,9,10)]$ 
\end{itemize}
This does not imply however that the sheaves are obtained in this manner.

\subsubsection{Three-block collection (7.1) of type $(2,2,1)$}
The first three-block exceptional collection (7.1) from \cite{karpov1998three} involves two rank-two 
sheaves $E^1$ and $E^2$. The Chern vectors of the projective and simple representations
are 
\be
\begin{array}{ccl}
 \gamma^1 &=& \[2,\(-2,1,1,1,1,1,1,1\),-\frac{3}{2}\] \\
 \gamma^2 &=& \[2,\(-1,0,0,0,1,1,1,1\),-\frac{3}{2}\] \\
 \gamma^3 &=& \[1,\(0,0,0,0,1,0,0,0\),-\frac{1}{2}\] \\
 \gamma^4 &=& \[1,\(0,0,0,0,0,1,0,0\),-\frac{1}{2}\] \\
 \gamma^5 &=& \[1,\(0,0,0,0,0,0,1,0\),-\frac{1}{2}\] \\
 \gamma^6 &=& \[1,\(0,0,0,0,0,0,0,1\),-\frac{1}{2}\] \\
 \gamma^7 &=& \[1,\(3,-1,-1,-1,-1,-1,-1,-1\),1\] \\
 \gamma^8 &=& \[1,\(1,-1,0,0,0,0,0,0\),0\] \\
 \gamma^9 &=& \[1,\(1,0,-1,0,0,0,0,0\),0\] \\
 \gamma^{10} &=& \[1,\(1,0,0,-1,0,0,0,0\),0\] \\
\end{array}
\quad
\begin{array}{ccl}
 \gamma_1 &=& \[2,\(-2,1,1,1,1,1,1,1\),-\frac{3}{2}\] \\
 \gamma_2 &=& \[-6,\(9,-4,-4,-4,-4,-4,-4,-4\),\frac{11}{2}\] \\
 \gamma_3 &=& \[1,\(0,0,0,0,0,0,0,0\),0\] \\
 \gamma_4 &=& \[1,\(-2,0,1,1,1,1,1,1\),-1\] \\
 \gamma_5 &=& \[1,\(-2,1,0,1,1,1,1,1\),-1\] \\
 \gamma_6 &=& \[1,\(-2,1,1,0,1,1,1,1\),-1\] \\
 \gamma_7 &=& \[1,\(-2,1,1,1,0,1,1,1\),-1\] \\
 \gamma_8 &=& \[1,\(-2,1,1,1,1,0,1,1\),-1\] \\
 \gamma_9 &=& \[1,\(-2,1,1,1,1,1,0,1\),-1\] \\
 \gamma_{10} &=& \[1,\(-2,1,1,1,1,1,1,0\),-1\] \\
\end{array}
\ee
with slopes $1/2,3/2,2$ and $1/2,1/6,0$, respectively. The Euler matrix has three-block structure
\be
S=\left(
\begin{array}{ccc}
 1 & 4 & 3 \\
 0 & 1 & 1 \\
 0 & 0 &   \mathbb{1}_8  \\
\end{array}
\right), \ 
S^\vee=\left(
\begin{array}{ccc}
 1 & -4 & 1 \\
 0 & 1 & -1 \\
 0 & 0 &   \mathbb{1}_8  \\
\end{array}
\right),\ 
\kappa=\left(
\begin{array}{ccc}
 0 & 4 & -1 \\
 -4 & 0 & 1 \\
 1 & -1 &   \mathbb{0}_8  \\
\end{array}
\right)
\ee
A general Chern character $\gamma=[N,c_1,\ch_2]$ decomposes as $\gamma=
\sum_i n_i \gamma_i$ with  
\bea
 n_1 &=& 2 c_{1,1}+2 c_{1,2}+2 c_{1,3}+2 c_{1,4}+2 c_{1,5}+2 c_{1,6}+2
   c_{1,7}+5 c_{1,H}+2 \ch_2 \nn\\
 n_2 &=& c_{1,1}+c_{1,2}+c_{1,3}+2 c_{1,4}+2 c_{1,5}+2 c_{1,6}+2 c_{1,7}+4
   c_{1,H}+2 \ch_2 \nn\\
 n_3 &=& \frac{1}{2} (c_{1,1}+c_{1,2}+c_{1,3}+3
   c_{1,4}+c_{1,5}+c_{1,6}+c_{1,7}+3 c_{1,H})+ \ch_2 \nn\\
 n_4 &=& \frac{1}{2} (c_{1,1}+c_{1,2}+c_{1,3}+c_{1,4}+3
   c_{1,5}+c_{1,6}+c_{1,7}+3 c_{1,H})+ \ch_2 \nn\\
 n_5 &=& \frac{1}{2} (c_{1,1}+c_{1,2}+c_{1,3}+c_{1,4}+c_{1,5}+3
   c_{1,6}+c_{1,7}+3 c_{1,H})+ \ch_2 \nn\\
 n_6 &=& \frac{1}{2} (c_{1,1}+c_{1,2}+c_{1,3}+c_{1,4}+c_{1,5}+c_{1,6}+3
   c_{1,7}+3 c_{1,H})+ \ch_2 \nn\\
 n_7 &=& \frac{1}{2}
   (-c_{1,1}-c_{1,2}-c_{1,3}-c_{1,4}-c_{1,5}-c_{1,6}-c_{1,7}-3
   c_{1,H})+\ch_2+N \nn\\
 n_8 &=& \frac{1}{2}
   (-c_{1,1}+c_{1,2}+c_{1,3}+c_{1,4}+c_{1,5}+c_{1,6}+c_{1,7}+c_{1,H})+\ch_2 \nn\\
 n_9 &=& \frac{1}{2}
   (c_{1,1}-c_{1,2}+c_{1,3}+c_{1,4}+c_{1,5}+c_{1,6}+c_{1,7}+c_{1,H})+2  \ch_2 \nn\\
 n_{10} &=& \frac{1}{2}
   (c_{1,1}+c_{1,2}-c_{1,3}+c_{1,4}+c_{1,5}+c_{1,6}+c_{1,7}+c_{1,H})+2 \ch_2 
\eea
or conversely
\bea
 N &=& 2 n_1+2 n_2-3 n_3-3 n_4-3 n_5-3 n_6+n_7+n_8+n_9+n_{10} \nn\\
 c_{1,H} &=& -2 n_1-n_2+3 n_3+3 n_4+3 n_5+3 n_6-2 n_8-2 n_9-2 n_{10} \nn\\
 c_{1,1} &=& n_1-n_3-n_4-n_5-n_6+n_9+n_{10} \nn\\
 c_{1,2} &=& n_1-n_3-n_4-n_5-n_6+n_8+n_{10} \nn\\
 c_{1,3} &=& n_1-n_3-n_4-n_5-n_6+n_8+n_9 \nn\\
 c_{1,4} &=& n_1+n_2-n_3-2 n_4-2 n_5-2 n_6+n_8+n_9+n_{10} \nn\\
 c_{1,5} &=& n_1+n_2-2 n_3-n_4-2 n_5-2 n_6+n_8+n_9+n_{10} \nn\\
 c_{1,6} &=& n_1+n_2-2 n_3-2 n_4-n_5-2 n_6+n_8+n_9+n_{10} \nn\\
 c_{1,7} &=& n_1+n_2-2 n_3-2 n_4-2 n_5-n_6+n_8+n_9+n_{10} \nn\\
 \ch_2 &=& \frac{1}{2} (-3 n_1-3 n_2+5 n_3+5 n_4+5 n_5+5 n_6)- n_8- n_9-n_{10}
\eea
As usual, we take $\vec N=-\vec n$. 
The total dimensions in each block are then 
\bea
\cN_1 &=& -2 c_{1,1}-2 c_{1,2}-2 c_{1,3}-2 c_{1,4}-2 c_{1,5}-2 c_{1,6}-2
   c_{1,7}-5 c_{1,H}-2
   \ch_2\nn\\
  \cN_2 &=&-c_{1,1}-c_{1,2}-c_{1,3}-c_{1,4}-c_{1,5}-c_{1,6}-c_{1,7}-2
   c_{1,H}-2 \ch_2,\\
   \cN_3 &=&-2 c_{1,1}-2 c_{1,2}-2 c_{1,3}-2 c_{1,4}-2 c_{1,5}-2
   c_{1,6}-2 c_{1,7}-2 c_{1,H}-8 \ch_2-N,\nn
   %\\
%   \cN &=& -5 c_{1,1}-5 c_{1,2}-5
%   c_{1,3}-5 c_{1,4}-5 c_{1,5}-5 c_{1,6}-5 c_{1,7}-9 c_{1,H}-12
%   \ch_2-N
\eea
The stability parameters $\vec\zeta=\vec\zeta^c$ in \eqref{zetacan}  for $J\propto c_1(S)$ are 
\bea
 \zeta_1 &=& (N-2 \deg ) \rho  + 2 \ch_2+\frac{3}{2} N\nn\\
 \zeta_2 &=& (6 \deg -N) \rho   -6 \ch_2-\frac{11}{2} N \nn\\
 \zeta_3 &=& -\deg  \rho  + \ch_2 \ ,\quad 
 \zeta_{4,\dots, 10} = -\deg  \rho  + \ch_2+N 
 \eea
The dimension agrees with \eqref{dimM} in the chamber where $\Phi_{i1}=0$ for $i=3,\dots 10$. 
This requires $\zeta_{1}\geq  0$, $\zeta_{3,4,5,6,7,8,9,10}\leq  0$ hence
\be
0 \leq \deg  < \frac{N}{2}
\ee
in agreement with the range of slopes of simple representations. 
\medskip

 The vanishing of the attractor
indices follows from the arguments in \S\ref{sec_threeblockatt}. 
For $\gamma=[1;-1,0,0,0,0,1,1,1;0]$ in the canonical chamber 
we find from the flow tree formula 
\be
\Omega(1,1,0,0,0,0,0,1,1,1,\vec\zeta^c)=1
\ee  as expected.
For  $\gamma=[1;0,0,0,0,0,0,0,0;1]$ one would expect 
$\Omega(2,2,0,1,1,1,1,1,1,1,\vec\zeta^c)=y^2+8+1/y^2$ but the height of the dimension vector is too high to check this directly. 

\medskip

In the blow-up chamber, the stability parameters are instead
\be
\vec\zeta^H = (-2 c_{1,H}-2 N ; 6 c_{1,H}+9 N ; -c_{1,H},
-c_{1,H}-2   N, \dots ,-c_{1,H}-2 N)+\epsilon \vec\zeta^c
\ee
This is never consistent with $\zeta_1\geq  0, \zeta_{3,\dots,10}$ unless $N=c_{1,H}=0$.

\subsubsection{Three-block collection (7.2) of type $(2,1,1)$}

The second strong exceptional collection (7.2) from \cite{karpov1998three} again
 involves two rank-two 
sheaves $E^1$ and $E^2$. 
%\be
%\begin{split}
%F,\quad F', & \quad  \cO(C_4), \quad \cO(C_5), \quad \cO(C_6), \quad \cO(C_7), \quad  \\
%& \cO(3H-C_1-\dots -C_7), \quad \cO(H-C_1), \quad  \cO(H-C_2), \quad  \cO(H-C_3)
%\end{split}
%\ee
%where $F,F'$ are rank 2 bundles with Chern characters $[2,-2H+C_1+\dots +C_7,-3/2]$ 
%and $[2,-H+C_4+C_5+C_6+C_7,-3/2 ]$. 
The Chern vectors of the projective and simple representations
are 
\be
\begin{array}{ccl}
 \gamma^1 &=& \[2,\(-2,1,1,1,1,1,1,1\),-\frac{3}{2}\] \\
 \gamma^2 &=& \[2,\(-1,0,0,0,1,1,1,1\),-\frac{3}{2}\] \\
 \gamma^3 &=& \[1,\(0,0,0,0,1,0,0,0\),-\frac{1}{2}\] \\
 \gamma^4 &=& \[1,\(0,0,0,0,0,1,0,0\),-\frac{1}{2}\] \\
 \gamma^5 &=& \[1,\(0,0,0,0,0,0,1,0\),-\frac{1}{2}\] \\
 \gamma^6 &=& \[1,\(0,0,0,0,0,0,0,1\),-\frac{1}{2}\] \\
 \gamma^7 &=& \[1,\(3,-1,-1,-1,-1,-1,-1,-1\),1\] \\
 \gamma^8 &=& \[1,\(1,-1,0,0,0,0,0,0\),0\] \\
 \gamma^9 &=& \[1,\(1,0,-1,0,0,0,0,0\),0\] \\
 \gamma^{10} &=& \[1,\(1,0,0,-1,0,0,0,0\),0\] \\
\end{array}
\ 
\begin{array}{ccl}
 \gamma_1 &=& \[2,\(-2,1,1,1,1,1,1,1\),-\frac{3}{2}\] \\
 \gamma_2 &=& \[2,\(-1,0,0,0,1,1,1,1\),-\frac{3}{2}\] \\
 \gamma_3 &=& \[-3,\(3,-1,-1,-1,-1,-2,-2,-2\),\frac{5}{2}\] \\
 \gamma_4 &=& \[-3,\(3,-1,-1,-1,-2,-1,-2,-2\),\frac{5}{2}\] \\
 \gamma_5 &=& \[-3,\(3,-1,-1,-1,-2,-2,-1,-2\),\frac{5}{2}\] \\
 \gamma_6 &=& \[-3,\(3,-1,-1,-1,-2,-2,-2,-1\),\frac{5}{2}\] \\
 \gamma_7 &=& \[1,\(0,0,0,0,0,0,0,0\),0\] \\
 \gamma_8 &=& \[1,\(-2,0,1,1,1,1,1,1\),-1\] \\
 \gamma_9 &=& \[1,\(-2,1,0,1,1,1,1,1\),-1\] \\
 \gamma_{10} &=& \[1,\(-2,1,1,0,1,1,1,1\),-1\] \\
\end{array}
\ee
with slope $1/2,1,2 $ and $4/3,6/5,1$, respectively. The Euler matrix has 
a three-block structure,
\be
S=\left(
\begin{array}{ccc}
   \mathbb{1}_2  & 1 & 3 \\
 0 &   \mathbb{1}_4  & 1 \\
 0 & 0 &   \mathbb{1}_4  \\
\end{array}
\right),\ 
S^\vee=\left(
\begin{array}{ccc}
   \mathbb{1}_2  & -1 & 1 \\
 0 &   \mathbb{1}_4  & -1 \\
 0 & 0 &   \mathbb{1}_4  \\
\end{array}
\right),\ 
\kappa=\left(
\begin{array}{ccc}
   \mathbb{0}_2 & 1 & -1 \\
 -1 &   \mathbb{0}_4  & 1 \\
 1 & -1 &   \mathbb{0}_4 \\
\end{array}
\right)
\ee
A general Chern character $\gamma=[N,c_1,\ch_2]$ decomposes as $\gamma=
\sum_i n_i \gamma_i$ with  
\bea
 n_1 &=& 2 c_{1,1}+2 c_{1,2}+2 c_{1,3}+2 c_{1,4}+2 c_{1,5}+2 c_{1,6}+2
   c_{1,7}+5 c_{1,H}+2 \ch_2 \nn\\
 n_2 &=& c_{1,1}+c_{1,2}+c_{1,3}+2 c_{1,4}+2 c_{1,5}+2 c_{1,6}+2 c_{1,7}+4
   c_{1,H}+2 \ch_2 \nn\\
 n_3 &=& \frac{1}{2} (c_{1,1}+c_{1,2}+c_{1,3}+3
   c_{1,4}+c_{1,5}+c_{1,6}+c_{1,7}+3 c_{1,H}) + \ch_2 \nn\\
 n_4 &=& \frac{1}{2} (c_{1,1}+c_{1,2}+c_{1,3}+c_{1,4}+3
   c_{1,5}+c_{1,6}+c_{1,7}+3 c_{1,H})+ \ch_2 \nn\\
 n_5 &=& \frac{1}{2} (c_{1,1}+c_{1,2}+c_{1,3}+c_{1,4}+c_{1,5}+3
   c_{1,6}+c_{1,7}+3 c_{1,H})+ \ch_2 \nn\\
 n_6 &=& \frac{1}{2} (c_{1,1}+c_{1,2}+c_{1,3}+c_{1,4}+c_{1,5}+c_{1,6}+3
   c_{1,7}+3 c_{1,H})+\ch_2 \nn\\
 n_7 &=& \frac{1}{2}
   (-c_{1,1}-c_{1,2}-c_{1,3}-c_{1,4}-c_{1,5}-c_{1,6}-c_{1,7}-3
   c_{1,H})+\ch_2+N \nn\\
 n_8 &=& \frac{1}{2}
   (-c_{1,1}+c_{1,2}+c_{1,3}+c_{1,4}+c_{1,5}+c_{1,6}+c_{1,7}+c_{1,H})+
   \ch_2 \nn\\
 n_9 &=& \frac{1}{2}
   (c_{1,1}-c_{1,2}+c_{1,3}+c_{1,4}+c_{1,5}+c_{1,6}+c_{1,7}+c_{1,H})+
   \ch_2 \nn\\
 n_{10} &=& \frac{1}{2}
   (c_{1,1}+c_{1,2}-c_{1,3}+c_{1,4}+c_{1,5}+c_{1,6}+c_{1,7}+c_{1,H})+
   \ch_2
\eea
or conversely
\bea
 N &=& 2 n_{1}+2 n_{2}-3 n_{3}-3 n_{4}-3 n_{5}-3 n_{6}+n_{7}+n_{8}+n_{9}+n_{10} \nn\\
 c_{1,H} &=& -2 n_{1}-n_{2}+3 n_{3}+3 n_{4}+3 n_{5}+3 n_{6}-2 n_{8}-2 n_{9}-2 n_{10} \nn\\
 c_{1,1} &=& n_{1}-n_{3}-n_{4}-n_{5}-n_{6}+n_{9}+n_{10} \nn\\
 c_{1,2} &=& n_{1}-n_{3}-n_{4}-n_{5}-n_{6}+n_{8}+n_{10} \nn\\
 c_{1,3} &=& n_{1}-n_{3}-n_{4}-n_{5}-n_{6}+n_{8}+n_{9} \nn\\
 c_{1,4} &=& n_{1}+n_{2}-n_{3}-2 n_{4}-2 n_{5}-2 n_{6}+n_{8}+n_{9}+n_{10} \nn\\
 c_{1,5} &=& n_{1}+n_{2}-2 n_{3}-n_{4}-2 n_{5}-2 n_{6}+n_{8}+n_{9}+n_{10} \nn\\
 c_{1,6} &=& n_{1}+n_{2}-2 n_{3}-2 n_{4}-n_{5}-2 n_{6}+n_{8}+n_{9}+n_{10} \nn\\
 c_{1,7} &=& n_{1}+n_{2}-2 n_{3}-2 n_{4}-2 n_{5}-n_{6}+n_{8}+n_{9}+n_{10} \nn\\
 \ch_2 &=& \frac{1}{2} \left(5 n_{4}+5 n_{5}+5 n_{6}-3 n_{1}-3 n_{2}+5 n_{3}\right)- n_{8}-n_{9}-n_{10}
\eea
A usual we set $\vec N=-\vec n$.
 The total dimensions for the three blocks
 and degree are then 
 \bea
 \cN_1 &=&-3 c_{1,1}-3 c_{1,2}-3 c_{1,3}-4 c_{1,4}-4 c_{1,5}-4 c_{1,6}-4
   c_{1,7}-9 c_{1,H}-4 \ch_2\nn\\
  \cN_2 &=&-2 c_{1,1}-2 c_{1,2}-2 c_{1,3}-3 c_{1,4}-3
   c_{1,5}-3 c_{1,6}-3 c_{1,7}-6 c_{1,H}-4
   \ch_2 ,\nn \\
   \cN_3 &=&-c_{1,4}-c_{1,5}-c_{1,6}-c_{1,7}-4 \ch_2-N\nn\\
   %\\
% \cN &=& -5
%   c_{1,1}-5 c_{1,2}-5 c_{1,3}-8 c_{1,4}-8 c_{1,5}-8 c_{1,6}-8 c_{1,7}-15
%   c_{1,H}-12 \ch_2-N
% \eea
%and the degree is
%\be
\deg&=& 3c_{1,H}+\sum_i c_{1,i} = N_4+N_5+N_6-N_1-N_2+N_3
\eea
The stability parameters  $\vec\zeta=\vec\zeta^c$ in \eqref{zetacan}  for $J\propto c_1(S)$ are 
\bea
 \zeta_{1,2} &=& -(2 \deg -N) \rho  +2\ch_2+\frac32 N \ ,\quad 
 \zeta_{3,4,5,6} = (3 \deg -N) \rho -3\ch_2-\frac52 N \nn\\
 \zeta_7 &=& -\deg\, \rho  +\ch_2\ , \quad 
  \zeta_{8,9,10} = -\deg  \rho +\ch_2 + N
\eea
The dimension agrees with \eqref{dimM} in the chamber where $\Phi_{ij}=0$ for $i=1,2$ and $j=7,8,9,10$.
This requires $\zeta_{1,2}\geq  0$, $\zeta_{7,8,9,10}\leq  0$ hence
\be
0 \leq \deg  \leq \frac{N}{2}
\ee
in agreement with the range of slopes of simple representations.

\medskip

The vanishing of the attractor
indices follows from the arguments in \S\ref{sec_threeblockatt}. 
For $\gamma=[1;-1,0,0,0,0,1,1,1;0]$ we find from 
the flow tree formula $\Omega(1,0,1,0,0,0,0,0,0,0,\vec\zeta^c)=1$ as expected.
For $\gamma=[1;0,0,0,0,0,0,0,0;1]$ one would expect 
$\Omega(2,2,1,1,1,1,0,1,1,1,\vec\zeta^c)=y^2+8+1/y^2$ 
but the height of the dimension vector is too high to
check this directly. 

\medskip

In the blow-up chamber, the stability parameters are instead
\be
\begin{split} 
\vec\zeta^H=& (-2 c_{1,H} - 2 N, -2 c_{1,H} - N; 
3 c_{1,H} + 3 N, 3 c_{1,H} + 3 N, 3 c_{1,H} + 3 N,  3 c_{1,H} + 3 N; \\ 
&-c_{1,H}, -c_{1,H} - 2 N, -c_{1,H} - 2 N, -c_{1,H} - 2 N)+\epsilon \vec\zeta^c
\end{split}
\ee
This is never consistent with $\zeta_{1,2}\geq  0, \zeta_{5,\dots,10}\leq  0$ unless $N=c_{1,H}=0$.

\subsubsection{Three-block collection (7.3) of type $(3,1,1)$}
The third three-block strong exceptional collection (7.3) from \cite{karpov1998three} involves
a rank three sheaf $E^1$. The Chern vectors of the projective and simple representations
are 
\be
\begin{array}{ccl}
 \gamma^1 &=& \[3,\(0,0,0,0,1,1,1,1\),-2\] \\
 \gamma^2 &=& \[1,\(1,-1,0,0,0,0,0,0\),0\] \\
 \gamma^3 &=& \[1,\(1,0,-1,0,0,0,0,0\),0\] \\
 \gamma^4 &=& \[1,\(1,0,0,-1,0,0,0,0\),0\] \\
 \gamma^5 &=& \[1,\(1,0,0,0,0,0,0,0\),\frac{1}{2}\] \\
 \gamma^6 &=& \[1,\(2,-1,-1,-1,0,0,0,0\),\frac{1}{2}\] \\
 \gamma^7 &=& \[1,\(3,-1,-1,-1,-1,-1,-1,0\),\frac{3}{2}\] \\
 \gamma^8 &=& \[1,\(3,-1,-1,-1,-1,-1,0,-1\),\frac{3}{2}\] \\
 \gamma^9 &=& \[1,\(3,-1,-1,-1,-1,0,-1,-1\),\frac{3}{2}\] \\
 \gamma^{10} &=& \[1,\(3,-1,-1,-1,0,-1,-1,-1\),\frac{3}{2}\] \\
\end{array}
\quad
\begin{array}{ccl}
 \gamma_1 &=& \[3,\(0,0,0,0,1,1,1,1\),-2\] \\
 \gamma_2 &=& \[-5,\(1,-1,0,0,-2,-2,-2,-2\),4\] \\
 \gamma_3 &=& \[-5,\(1,0,-1,0,-2,-2,-2,-2\),4\] \\
 \gamma_4 &=& \[-5,\(1,0,0,-1,-2,-2,-2,-2\),4\] \\
 \gamma_5 &=& \[1,\(-2,1,1,1,1,1,1,1\),-\frac{3}{2}\] \\
 \gamma_6 &=& \[1,\(-1,0,0,0,1,1,1,1\),-\frac{3}{2}\] \\
 \gamma_7 &=& \[1,\(0,0,0,0,0,0,0,1\),-\frac{1}{2}\] \\
 \gamma_8 &=& \[1,\(0,0,0,0,0,0,1,0\),-\frac{1}{2}\] \\
 \gamma_9 &=& \[1,\(0,0,0,0,0,1,0,0\),-\frac{1}{2}\] \\
 \gamma_{10} &=& \[1,\(0,0,0,0,1,0,0,0\),-\frac{1}{2}\] \\
\end{array}
\ee
with slopes $4/3,2,$ and $4/3, 6/5,1$, respectively. 
The Euler matrix has a three-block form,
\be
S=\left(
\begin{array}{ccc}
 1 & 2 & 5 \\
 0 &   \mathbb{1}_3  & 1 \\
 0 & 0 &   \mathbb{1}_6  \\
\end{array}
\right),\ 
S^\vee=\left(
\begin{array}{ccc}
 1 & -2 & 1 \\
 0 &   \mathbb{1}_3  & -1 \\
 0 & 0 &   \mathbb{1}_6  \\
\end{array}
\right),\ 
\kappa=\left(
\begin{array}{ccc}
 0 & 2 & -1 \\
 -2 &   \mathbb{0}_3 & 1 \\
 1 & -1 &   \mathbb{0}_6  \\
\end{array}
\right)
\ee
A general Chern character $\gamma=[N,c_1,\ch_2]$ decomposes as $\gamma=
\sum_i n_i \gamma_i$ with  
\bea
 n_1 &=& \frac{1}{2} (3 c_{1,1}+3 c_{1,2}+3 c_{1,3}+5 c_{1,4}+5 c_{1,5}+5
   c_{1,6}+5 c_{1,7}+9 c_{1,H})+3 \ch_2- N \nn\\
 n_2 &=& \frac{1}{2}
   (-c_{1,1}+c_{1,2}+c_{1,3}+c_{1,4}+c_{1,5}+c_{1,6}+c_{1,7}+c_{1,H})+2 \ch_2 \nn\\
 n_3 &=& \frac{1}{2}
   (c_{1,1}-c_{1,2}+c_{1,3}+c_{1,4}+c_{1,5}+c_{1,6}+c_{1,7}+c_{1,H})+\ch_2 \nn\\
 n_4 &=& \frac{1}{2}
   (c_{1,1}+c_{1,2}-c_{1,3}+c_{1,4}+c_{1,5}+c_{1,6}+c_{1,7}+c_{1,H})+  \ch_2 \nn\\
 n_5 &=& \frac{1}{2}
   (c_{1,1}+c_{1,2}+c_{1,3}+c_{1,4}+c_{1,5}+c_{1,6}+c_{1,7}+c_{1,H})+ \ch_2 \nn\\
 n_6 &=& \frac{1}{2}
   (-c_{1,1}-c_{1,2}-c_{1,3}+c_{1,4}+c_{1,5}+c_{1,6}+c_{1,7}-c_{1,H})+  \ch_2 \nn\\
 n_7 &=& \frac{1}{2}
   (-c_{1,1}-c_{1,2}-c_{1,3}-c_{1,4}-c_{1,5}-c_{1,6}+c_{1,7}-3
   c_{1,H})+ \ch_2+ N \nn\\
 n_8 &=& \frac{1}{2}
   (-c_{1,1}-c_{1,2}-c_{1,3}-c_{1,4}-c_{1,5}+c_{1,6}-c_{1,7}-3
   c_{1,H}) + \ch_2+ N \nn\\
 n_9 &=& \frac{1}{2}
   (-c_{1,1}-c_{1,2}-c_{1,3}-c_{1,4}+c_{1,5}-c_{1,6}-c_{1,7}-3
   c_{1,H})+ \ch_2+N \nn\\
 n_{10} &=& \frac{1}{2}
   (-c_{1,1}-c_{1,2}-c_{1,3}+c_{1,4}-c_{1,5}-c_{1,6}-c_{1,7}-3
   c_{1,H})+\ch_2+N 
\eea
or conversely
\bea
 N &=& 3 n_{1}-5 n_{2}-5 n_{3}-5 n_{4}+n_{5}+n_{6}+n_{7}+n_{8}+n_{9}+n_{10} \nn\\
 c_{1,H} &=& n_{2}+n_{3}+n_{4}-2 n_{5}-n_{6} \nn\\
 c_{1,1} &=& n_{5}-n_{2} \ ,\quad 
 c_{1,2} = n_{5}-n_{3} \ ,\quad 
 c_{1,3} = n_{5}-n_{4} \nn\\
 c_{1,4} &=& n_{1}-2 n_{2}-2 n_{3}-2 n_{4}+n_{5}+n_{6}+n_{10} \nn\\
 c_{1,5} &=& n_{1}-2 n_{2}-2 n_{3}-2 n_{4}+n_{5}+n_{6}+n_{9} \nn\\
 c_{1,6} &=& n_{1}-2 n_{2}-2 n_{3}-2 n_{4}+n_{5}+n_{6}+n_{8} \nn\\
 c_{1,7} &=& n_{1}-2 n_{2}-2 n_{3}-2 n_{4}+n_{5}+n_{6}+n_{7} \nn\\
 \ch_2 &=&4 n_{2}+4 n_{3}+4 n_{4}-2 n_{1} - \frac{1}{2} (3 n_{5}+3 n_{6}+n_{7}+n_{8}+n_{9}+n_{10})
\eea
As usual, we take $\vec N=-\vec n$. 
The total dimensions in each block are then 
\bea
\cN_1 &=&
-\frac12\left( 3 c_{1,1}+3 c_{1,2}+3 c_{1,3}+5c_{1,4}+5c_{1,5}+5c_{1,6}+5c_{1,7}+9 c_{1,H}\right)
   -3 \ch_2+N \nn\\
  \cN_2 &=&
  -\frac12\left( c_{1,1}+c_{1,2}+c_{1,3}+3c_{1,4}+3c_{1,5}+3c_{1,6}+3c_{1,7}+3c_{1,H}\right)
   -3 \ch_2
  \nn \\
   \cN_3 &=&2 c_{1,1}+2 c_{1,2}+2 c_{1,3}+6
   c_{1,H}-6 \ch_2-4 N
%   \nn\\
% \cN &=&  -4 c_{1,4}-4 c_{1,5}-4 c_{1,6}-4 c_{1,7}-12
%   \ch_2-3 N
\eea
The relevant stability parameters $\vec\zeta=\vec\zeta^c$ in \eqref{zetacan}
in the canonical chamber $J\propto c_1(S)$ are 
\bea
 \zeta_1 &=& (4 N-3 \deg ) \rho  + 3 \ch_2+2 N \ ,\quad 
 \zeta_{2,3,4} = (5 \deg -6 N) \rho  -5 \ch_2-4 N \nn\\
 \zeta_{5,6} &=& (N-\deg ) \rho  + \ch_2+\frac{3}{2} N \ ,\quad 
 \zeta_{7,8,9,10} = (N-\deg ) \rho  + \ch_2+\frac{1}{2} N 
\eea
The dimension agrees with \eqref{dimM} in the chamber where $\Phi_{i1}=0$ for $i=5,\dots 10$. 
This requires $\zeta_{1}\geq  0$, $\zeta_{5,6,7,8,9,10}\leq  0$ hence
\be
N \leq  \deg   \leq \frac{4}{3} N
\ee
in agreement with the range of slopes of simple representations.

\medskip

The vanishing of the attractor
indices follows from the arguments in \S\ref{sec_threeblockatt}. 
For $\gamma=\[ 1; 0,1,0,0,0,0,0,0;0\]$ 
we get from  the flow tree formula $\Omega(0,1,0,0,0,1,1,0,0,0,\vec\zeta^c)=1$
as expected. 
For $\gamma=[1;0,0,0,0,0,0,0,1;1]$
we expect $\Omega(3, 1, 1, 1, 1, 1, 0, 1, 1, 1,\vec\zeta^c)=y^2+8+1/y^2$
but the height of the dimension vector is too high to
check this directly.

\medskip

In the blow-up chamber, the stability parameters are instead
\be
\begin{split}
\vec\zeta^H=& (-3 c_{1,H} ; 
5 c_{1,H}+N,5 c_{1,H}+N,5   c_{1,H}+N;-c_{1,H}-2   N,-c_{1,H}-N, \\ 
& -c_{1,H},-c_{1,H},-c_{1,H},-c_{1,H}) 
 +\epsilon \vec\zeta^c
\end{split}
\ee
This is consistent with $\zeta_{1}\geq  0, \zeta_{5,\dots,10}\leq 0$ only when $c_{1,H}=0$.
From the flow tree formula we get
\be
\begin{array}{|c|c|l|}
\hline
\[N; c_1 ;c_2\] & \vec N &\Omega(\vec N, \vec\zeta^H)  \\ \hline
\[ 1; 0,1,0,0,0,0,0,0;0\] & (0,1,0,0,0,1,1,0,0,0) & 1 \nn\\
\[ 2; 0, -1, 0, 0, 0, 1, 1, 1; 0\]  & (2, 0, 1, 1, 1, 0, 0, 0, 0, 1) & -y-1/y \nn\\
\[ 3;0,1,0,0,1,1,1,0;0\] & (0,1,0,0,0,1,1,0,0,0) & 1  \nn\\
\[3; 0, 0, 0, -1, 1, 1, 1, 1; 0\] & (2, 1, 1, 0, 1, 0, 0, 0, 0, 0) & y^2+1+1/y^2   \nn\\
\[ 3;0,1,-1,-1,1,1,1,1;-1\] & (2,2,0,0,1,0,0,0,0,0) & 0 
    \\ \hline
    \end{array}
\ee
in agreement with the analysis in  \S\ref{sec_VWdP}.

\subsection{$dP_8$}
The authors of  \cite{karpov1998three} provide four distinct three-block exceptional 
collections on $dP_8$. As noted in  \cite{Wijnholt:2002qz}, the quivers for the collections
$(8.1)$, $(8.3)$, $(8.4)$ can be obtained from the one for $(8.2)$ by applying a sequence of Seiberg dualities, 
\begin{itemize}
\item  $(8.1) \simeq SD[ (8.2), \{4,1,2,3\},(2,3,5,6,7,8,9,10,11,1,4)]$  
\item  $(8.3) \simeq SD[ (8.2), \{4,5,1\},(4,5,1,2,3,6,7,8,9,10,11)]$ 
\item  $(8.4) \simeq SD[ (8.2), \{4,5,6,1,4,5,6\}, {\rm Id} ]$ 
\end{itemize}

\subsubsection{Three-block collection (8.1) of type $(3,3,1)$}
As noted in \cite{Wijnholt:2002qz},  the three-block exceptional collection (8.1) from \cite{karpov1998three} appears to be  invalid in the form stated there. A valid collection
of the same type can be obtained from  the collection of type $(8.2)$ below 
by applying the sequence of Seiberg dualities indicated above.\footnote{{\bp See \cite{Verlinde:2005jr}
for an alternative collection of type (8.1).}}  In this way we find the Chern vectors 
\be
\begin{array}{ccl}
 \gamma^1 &=& \[3,\(0,0,0,0,0,1,1,1,1\),-2\] \\
 \gamma^2 &=& \[3,\(3,-1,-1,-1,-1,0,0,0,0\),-\frac{1}{2}\] \\
 \gamma^3 &=& \[1,\(2,-1,-1,-1,-1,0,0,0,0\),0\] \\
 \gamma^4 &=& \[1,\(1,0,0,0,-1,0,0,0,0\),0\] \\
 \gamma^5 &=& \[1,\(3,-1,-1,-1,-1,0,-1,-1,-1\),1\] \\
 \gamma^6 &=& \[1,\(3,-1,-1,-1,-1,-1,0,-1,-1\),1\] \\
 \gamma^7 &=& \[1,\(3,-1,-1,-1,-1,-1,-1,0,-1\),1\] \\
 \gamma^8 &=& \[1,\(3,-1,-1,-1,-1,-1,-1,-1,0\),1\] \\
 \gamma^9 &=& \[1,\(1,-1,0,0,0,0,0,0,0\),0\] \\
 \gamma^{10} &=& \[1,\(1,0,-1,0,0,0,0,0,0\),0\] \\
 \gamma^{11} &=& \[1,\(1,0,0,-1,0,0,0,0,0\),0\] \\
\end{array}
\ 
\begin{array}{ccl}
 \gamma _1 &=& \[3,\(0,0,0,0,0,1,1,1,1\),-2\] \\
 \gamma _2 &=& -\[6,\(-3,1,1,1,1,3,3,3,3\),-\frac{11}{2}\] \\
 \gamma _3 &=& \[1,\(-1,0,0,0,0,1,1,1,1\),-\frac{3}{2}\] \\
 \gamma _4 &=& \[1,\(-2,1,1,1,0,1,1,1,1\),-\frac{3}{2}\] \\
 \gamma _5 &=& \[1,\(0,0,0,0,0,1,0,0,0\),-\frac{1}{2}\] \\
 \gamma _6 &=& \[1,\(0,0,0,0,0,0,1,0,0\),-\frac{1}{2}\] \\
 \gamma _7 &=& \[1,\(0,0,0,0,0,0,0,1,0\),-\frac{1}{2}\] \\
 \gamma _8 &=& \[1,\(0,0,0,0,0,0,0,0,1\),-\frac{1}{2}\] \\
 \gamma _9 &=& \[1,\(-2,0,1,1,1,1,1,1,1\),-\frac{3}{2}\] \\
 \gamma _{10} &=& \[1,\(-2,1,0,1,1,1,1,1,1\),-\frac{3}{2}\] \\
 \gamma _{11} &=& \[1,\(-2,1,1,0,1,1,1,1,1\),-\frac{3}{2}\] \\
\end{array}
\ee
with slopes $4/3,5/3,2$ and $4/3,7/6,1$, respectively. The Euler matrix has a three-block structure
\be
S=\left(
\begin{array}{ccc}
 1 & 3 & 2 \\
 0 & 1 & 1 \\
 0 & 0 &   \mathbb{1}_9  \\
\end{array}
\right),\ 
S^\vee=\left(
\begin{array}{ccc}
 1 & -3 & 1 \\
 0 & 1 & -1 \\
 0 & 0 &   \mathbb{1}_9 \\
\end{array}
\right),\ 
\kappa=\left(
\begin{array}{ccc}
 0 & 3 & -1 \\
 -3 & 0 & 1 \\
 1 & -1 &   \mathbb{0}_9  \\
\end{array}
\right)
\ee
A general Chern character $\gamma=[N,c_1,\ch_2]$ decomposes as $\gamma=
\sum_i n_i \gamma_i$ with  
\bea
 n_1 &=& \frac{1}{2} (3 c_{1,1}+3 c_{1,2}+3 c_{1,3}+3 c_{1,4}+5 c_{1,5}+5
   c_{1,6}+5 c_{1,7}+5 c_{1,8}+9 c_{1,H})+3 \ch_2- N \nn\\
 n_2 &=& \frac{1}{2} (c_{1,1}+c_{1,2}+c_{1,3}+c_{1,4}+3 c_{1,5}+3 c_{1,6}+3
   c_{1,7}+3 c_{1,8}+3 c_{1,H})+3 \ch_2 \nn\\
 n_3 &=& \frac{1}{2}
   (-c_{1,1}-c_{1,2}-c_{1,3}-c_{1,4}+c_{1,5}+c_{1,6}+c_{1,7}+c_{1,8}-c_{1,H})+ \ch_2 \nn\\
 n_4 &=& \frac{1}{2}
   (c_{1,1}+c_{1,2}+c_{1,3}-c_{1,4}+c_{1,5}+c_{1,6}+c_{1,7}+c_{1,8}+c_{1,H})+ \ch_2 \nn\\
 n_5 &=& \frac{1}{2}
   (-c_{1,1}-c_{1,2}-c_{1,3}-c_{1,4}+c_{1,5}-c_{1,6}-c_{1,7}-c_{1,8}-3
   c_{1,H})+ \ch_2+ N \nn\\
 n_6 &=& \frac{1}{2}
   (-c_{1,1}-c_{1,2}-c_{1,3}-c_{1,4}-c_{1,5}+c_{1,6}-c_{1,7}-c_{1,8}-3
   c_{1,H})+ \ch_2+ N \nn\\
 n_7 &=& \frac{1}{2}
   (-c_{1,1}-c_{1,2}-c_{1,3}-c_{1,4}-c_{1,5}-c_{1,6}+c_{1,7}-c_{1,8}-3
   c_{1,H})+ \ch_2+ N \nn\\
 n_8 &=& \frac{1}{2}
   (-c_{1,1}-c_{1,2}-c_{1,3}-c_{1,4}-c_{1,5}-c_{1,6}-c_{1,7}+c_{1,8}-3
   c_{1,H})+ \ch_2+N \nn\\
 n_9 &=& \frac{1}{2}
   (-c_{1,1}+c_{1,2}+c_{1,3}+c_{1,4}+c_{1,5}+c_{1,6}+c_{1,7}+c_{1,8}+c_{1,H})+\ch_2 \nn\\
 n_{10} &=& \frac{1}{2}
   (c_{1,1}-c_{1,2}+c_{1,3}+c_{1,4}+c_{1,5}+c_{1,6}+c_{1,7}+c_{1,8}+c_{1,H})+ \ch_2 \nn\\
 n_{11} &=& \frac{1}{2}
   (c_{1,1}+c_{1,2}-c_{1,3}+c_{1,4}+c_{1,5}+c_{1,6}+c_{1,7}+c_{1,8}+c_{1,H})+\ch_2
\eea
or conversely
\bea
 N &=& 3 n_{1}-6 n_{2}+n_{3}+n_{4}+n_{5}+n_{6}+n_{7}+n_{8}+n_{9}+n_{10}+n_{11} \nn\\
 c_{1,H} &=& 3 n_{2}-n_{3}-2 n_{4}-2 n_{9}-2 n_{10}-2 n_{11} \nn\\
 c_{1,1} &=& -n_{2}+n_{4}+n_{10}+n_{11} \nn\\
 c_{1,2} &=& -n_{2}+n_{4}+n_{9}+n_{11} \nn\\
 c_{1,3} &=& -n_{2}+n_{4}+n_{9}+n_{10} \nn\\
 c_{1,4} &=& -n_{2}+n_{9}+n_{10}+n_{11} \nn\\
 c_{1,5} &=& n_{1}-3 n_{2}+n_{3}+n_{4}+n_{5}+n_{9}+n_{10}+n_{11} \nn\\
 c_{1,6} &=& n_{1}-3 n_{2}+n_{3}+n_{4}+n_{6}+n_{9}+n_{10}+n_{11} \nn\\
 c_{1,7} &=& n_{1}-3 n_{2}+n_{3}+n_{4}+n_{7}+n_{9}+n_{10}+n_{11} \nn\\
 c_{1,8} &=& n_{1}-3 n_{2}+n_{3}+n_{4}+n_{8}+n_{9}+n_{10}+n_{11} \nn\\
 \ch_2 &=& \frac{1}{2} (-4 n_{1}+11 n_{2}-3 n_{3}-3 n_{4}-n_{5}-n_{6}-n_{7}-n_{8}-3 n_{9}-3 n_{10}-3
   n_{11})
 \eea
 The total dimensions in each block are then 
 \bea
 \cN_1 &=&
 -\frac12\left(3 c_{1,1}+3 c_{1,2}+3 c_{1,3}+3 c_{1,4}+5 c_{1,6}+5 c_{1,7}+5 c_{1,8}+9c_{1,H}\right)
   -3 \ch_2+N\nn\\
     \cN_2 &=&
    -\frac12\left(c_{1,1}+c_{1,2}+ c_{1,3}+c_{1,4}+3 c_{1,6}+3 c_{1,7}+3 c_{1,8}+3c_{1,H}\right)
   -3 \ch_2
   \\
    \cN_3 &=&
  \frac12\left(3 c_{1,1}+3 c_{1,2}+3 c_{1,3}+3 c_{1,4}-3 c_{1,6}-3 c_{1,7}-3 c_{1,8}+9c_{1,H}\right)
   -9 \ch_2-4 N
  \nn
%    \cN &=&-\frac{c_{1,1}}{2}-\frac{c_{1,2}}{2}-\frac{c_{1,3}}{2}-\frac{c_{1,4}}
%   {2}-\frac{11 c_{1,5}}{2}-\frac{11 c_{1,6}}{2}-\frac{11 c_{1,7}}{2}-\frac{11
%   c_{1,8}}{2}-\frac{3 c_{1,H}}{2}-15 \ch_2-3 N
   \eea
The stability vector $\vec\zeta=\vec\zeta^c$ in \eqref{zetacan}
in the canonical chamber $J\propto c_1(S)$ is
\bea
 \zeta _1 &=& (4 N-3 \deg ) \rho  + 3 \ch_2+2 N \nn\\
 \zeta _2 &=& (6 \deg -7 N) \rho   -6 \ch_2-\tfrac{11}{2}N \nn\\
 \zeta _{3,4} &=& (N-\deg ) \rho  + \ch_2+\tfrac{3 }{2}N \nn\\
 \zeta _{5,6,7,8} &=& (N-\deg ) \rho  + \ch_2+\tfrac{1}{2}N \nn\\
 \zeta _{9,10,11} &=& (N-\deg ) \rho  + \ch_2+\tfrac{3}{2}N 
\eea
The dimension agrees with \eqref{dimM} in the chamber where $\Phi_{i,1}=0$ with $i=3,\dots 11$. 
This requires $\zeta_{1}\geq  0$, $\zeta_{3,\dots 11}\leq  0$ hence
\be
N \leq  \deg   \leq \frac43 N 
\ee

\medskip
The vanishing of the attractor
indices follows from the arguments in \S\ref{sec_threeblockatt}. 
For $\gamma=[1;0,1,0,0,0,0,0,0,0,0,0;0]$ we get $\Omega(1,0,1,0,0,0,0,0,0,1,0,\vec\zetastar)=1$ as expected. For $\gamma=[1;(0,-1,0,0,0,0,0,0,0,0);1]$
we expect $\Omega(7, 5, 1, 2, 0, 0, 0, 0, 1, 2, 2,\vec\zeta^c)
=y^2+9+1/y^2$ but the height of the dimension
vector is too high for a direct check.

\medskip

In the blow-up chamber the stability parameters are instead
\be
\begin{split}
\vec\zeta^H=(-3 c_{1,H}; \  & 6 c_{1,H}+3 N; 
 -c_{1,H}-N,-c_{1,H}-2
   N,-c_{1,H}, \nn\\ 
   & -c_{1,H},-c_{1,H},-c_{1,H},-c_{1,H}-2
   N,-c_{1,H}-2 N,-c_{1,H}-2 N)
   \end{split}
\ee
This is consistent with $\zeta_1\geq  0, \zeta_{3,\dots,11}\leq  0$ for $c_{1,H}=0$ only. From the
tree flow formula we get
\be
\begin{array}{|c|c|l|}
\hline
\[N; c_1 ;c_2\] & \vec N &\Omega(\vec N, \vec\zeta^H)  \\ \hline
 \[2;  0,0,0,0,-1,0,1,1,1,1;0\] &(2, 2, 0, 0,1, 0, 0, 0, 1,1,1) & -y-1/y\nn \\
 \[3;0,0,0,0,1,0,1,1,1;0\] &(0,1, 1, 1, 1, 0, 0, 0, 0, 0,0) & 1 \nn \\
 \[3;0,0,0,-1,0,1,1,1,1;0\] & (2,2,0,1,0,0,0,0,1,1,0) & y^2+1+1/y^2 
    \\ \hline
    \end{array}
\ee
in agreement with the analysis in \S\ref{sec_VWdP}.

\subsubsection{Three-block collection (8.2) of type $(4,2,1)$}

The second strong exceptional collection (8.2) from \cite{karpov1998three}, also studied in \cite{Herzog:2003dj}, involves sheaves of rank $4,2,1$:
\be
\begin{array}{ccl}
 \gamma^1 &=& \[4,\(0,0,0,0,1,1,1,1,1\),-\frac{5}{2}\] \\
 \gamma^2 &=& \[2,\(1,0,0,0,0,0,0,0,0\),-\frac{1}{2}\] \\
 \gamma^3 &=& \[2,\(2,-1,-1,-1,0,0,0,0,0\),-\frac{1}{2}\] \\
 \gamma^4 &=& \[1,\(3,-1,-1,-1,0,-1,-1,-1,-1\),1\] \\
 \gamma^5 &=& \[1,\(3,-1,-1,-1,-1,0,-1,-1,-1\),1\] \\
 \gamma^6 &=& \[1,\(3,-1,-1,-1,-1,-1,0,-1,-1\),1\] \\
 \gamma^7 &=& \[1,\(3,-1,-1,-1,-1,-1,-1,0,-1\),1\] \\
 \gamma^8 &=& \[1,\(3,-1,-1,-1,-1,-1,-1,-1,0\),1\] \\
 \gamma^9 &=& \[1,\(1,-1,0,0,0,0,0,0,0\),0\] \\
 \gamma^{10} &=& \[1,\(1,0,-1,0,0,0,0,0,0\),0\] \\
 \gamma^{11} &=& \[1,\(1,0,0,-1,0,0,0,0,0\),0\] \\
\end{array}
\quad 
\begin{array}{ccl}
 \gamma_1 &=& \[4,\(0,0,0,0,1,1,1,1,1\),-\frac{5}{2}\] \\
 \gamma_2 &=& \[-6,\(1,0,0,0,-2,\dots, -2\),\frac{9}{2}\] \\
 \gamma_3 &=& \[-6,\(2,-1,-1,-1,-2,\dots,-2\),\frac{9}{2}\] \\
 \gamma_4 &=& \[1,\(0,0,0,0,1,0,0,0,0\),-\frac{1}{2}\] \\
 \gamma_5 &=& \[1,\(0,0,0,0,0,1,0,0,0\),-\frac{1}{2}\] \\
 \gamma_6 &=& \[1,\(0,0,0,0,0,0,1,0,0\),-\frac{1}{2}\] \\
 \gamma_7 &=& \[1,\(0,0,0,0,0,0,0,1,0\),-\frac{1}{2}\] \\
 \gamma_8 &=& \[1,\(0,0,0,0,0,0,0,0,1\),-\frac{1}{2}\] \\
 \gamma_9 &=& \[1,\(-2,0,1,1,1,1,1,1,1\),-\frac{3}{2}\] \\
 \gamma_{10} &=& \[1,\(-2,1,0,1,1,1,1,1,1\),-\frac{3}{2}\] \\
 \gamma_{11} &=& \[1,\(-2,1,1,0,1,1,1,1,1\),-\frac{3}{2}\] \\
\end{array}
\ee
with slope $5/4,3/2,2 $ and $5/4,7/6,1$, respectively. The Euler matrix has a three-block form, 
\be
S=\left(
\begin{array}{ccc}
 1 & 2 & 3 \\
 0 &   \mathbb{1}_2  & 1 \\
 0 & 0 &   \mathbb{1}_8  \\
\end{array}
\right),\ 
S^\vee=\left(
\begin{array}{ccc}
 1 & -2 & 1 \\
 0 &   \mathbb{1}_2  & -1 \\
 0 & 0 &   \mathbb{1}_8  \\
\end{array}
\right),\ 
\kappa=\left(
\begin{array}{ccc}
 0 & 2 & -1 \\
 -2 &   \mathbb{0}_2 & 1 \\
 1 & -1 &   \mathbb{0}_8  \\
\end{array}
\right)
\ee
A general Chern character $\gamma=[N,c_1,\ch_2]$ decomposes as $\gamma=
\sum_i n_i \gamma_i$ with  
\bea
 n_1 &=& 2 c_{1,1}+2 c_{1,2}+2 c_{1,3}+3 c_{1,4}+3 c_{1,5}+3 c_{1,6}+3
   c_{1,7}+3 c_{1,8}+6 c_{1,H}+4 \ch_2-N \nn\\
 n_2 &=&
   c_{1,1}+c_{1,2}+c_{1,3}+c_{1,4}+c_{1,5}+c_{1,6}+c_{1,7}+c_{1,8}+2
   c_{1,H}+2 \ch_2 \nn\\
 n_3 &=& c_{1,4}+c_{1,5}+c_{1,6}+c_{1,7}+c_{1,8}+c_{1,H}+2 \ch_2 \nn\\
 n_4 &=& \frac{1}{2}
   (-c_{1,1}-c_{1,2}-c_{1,3}+c_{1,4}-c_{1,5}-c_{1,6}-c_{1,7}-c_{1,8}-3
   c_{1,H})+ \ch_2+ N \nn\\
 n_5 &=& \frac{1}{2}
   (-c_{1,1}-c_{1,2}-c_{1,3}-c_{1,4}+c_{1,5}-c_{1,6}-c_{1,7}-c_{1,8}-3
   c_{1,H})+ \ch_2+N \nn\\
 n_6 &=& \frac{1}{2}
   (-c_{1,1}-c_{1,2}-c_{1,3}-c_{1,4}-c_{1,5}+c_{1,6}-c_{1,7}-c_{1,8}-3
   c_{1,H})+ \ch_2+N \nn\\
 n_7 &=& \frac{1}{2}
   (-c_{1,1}-c_{1,2}-c_{1,3}-c_{1,4}-c_{1,5}-c_{1,6}+c_{1,7}-c_{1,8}-3
   c_{1,H})+ \ch_2+N \nn\\
 n_8 &=& \frac{1}{2}
   (-c_{1,1}-c_{1,2}-c_{1,3}-c_{1,4}-c_{1,5}-c_{1,6}-c_{1,7}+c_{1,8}-3
   c_{1,H})+ \ch_2+N \nn\\
 n_9 &=& \frac{1}{2}
   (-c_{1,1}+c_{1,2}+c_{1,3}+c_{1,4}+c_{1,5}+c_{1,6}+c_{1,7}+c_{1,8}+c_{1,H})+ \ch_2 \nn\\
 n_{10} &=& \frac{1}{2}
   (c_{1,1}-c_{1,2}+c_{1,3}+c_{1,4}+c_{1,5}+c_{1,6}+c_{1,7}+c_{1,8}+c_{1,H})+ \ch_2 \nn\\
 n_{11} &=& \frac{1}{2}
   (c_{1,1}+c_{1,2}-c_{1,3}+c_{1,4}+c_{1,5}+c_{1,6}+c_{1,7}+c_{1,8}+c_{1,H})+ \ch_2
\eea
or conversely
\bea
 N &=& 4 n_{1}-6 n_{2}-6 n_{3}+n_{4}+n_{5}+n_{6}+n_{7}+n_{8}+n_{9}+n_{10}+n_{11} \nn\\
 c_{1,H} &=& n_{2}+2 n_{3}-2 n_{9}-2 n_{10}-2 n_{11} \nn\\
 c_{1,1} &=& -n_{3}+n_{10}+n_{11} \ ,\quad
 c_{1,2} = -n_{3}+n_{9}+n_{11} \ ,\quad 
  c_{1,3} = -n_{3}+n_{9}+n_{10} \nn\\
 c_{1,4} &=& n_{1}-2 n_{2}-2 n_{3}+n_{4}+n_{9}+n_{10}+n_{11} \nn\\
 c_{1,5} &=& n_{1}-2 n_{2}-2 n_{3}+n_{5}+n_{9}+n_{10}+n_{11} \nn\\
 c_{1,6} &=& n_{1}-2 n_{2}-2 n_{3}+n_{6}+n_{9}+n_{10}+n_{11} \nn\\
 c_{1,7} &=& n_{1}-2 n_{2}-2 n_{3}+n_{7}+n_{9}+n_{10}+n_{11} \nn\\
 c_{1,8} &=& n_{1}-2 n_{2}-2 n_{3}+n_{8}+n_{9}+n_{10}+n_{11} \nn\\
 \ch_2 &=& \frac{1}{2} \left(-5 n_{1}+9 n_{2}+9 n_{3}-n_{4}-n_{5}-n_{6}-n_{7}-n_{8}-3 n_{9}-3 n_{10}-3
   n_{11}\right)
\eea
As usual we set $\vec N=-\vec n$. The total dimensions for the three blocks are then 
 \bea
 \cN_1 &=&-2 c_{1,1}-2 c_{1,2}-2 c_{1,3}-3 c_{1,4}-3 c_{1,5}-3 c_{1,6}-3
   c_{1,7}-3 c_{1,8}-6 c_{1,H}-4
   \ch_2+N \nn\\
  \cN_2 &=&-c_{1,1}-c_{1,2}-c_{1,3}-2 c_{1,4}-2 c_{1,5}-2
   c_{1,6}-2 c_{1,7}-2 c_{1,8}-3 c_{1,H}-4 \ch_2\nn\\
   \cN_3 &=& 2 c_{1,1}+2 c_{1,2}+2
   c_{1,3}+6 c_{1,H}-8 \ch_2-5 N %\\
   %,\nn\\
% \cN &=&   -c_{1,1}-c_{1,2}-c_{1,3}-5
%   c_{1,4}-5 c_{1,5}-5 c_{1,6}-5 c_{1,7}-5 c_{1,8}-3 c_{1,H}-16 \ch_2-4
%   N
%\deg&=& 3c_{1,H}+\sum_i c_{1,i} = 7 N_2+7 N_3 -5 N_1+-N_4-N_5-N_6-N_7-N_8-N_9-N_{10}-N_{11}
%\nn
\eea
The stability parameters  $\vec\zeta=\vec\zeta^c$ in \eqref{zetacan}  for $J\propto c_1(S)$ are 
\bea
 \zeta_1 &=& -(4 \deg -5 N) \rho  + 4\ch_2 + \frac52 N \ ,\quad 
 \zeta_{2,3} = (6 \deg -7 N) \rho -6\ch_2-\frac92 N \nn\\
 \zeta_{4,5,6,7,8} &=&
   -(\deg-N) \rho +\ch_2+\frac12 N  \ ,\quad 
 \zeta_{9,10,11} =  -(\deg-N) \rho +\ch_2 + \frac32 N 
\eea
The dimension agrees with \eqref{dimM} in the chamber where $\Phi_{1j}=0$
with $j=4\dots 11$. This requires $\zeta_1\geq  0,\zeta_{4,\dots 11}\leq  0$ hence
\be
N \leq  \deg   \leq \frac54 N
\ee
in agreement with the range of slopes of stable objects.

\medskip
The vanishing of the attractor
indices follows from the arguments in \S\ref{sec_threeblockatt}. 
For $\gamma= \[1;  0,1,0,0,0,0,0,0,0,0;0\]$ we find
from the flow tree formula $\Omega(1, 0, 1, 0,0, 0, 0, 0, 0,1,0,\zeta^c) =1$ 
as expected.
For $\gamma=[1;0,0,0,0,0,0,0,0,1;1]$ one would expect
$\Omega(4, 2, 2, 1, 1, 1, 1, 0, 1, 1, 1)=y^2+9+1/y^2$, however the height of the
dimension vector is too high for a direct check. 

\medskip

In the blow-up chamber the stability parameters are instead
\be
\begin{split} 
\vec\zeta^H=&(-4 c_{1,H}; 
6 c_{1,H}+N,6 c_{1,H}+2N; \\
&-c_{1,H},-c_{1,H},-c_{1,H},-c_{1,H},-c_{1,H},-c_{1,H}-2
   N,-c_{1,H}-2 N,-c_{1,H}-2 N) + \epsilon \vec\zeta^c
\end{split} 
\ee
This is consistent with $\zeta_1\geq  0, \zeta_{4,\dots,11}\leq  0$ for $c_{1,H}=0$ only. From the flow tree formula we get
\be
\begin{array}{|c|c|l|}
\hline
\[N; c_1 ;c_2\] & \vec N &\Omega(\vec N, \vec\zeta^H)  \\ \hline
 \[1;  0,1,0,0,0,0,0,0,0,0;0\] &(1, 0, 1, 0,0, 0, 0, 0, 0,1,0) & 1 \nn \\
 \[3;0,0,-1,-1,1,1,1,1,1;0\] & (2,2,0,0,0,0,0,0,0,1,0) & 0 
    \\ \hline
    \end{array}
\ee
in agreement with the analysis in \S\ref{sec_VWdP}.

\subsubsection{Three-block collection (8.3) of type  $(3,2,1)$}
The third strong  exceptional collection (8.3) from \cite{karpov1998three} involves
sheaves of rank 3,2,1:
\be
\begin{array}{ccl}
 \gamma^1 &=& \[3,\(0,0,0,0,1,1,1,1,0\),-2\] \\
 \gamma^2 &=& \[3,\(0,0,0,0,1,1,1,0,1\),-2\] \\
 \gamma^3 &=& \[2,\(1,0,0,0,0,0,0,0,0\),-\frac{1}{2}\] \\
 \gamma^4 &=& \[2,\(2,-1,-1,-1,0,0,0,0,0\),-\frac{1}{2}\] \\
 \gamma^5 &=& \[2,\(0,0,0,0,1,1,1,0,0\),-\frac{3}{2}\] \\
 \gamma^6 &=& \[1,\(3,-1,-1,-1,0,-1,-1,-1,-1\),1\] \\
 \gamma^7 &=& \[1,\(3,-1,-1,-1,-1,0,-1,-1,-1\),1\] \\
 \gamma^8 &=& \[1,\(3,-1,-1,-1,-1,-1,0,-1,-1\),1\] \\
 \gamma^9 &=& \[1,\(1,-1,0,0,0,0,0,0,0\),0\] \\
 \gamma^{10} &=& \[1,\(1,0,-1,0,0,0,0,0,0\),0\] \\
 \gamma^{11} &=& \[1,\(1,0,0,-1,0,0,0,0,0\),0\] \\
\end{array}
\ 
\begin{array}{ccl}
 \gamma_1 &=& \[3,\(0,0,0,0,1,1,1,1,0\),-2\] \\
 \gamma_2 &=& \[3,\(0,0,0,0,1,1,1,0,1\),-2\] \\
 \gamma_3 &=&- \[4,\(-1,0,0,0,2,2,2,1,1\),-\frac{7}{2}\] \\
 \gamma_4 &=& -\[4,\(-2,1,1,1,2,2,2,1,1\),-\frac{7}{2}\] \\
 \gamma_5 &=& -\[4,\(0,0,0,0,1,1,1,1,1\),-\frac{5}{2}\] \\
 \gamma_6 &=& \[1,\(0,0,0,0,1,0,0,0,0\),-\frac{1}{2}\] \\
 \gamma_7 &=& \[1,\(0,0,0,0,0,1,0,0,0\),-\frac{1}{2}\] \\
 \gamma_8 &=& \[1,\(0,0,0,0,0,0,1,0,0\),-\frac{1}{2}\] \\
 \gamma_9 &=& \[1,\(-2,0,1,1,1,1,1,1,1\),-\frac{3}{2}\] \\
 \gamma_{10} &=& \[1,\(-2,1,0,1,1,1,1,1,1\),-\frac{3}{2}\] \\
 \gamma_{11} &=& \[1,\(-2,1,1,0,1,1,1,1,1\),-\frac{3}{2}\] \\
\end{array}
\ee
with slope $4/3,3/2,2 $ and $4/3,5/4,1$, respectively. The Euler matrix has a three-block form,
\be
S=\left(
\begin{array}{ccc}
   \mathbb{1}_2 & 1 & 2 \\
 0 &   \mathbb{1}_3  & 1 \\
 0 & 0 &   \mathbb{1}_6  \\
\end{array}
\right),\ 
S^\vee=\left(
\begin{array}{ccc}
   \mathbb{1}_2 & -1 & 1 \\
 0 &   \mathbb{1}_3  & -1 \\
 0 & 0 &   \mathbb{1}_6  \\
\end{array}
\right),\ 
\kappa=\left(
\begin{array}{ccc}
   \mathbb{0}_2  & 1 & -1 \\
 -1 &   \mathbb{0}_3  & 1 \\
 1 & -1 &   \mathbb{0}_6 \\
\end{array}
\right)
\ee
A general Chern character $\gamma=[N,c_1,\ch_2]$ decomposes as $\gamma=
\sum_i n_i \gamma_i$ with 
\bea
 n_1 &=& \frac{1}{2} (3 c_{1,1}+3 c_{1,2}+3 c_{1,3}+5 c_{1,4}+5 c_{1,5}+5
   c_{1,6}+5 c_{1,7}+3 c_{1,8}+9 c_{1,H})+3 \ch_2-N \nn\\
 n_2 &=& \frac{1}{2} (3 c_{1,1}+3 c_{1,2}+3 c_{1,3}+5 c_{1,4}+5 c_{1,5}+5
   c_{1,6}+3 c_{1,7}+5 c_{1,8}+9 c_{1,H})+3 \ch_2- N \nn\\
 n_3 &=&
   c_{1,1}+c_{1,2}+c_{1,3}+c_{1,4}+c_{1,5}+c_{1,6}+c_{1,7}+c_{1,8}+2
   c_{1,H}+2 \ch_2 \nn\\
 n_4 &=& c_{1,4}+c_{1,5}+c_{1,6}+c_{1,7}+c_{1,8}+c_{1,H}+2 \ch_2 \nn\\
 n_5 &=& c_{1,1}+c_{1,2}+c_{1,3}+2 c_{1,4}+2 c_{1,5}+2
   c_{1,6}+c_{1,7}+c_{1,8}+3 c_{1,H}+2 \ch_2-N \nn\\
 n_6 &=& \frac{1}{2}
   (-c_{1,1}-c_{1,2}-c_{1,3}+c_{1,4}-c_{1,5}-c_{1,6}-c_{1,7}-c_{1,8}-3
   c_{1,H})+\ch_2 + N \nn\\
 n_7 &=& \frac{1}{2}
   (-c_{1,1}-c_{1,2}-c_{1,3}-c_{1,4}+c_{1,5}-c_{1,6}-c_{1,7}-c_{1,8}-3
   c_{1,H})+\ch_2 + N \nn\\
 n_8 &=& \frac{1}{2}
   (-c_{1,1}-c_{1,2}-c_{1,3}-c_{1,4}-c_{1,5}+c_{1,6}-c_{1,7}-c_{1,8}-3
   c_{1,H})+\ch_2 + N \nn\\
 n_9 &=& \frac{1}{2}
   (-c_{1,1}+c_{1,2}+c_{1,3}+c_{1,4}+c_{1,5}+c_{1,6}+c_{1,7}+c_{1,8}+c_{1,H})+\ch_2 \nn\\
 n_{10} &=& \frac{1}{2}
   (c_{1,1}-c_{1,2}+c_{1,3}+c_{1,4}+c_{1,5}+c_{1,6}+c_{1,7}+c_{1,8}+c_{1,H})+\ch_2 \nn\\
 n_{11} &=& \frac{1}{2}
   (c_{1,1}+c_{1,2}-c_{1,3}+c_{1,4}+c_{1,5}+c_{1,6}+c_{1,7}+c_{1,8}+c_{1,H})+\ch_2 
\eea
or conversely
\bea
 N &=& 3 n_{1}+3 n_{2}-4 n_{3}-4 n_{4}-4 n_{5}+n_{6}+n_{7}+n_{8}+n_{9}+n_{10}+n_{11} \nn\\
 c_{1,H} &=& n_{3}+2 n_{4}-2 n_{9}-2 n_{10}-2 n_{11} \nn\\
 c_{1,1} &=& -n_{4}+n_{10}+n_{11} \ ,\quad 
  c_{1,2} = -n_{4}+n_{9}+n_{11} \ ,\quad
 c_{1,3} = -n_{4}+n_{9}+n_{10} \nn\\
 c_{1,4} &=& n_{1}+n_{2}-2 n_{3}-2 n_{4}-n_{5}+n_{6}+n_{9}+n_{10}+n_{11} \nn\\
 c_{1,5} &=& n_{1}+n_{2}-2 n_{3}-2 n_{4}-n_{5}+n_{7}+n_{9}+n_{10}+n_{11} \nn\\
 c_{1,6} &=& n_{1}+n_{2}-2 n_{3}-2 n_{4}-n_{5}+n_{8}+n_{9}+n_{10}+n_{11} \nn\\
 c_{1,7} &=& n_{1}-n_{3}-n_{4}-n_{5}+n_{9}+n_{10}+n_{11} \nn\\
 c_{1,8} &=& n_{2}-n_{3}-n_{4}-n_{5}+n_{9}+n_{10}+n_{11} \nn\\
 \ch_2 &=& \frac{1}{2} (-4 n_{1}-4 n_{2}+7 n_{3}+7 n_{4}+5 n_{5}-n_{6}-n_{7}-n_{8}-3 n_{9}-3 n_{10}-3
   n_{11})
\eea
As usual, we take $\vec N=-\vec n$. 
The total dimensions in each block are then 
\bea
\cN_1 &=& -3 c_{1,1}-3 c_{1,2}-3 c_{1,3}-5 c_{1,4}-5 c_{1,5}-5 c_{1,6}-4
   c_{1,7}-4 c_{1,8}-9 c_{1,H}-6 \ch_2+2 N\nn\\
\cN_2 &=&   -2 c_{1,1}-2 c_{1,2}-2
   c_{1,3}-4 c_{1,4}-4 c_{1,5}-4 c_{1,6}-3 c_{1,7}-3 c_{1,8}-6 c_{1,H}-6
   \ch_2+N \nn \\
\cN_3 &=&c_{1,1}+c_{1,2}+c_{1,3}-c_{1,4}-c_{1,5}-c_{1,6}+3
   c_{1,H}-6 \ch_2-3 N 
%   \\
%\cN&=&   -4 c_{1,1}-4 c_{1,2}-4 c_{1,3}-10
%   c_{1,4}-10 c_{1,5}-10 c_{1,6}-7 c_{1,7}-7 c_{1,8}-12 c_{1,H}-18
%   \ch_2
   \eea
The stability vector  $\vec\zeta=\vec\zeta^c$ in \eqref{zetacan}
in the canonical chamber $J\propto c_1(S)$ is
\bea
 \zeta_{1,2} &=& (4 N-3 \deg ) \rho  + 3 \ch_2+2 N \nn\\
 \zeta_{3,4,5} &=& (4 \deg -5 N) \rho   -4 \ch_2-\tfrac{7}{2} N\nn\\
 \zeta_{6,7,8} &=& (N-\deg ) \rho  + \ch_2+\tfrac{1}{2}N \nn\\
 \zeta_{9,10,11} &=& (N-\deg) \rho  + \ch_2+\tfrac{3 }{2} N \nn\\
\eea
The dimension agrees with \eqref{dimM} in the chamber where $\Phi_{ij}=0$ for $i=6,\dots 11$, $j=1,2$. 
This requires $\zeta_{1,2}\geq  0$, $\zeta_{6,7,8,9,10,11}\leq  0$ hence
\be
N \leq  \deg   \leq \frac{4}{3} N 
\ee

\medskip

The vanishing of the attractor
indices follows from the arguments in \S\ref{sec_threeblockatt}. 
For $\gamma=[1;0,0,0,0,0,0,0,1,0;0]$ in the canonical chamber 
we find from the flow tree formula 
\be
\Omega(0,1,0,0,1,0,0,0,0,0,0,\vec\zeta^c)=1
\ee
 as expected. 
For $\gamma=[1;0,0,0,0,0,0,0,0,0,1;1]$
we expect $\Omega(4, 3, 2, 2, 3, 1, 1, 1, 1, 1, 1,\vec\zeta^c)=y^2+9+1/y^2$ 
but the height is too high for a direct check.

\medskip

In the blow-up chamber the stability parameters are instead
\be
\begin{split} 
\vec\zeta^H=& (-3 c_{1,H},-3 c_{1,H};
4 c_{1,H}+N,4 c_{1,H}+2 N,4 c_{1,H}; \\
&-c_{1,H},-c_{1,H},-c_{1,H},-c_{1,H}-2 N,-c_{1,H}-2
   N,-c_{1,H}-2 N) + \epsilon\zeta^c
   \end{split}
\ee
This is consistent with $\zeta_{1,2}\geq  0, \zeta_{6,\dots,11}\leq  0$ for $c_{1,H}=0$ only. 
From the flow tree formula we get
\be
\begin{array}{|c|c|l|}
\hline
\[N; c_1 ;c_2\] & \vec N &\Omega(\vec N, \vec\zeta^H)  \\ \hline
 \[3;  0,1,0,0,1,1,1,0,0;0\] &(0,0, 0, 1, 0, 0, 0, 0,1,0,0) & 1 \nn \\
  \[3;  0,1,0,0,0,1,1,1,0;0\] &(0,1, 0, 1, 1, 1, 0, 0,1,0,0)  & 1 \nn \\
 \[3;0,0,0,-1,1,2,2,0,0;0\] &(1,1,2,1,0,1,0,0,1,1, 0) & 1
    \\ \hline
    \end{array}
\ee
in agreement with the analysis in \S\ref{sec_VWdP}.

\subsubsection{Three-block collection (8.4) of type  $(5,2,1)$}
The fourth and last  strong exceptional collection (8.4) from \cite{karpov1998three}
involves sheaves of rank 5,2,1:
\be
\begin{array}{ccl}
 \gamma^1 &=& \[5,\(6,-2,-2,-2,0,0,0,0,0\),0\] \\
 \gamma^2 &=& \[2,\(3,-1,-1,-1,-1,0,0,0,0\),\frac{1}{2}\] \\
 \gamma^3 &=& \[2,\(3,-1,-1,-1,0,-1,0,0,0\),\frac{1}{2}\] \\
 \gamma^4 &=& \[2,\(3,-1,-1,-1,0,0,-1,0,0\),\frac{1}{2}\] \\
 \gamma^5 &=& \[2,\(3,-1,-1,-1,0,0,0,-1,0\),\frac{1}{2}\] \\
 \gamma^6 &=& \[2,\(3,-1,-1,-1,0,0,0,0,-1\),\frac{1}{2}\] \\
 \gamma^7 &=& \[1,\(1,0,0,0,0,0,0,0,0\),\frac{1}{2}\] \\
 \gamma^8 &=& \[1,\(2,-1,-1,-1,0,0,0,0,0\),\frac{1}{2}\] \\
 \gamma^9 &=& \[1,\(4,-2,-1,-1,-1,-1,-1,-1,-1\),\frac{5}{2}\] \\
 \gamma^{10} &=& \[1,\(4,-1,-2,-1,-1,-1,-1,-1,-1\),\frac{5}{2}\] \\
 \gamma^{11} &=& \[1,\(4,-1,-1,-2,-1,-1,-1,-1,-1\),\frac{5}{2}\] \\
\end{array}
\ 
\begin{array}{ccl}
 \gamma_1 &=& \[5,\(6,-2,-2,-2,0,0,0,0,0\),0\] \\
 \gamma_2 &=& \[-3,\(-3,1,1,1,-1,0,0,0,0\),\frac{1}{2}\] \\
 \gamma_3 &=& \[-3,\(-3,1,1,1,0,-1,0,0,0\),\frac{1}{2}\] \\
 \gamma_4 &=& \[-3,\(-3,1,1,1,0,0,-1,0,0\),\frac{1}{2}\] \\
 \gamma_5 &=& \[-3,\(-3,1,1,1,0,0,0,-1,0\),\frac{1}{2}\] \\
 \gamma_6 &=& \[-3,\(-3,1,1,1,0,0,0,0,-1\),\frac{1}{2}\] \\
 \gamma_7 &=& \[1,\(-2,1,1,1,1,1,1,1,1\),-2\] \\
 \gamma_8 &=& \[1,\(-1,0,0,0,1,1,1,1,1\),-2\] \\
 \gamma_9 &=& \[1,\(1,-1,0,0,0,0,0,0,0\),0\] \\
 \gamma_{10} &=& \[1,\(1,0,-1,0,0,0,0,0,0\),0\] \\
 \gamma_{11} &=& \[1,\(1,0,0,-1,0,0,0,0,0\),0\] \\
\end{array}
\ee
with slopes $12/5,5/2,3$ and $12/5,7/3,2$, respectively. 
The Euler matrix has a three-block  form,
\be
S=\left(
\begin{array}{ccc}
 1 & 1 & 3 \\
 0 &   \mathbb{1}_5  & 1 \\
 0 & 0 &  \mathbb{1}_5  \\
\end{array}
\right),\ 
S^\vee=\left(
\begin{array}{ccc}
 1 & -1 & 2 \\
 0 &  \mathbb{1}_5  & -1 \\
 0 & 0 &  \mathbb{1}_5  \\
\end{array}
\right),\ 
\kappa=\left(
\begin{array}{ccc}
 0 & 1 & -2 \\
 -1 &  \mathbb{0}_5  & 1 \\
 2 & -1 &  \mathbb{0}_5  \\
\end{array}
\right)\
\ee
A general Chern character $\gamma=[N,c_1,\ch_2]$ decomposes as $\gamma=
\sum_i n_i \gamma_i$ with  
\bea
 n_1 &=& \frac{1}{2} (c_{1,1}+c_{1,2}+c_{1,3}+5 c_{1,4}+5 c_{1,5}+5 c_{1,6}+5
   c_{1,7}+5 c_{1,8}+3 c_{1,H})+5 \ch_2- N \nn\\
 n_2 &=& c_{1,5}+c_{1,6}+c_{1,7}+c_{1,8}+2 \ch_2 \nn\\
 n_3 &=& c_{1,4}+c_{1,6}+c_{1,7}+c_{1,8}+2 \ch_2 \nn\\
 n_4 &=& c_{1,4}+c_{1,5}+c_{1,7}+c_{1,8}+2 \ch_2 \nn\\
 n_5 &=& c_{1,4}+c_{1,5}+c_{1,6}+c_{1,8}+2 \ch_2 \nn\\
 n_6 &=& c_{1,4}+c_{1,5}+c_{1,6}+c_{1,7}+2 \ch_2 \nn\\
 n_7 &=& \frac{1}{2}
   (c_{1,1}+c_{1,2}+c_{1,3}+c_{1,4}+c_{1,5}+c_{1,6}+c_{1,7}+c_{1,8}+c_{1,H})+ \ch_2 \nn\\
 n_8 &=& \frac{1}{2}
   (-c_{1,1}-c_{1,2}-c_{1,3}+c_{1,4}+c_{1,5}+c_{1,6}+c_{1,7}+c_{1,8}-c_{1,H})+ \ch_2 \nn\\
 n_9 &=& \frac{1}{2} (-3
   c_{1,1}-c_{1,2}-c_{1,3}-c_{1,4}-c_{1,5}-c_{1,6}-c_{1,7}-c_{1,8}-5  c_{1,H})+ \ch_2+2 N \nn\\
 n_{10} &=& \frac{1}{2} (-c_{1,1}-3
   c_{1,2}-c_{1,3}-c_{1,4}-c_{1,5}-c_{1,6}-c_{1,7}-c_{1,8}-5 c_{1,H})+ \ch_2+2 N \nn\\
 n_{11} &=& \frac{1}{2} (-c_{1,1}-c_{1,2}-3
   c_{1,3}-c_{1,4}-c_{1,5}-c_{1,6}-c_{1,7}-c_{1,8}-5 c_{1,H})+  \ch_2+2N 
\eea
or conversely
\bea
 N &=& 5 n_{1}-3 n_{2}-3 n_{3}-3 n_{4}-3 n_{5}-3 n_{6}+n_{7}+n_{8}+n_{9}+n_{10}+n_{11} \nn\\
 c_{1,H} &=& 6 n_{1}-3 n_{2}-3 n_{3}-3 n_{4}-3 n_{5}-3 n_{6}-2 n_{7}-n_{8}+n_{9}+n_{10}+n_{11} \nn\\
 c_{1,1} &=& -2 n_{1}+n_{2}+n_{3}+n_{4}+n_{5}+n_{6}+n_{7}-n_{9} \nn\\
 c_{1,2} &=& -2 n_{1}+n_{2}+n_{3}+n_{4}+n_{5}+n_{6}+n_{7}-n_{10} \nn\\
 c_{1,3} &=& -2 n_{1}+n_{2}+n_{3}+n_{4}+n_{5}+n_{6}+n_{7}-n_{11} \nn\\
 c_{1,4} &=& -n_{2}+n_{7}+n_{8} \nn\\
 c_{1,5} &=& -n_{3}+n_{7}+n_{8} \nn\\
 c_{1,6} &=& -n_{4}+n_{7}+n_{8} \nn\\
 c_{1,7} &=& -n_{5}+n_{7}+n_{8} \nn\\
 c_{1,8} &=& -n_{6}+n_{7}+n_{8} \nn\\
 \ch_2 &=& \frac{1}{2} (n_{2}+n_{3}+n_{4}+n_{5}+n_{6}-4 n_{7}-4 n_{8}
\eea
The total dimensions in each block are then 
\bea
\cN_1 &=& -\frac12\left( c_{1,1}+c_{1,2}+c_{1,3}+5c_{1,4}+5c_{1,5}+5c_{1,6}+5c_{1,7}+5c_{1,8}
+3c_{1,H}\right)
   -5 \ch_2+N, \nn\\
\cN_2 &=&  -4 c_{1,4}-4 c_{1,5}-4
   c_{1,6}-4 c_{1,7}-4 c_{1,8}-10 \ch_2, \\
\cN_3 &=&  
\frac12\left( 5c_{1,1}+5c_{1,2}+5c_{1,3}+c_{1,4}+c_{1,5}+c_{1,6}+c_{1,7}+c_{1,8}
+15c_{1,H}\right)
   -5 \ch_2-6 N
   \nn
%\cN&=&   2
%   c_{1,1}+2 c_{1,2}+2 c_{1,3}-6 c_{1,4}-6 c_{1,5}-6 c_{1,6}-6 c_{1,7}-6
%   c_{1,8}+6 c_{1,H}-20 \ch_2-5 N
   \eea
 The stability vector is 
\bea
 \zeta_1 &=& (12 N-5 \deg +) \rho  +5 \ch_2 \nn\\
 \zeta_{2,\dots,6} &=& (3 \deg -7 N) \rho  -3 \ch_2-\frac{N}{2} \nn\\
 \zeta_{7,8,9} &=& (2 N-\deg ) \rho + \ch_2+2 N \nn\\
  \zeta_{10,11} &=& (2 N-\deg) \rho + \ch_2 
\eea
The dimension agrees with \eqref{dimM} in the chamber where $\Phi_{i1}=0$ for $i=9,10, 11$. 
This requires $\zeta_{1}\geq  0$, $\zeta_{7,8,9,10,11}\leq  0$ hence
\be
2N \leq  \deg   \leq \frac{12}{5} N 
\ee

\medskip

The vanishing of the attractor
indices follows from the arguments in \S\ref{sec_threeblockatt}. 
For $\gamma= \[1;  1,0,0,0,0,0,0,0,-1;0\]$ 
we get from the flow tree formula $\Omega(2,1, 1, 1, 1,0, 0, 1,0,0,0,\vec\zeta^c)=1$
as expected. 
For $\gamma=[1;1,-1,0,0,0,0,0,0,0,0;1]$
we expect $\Omega(6, 2, 2, 2, 3, 3, 1, 1, 1, 1, 1,\vec\zeta^c)=
y^2+9+1/y^2$ but the height is too high for a direct check.

\medskip

In the blow-up chamber the stability parameters are instead
\be
\begin{split}
\vec \zeta^H =& (6 N-5 c_{1,H}; 
3 c_{1,H}-3 N,3 c_{1,H}-3 N,3 c_{1,H}-3N,3 c_{1,H}-3 N,3 c_{1,H}-3 N; \\
& -c_{1,H}-2 N,-c_{1,H}-N,N-c_{1,H},N-c_{1,H},N-c_{1,H} ) + \epsilon \vec\zeta^c
\end{split} 
\ee
This is consistent with $\zeta_{1}\geq  0, \zeta_{7,8,9,10,11}\leq  0$ for $N\leq c_{1,H}\leq \frac65 N$. 
From the flow tree formula we get
\be
\begin{array}{|c|c|l|}
\hline
\[N; c_1 ;c_2\] & \vec N &\Omega(\vec N, \vec\zeta^H)  \\ \hline
  \[1;  1,0,0,0,0,0,0,0,-1;0\] &(2,1, 1, 1, 1,0, 0, 1,0,0,0)  & 1   \nn \\
   \[3;  3,-1,-1,-1,1,0,0,0,0;3\] &(0,1, 0, 0, 0, 0, 0, 0,0,0,0) & 1 \nn \\
    \[3;  3,0,-1,-1,0,0,0,0,0;4\] &(2,1, 1, 1, 1, 1, 0, 1,1,0,0) & y^2+7+1/y^2 \nn 
    \\ \hline
    \end{array}
\ee
in agreement with the analysis in \S\ref{sec_VWdP}. 

\section{Summary and discussion \label{sec_discuss}}

{\bp In this work, we have proposed a general method for computing Vafa-Witten invariants 
$c_{\gamma,J}$ on any complex surface $S$  which admits  a strong, full, 
cyclic exceptional collection $\cC$. This assumes that vanishing theorems apply such that
Vafa-Witten invariants on $S$ are solely given by contributions from the instanton branch. }
The method involves 1) constructing the quiver $Q$ 
and its superpotential $W$
from the dual collection $\cC^\vee$; 2) determining  the dimension vector $\vec N$ and stability parameter $\vec\zeta$ 
for given Chern vector $\gamma$ and polarization $J$; 3) computing the attractor indices
$\Omstar(\vec N):=\Omega(\vec N,\vec \zetastar)$ for all $\vec N$ and 4) evaluating the index $\Omega(\vec N,\vec \zeta)=
c_{\gamma,J}$  in the desired chamber using the flow tree formula.
The equality $\Omega(\vec N,\vec \zeta)=c_{\gamma,J}$ is  expected to hold for Chern classes such that the slope lies in a certain window $\nu_-\leq \nu_J(E) \leq \nu_+$,
typically determined by the slopes of the dual collection $\cC^\vee$, such that stable representations are supported on the Beilinson subquiver.  
For the canonical polarization $J\propto c_1(S)$, we showed that the relevant stability vector $\vec\zeta^c$ is  {\it opposite} to the attractor value $\vec\zetastar$, up to a small perturbation.

\medskip 

We  validated
the method outlined above in the case of 
Fano surfaces, where suitable exceptional collections are known.
For all considered cases, we could show that the  attractor indices
$\Omstar(\vec N)$ vanish except for dimension vectors associated to 
simple representations and (possibly) pure D0-branes. Since
pure D0-branes do not contribute to the flow tree formula for Chern vectors associated
to torsion-free sheaves, we were able to compute the index $\Omega(\vec N,\vec \zeta)$
for a large variety of Chern vectors, and successfully match with results obtained by combining 
the blow-up and wall-crossing formulae. It would be very  interesting to see if our method
can be extended to torsion sheaves (i.e. D2-D0 brane bound states), and provide a new way of computing Gopakumar-Vafa invariants on these surfaces, some which are not toric.

\medskip

We also found evidence that this method extends to weak Fano surfaces ($\IF_2$ and the pseudo del Pezzo surfaces $PdP_n$) where VW invariants turn out to be related to those of Fano surfaces of the same degree, despite the fact that the quiver sometimes involves bi-directional arrows. 
For $\IF_m$ with $m\geq 3$, we have not identified the full quiver description, although we expect that the invariants can still be related to those of $\IF_0$ or $\IF_1$, depending on the parity of $m$. It would be very interesting to extend our method to more general surfaces, such as K3 and surfaces of general type.

\medskip 

 In order to prove that the attractor indices $\Omstar(\vec N,y)$ vanish for generic dimension vector, we showed  that the expected dimension in the attractor chamber is strictly negative, by producing a positive quadratic form $\cQ(\vec N)$ such that the dimension 
can be written as in \eqref{dstareps}. It would be useful to understand the geometric origin
of the quadratic form $\cQ(\vec N)$, which we have found by {\it ad hoc} methods.  
{\bp We note that our proof is not mathematically rigorous, since it depends on the unproven assertion that for any semi-stable representation with generic dimension vector, there exists a cut $Q_1'\subset Q_1$ of the superpotential
such that all arrows in $Q'_1$ vanish.  For toric cases, the superpotential 
can be easily extracted from the brane tiling, and the possible cuts correspond to perfect
matchings. It would be interesting to determine the superpotential in non-toric cases
(including its dependence on complex structure moduli) and classify their possible cuts.}

\medskip

{\bp We emphasize that the vanishing of attractor indices is a special property of quivers
associated to del Pezzo surfaces, and is not be expected to hold in general.
The vanishing
of attractor indices (or initial data in the language of \cite{Kontsevich:2013rda}) is known to hold
for  acyclic quivers and  for quivers in the class $\cP$ defined in \cite[\S 8.4]{ks},
but del Pezzo quivers do not belong to  these classes of examples.\footnote{We thank Tom Bridgeland, Maxim Kontsevich and Yan Soibelman for discussions on this issue.}
The vanishing of attractor invariants for $S=\IP^2$ seems to be consistent with the 
scattering diagram formalism developped in  \cite{gross2010tropical,
Gross:2014fwa,bridgeland2016scattering,bousseau2019scattering},
and it would be interesting to understand the relation 
between scattering diagrams and  attractor flow trees.
 A property shared by all quivers for any complex surface is that the adjacency matrix has rank 2, but this in itself is not a sufficient condition for the vanishing
of attractor indices, as shown by the example of three-node quivers with generic values of $(a,b,c)$ outside the list in below \eqref{dtotP2gencond}. 
It seems plausible that the vanishing
of attractor indices should hold for all rational surfaces, although we have little to back 
up this conjecture. }

\medskip 

In addition to the vanishing of the attractor indices, we have found circumstancial 
evidence that the single centered indices  $\OmS(\vec N,y)$ coincide with attractor indices, and therefore vanish under the same conditions on $\vec N$ (for the same quiver $Q$). 
The conjectural 
identity $\OmS(\vec N,y)= \Omstar(\vec N,y)$ follows from the observation 
that in evaluating the Coulomb branch formula \eqref{CBform} in the attractor chamber $\vec \zeta=\vec\zetastar$, all contributions from decompositions 
$\gamma=\sum_j \alpha_j$ where all $\alpha_j$'s are associated to simple representations 
appear to have a vanishing coefficient $\widehat{g}_C(\{\alpha_j,m_j, c_j\},y)$. This does {\it not} mean that collinear black hole configurations with charges $\{\alpha_j\}$ do not exist at the attractor point, but rather that
they carry vanishingly small angular momentum, such that  their contributions are cancelled by the minimal modification prescription of \cite{Manschot:2011xc}. Physically, black holes 
corresponding to the exceptional D-branes $E_i^\vee$ may {\it  classically} form scaling configurations, but not quantum mechanically. 
This seems to be a remarkable property of the adjacency matrix coming
from strong exceptional collections, though it does not rely on the specific form of the superpotential. 
Moreover, the vanishing of  $\OmS(\vec N,y)$ for almost all dimension vectors is consistent with the general expectation that  there should exist no single-centered black holes with large entropy in local Calabi-Yau geometries. 
For dimension vectors corresponding to pure D0-branes on $X=\mathrm{Tot}(K_S)$,  
single-centered and attractor indices no longer coincide, rather 
\be 
\Omstar(\vec N,y) = -y-1/y + \OmS(\vec N,y) + \dots
\ee
where the dots indicate contributions proportional
to $\OmS(\alpha_j)$ with generic dimension vectors, which conjecturally vanish.
 In this case it is clear that 
$\Omstar(\vec N,y)$ and $\OmS(\vec N,y)$ cannot both vanish, and it would be interesting to determine either of them. Due to the fact that $X$ is non-compact, these indices are not expected to be invariant under $y\to 1/y$.

\medskip

The fact that the stability vector $\vec \zeta^c$ relevant for the canonical polarization $J\propto c_1(S)$ is {\it opposite} to the attractor value  $\vec \zetastar$ raises an interesting paradox. As emphasized in 
\cite{Alexandrov:2019rth}, the canonical polarization corresponds to the large volume attractor point $z_\gamma$ for a D4-brane wrapped on $S$. Generally, at the attractor point 
one expects only single-centered black holes to contribute, possibly along with scaling solutions. In contrast, for $\vec\zeta=\vec\zeta^c$ the spectrum of BPS bound states in quiver quantum mechanics is in some sense at its richest point  (while being at its poorest for $\vec\zeta=-\vec\zeta^c$). 
The resolution of this paradox is presumably that the BPS spectrum in the large volume attractor point (counted
by the so-called MSW index) includes BPS bound states of D6-anti D6-branes which are well described 
by the quiver quantum mechanics in the canonical chamber $\vec\zeta=\vec\zeta^c$. These bound
states decay at finite volume on the way to the true attractor point, consistently with the absence of 
bound states in the quiver quantum mechanics for $\vec\zeta=\vec\zetastar$.

\medskip

Another property of the canonical polarization is that the (mock) modular properties and holomorphic anomalies of the generating function of VW invariants  with fixed rank and first Chern class are expected to be simpler than for generic $J$ \cite{Manschot:2011dj}. It would clearly be of great interest to understand the appearance of these modular properties from the quiver description. In analogy with the case of framed sheaves on ALE spaces \cite{nakajima1994instantons} or Hilbert scheme of points on surfaces \cite{nakajima1999lectures}, one may hope to identify the generating function of quiver indices with the character of an affine or possibly more exotic vertex operator algebra, possibly in the spirit of \cite{Li:2020rij}, relating the cohomologies of quiver moduli spaces with different dimension vectors. We hope to return to this in the future.

%\enlargethispage{1cm}

 \medskip
 
 \noindent  {\bf Acknowledgements:}  We are grateful to Sergey Alexandrov, Pierrick Bousseau, 
  Emmanuel Diaconescu, Amihay Hanany, Yang-Hui He, Amir Kashani-Poor, 
 Sergey Mozgovoy, Olivier Schiffmann, Ashoke Sen, 
 and especially Markus Perling for helpful discussions or communications in the recent or more distant past. We also thank Tom Bridgeland, Sebastian Franco, Maxim Kontsevitch, Richard Thomas and Yan Soibelman 
 for useful comments on an earlier version of this work.
 The research of J.M. is
supported by IRC Laureate Award 15175 ``Modularity in Quantum Field
Theory and Gravity''.

\medskip
\medskip

%\newpage

\appendix

\section{Generating functions of  VW invariants \label{sec_genVW}}
In this appendix, we collect known results about generating functions of VW invariants 
for various rational surfaces, in order to provide checks for the
quiver description in the body of this paper. As in \S\ref{sec_VW}, let
$P(\cM^S_{\gamma,J},y)$ be the Poincar\'e polynomial of
the moduli space of Gieseker semi-stable sheaves with Chern character $\gamma=[N;\mu;n]$. This is a Laurent
polynomial with positive integer coefficients, symmetric under $y\leftrightarrow y^{-1}$. The rational
invariant $c_{\gamma,J}^{\rm ref}(y)$ is defined by
(\ref{defcref}), and coincides with $P(\cM^S_{\gamma,J},y)$ when $\gamma$ is primitive. 
It is useful to further introduce  the stack invariant \be  
\label{Itoc}
\mathcal{I}_{\gamma, J}(y) :=\sum_{\substack{\gamma_1+\dots+\gamma_\ell=\gamma
\\ \nu_J(\gamma_j)=\nu_J(\gamma)\,\,\forall j  \\
\frac{\ch_2}{\rk} (\gamma_j) = \frac{\ch_2}{\rk} (\gamma) \, \forall i
}}
\frac{1}{\ell !\,(y-y^{-1})^{\ell}}\, \prod_{j=1}^\ell\, c_{\gamma_j,J}^{\rm ref}(y),
\ee
which has simpler transformation properties under wall-crossing. 
Conversely, the rational invariant can be expressed in terms of stack invariants of the moduli space of Gieseker semi-stable sheaves
  through
\be
\label{ctoI}
c_{\gamma_j,J}^{\rm ref}(y)=(y-y^{-1})\sum_{\substack{\gamma_1+\dots+\gamma_\ell=\gamma
\\ \nu_J(\gamma_j)=\nu_J(\gamma)\,\,\forall j  \\
\frac{\ch_2}{N} (\gamma_j) = \frac{\ch_2}{N} (\gamma) \, \forall j
}}
\frac{(-1)^{\ell-1}}{\ell }\, \prod_{j=1}^\ell\mathcal{I}_{\gamma_j,J}(y).
\ee
In order to make use of the blow-up formula, we shall also need the stack invariants 
$\mathcal{I}_{\gamma, J}^{\nu}(y)$ 
of the moduli space $\cM^{S,\nu}_{\gamma,J}$ of {\it slope} semi-stable sheaves. For a generic polarization $J$, these are related to the stack invariants $\mathcal{I}_{\gamma, J}(y)$   for
Gieseker stability via 
\be
\label{IMI}
\mathcal{I}_{\gamma, J}^{\nu}(y)=\sum_{{\gamma_1+\dots+\gamma_\ell=\gamma
  \atop \nu_J(\gamma_j)=\nu_J(\gamma)\, \forall j } \atop
\frac{\ch_{2}}{N}(\gamma_j)< \frac{\ch_{2}}{N}(\gamma_{j+1})} \prod_{j=1}^\ell \mathcal{I}_{\gamma_j, J}(y),
\ee 
where $\nu_J$ is the slope (\ref{defnuE}). Similar to \eqref{defhVWref}, we define the 
 generating function of the stack invariants for slope-stability as
\be
\label{defHS}
H^{S}_{N,c_1,J}(\tau,w)=\sum_n \mathcal{I}^{\nu}_{\gamma, J}(y)\,\q^{n-\frac{N-1}{2N}\mu^2-\frac{N\chi(S)}{24}}.
\ee
Using (\ref{ctoI}) and (\ref{IMI}), one can express the generating function 
of rational invariants  (\ref{defhVWref}) in
terms of \eqref{defHS} for a generic polarization $J$,
\be
\label{hSHS}
h^{S}_{N,c_1,J}(\tau,w) =\sum_{\sum (N_i,c_{1,i})=(N,c_1)\atop
  \nu_J(N_i,c_{1,i})=\nu_J(N,c_{1})} \frac{(-1)^\ell}{\ell} \prod_{j=1}^\ell H^{S}_{N_i,c_{1,i},J}.
\ee
To list generating functions in a more concise fashion, we further 
introduce the generating function $\hiVW^S_{N,\mu,J}$ of Poincar\'e polynomials of the moduli spaces of  Gieseker semi-stable sheaves,
\be
\hiVW^S_{N,\mu,J} (\tau,w) = 
\sum_{n\geq 0}
P(\cM^S_{[N;\mu;n],J},y)\, \q^{n} \ .
\ee 
When $(N,\mu)$ is primitive, this is related to the generating function \eqref{defhVWref} by
\be
\hiVW^S_{N,\mu,J} (\tau,w) = \q^{\tfrac{N-1}{2N} \mu^2 +  \tfrac{N \chi(S)}{24}}\,  (y-1/y)\, 
\hVW_{N,\mu,J}(\tau,w),
\ee 
whereas for generic $(N,\mu)$, $\hiVW^S_{N,\mu,J}$ follows  from the generating function 
$\hVW_{N,\mu,J}$ of rational invariants
 by inverting (\ref{defcref}). 
In all cases we omit the dependence on $J$ when it is clear from the context. 

\medskip

For rank $N=1$, VW invariants are independent of $J$ and $\mu$
and given by Goettsche's formula  \eqref{h10anyS} for any smooth, simply connected 
4-manifold. Expanding at the first few orders in $\q$ we get 
\be
\label{VW1}
\begin{split}
%q^{\frac{\chi(S)}{24}}(y-1/y) h_{1,0} &=
\hiVW^S_{1} &=
1 + (y^2 + b_2 + 1/y^2) \q+ \left(y^4 + (b_2+1) y^2 + \frac12(b_2+1)(b_2+2) + \dots \right) \q^2 \\
&+ \left(y^6 + (b_2+1) y^4 + 
\frac12(b_2+1)b_2+4)  y^2 + \left(2 + \frac16 b_2 (b_2+2) (b_2+7)\right)+ \dots  \right) \q^3 + \dots ,
\end{split}
\ee
where $b_2=b_2(S)$  and the dots inside the brackets denote terms required by 
invariance under $y\to 1/y$.
The order $\q^n$  term in this expression is the Poincar\'e polynomial of the Hilbert scheme $\cM^S_{[1;\mu;n],J}=S^{[n]}$,
rescaled by a power of $y^{-\dim_\IC \cM}$. 
{\bp It is interesting to note that the Betti numbers  $b_{2k}(S^{[n]})$ stabilize for $n\geq 2k$ to 
a value $\tilde b_{2k}$, the generating function of which is given by \cite{Gottsche:1990}}
\be
\label{stableBetti}
\sum_{k=0}^{\infty} \tilde b_{2k}\,y^{2k} = 
\frac{(1-y^2)}{\prod_{k=1}^{\infty} (1-y^{2k})^{b_2+2}}
=1 + (b_2+1)y^2+ \frac12 (b_2^2+5b_2+6) y^4 + \dots 
\ee
Moreover, they reach this value from below, $b_{2k}(S^{[n]})\leq \tilde b_{2k}$ for all $k\geq 0$.

\medskip

As explained in a series of works \cite{Yoshioka:1994, Yoshioka:1995, yoshioka1999euler, Gottsche:1999ix, Manschot:2010nc, Manschot:2011ym, Manschot:2014cca}, a general strategy for obtaining VW invariants for rational surfaces at arbitrary rank $N\geq 2$ is to first compute generating functions for Hirzebruch surfaces $\IF_m$, and then apply the blow-up formula
to reach $\IP^2$ (by blowing down one point on $\IF_1$) or $dP_k$ (by blowing up $k-1$ points
on $\IF_1$). The blow-up formula relates the generating functions of stack invariants for slope stability on the surface $S$ to those on the blow-up $\pi:\hat S\to S$ \cite{Yoshioka:1996, Li_1999, Gottsche:1999ix}. We have in particular,  
\be
\label{HtoHhat}
H^{\hat S}_{N,\pi_*(c_1)-k C,\pi_*(J)}(\tau,w)= B_{N,k}(\tau,w)\, H^{S}_{N,c_1,J}(\tau,w),
\ee
where $C$ is the exceptional curve and      
\be 
\label{defblowB}
B_{N,k}(\tau,w) = \frac{1}{\eta(\tau)^N} \sum_{\sum_{i=1}^N a_i=0 \atop a_i\in\IZ+\frac{k}{N}}
\q^{-\sum_{i<j} a_i a_j}\, y^{\sum_{i<j}(a_i-a_j)},
\ee
is a Jacobi form of weight $-\frac12$ and index $\frac16(N^3-N)$,
We thus start by explaining how to compute stack invariants and VW invariants on
$\IF_m$, and then go on to discuss $\IP^2$ and $dP_k$ for $k\leq 8$. 
{\bp It is interesting to observe that in all cases,
the Betti numbers $b_{2k}(\cM^S_{[N;\mu;n],J})$ for canonical polarization stabilize for large enough $n$ to the same value  $\tilde b_{2k}$ as for the Hilbert scheme of points on $S$ \eqref{stableBetti}, in agreement with the conjecture \cite{Coskun:2020}. Moreover, they appear
to increase monotonically for any $n$ towards the asymptotic value $\tilde b_{2k}$. }

\subsection{Hirzebruch surfaces \label{sec_VWHirz}}

As in \S\ref{sec_Hirz}, we denote by $C$ and $F$ the base and fiber of the projective bundle 
$\IF_m\to \IP_1$, such that  $C^2=-m, F^2=0, C\cdot F=1$. We denote by $c_1(E)=\beta C-\alpha F$ the first Chern class, and  parametrize the polarization $J$  by  
$J_{m_1,m_2}=m_1(C+mF)+m_2 F$. For the polarization
$J_{0,1}$ at the boundary of the  \kahler 
 cone, the generating functions
$H_{N,\mu,J_{0,1}}^{\mathbb{F}_m} $ are independent of $m$, and   
given by (\cite[Conjecture 4.1]{Manschot:2011ym}, proven in \cite{Mozgovoy:2013zqx} by making use of the Hall algebra of $\IP^1$)
\be 
\label{defHN} 
H_{N,\mu,J_{0,1}}^{\mathbb{F}_m}(\tau,w)= 
H_{N}(\tau,w) := \frac{\I (-1)^{N-1} \eta(\tau)^{2N-3}}
{\theta_1(\tau,2Nw)\, \prod_{k=1}^{N-1} \theta_1(\tau,2kw)^2},
\ee
for $\mu \cdot F=0 \mod N$, or 0 otherwise. Here $H_N(\tau,w)$ 
is a Jacobi form of weight $-1$ and index $-\frac13(4N^3+2N)$.

Let $J_{\epsilon,1}$ with $0<\epsilon\ll 1$ 
be a ``suitable polarization''. In other words,
$\epsilon$ is sufficiently small such that no walls of marginal
stability are crossed between $J_{\epsilon,1}$ and $J_{0,1}$. Then,
the stack invariants for $J_{\epsilon,1}$ can be obtained from those at $J_{0,1}$
by applying the Joyce wall-crossing formula \cite{joyce2008configurations}, leading to  \cite[Prop. 5.7]{Manschot:2016gsx}
\be 
\label{eq:Hrabeps}
H^{\IF_m}_{N,\beta C-\alpha F,J_{\epsilon,1}}(\tau,w)=\sum_{N_1+\dots+N_\ell=N, \, N_i\in \mathbb{N}^* }
\frac{y^{2\sum_{i=2}^N (N_i+N_{i-1}) \{\frac{\alpha}{N} \sum_{k=i}^N N_k \}}}
{\prod_{i=2}^{N-1}(1-y^{2 (N_i+N_{i-1})})}
\,\prod_{j=1}^\ell H_{N_j}(\tau,w),
\ee
when  $\beta=0\mod N$, 
or zero if $\beta\neq 0\mod N$, 
where $H_N$ was defined in \eqref{defHN}. Since $J_{\eps,1}$ is a
generic polarization,  Eq. \eqref{hSHS}
implies that the generating function of rational invariants with respect to Gieseker
stability are given by  
\be
\label{htoH}
\hrVW_{N,\mu}(\tau,w) = \sum_{d=d_1+\dots +d_\ell}
 \frac{(-1)^{\ell-1}}{\ell}  \prod_{i=1}^\ell H^{\IF_m}_{d_i N_0, d_i \mu_0,J_{\epsilon,1}}(\tau,w),
\ee
where $d$ is the largest integer such that $(N_0,\mu_0):=(N/d,\mu/d)$ is primitive,
and the sum runs over all ordered decompositions of $d$ with $d_i>0$. 
Contributions
with $\ell>1$ in \eqref{htoH} can be expressed in terms of $h_{N',\mu}$ with $N'<N$ by applying \eqref{htoH} recursively.

\medskip

For rank $N=2$, we get 
\be 
H^{\IF_m}_{2,\beta C-\alpha F,J_{\eps,1}}(\tau,w) = 
\begin{cases} 
0, \quad \mbox{if} \quad \beta=1\mod 2,
\\
 H_2  + \frac{y^2}{1-y^4} H_1^2,
 \quad \mbox{if} \quad (\alpha,\beta)=(1,0) \mod 2
 \\
 H_2 + \frac{1}{1-y^4}   H_1^2,
 \quad \mbox{if} \quad (\alpha,\beta)=(0,0) \mod 2
\end{cases}
\ee
hence, using (\ref{ctoI}) we find for $\beta=0\mod 2$ in the chamber $J_{\epsilon,1}$,
\bea
\label{hFmboundary2}
%(y-1/y) q^{1/3}&& \left( h_{2,0} + \frac12h_1(2\tau,2w) \right) 
\hiVW^{\IF_m}_{2,0} &=&
 \left( y^5 + 2 y^3 + 3 y+ \dots\right)\q^2 %\nn\\&& 
 + \left(y^9 + 3 y^7 + 8 y^5 + 16 y^3 + 20 y+\dots\right) \q^3 + \dots
\nn\\
%(y-1/y) \q^{1/3}&& h_{2,F} =
\hiVW^{\IF_m}_{2,F} &=&
(y+1/y) \q + (y^5+3y^3+7y+\dots) \q^2 
%\nn\\&& 
+  \left(  y^9+  3 y^7 + 10 y^5 +22 y^3 + 36 y +\dots\right) \q^3 + \dots \nn 
\\
\eea 
and $0$ for $\beta\neq 0\mod 2$. 

\medskip

For rank $N=3$,
\be
H^{\IF_m}_{3,\beta C-\alpha F, J_{\eps,1}} = 
\begin{cases} 
0, \quad \mbox{if} \quad \beta\neq 0\mod 3
\\
 H_3 + \frac{1}{1-y^6} H_1\,  H_2 + \frac{1}{(1-y^4)^2}  H_1^3,
 \quad \mbox{if} \quad (\alpha,\beta)=(0,0) \mod 3,
 \\
 H_3 + \frac{y^2+y^4}{1-y^6} H_1\,  H_2 + \frac{y^4}{(1-y^4)^2}  H_1^3,
 \quad \mbox{if} \quad (\alpha,\beta)=(\pm 1,0) \mod 3
\end{cases}
\ee
hence for 
$\beta=0\mod 3$ in the chamber $J_{\epsilon,1}$,
\bea
\label{hFmboundary3}
%(y-1/y) \q^{1/2} &&\left( h_{3,0} - \frac13 h_1(3\tau,3w)  \right) =
\hiVW^{\IF_m}_{3,0} &=&
\left(y^{10} + 2 y^8 + 5 y^6 + 8 y^4 + 9 y^2 + 10 +\dots\right) \q^3 + \dots \nn\\
\hiVW^{\IF_m}_{3,\pm F} &=&
%(y-1/y) \q^{1/2} &&h_{3,\pm F} 
\left( y^4 + 2 y^2 + 4+\dots \right) \q^2  %\nn\\&& 
+ \left(y^{10} + 3 y^8 + 9 y^6 + 20 y^4 + 34 y^2 + 42+\dots\right) \q^3 + \dots \nn\\ 
\eea
and $0$ for $\beta\neq 0\mod 3$. 

\medskip

For rank $N=4$, we get $H^{\IF_m}_{4,\beta C-\alpha F, J_{\eps,1}}=$ 
\be
%H_{4,\beta C-\alpha F}(\tau,w, J_{\eps,1}) = 
\begin{cases} 
0, \quad \mbox{if} \quad \beta\neq 0\mod 4
\\
 H_4 + \frac{2}{1-y^8} H_1\,  H_3 
+\frac{1}{1-y^8}   H_2^2 
+\frac{3+3y^2+2y^4}{(1+y^2)(1-y^6)^2}   H_1^2  H_2 
+ \frac{1}{(1-y^4)^3}
  H_1^4
 \quad \mbox{if }(\alpha,\beta)=(0,0) \mod 4
 \\
 H_4 + \frac{y^2+y^6}{1-y^8} H_1\,  H_3 
+\frac{y^4}{1-y^8}   H_2^2 
+ \frac{y^4(1+y^2)^2}{(1-y^6)^2}  H_1^2  H_2 
+\frac{y^6}{(1-y^4)^3}   H_1^4
 \quad \mbox{if } (\alpha,\beta)=(\pm 1,0) \mod 4
\\
 H_4 + \frac{2y^4}{1-y^8} H_1\,  H_3 
+\frac{1}{1-y^8}   H_2^2 
+\frac{y^2(2+2y^2+3y^4+y^6)}{(1+y^2)(1-y^6)^2}    H_1^2  H_2 
+\frac{y^4}{(1-y^4)^3}  H_1^4
 \quad \mbox{if } (\alpha,\beta)=(2,0) \mod 4
\end{cases}
\ee
hence for $\beta=0\mod 4$ in the chamber $J_{\epsilon,1}$,
\bea
\label{hFmboundary4}
%(y-1/y) \q^{2/3}&&  \left( h_{4,0} - \frac12 h_{2,0}(2\tau,2w)  \right)=
\hiVW^{\IF_m}_{4,0} &=&
(y^{17} + 2 y^{15} + 6 y^{13} + 12 y^{11} + 23 y^9 + 32 y^7 + 43 y^5 
% \nn\\&& + 
 49 y^3 + 52 y+\dots) \q^4 + \dots
\nn \\
\hiVW^{\IF_m}_{4,\pm F} &=&
%(y-1/y) q^{2/3} &&h_{4,\pm F} =
(y^9 + 2 y^7 + 6 y^5 + 11 y^3 + 15 y+\dots) \q^3 \nn\\
&+&
(y^{17}+3 y^{15}+9 y^{13}+22 y^{11}+49 y^9+92 y^7+158 y^5+230
   y^3+282 y+\dots) \q^4 + \dots
\nn\\
\hiVW^{\IF_m}_{4,\pm 2F} &=&
%(y-1/y) \q^{2/3}&& \left( h_{4,\pm 2F} - \frac12 h_{2,F}(2\tau,2w) \right)=
(y^9+3 y^7+7 y^5+13 y^3+16
   y+\dots) \q^3
  \nn\\&+&
\left( y^{17}+3 y^{15}+10 y^{13}+24 y^{11}+55 y^9+103 y^7+173
   y^5+243 y^3+288 y+\dots\right) \q^4 + \dots\nn \\
   \eea
and $0$ for $\beta\neq 0\mod 4$.

\medskip

The $H^{\IF_m}_{2,c_1,J}$ for other chambers can be obtained by again applying the Joyce wall-crossing formula. For $N=2$ and arbitrary $\IF_m$ , one finds 
 \cite[(5.14)]{Manschot:2016gsx}
 \bea
\label{VWFm2wc}
&&H^{\IF_m}_{2,\beta C-\alpha F,J_{m_1,m_2}}   =
H^{\IF_m}_{2,\beta C-\alpha F,J_{\epsilon,1}}
\\
\quad && +\frac12 (h^{\IF_m}_{1})^2 
\sum_{a\in\IZ+\frac12\alpha \atop b\in\IZ+\frac12\beta} 
\left[ \sgn(2b\, m_2-2 m_1 a + v) - \sgn(2b-2a\epsilon+v) \right]\, 
y^{-2(m-2)b-4a} \, \q^{m b^2 +2a b}.\nn
\eea
for 
 $0<v\ll \epsilon \ll 1$. 
For $N=3$ we get instead \cite[\S 5]{Manschot:2011ym}
\bea
\label{VWFm3wc}
&&H^{\IF_m}_{3,\beta C-\alpha F,J_{m_1,m_2}}   =
H^{\IF_m}_{3,\beta C-\alpha F,J_{\epsilon,1}}
\nn\\
\quad && + h^{\IF_m}_{1}
\sum_{a\in\IZ-\frac23\alpha \atop b\in\IZ-\frac23\beta} 
\left[ \sgn(3b\, m_2-3 m_1 a + v) - \sgn(3b-3a\epsilon+v) \right]\, 
y^{-3(m-2)b-6a} \, \q^{\frac34 m b^2 +\frac32 a b}\nn\\&& \times 
H^{\IF_m}_{2,(b+\frac23\beta) C-(a+\frac23\alpha) F,J_{|b|,|a|}} 
\eea
In the next subsections we consider the cases $m=0,1,2$ in more detail. 
We shall be interested in particular in the canonical chamber $J= c_1(S)$, where the modular properties of the generating functions are expected to be simpler, and for $m=1$ in the 
chamber $J=H_\epsilon=C+F+\epsilon(2C+3F)$, where the VW invariants of $\IF_1$ are related to those of $\IP^2$ by the
blow-up formula.

\subsubsection{$\IF_0$ \label{VWF0}}

For $m=0$ in the chamber $J=c_1(S)$, i.e. $(m_1,m_2)=(2,2)$, we find \cite[Table 6]{Manschot:2016gsx}\cite[(4.36)]{Haghighat:2012bm}
\bea
\label{F0mod2} 
\hiVW^{\IF_0}_{2,(0,0)}&=&  
%\q^{\frac13} (y-1/y) &  \left[ h_{2,(0,0)}(\tau,w) + \frac12 h_1(2\tau,2w) \right] 
(y^5+2 y^3+3
   y+\dots) \q^2 %\\ & 
   + (y^9+3 y^7+8 y^5+16 y^3+20y+\dots) \q^3 \nn  \\
&&   + ( y^{13}+3 y^{11}+10 y^9+24 y^7+51 y^5+83 y^3+104
   y +\dots) \q^4 +\dots \nn \\
\hiVW^{\IF_0}_{2,(0,1)}&=& 
%\q^{\frac13} (y-1/y) h_{2,(1,0)}(\tau,w) =& \q^{1/3} (y-1/y) h_{2,(0,1)}(\tau,w) \\
(y+1/y)\, \q + (y^5+3 y^3+7y+\dots)\q^2
\nn\\ && + (y^9+3 y^7 +10 y^5 +22 y^3 + 37 y+\dots) \q^3+\dots
\nn\\
\hiVW^{\IF_0}_{2,(1,1)}&=& 
%\q^{\frac56} (y-1/y) h_{2,(1,1)}(\tau,z) =&
 (y^3+y+\dots) \q^2 + (y^7+3y^5+7y^3+9y+\dots) \q^3 +\dots
\eea
Away from the canonical chamber, we find from \eqref{VWFm2wc} the following wall-crossing phenomena as a function of $\eta=m_1/m_2$ (assuming $m_1,m_2>0$):
\begin{itemize}
\item $(\alpha,\beta)=(0,0)$: for $c_2=1,2,3$, the result above holds for any $\eta>0$.
For $c_2=4$, it holds for $1/2<\eta<2$, while the index equals $y^{13}+3 y^{11}+10 y^9+24 y^7+51 y^5+82 y^3+103
   y +\dots$  outside this range; 
 \item $(\alpha,\beta)=(1,0)$: for $c_2=1$, the index is $y+1/y$ for $\eta<2$, vanishes otherwise;
 for $c_2=2$, the result above holds for $0<\eta<2$; for $2<\eta<4$,
 the index jumps to $y^5+y^3+y+\dots$, and vanishes for $\eta>4$. For
 $c_2=3$,  the result above holds for $2/3<\eta<2$, for $\eta<2/3$ it
 jumps to $y^9+3 y^7+10 y^5+22 y^3+36y+\dots$; for $2<\eta<4$ to
 $y^9+3 y^7+7 y^5+ 9y^3+9 y+\dots$; for $4<\eta<6$, $y^9+y^7+y^5+
 y^3+y+\dots$; for $\eta>6$ it vanishes;  
  \item $(\alpha,\beta)=(1,1)$: for $c_2=2$ the result above holds for
    $1/3<\eta<3$, while the index vanishes outside this window; for
    $c_2=3$, the index vanishes for $\eta<1/5$ or $\eta>5$, is given
    by the result above for $1/3<\eta<3$, and by $y^7+y^5+y^3+y+\dots$
    for $1/5<\eta<1/3$ or $3<\eta<5$. 
\end{itemize}

For rank 3 in the chamber $J=c_1(S)$, from \eqref{VWFm3wc} we get
\bea
\label{F0mod3}
\hiVW^{\IF_0}_{3,(0,0)}
%\q^{\frac12} (y-1/y) &  \left[ h_{3,(0,0)}(\tau,w) - \frac13 h_1(3\tau,3w) \right] 
&=& (y^{10}+2 y^8+5   y^6 +8 y^4 + 9 y^2 + 10 +\dots) \q^3 \nn \\ 
&&+ (y^{16}+3 y^{14}+9 y^{12}+ 21 y^{10} + 44 y^8 + 74 y^6 +106 y^4+ 124 y^2 + 133 +\dots) \q^4
+\dots \nn
\\
\hiVW^{\IF_0}_{3,(\pm 1,0)}&=&(y^4+2y^2+4+\dots) \q^2 + (y^{10}+3 y^8+9 y^6+20 y^4+35 y^2+44+\dots) \q^3+\dots\nn\\
\hiVW^{\IF_0}_{3,(\pm 1,\pm 1)}&=&\q^2+ (y^6+3y^4+7y^2+10+\dots) \q^3 +\dots\nn\\
\hiVW^{\IF_0}_{3,(\pm 1,\mp 1)}&=&(y^8+2y^6+4y^4+4y^2+5+\dots) \q^2 \nn\\&& 
+(y^{14} + 3 y^{12} + 9 y^{10} + 20 y^8 + 37 y^6 + 54 y^4 + 64 y^2 + 68 +\dots) \q^3+  \dots
\eea
The first line is in agreement with
\cite[Table 13]{Manschot:2016gsx}.

\subsubsection{$\IF_1$ \label{VWF1}}

For  $N=2$, $c_1=(c_{1,H},c_{1,C})$ in the canonical chamber  we find
from \eqref{VWFm2wc} 
\bea
\label{F1mod2}
%\label{F1mod200}
%q^{\frac13} (y-1/y) && \left[ h_{2,(0,0)}(\tau,w) + \frac12 h_1(2\tau,2w)\right]=
\hiVW^{\IF_1}_{2,(0,0)} &=& 
(y^5+2 y^3+3   y+\dots) \q^2  + (y^9+3 y^7+8 y^5+16 y^3+21y+\dots) \q^3\nn\\
&&   + ( y^{13}+3 y^{11}+10 y^9+24 y^7+51 y^5+84 y^3+109  y+\dots ) \q^4 +\dots,
\nn\\
%\label{F1mod201}
%\q^{\frac{1}{12}} (y-1/y) h_{2,(0,1)}(\tau,w) &=&
\hiVW^{\IF_1}_{2,(0,1)} &=& 
(y^2+1+1/y^2) \q + 
(y^6 + 3 y^4 + 7 y^2 + 9+ \dots) \q^2  \nn \\
&& + (y^{10}+3
   y^8+10 y^6+22 y^4+40
   y^2+47+\dots) \q^3\nn \\
&&   + (y^{14}+3
   y^{12}+10 y^{10}+26 y^8+60 y^6+114 y^4+177
   y^2+205+\dots ) \q^4 +\dots,
\nn\\
%(y-1/y) \q^{\frac{7}{12}} h_{2,(1,0)} &=&
\hiVW^{\IF_1}_{2,(1,0)} &=& 
\q + \left( y^4 + 3 y^2 + 5 + \dots\right)\q^2 + \left(y^8+3y^6+10y^4+19y^2+27+\dots\right) \q^3 + \dots,
\nn\\
%(y-1/y) \q^{\frac13} h_{2,(1,1)} &=&
\hiVW^{\IF_1}_{2,(1,1)} &=& 
(y+1/y) \q + (y^5+3y^3+7y+\dots) \q^2 
\nn\\&& + 
 \left(  y^9+  3 y^7 + 10 y^5 +22 y^3 + 36 y +\dots\right) \q^3 + \dots
\eea

\medskip 

Away from the canonical chamber, we find the following wall-crossing phenomena from \eqref{VWFm2wc}
(with $\eta=m_1/m_2$
assuming $m_1,m_2>0$):
\begin{itemize}
\item No $\eta$ dependence for $(\alpha,\beta)=(1,0)$ i.e. $(c_{1,H},c_{1,C})=(1,1) \mod 2$, $c_2=1,2,3$.
\item For $(\alpha,\beta)=(0,1)$ i.e. $(c_{1,H},c_{1,C})=(0,1) \mod 2$: for $c_2=1$, the index is $y^2+1+1/y^2$ for $\eta>1/2$, vanishes otherwise; for $c_2=2$, the index is $y^6+3y^4+7y^2+9+\dots$ for $\eta>1/2$, $y^6+y^4+y^2+1+\dots$ for $1/4<\eta<1/2$, 0 otherwise;
\item For $(\alpha,\beta)=(1,1)$ $(c_{1,H},c_{1,C})=(1,0) \mod 2$: for $c_2=1$, the index is $1$ for $\eta>1$, vanishes otherwise; for $c_2=2$, the index is $y^4+3y^2+5+\dots$ for $\eta>1$, 
$y^4+y^2+1+\dots$ for $1/3<\eta<1$, 0 otherwise;
\item For $(\alpha,\beta)=(0,0)$: for $c_2=1,2$, the index is
  constant; for  $c_2=3$, the index is $y^9+3y^7+8y^5+16y^3+21y+\dots$
  for $\eta>1$, $y^9+3y^7+8y^5+16y^3+20y+\dots$ for $0<\eta<1$. 
\end{itemize} 
 
 At higher rank, the VW invariants can be obtained from those on $\IP^2$ given in the next
 section using the blow-up formula \eqref{HtoF1}. 
 For  $N=3$  in the blow-up chamber $J\propto H_\epsilon$, we get 
%(see  \cite[Table 14]{Manschot:2016gsx} for the first case $c_1=0$)
\bea
\label{F1mod300}
\hiVW^{\IF_1}_{3,(0,0)} &=& 
% \q^{\frac12} (y-1/y) & \left[ h_{3,(0,0)}(\tau,w) - \frac13 h_1(3\tau,3w)\right]=
(y^{10}+2 y^8+5   y^6 +8 y^4 + 10 y^2 + 11 +\dots) \q^3 \\ && 
+ (y^{16}+3 y^{14}+9 y^{12}+ 21 y^{10} + 44 y^8 + 75 y^6 +111 y^4+ 137 y^2 + 149 +\dots) \q^4
+\dots \nn\\
 \hiVW^{\IF_1}_{3,(0,\pm 1)}   &=& (y^6+2 y^4+4
   y^2+4+\dots) \q^2 \nn\\ && 
   + (y^{12}+3 y^{10}+9 y^8+20 y^6+37 y^4+53
   y^2+59+\dots) \q^3 + \dots
\nn\\
%(y-1/y) \q^{\frac56} h_{3,(1,0)} &=&
\hiVW^{\IF_1}_{3,(\pm 1,0)} &=& 
(y^2+1+1/y^2) \q^2 + \left(y^8+3y^6 + 8 y^4+14 y^2+17+\dots\right) \q^3 + \dots
\nn\\
%(y-1/y) \q^{\frac12} h_{3,(1,1)} &=&
\hiVW^{\IF_1}_{3,(\pm 1, \pm 1)} &=& 
 \left( y^4 + 2 y^2 + 3+\dots \right) \q^2 %\nn\\&& 
 + \left(y^{10} + 3 y^8 + 9 y^6 + 19 y^4 + 31 y^2 + 36+\dots\right) \q^3 + \dots
\nn\\
%(y-1/y) \q^{-\frac12} h_{3,(1,2)} &=&
\hiVW^{\IF_1}_{3,(\pm 1,\mp 1)} &=& 
 \left( y^4 + 2 y^2 + 3+\dots \right) \q  %\nn\\&& 
 + \left(y^{10} + 3 y^8 + 9 y^6 + 19 y^4 + 31 y^2 + 36+\dots\right) \q^2 + \dots \nn
    \eea
For $N=4$ in the chamber $J\propto H_\epsilon$, 
\bea
%(y-1/y) \q^{\frac{25}{24}} &&h_{4,(1,0)} =
\hiVW^{\IF_1}_{4,(1,0)} &=& 
(y^6 + y^4 + 3 y^2 + 3 +\dots) \q^3 \nn\\
&&+( y^{14} + 3 y^{12} + 8 y^{10} + 18 y^8 + 34 y^6 + 53 y^4 + 71 y^2 + 78+\dots) \q^4 + \dots
\nn\\
%(y-1/y) \q^{\frac{2}{3}} &&h_{4,(1,\pm 1)} =
\hiVW^{\IF_1}_{4,(1,\pm 1 )} &=& 
(y^9 + 2 y^7 + 5 y^5 + 8 y^3 + 10 y +\dots) \q^3 \nn\\
&&+
(y^{17} + 3 y^{15} + 9 y^{13} + 21 y^{11} + 45 y^9 + 81 y^7 + 130 y^5 + 
 177 y^3 + 207 y+\dots ) \q^4 + \dots
\nn\\
%(y-1/y) \q^{-\frac{11}{24}} &&h_{4,(1,2)} =
\hiVW^{\IF_1}_{4,(1,2)} &=& 
(y^{10} + 2 y^8 + 6 y^6 + 9 y^4 + 14 y^2 + 14 +\dots) \q^2+ \left( y^{18}+3 y^{16}
+9 y^{14}\right.  \nn\\
&&\left. 
+22 y^{12}+47 y^{10}+89 y^8+149
   y^6+214 y^4 +268 y^2+288+\dots\right) \q^3 + \dots
\eea

\subsubsection{Higher $\IF_m$'s \label{sec_VWFm}}
For $m$ even, changing variables from $a$ to $a'=a+\frac{m}{2}b$ in the sum in \eqref{VWFm2wc}, we get 
\be
\sum_{a'\in\IZ+\frac12\alpha' \atop b\in\IZ+\frac12\beta} 
\left[ \sgn(2b\, m_2' -2 m_1 a' + v) - \sgn((2-m\epsilon)b-2a'\epsilon+v) \right]\, 
y^{4(b-a')} \, \q^{2a' b},
\ee
which is the same sum as for $\IF_0$, with shifted first Chern class $\alpha'=\alpha+\frac{m}{2}\beta$ and \kahler class $m_2'=m_2+\frac{m}{2} m_1$. Similarly, for $m$ odd, changing variables from $a$ to $a'=a+\frac{m-1}{2}b$ in the sum, we get 
\be
\sum_{a'\in\IZ+\frac12\alpha' \atop b\in\IZ+\frac12\beta} 
\left[ \sgn(2b\, m_2' -2 m_1 a' + v)- \sgn((2-(m-1)\epsilon)b-2a'\epsilon+v) \right]\, 
y^{2b-4a'} \, \q^{b^2+ 2a' b},
\ee
which is the same sum as for $\IF_1$, with shifted first Chern class $\alpha'=\alpha+\frac{m-1}{2}\beta$ and \kahler class $m_2'=m_2+\frac{m-1}{2} m_1$. Thus, the rank 2 VW invariants for 
$\IF_{m}$ depend only on $m$ modulo 2, provided the polarization is suitably adjusted.
The same dependance on $m$ modulo 2  presumably holds for the rank 3 invariants in  \eqref{VWFm3wc}, although
the dependence of the last term on $J_{|b|,|a|}$ makes it harder to check explicitely.

\medskip

For example, for $m=2$, we get as a function of $\eta>0$ (with $\eta=+\infty$ the canonical chamber)
\begin{itemize}
\item  No dependence for $(\alpha,\beta)=(0,0)$, $c_2\leq 3$:
\be
\hiVW^{\IF_2}_{2,(0,0)} =  (y^5+2 y^3+5 y+\dots) \q^2 + (y^9+3y^7+8y^5+16 y^3+20 y+\dots) \q^3 +\dots
\ee
which coincides with $h_{2,(0,0)}$ for $\IF_0$ in \eqref{F0mod2}. 
\item For $(\alpha,\beta)=(1,0)$, $\hiVW^{\IF_2}_{2,(1,0)} = $
\be
\hspace*{-1cm}
\begin{cases}
\eta>2 : & 
(y+1/y) \q + (y^5+3 y^3+7 y+\dots) \q^2 + (y^9+3y^7+10y^5+22 y^3+37 y+\dots) \q^3 +\dots \\
\eta<2: & 
(y+1/y) \q + (y^5+3 y^3+7 y+\dots) \q^2 + (y^9+3y^7+10y^5+22 y^3+36 y+\dots) \q^3 + \dots
\end{cases}
\ee
which for $\eta>2$ coincides with  $h_{2,(1,0)}$ for $\IF_0$ in \eqref{F0mod2}. 
\item For $(\alpha,\beta)=(0,1)$, $\hiVW^{\IF_2}_{2,(0,1)} = $
\be
\hspace*{-1cm}
\begin{cases}
\eta>\frac12 : & 
(y^3+y+\dots) \q + (y^7+3 y^5+7 y^3+9y+\dots) \q^2 \\&  +
 (y^{11}+3y^9+10y^7+22y^5+40 y^3+50 y+\dots) \q^3 +\dots \\
\frac14<\eta<\frac12 : & 
%0\, q +
 (y^7+y^5+y^3+y+\dots) \q^2 +
 (y^{11}+3y^9+7y^7+9y^5+9 y^3+9 y+\dots) \q^3 +\dots \\
\frac16< \eta<\frac14 : & 
%0\, \q + 0\, \q^2 +
 (y^{11}+y^9+y^7+y^5+y^3+ y+\dots) \q^3 +\dots \\
 \eta<\frac16 & 0\, \q + 0\, \q^2 +0 \q^3 +\dots
\end{cases}
\ee
which for $\eta>1/2$ coincides with  $h_{2,(1,1)}$ for $\IF_0$ in \eqref{F0mod2}. 
\item For $(\alpha,\beta)=(1,1)$, $\hiVW^{\IF_2}_{2,(1,1)} = $
\be
\hspace*{-1cm}
\begin{cases}
\eta>1 : & 
(y+1/y) \q + (y^5+3 y^3+7 y+\dots) \q^2  \\ & 
+ (y^9+3y^7+10y^5+22 y^3+37 y+\dots) \q^3 +\dots \\
\frac13<\eta<2: & 
%0\, q +
 (y^5+ y^3+y+\dots) \q^2 + (y^9+3y^7+7y^5+9 y^3+9 y+\dots) \q^3 +\dots\\
\frac15<\eta<2: & 
%0\, \q + 0\, \q^2 +
 (y^9+y^7+y^5+y^3+y+\dots) \q^3 +\dots \\
\eta<\frac15 & 0\, \q + 0\, \q^2 +0 \q^3 +\dots
\end{cases}
\ee
which for $\eta>2$ coincides with  $h_{2,(0,1)}$ for $\IF_0$ in \eqref{F0mod2}. 
\end{itemize}

\subsection{$\IP^2$  \label{sec_VWP2}}

In order to compute VW invariants of $\IP^2$, one strategy is   to first
compute the VW invariants of the Hirzebruch
surface $\mathbb{F}_1$ in a suitable chamber $J_{1,0}$ (where $J_{m_1,m_2}=m_1(C+F)+m_2 F=(m_1+m_2)H-m_2 C$) and then apply the blow-up formula \eqref{HtoHhat} 
in the form 
\be
\label{HtoF1}
H^{\IP_2}_{N,\mu}(\tau,w)= \frac{H_{N,(-\mu,\mu),J_{1,0}}^{\mathbb{F}_1} (\tau,w)}
{B_{N,0}(\tau,w)}.
\ee
The stack invariants of $\mathbb{F}_1$
in the chamber $J_{1,0}$ are  obtained from those in the chamber $J_{\epsilon,1}$
by applying the wall-crossing formula. This results in  \cite[(4.7)]{Manschot:2014cca}
\be
\label{eq:Hrab}
H^{\mathbb{F}_1} _{N,-aF+bC,J_{1,0}}(\tau,w)=\sum_{N_1+\dots+N_\ell=N, \, N_i\in \mathbb{N}^* }
 \Psi_{N_1,\dots,N_\ell}^{a,b}(\tau,w)\,\prod_{j=1}^\ell H_{N_j}(\tau,w),
\ee
where the the  generating functions $H_N$ are defined in \eqref{defHN} and the generalized Appell functions $\Psi$ are defined as follows \cite[Prop. 4.1]{Manschot:2014cca}
\bea
\label{defPsi}
\Psi^{a,b}_{N_1,\dots,N_\ell}(\tau,w)&=&\sum_{N_1k_1+\dots+N_\ell k_\ell=b, \atop
k_i\in \mathbb{Z}} 
\frac{y^{\sum_{j<i} N_iN_j(k_i-k_j) +2\sum_{i=2}^{\ell}
(N_i+N_{i-1})  \{ \frac{a}{N} \sum_{k=i}^\ell N_k \} }}{\prod_{i=2}^{\ell} \left(1-y^{2(N_{i}+N_{i-1})} \q^{k_{i-1}-k_{i}}\right)} 
\non \\
&& \times\, \q^{\sum_{i=1}^\ell \frac{N_i(N-N_i)}{2N}k_i^2 
- \frac{1}{N} \sum_{i<j} N_iN_j k_ik_j + \sum_{i=2}^\ell (k_{i-1}-k_{i})\{ \frac{a}{N} \sum_{k=i}^\ell N_k \}  },
\eea
where $N=N_1+\dots +N_\ell$. 
Note that when $\mu=0 \mod N$, there is a contribution $H_N$ from $\ell=1$.
Applying these formulae, one finds 
\be
h^{\IP^2}_{2,0} = \frac{H_2 + \Psi_{1,1}^{0,0} H_1^2}{B_{2,0}}- \frac12 (h_1^{\IP^2})^2\ ,\quad 
h_{2,1}^{\IP^2}= \frac{\Psi_{1,1}^{-1,1} H_1^2}{B_{2,0}}\ ,\quad 
\ee
leading to the generating functions of rank 2 VW invariants
\cite{Yoshioka:1994, Yoshioka:1995}
\bea
\hiVW_{2,0}^{\IP^2} &=&  \q^{1/4} ( y-1/y) 
 \left( h^{\IP^2}_{2,0}(\tau,w) +\frac12 h^{\IP^2}_{1}(2\tau,2w) \right) \nn\\&&
 =
\left( y^5 + y^3 + y + y^{-1} +y^{-3} + y^{-5} \right) \q^2 
  + \left( y^9+2 y^7+4 y^5+6 y^3+6  y+\dots \right) \q^3 
\nn  \\ &&  + 
   \left( y^{13}+2 y^{11}+6 y^9+11 y^7+19 y^5+24 y^3+27y+\dots \right) \q^4 + 
   \dots
   \nn\\
\hiVW_{2,1}^{\IP^2} &=& \q^{1/2} ( y-1/y) h_{2,1}^{\IP^2} \nn\\&&
 =    \q+\left(y^4+2y^2+3+\dots\right) \q^2 +\q^3 
\left(y^8+2   y^6+6 y^4+9 y^2+12+\dots\right) + \dots
\eea
For $N=3$,
\bea
h^{\IP^2} _{3,0} &=& \frac{H_3 + \left( \Psi_{1,2}^{0,0} + \Psi_{2,1}^{0,0} \right) H_1 H_2+ \Psi_{1,1,1}^{0,0} H_1^3}{B_{3,0}} - h^{\IP^2} _{2,0} h^{\IP^2} _1 - \frac16 (h^{\IP^2} _1)^3\ ,\quad \nn\\
h^{\IP^2} _{3,1} &=&  \frac{ \left( \Psi_{1,2}^{-1,1} + \Psi_{2,1}^{-1,1} \right) H_1 H_2+ \Psi_{1,1,1}^{-1,1} H_1^3}{B_{3,0}} ,
\eea
leading to the generating functions of rank 3 VW invariants
\bea
\hiVW_{3,0}^{\IP^2} &=&
\q^{3/8}  (y-1/y) \left( h^{\IP^2} _{3,0}(\tau,w)-\frac13 h^{\IP^2} _1(3\tau,3w) \right) \nn\\
&=& 
\left( y^{10}+y^8+2 y^6+2 y^4+2   y^2+2+\dots \right) \q^3
 \nn\\  &&  
 + \left( y^{16}+2 y^{14}+5 y^{12}+9 y^{10}+15 y^8+19 y^6+22 y^4+23
   y^2+24  +\dots \right) \q^4 + \dots
   \nn \\
\hiVW_{3,1}^{\IP^2} &=& \q^{17/24}  (y-1/y) h^{\IP^2} _{3,\pm 1} \nn\\
&=&    
\q^2 \left(y^2+1+1/y^2\right)+\q^3 \left(y^8+2  y^6+5 y^4+8 y^2+10
+\dots \right)
  \nn \\&&
%\hspace*{-2cm}
+\q^4 \left( y^{14}+2 y^{12}+6 y^{10}+12
   y^8+24 y^6+38 y^4+54   y^2+59+ \dots\right)
  \nn \\&&
 % \hspace*{-2cm} 
 +\q^5 \left( y^{20}+2 y^{18}+6 y^{16}+13
   y^{14}+28 y^{12}+52 y^{10}+94 y^8+149 y^6+217 y^4 \right. \nn\\ && 
   \left. +273
   y^2+298+\dots \right) + \dots
\eea
For $N=4$,
\bea
h^{\IP^2}_{4,0} &=&  \frac{1}{B_{4,0}} \left[ 
H_4 + 2 \Psi_{3,1}^{0,0} \, H_1 \, H_3 
+ \Psi_{2,2}^{0,0}\, H_2^2  
+ \left( \Psi_{1,2,1}^{0,0}  + 2\Psi_{2,1,1}^{0,0}  \right) H_1^2\, H_2
+ \Psi_{1,1,1,1}^{0,0} H_1^4 \right]
\nn\\
&&  - \frac12 (h^{\IP^2}_1)^2\,h^{\IP^2}_{2,0} - h^{\IP^2}_1 \, h^{\IP^2}_{3,0} 
-\frac12 (h^{\IP^2}_2)^2-\frac1{24} (h^{\IP^2}_1)^4,
\nn \\
h^{\IP^2}_{4,1} &=&  \frac{1}{B_{4,0}} \left[ 
\left( \Psi_{3,1}^{-1,1} +\Psi_{1,3}^{-1,1}  \right)\, H_1 \, H_3 
%+ \Psi_{2,2}^{-1,1}\, H_2^2  
+ \left( \Psi_{2,1,1}^{-1,1}  + \Psi_{1,2,1}^{-1,1}  + \Psi_{1,1,2}^{-1,1}  \right) H_1^2\, H_2
+ \Psi_{1,1,1,1}^{-1,1} H_1^4 \right],
\nn
\\
h^{\IP^2}_{4,2} &=&  \frac{1}{B_{4,0}} \left[ 
2 \Psi_{3,1}^{-2,2}\, H_1 \, H_3 
+ \Psi_{2,2}^{-2,2}\, H_2^2  
+ \left( 2 \Psi_{2,1,1}^{-2,2}  + \Psi_{1,2,1}^{-2,2}   \right) H_1^2\, H_2
+ \Psi_{1,1,1,1}^{-2,2} H_1^4 \right] -\frac12 (h^{\IP^2}_{2,1})^2,
\nn\\
\eea
leading to the generating functions of rank 4 VW invariants
\bea
\hiVW^{\IP^2}_{4,0}&=&\q^{1/2} (y-1/y) \left( h^{\IP^2}_{4,0} -\frac12 h^{\IP^2}_{2,0}(2\tau,2w) 
\right)  \nn\\
&=& \q^4 
\left( y^{17}+y^{15}+3 y^{13}+4 y^{11}+6 y^9+6 y^7+7 y^5+7 y^3+7
   y+\dots \right)
  \nn \\
&&+ \q^5 \left( y^{25}+2 y^{23}+5 y^{21}+10
   y^{19}+19 y^{17}+32 y^{15}+49 y^{13}+68 y^{11}+85 y^9+98 y^7+107 y^5\right. \nn \\
   && \left.+112
   y^3+114 y+ \dots\right) + \dots
 \nn  \\ 
\hiVW^{\IP^2}_{4,1}&=& \q^3 \left(y^6+y^4+3   y^2+3+\dots\right)
   \nn\\ &&
   +\q^4 \left( y^{14}+2
   y^{12}+5 y^{10}+10 y^8+18 y^6+28 y^4 +38
   y^2+42+ \dots\right)  
   \nn \\  && 
   +\q^5\left( y^{22}+2 y^{20}+6 y^{18}+12
   y^{16}+26 y^{14}+46 y^{12}+83 y^{10}+131 y^8+200 y^6 \right. \nn\\&&
   \left. +268 y^4+332
   y^2+351+\dots\right) + \dots
   \nn\\
\hiVW^{\IP^2}_{4,2}&=& \q^{2} (y-1/y) \left[    
 h^{\IP^2}_{4,2}+\frac12 h^{\IP^2}_{2,1}(2\tau,2w) \right] \nn\\
  &=&  
 \q^4 \left(y^5+y^3+y+\dots \right)
 %  \nn\\ &&
   +\q^5\left( y^{13}+2   y^{11}+6 y^9+10 y^7+17 y^5+21 y^3+24
   y+ \dots \right)       \nn \\
   &&
   +\q^6\left(   y^{21}+2 y^{19}+6 y^{17}+13
   y^{15}+27 y^{13}+49 y^{11}+84 y^9+126 y^7+173 y^5 \right.\nn\\&&\left. +211 y^3+231
   y+\dots \right) 
+ \dots
\eea
Note that  $\hiVW^{\IP^2}_{4,2}$ is consistent with the results in  \cite[Table 1]{Manschot:2014cca}.

For $N=5$,
\bea
h^{\IP^2}_{5,0} &=&  \frac{1}{B_{5,0}} \left[ 
H_5 + 2 \Psi_{1,4}^{0,0} \, H_1 \, H_4  + 2 \Psi_{3,2}^{0,0} \, H_2 \, H_3 
+  \left( 2 \Psi_{2,2,1}^{0,0}+\Psi_{2,1,2}^{0,0}\right) \, H_2^2 H_1
\right. \nn \\
&& \left. 
+ \left( 2 \Psi_{2,1,1,1}^{0,0}  + 2\Psi_{1,2,1,1}^{0,0}  \right) H_1^3\, H_2
+ \left( 2 \Psi_{3,1,1}^{0,0}  + \Psi_{1,3,1}^{0,0}  \right) H_1^2\, H_3
+ \Psi_{1,1,1,1,1}^{0,0} H_1^5 \right]
\nn\\
&& -\frac{1}{120} (h^{\IP^2}_{1})^5-\frac{1}{6} h^{\IP^2}_{2,0}
   (h_1^{\IP^2}) ^3-\frac{1}{2} h^{\IP^2}_{3,0} (h^{\IP^2}_1)^2-\frac{1}{2} (h^{\IP^2}_{2,0})^2
   h^{\IP^2}_1-h^{\IP^2}_{4,0} h^{\IP^2}_1-h^{\IP^2}_{2,0}  h^{\IP^2}_{3,0} ,
\nn \\
h^{\IP^2}_{5,1} &=& \frac{1}{B_{5,0}} \left[ 
H_5 + 2 \Psi_{1,4}^{-1,1} \, H_1 \, H_4  + 2 \Psi_{3,2}^{-1,1} \, H_2 \, H_3 
+  \left( 2 \Psi_{2,2,1}^{-1,1}+\Psi_{2,1,2}^{-1,1}\right) \, H_2^2 H_1    \right. \\
&& \left. 
+ \left( 2 \Psi_{2,1,1,1}^{-1,1}  + 2\Psi_{1,2,1,1}^{-1,1}  \right) H_1^3\, H_2
+ \left( 2 \Psi_{3,1,1}^{-1,1}  + \Psi_{1,3,1}^{-1,1}  \right) H_1^2\, H_3
+ \Psi_{1,1,1,1,1}^{-1,1} H_1^5 \right], \nn
\nn
\\
h^{\IP^2}_{5,2} &=&  \frac{1}{B_{5,0}} \left[ 
H_5 + 2 \Psi_{1,4}^{-2,2} \, H_1 \, H_4  + 2 \Psi_{3,2}^{-2,2} \, H_2 \, H_3 
+  \left( 2 \Psi_{2,2,1}^{-2,2}+\Psi_{2,1,2}^{-2,2}\right) \, H_2^2
H_1    \right. \nn \\
&& \left. 
+ \left( 2 \Psi_{2,1,1,1}^{-2,2}  + 2\Psi_{1,2,1,1}^{-2,2}  \right) H_1^3\, H_2
+ \left( 2 \Psi_{3,1,1}^{-2,2}  + \Psi_{1,3,1}^{-2,2}  \right) H_1^2\, H_3
+ \Psi_{1,1,1,1,1}^{-2,2} H_1^5 \right],
\nn
\eea
leading to 
\bea
\hiVW^{\IP^2}_{5,0} &=& \q^{5/8} (y-1/y) \, \left[ h^{\IP^2}_{5,0} -\frac15 h_{1}(5\tau,5w) \right]
\nn\\
&=& \q^5
\left( y^{26}+y^{24}+3 y^{22}+5 y^{20}+9 y^{18}+13 y^{16}+18 y^{14}+22 y^{12}+26
   y^{10} \right. \nn \\
   && \left. +28 y^8+30 y^6+30 y^4+31 y^2+31+ \dots \right)\nn\\&&
+ \q^6 \left(y^{36}+2
   y^{34}+5 y^{32}+10 y^{30}+20 y^{28}+35 y^{26}+61 y^{24}+96 y^{22}+148 y^{20}+212 y^{18}
   \right. \nn \\
   && \left. +289
   y^{16}+368 y^{14}+446 y^{12}+509 y^{10}+561 y^8+596 y^6+620 y^4+632
   y^2+638+ \dots\right) \nn\\&& +\dots
   \nn\\
 \hiVW^{\IP^2}_{5,1}&=& \q^4 \left(y^{12}+y^{10}+3 y^8+5y^6+8y^4+10y^2+12+\dots\right)
+\q^5 \left( y^{22}+2 y^{20}+ 5 y^{18}  \right. \nn \\
   &&  \left.
   + 10 y^{16}+20  y^{14}   +34  y^{12}+57 y^{10}+87 y^8 + 126 y^6+165 y^4+198 y^2+210+\dots
    \right)  \nn \\ 
    && + \q^6 \left( y^{32}+2 y^{30}+6 y^{28}+12 y^{26}+26
   y^{24}+48 y^{22}+89 y^{20}+150 y^{18}+251 y^{16}+393 y^{14} \right. \nn \\
   && \left.  +600 y^{12} +865 y^{10}+1201
   y^8+1564 y^6+1921 y^4+2177 y^2+2280+ \dots \right)+\dots
   \nn\\
\hiVW^{\IP^2}_{5,2}&=& \q^4 + \q^5  \left(y^{10}+2 y^8+5 y^6+8 y^4+13
   y^2+14+\dots\right) +
\q^6 \left(y^{20}+2 y^{18}+6 y^{16}   \right.  \nn \\
   && \left.  +12 y^{14}+25
   y^{12}+44 y^{10}+76 y^8+114 y^6+161 y^4+196
   y^2+214+ \dots \right)+\dots .
\eea

\subsection{Higher del Pezzo surfaces \label{sec_VWdP}}
We now turn to the del Pezzo surfaces $dP_k$ with $k\geq 2$. Since $dP_k$ is the blow-up of 
$dP_{k-1}$ at one point, the generating function of stack invariants can be obtained by applying the blow-up formula \eqref{HtoHhat} iteratively: 
\be
\label{dPkblow}
H^{dP_k}_{N,\mu_0 H - \mu_i C_i,H}(\tau,w)=  H^{\IP^2}_{N,\mu_0}(\tau,w)\, \prod_{i=1}^k B_{N,\mu_i}(\tau,w)\,
\ee
where we identified the hyperplane class $H$ on $\IP^2$ with its pull-back $\pi_*(H)$ to $dP_k$. 
In general however, the polarization $J=H$ is not generic, and we
choose to deform it to 
$J=H_\epsilon:=H+\epsilon c_1(S)$ where $0<\epsilon\ll 1$. For $\mu_0\neq 0 \mod N$, the stack invariants do not change under this deformation, but for $\mu_0=0 \mod N$, there are sheaves
which are strictly semi-stable for $J$ but unstable for $J_\epsilon$ and their contribution must be subtracted from \eqref{dPkblow}. In addition, even at $J_\epsilon$ there are strictly semi-stable sheaves which must be taken into account in obtaining the rational invariants from the stack invariants. We shall now spell out this procedure for $N=2$ and $N=3$, before tabulating
results for all del Pezzo surfaces with $k\leq 8$.
 
\subsubsection{Rank $N=2$ and $3$ for $\mu_0 \neq 0\mod N$}
For $N=2$ with $\mu_0\neq 0 \mod 2$, we get for any $k\geq 1$ and $J= H_\epsilon$
\bea
%\q^{\frac{k+6}{12}} &&(y-1/y) h^{dP_k}_{2,H} = 
\hiVW^{dP_k}_{2,H} &=&
\q + \left( y^4 + (k+2) y^2 + (2k+3) + \dots \right) \q^2\nn\\
&& + 
\left( y^8 + (k+2) y^6 + \tfrac12(k+3)(k+4) y^4 + (9+8k+2k^2) y^2 + 3 (k+2)^2 + \dots \right) \q^3\nn\\
&&+\dots
\nn\\
%q^{\frac{k+3}{12}} &&(y-1/y) h^{dP_k}_{2,H-C_1} = 
\hiVW^{dP_k}_{2,H-C_1} &=&
(y+1/y)\, \q + 
\left( y^5 + (k+2) y^3 + (3k+4) + \dots \right) \q^2\nn\\
&&+ 
\left( y^9 + (k+2) y^7 + \tfrac12(k+3)(k+4) y^5 + \tfrac12 (22+17k+5k^2) y^3 \right. \nn\\
&&+
\left.  (15+16k+5k^2)  y + \dots \right) \q^3+\dots
\nn\\
%\q^{\frac{k}{12}} &&(y-1/y) h^{dP_k}_{2,H-C_1-C_2} 
\hiVW^{dP_k}_{2,H-C_1-C_2} &=&
 (y^2+2+1/y^2)\, \q + 
\left( y^6 + (k+2) y^4 + (4k+4) y^2+ 6k+6+\dots \right) \q^2\nn\\
&&+  
\left( y^{10} + (k+2) y^8 + \tfrac12(k+3)(k+4) y^6 + (13+8k+3k^2) y^4  \right. \nn\\ 
&&+ \left.  \tfrac12(28+35k+15k^2)  y^2  + 
(22+24k+10k^2)\dots \right) \q^3+\dots
\eea
More generally,
\bea
\hiVW^{dP_k}_{2,H-C_1-\dots-C_\ell} =&& (y+1/y)^{\ell}  \left[ 
\q +  
 (y^4+(k-\ell+2) y^2+(2
   k+3)+\dots) \q^2  \right. \nn\\
  &&+ \left. 
 \left(y^8+(k-\ell +2) y^6+\frac{1}{2}
   \left(k^2-2 \ell k+7 k+\ell^2-3 \ell+12\right) y^4 \right.\right. \nn\\
\hspace*{-5mm}  &&+ 
\left(2 k^2-2 \ell k+8 k-3 \ell+9\right) y^2 \nn\\ 
&&   \left. \left. 
   +\left(3 k^2-2 \ell k+12 k+\ell^2-4\ell+12\right) +\dots \right) \q^3 + \dots  \right] 
\eea

%\subsubsection{Rank 3}
For $N=3$, $\mu_0\neq 0 \mod N$, we get for any $k\geq 1$ and $J=H_\epsilon$
\bea
%\q^{\frac{3k+17}{24}} &&(y-1/y) H^{dP_k}_{3,H} = 
   \hiVW^{dP_k}_{3,H} &=& 
(y^2+1+1/y^2)\, \q^2 \nn\\&& + 
\left( y^8 + (k+2) y^6 + (3k+5) y^4 + (6k+8) y^2 + 7k+10 + \dots \right) \q^3+\dots
\nn\\
%q^{\frac{k+3}{8}} &&(y-1/y) H^{dP_k}_{3,H-C_1} = 
   \hiVW^{dP_k}_{3,H-C_1} &=& 
(y^4+2y^2+3+\dots)\, \q^2  + 
\left( y^{10} + (k+2) y^8 + (4k+5) y^6  \right.\nn\\&&\left.  
+(10k+9) y^4 + (16k+15)  y^2 + 19k+17+ \dots \right) \q^3+\dots
\nn\\
%q^{\frac{k-5}{8}} &&(y-1/y) H^{dP_k}_{3,H-2C_1} = 
   \hiVW^{dP_k}_{3,H-2C_1} &=& 
(y^4+2y^2+3+\dots)\, \q 
\left( y^{10} + (k+2) y^8 + (4k+5) y^6   \right.\nn\\&&\left.  +(10k+9) y^4 + (16k+15)  y^2 + 19k+17+ \dots \right) \q^2+\dots
\nn\\
%\q^{\frac{3k+1}{24}} &&(y-1/y) H^{dP_k}_{3,H-C_1-C_2} =
   \hiVW^{dP_k}_{3,H-C_1-C_2} &=& 
 (y^6+3y^4+6y^2+7+\dots)\, \q^2 + 
\left( y^{12} + (k+2) y^{10} + (4k+5) y^8 +(15k+6) y^6   \right.\nn\\&&\left. 
+ (30k+12) y^4 + (45k+18)  y^2 + 51k+22+ \dots \right) \q^3+\dots
\nn\\
%\q^{\frac{3k-23}{24}} &&(y-1/y) H^{dP_k}_{3,H-2C_1-C_2} =
   \hiVW^{dP_k}_{3,H-2C_1-C_2} &=& 
 (y^6+3y^4+6y^2+7+\dots)\, \q + 
\left( y^{12} + (k+2) y^{10} + (4k+5) y^8 +(15k+6) y^6  \right.\nn\\&&\left.  + (30k+12) y^4 + (45k+18)  y^2 + 51k+22+ \dots \right) \q^2+\dots
\eea

\subsubsection{Rank $N=2$ and $3$ for $\mu_0=0\mod N$}
We next consider $N=2$ and $\mu_0=0 \mod 2$.
To determine the generating function of rational invariants 
$h^{dP_k}_{r,\mu,J}(\tau,w)$, we have to subtract from
$H_{2,c_1}$ contributions from sheaves which are strictly
semi-stable for $J=H$ but unstable for $J=H_\varepsilon$. 
Such sheaves exist only for $\mu_0$ even, and have an
Harder-Narasimhan (HN) filtration of the form 
$0=F_0\subset F_1 \subset F_2=F$, with
quotients $E_i=F_i/F_{i-1}$ of rank 1. 
The stability conditions imply for these filtrations imply
\be
\begin{split}
&{\rm 1)\,\,\mbox{strictly semi-stable}\,\, for\,\,} J=H:\qquad \nu_H(E_1)-\nu_H(E_2)=0, \\
&{\rm 2)\,\,unstable\,\, for\,\,} J=H_\epsilon:\qquad
\nu_{H_\epsilon}(E_1)-\nu_{H_\epsilon}(E_2)>0.
\end{split}
\ee
Subtracting the two equations gives
\be 
\label{E1E2c1}
\left(\frac{c_1(E_1)}{N_1}-\frac{c_1(E_2)}{N_2}\right)\cdot c_1(S)>0\ . 
\ee
The contribution of such a filtration to  the stack invariant 
$\mathcal{I}^\nu_{\gamma_1+\gamma_2,H}(y)$ is 
\be
\label{yII}
y^{-(N_2c_1(E_1)-N_1c_1(E_2))\cdot c_1(S)}\,
\mathcal{I}^\nu_{\gamma_1,H}(y)\,\mathcal{I}^\nu_{\gamma_2,H}(y),
\ee
 which should be subtracted in order to arrive
at $\mathcal{I}^\nu_{\gamma_1+\gamma_2,H_\epsilon}(y)$. To determine the generating series, we denote
$c_{1}(E_2)=(0,m_1,\dots,m_k)$ and $c_1:=c_1(E)=a_1 C_1+\dots + a_k
C_k$, such that (\ref{E1E2c1}) becomes $\sum_{i}(2m_i-a_{i})>0$.
As a result, the $ H^{dP_k}_{2,c_1,J}$ are given for the polarization $H_\epsilon$ by 
\bea 
 H^{dP_k}_{2,a_1 C_1+\dots + a_k C_k,H_\epsilon}(\tau,w) &=& 
 H^{dP_k}_{2,a_1 C_1+\dots + a_k C_k,H}(\tau,w)
- (h^{dP_k}_{1})^2 \, 
\sum_{b_i\in \IZ, \sum_i b_i>0 \atop b_i=a_i \mod 2}   y^{-\sum_i b_i} \q^{\frac14 \sum_i b_i^2}\nn\\
\eea
where we have set $b_i=2m_i-a_i$. 

We can then obtain $h^{dP_k}_{2,c_1}$ by using (\ref{hSHS}). The polarization $H_\epsilon$ is not generic for 
$dP_k$ with $k>1$. The Chern classes which have the same
slope for $J=H$ and for $J=H_\epsilon$ differ by elements of
$H^2(S)$ which are orthogonal to $c_1(S)$ and correspond to vectors $b_i$
such that $\sum_i b_i=0$. Altogether, we arrive at
\bea
h^{dP_k}_{2,a_1 C_1+\dots + a_k C_k,H_\epsilon}(\tau,w) &=& 
 H^{dP_k}_{2,a_1 C_1+\dots + a_k C_k,H_\epsilon}(\tau,w)
- \frac12 (h^{dP_k}_{1})^2 \, 
\sum_{b_i\in \IZ, \sum_i b_i=0 \atop b_i=a_i \mod 2} \q^{\frac14 \sum_i b_i^2} \nn\\
\eea

For $\mu_0=0 \mod 3$, there are similarly strictly
semi-stable sheaves for $J=H$, which become unstable for $J=H_\varepsilon$.
Their HN filtration is either of the previous form with $(E_1,E_2)$ of rank $(1,2)$ or $(2,1)$,
or $0=F_0\subset F_1 \subset F_2\subset F_3=F$, with
quotients $E_i=F_i/F_{i-1}$ of rank 1. Similar reasoning to the above leads to the relations
\be
\begin{split}
&H^{dP_k}_{3,(0,c_i),H_\epsilon}=H^{dP_k}_{3,(0,c_i),H} \\
&- h^{dP_k}_1\, \left[ 
\sum_{b_i\in \mathbb{Z}-\frac13 c_{i} \atop\sum_i b_{i}>0} H^{dP_k}_{2,(0,b_i-\frac23c_{i}),H}\,  y^{-3\sum_i b_i} \q^{\frac34 \sum_i b_i^2}
+
 \sum_{b_i\in \mathbb{Z}-\frac23 c_{i} \atop\sum_i b_{i}>0} H^{dP_k}_{2,(0,b_i+\frac23c_{i}),H}\,  y^{-3\sum_i b_i} \q^{\frac34 \sum_i b_i^2} \right] \\
&+ \left( h^{dP_k}_1 \right)^3 \, \sum_{b_{1,i},b_{2,i}\in \mathbb{Z}-\frac13 c_{i}
\atop \sum_i  b_{1,i}>0; \sum_i b_{1,i}+b_{2,i}>0} 
y^{-4\sum_i b_{1,i}-2\sum_i b_{2,i}} \q^{\sum_i (b_{1,i}^2+b_{2,i}^{2}+b_{1,i}b_{2,i})},
\end{split}
\ee
and 
\be
\begin{split}
&h^{dP_k}_{3,(0,c_i),H_\epsilon}=H^{dP_k}_{3,(0,c_i),H_\epsilon}\\
&-\frac{1}{2}h^{dP_k}_1 \left[ 
\sum_{b_i\in \mathbb{Z} -\frac13 c_{i} \atop\sum_i b_{i}=0} 
H^{dP_k}_{2,(0,b_i-\frac23 c_{i}),H_\epsilon}\,\q^{\frac34 \sum_i b_i^2} 
+
\sum_{b_i\in \mathbb{Z} -\frac23 c_{i}  \atop\sum_i b_{i}=0} 
H^{dP_k}_{2,(0,b_i+\frac23 c_{i}),H_\epsilon}\, \q^{\frac34 \sum_i b_i^2} \right]
\\
&+ \frac{1}{3} \left( h^{dP_k}_1 \right)^3 \sum_{b_{1,i},b_{2,i}\in \mathbb{Z}-c_{i}/3\atop \sum_i
  b_{1,i}=0; \sum_i b_{1,i}+b_{2,i}=0}
\q^{\sum_i(b_{1,i}^2+b_{2,i}^{2}+b_{1,i}b_{2,i})}. 
\end{split}
\ee 
In the next subsections we list the first few terms in the generating functions of integer VW invariants with $\mu_0=0$ in the chamber $J'\propto H_\epsilon$ at rank 2 and 3. For $N=2$ (respectively $N=3$) we choose a representative with $c_i\in\{0,1\}$ (respectively $c_i\in \{-1,0,1\}$). The result is of course invariant under permutations of the $c_i$'s.

\subsubsection{$dP_2$}

\bea
%\q^{\frac{5}{12}} &&(y-1/y) h^{dP_2}_{2,0} 
\hiVW^{dP_2}_{2,000} &=&
\left(y^5+3 y^3+6 y+\dots \right) \q^2+\left(y^9+4
   y^7+13 y^5+32 y^3+51 y+\dots \right) \q^3 \nn \\ && +\left(y^{13}+4 y^{11}+ 
   15 y^9+43
   y^7+107 y^5+210 y^3+309 y+\dots\right)
   \q^4 + \dots \nn\\ 
%\q^{\frac{1}{6}} &&(y-1/y) h^{dP_2}_{2,C_1} =
\hiVW^{dP_2}_{2,010} &=&
   \left(y^2+2+\dots \right) \q+\left(y^6+4 y^4+12
   y^2+19+\dots \right) \q^2 \nn \\ &&
   +\left(y^{10}+4 y^8+15 y^6+41
   y^4+87 y^2+119+\dots \right) \q^3 \nn \\ &&
   +\left(
   y^{14}+4 y^{12}+15 y^{10}+45
   y^8+120 y^6+266 y^4+475 y^2+603+\dots\right)
   \q^4+ \dots \nn\\
%\q^{-\frac{1}{12}} &&(y-1/y) h^{dP_2}_{2,C_1+C_2} =   
\hiVW^{dP_2}_{2,011} &=&
\left(y^3+2 y+\dots\right) \q+\left(y^7+4 y^5+12
   y^3+21 y+\dots\right) \q^2 \nn \\ &&
   +\left(y^{11}+4 y^9+15 y^7+41 y^5+90
   y^3+138 y+\dots\right) \q^3  \nn \\ &&
   +\left(y^{15}+4 y^{13}+15 y^{11}+45 y^9+120
   y^7+270 y^5+504 y^3+713 y+\dots\right)
   \q^4 \nn \\ && + \dots
\eea

\bea
\hiVW^{dP_2}_{3,000} &=& ( y^{10}+3 y^8+9 y^6+19 y^4+28 y^2+33+ \dots) \q^3 + \dots \nn\\
\hiVW^{dP_2}_{3,0\pm0}   &=&   (y^6+3 y^4+8
   y^2+10+\dots) \q^2 \nn\\ &&
    + (y^{12}+4 y^{10}+14 y^8+38 y^6+84 y^4+145
   y^2+177+\dots) \q^3 + \dots \nn \\
\hiVW^{dP_2}_{3,0\pm\pm}   &=&   (y^2+1+\dots) \q + (y^8+4 y^6+12 y^4+25
   y^2+33+\dots) \q^2 + \nn\\ &&
    (y^{14}+4 y^{12}+15 y^{10}+42 y^8+105 y^6+212 y^4+350
   y^2+417+\dots) \q^3 + \dots\nn\\
\hiVW^{dP_2}_{3,0\pm\mp}   &=&   (y^8+3 y^6+8 y^4+12
   y^2+15+\dots) \q^2 \nn\\ &&
   + (y^{14}+4 y^{12}+14 y^{10}+38 y^8+86 y^6+158 y^4+228
   y^2+259+\dots) \q^3 \nn \\ && + \dots
\eea

\subsubsection{$dP_3$}

\bea
%\q^{\frac{1}{2}} &&(y-1/y) h^{dP_3}_{2,0} = 
\hiVW^{dP_3}_{2,0000} &=&
(y^5+4 y^3+10 y+\dots) \q^2+ (y^9+5 y^7+19 y^5+55
   y^3+102 y+\dots) \q^3 \nn\\ && 
   + (y^{13}+5 y^{11}+21 y^9+69 y^7+195  y^5+438 y^3+717 y+\dots) \q^4   + \dots \nn\\ 
%\q^{\frac{1}{4}} &&(y-1/y) h^{dP_3}_{2,C_1} =
\hiVW^{dP_3}_{2,0100} &=&
( y^2+3+\dots) \q + (y^6+5 y^4+18
   y^2+34+\dots) \q^2  \nn\\ && + (y^{10}+5 y^8+21 y^6+67 y^4+163
   y^2+251+\dots) \q^3 \nn\\&&+ (y^{14}+5 y^{12}+21 y^{10}+71 y^8+212
   y^6+532 y^4+1066 y^2+1470+\dots )\q^4
 + \dots \nn\\
%\q^{0} &&(y-1/y) h^{dP_3}_{2,C_1+C_2} =  
\hiVW^{dP_3}_{2,0110} &=&
 (y^3+3   y+\dots) \q + (y^7+5 y^5+18 y^3+38+\dots) \q^2 \nn\\&& + 
   ( y^{11}+5 y^9+21 y^7+67 y^5+169 y^3+295
   y + \dots) \q^3 \nn\\&&
   + (y^{15}+5 y^{13}+21 y^{11}+71 y^9+212
   y^7+540 y^5+1134 y^3+1769 y+\dots) \q^4  + \dots \nn\\
%\q^{-\frac14} &&(y-1/y) h^{dP_3}_{2,C_1+C_2+C_3} =
   \hiVW^{dP_3}_{2,0111} &=&
1+ (
y^4+5  y^2+ y^2+12+1/y^2) \q
% \nn\\&&
+( y^8+5 y^6+21 y^4+58
   y^2+100+\dots) \q^2 \nn\\&&
   + (y^{12}+5 y^{10}+21 y^8+71 y^6+200
   y^4+435 y^2+634+\dots) \q^3
   \nn\\&&
    + ( y^{16}+5 y^{14}+21 y^{12}+71
   y^{10}+217 y^8+582 y^6+1355 y^4+2536 y^2+3378 +\dots) \q^4  \nn\\ 
   &&  + \dots 
\eea

\bea
\hiVW^{dP_3}_{3,0000} &=&    
   (y^{10}+4 y^8+14 y^6+36 y^4+61 y^2+74+\dots) \q^3 + \dots\nn\\
   \hiVW^{dP_3}_{3,0\pm00}   &=&   (y^6+4 y^4+13
   y^2+20+\dots) \q^2 \nn\\&&
    + (y^{12}+5 y^{10}+20 y^8+63 y^6+160 y^4+317
   y^2+422+\dots) \q^3 + \dots \nn \\
\hiVW^{dP_3}_{3,0\pm\pm0}   &=&   (y^2+2+\dots) \q + (y^8+5 y^6+18 y^4+45
   y^2+68+\dots) \q^2 + \nn\\ &&
    (y^{14}+5 y^{12}+21 y^{10}+68 y^8+191 y^6+442 y^4+817
   y^2+1053+\dots) \q^3 + \dots\nn\\
\hiVW^{dP_3}_{3,0\pm\mp0}   &=& (y^8+4 y^6+13 y^4+24
   y^2+30+\dots) \q^2 + \nn\\ &&
    (y^{14}+5 y^{12}+20 y^{10}+63 y^8+164 y^6+347 y^4+557
   y^2+650+\dots) \q^3 + \dots\nn\\
\hiVW^{dP_3}_{3,0\pm\pm \pm}   &=& (y^4+2
   y^2+3+\dots) \q + (y^{10}+5 y^8+18 y^6+47 y^4+85
   y^2+102+\dots) \q^2 \nn\\ &&
   + (y^{16}+5 y^{14}+21 y^{12}+68 y^{10}+193 y^8+459 y^6+916 y^4+1418
   y^2+1641+\dots) \q^3 \nn \\ && + \dots \nn\\     
\hiVW^{dP_3}_{3,0\pm\pm \mp}   &=& (y^4+4
   y^2+8+\dots) \q + (y^{10}+5 y^8+20 y^6+58 y^4+129
   y^2+181+\dots) \q^2 \nn\\ &&
   + (y^{16}+5 y^{14}+21 y^{12}+70 y^{10}+204 y^8+515 y^6+1112 y^4+1930
   y^2+2423+\dots) \q^3 \nn \\ && + \dots
\eea

\subsubsection{$dP_4$}
\bea
%\q^{\frac{7}{12}} &&(y-1/y) h^{dP_4}_{2,0} = 
   \hiVW^{dP_4}_{2,00000} &=& 
 (y^5+5 y^3+15 y+\dots) \q^2 + 
(y^9+6 y^7+26 y^5+86  y^3+180 y+\dots) \q^3 \nn\\ &&
+ (   y^{13}+6 y^{11}+28 y^9+103 y^7+324 y^5+812 y^3+1452 y+\dots) \q^4
   + \dots \nn \\ 
%\q^{\frac{1}{3}} &&(y-1/y) h^{dP_4}_{2,C_1} = 
   \hiVW^{dP_4}_{2,01000} &=& 
(y^2+4+\dots) \q+ ( y^6+6 y^4+25
   y^2+54+\dots) \q^2 \nn\\ && + (y^{10}+6 y^8+28 y^6+101 y^4+275
   y^2+465+\dots) \q^3 \nn\\ && + (y^{14}+6 y^{12}+28 y^{10}+105
   y^8+345 y^6+958 y^4+2110 y^2+3119+\dots) \q^4
+ \dots \nn\\
%\q^{\frac{1}{12}} &&(y-1/y) h^{dP_3}_{2,C_1+C_2} = 
   \hiVW^{dP_4}_{2,011000} &=& 
(y^3+4 y+\dots) \q + (y^7+6 y^5+25 y^3+61
   y+\dots) \q^2  \nn\\ && + (y^{11}+6 y^9+28 y^7+101 y^5+286 y^3+553
   y+\dots) \q^3  \nn\\ && + (y^{15}+6 y^{13}+28 y^{11}+105 y^9+345
   y^7+973 y^5+2253 y^3+3806 y+\dots) \q^4 
      + \dots \nn \\
%\q^{-\frac16} &&(y-1/y) h^{dP_4}_{2,C_1+C_2+C_3} =
   \hiVW^{dP_4}_{2,01110} &=& 
1 + (y^4+6  y^2+17+\dots) \q + (y^8+6 y^6+28 y^4+89
   y^2+169+\dots) \q^2  \nn\\ && + ( y^{12}+6 y^{10}+28 
   \dots) \q^3 \nn\\ && 
   + (y^{16}+6 y^{14}+28
   y^{12}+105 y^{10}+350 y^8+1031 y^6+2630 y^4+5380 y^2+7606+\dots)
   \q^4 \nn\\ && + 
\dots \nn\\
%\q^{-\frac{5}{12}} &&(y-1/y) h^{dP_4}_{2,C_1+C_2+C_3+C_4} 
   \hiVW^{dP_4}_{2,01111} &=&  (y+1/y) + 
(y^5+6 y^3+19   y+\dots) \q
% \nn\\&&
   + ( y^9+6 y^7+28 y^5+92 y^3+200
   y+\dots) \q^2 \nn\\ && +(y^{13}+6 y^{11}+28 y^9+105 y^7+333 y^5+843
   y^3+1522 y+\dots) \q^3 \nn\\ && 
   + (y^{17}+6 y^{15}+28 y^{13}+105
   y^{11}+350 y^9+1036 y^7+2697 y^5+5825 y^3+9399 y+\dots) \q^4 \nn\\ && 
   + \dots
\eea

\bea
\hiVW^{dP_4}_{3,00000} &=&    
   (y^{10}+5 y^8+20 y^6+60 y^4+114 y^2+140+\dots) \q^3 + \dots\nn\\
   \hiVW^{dP_4}_{3,0\pm000}   &=&   (y^6+5 y^4+19
   y^2+34+\dots) \q^2 \nn\\&&
    + (y^{12}+6 y^{10}+27 y^8+96 y^6+272 y^4+597
   y^2+855+\dots) \q^3 + \dots \nn\\
   \hiVW^{dP_4}_{3,0\pm\pm00}   &=&   (y^2+3+\dots) \q + (y^8+6 y^6+25 y^4+71
   y^2+120+\dots) \q^2 + \nn\\ &&
    (y^{14}+6 y^{12}+28 y^{10}+102 y^8+316 y^6+810 y^4+1633
   y^2+2243+\dots) \q^3 + \dots\nn\\
   \hiVW^{dP_4}_{3,0\pm\mp00}   &=& (y^8+5 y^6+19 y^4+41
   y^2+51+\dots) \q^2 + \nn\\ &&
    (y^{14}+6 y^{12}+27 y^{10}+96 y^8+280 y^6+660 y^4+1151
   y^2+1363+\dots) \q^3 + \dots\nn\\
\hiVW^{dP_4}_{3,0\pm\pm \pm0}   &=& (y^4+3
   y^2+4+\dots) \q + (y^{10}+6 y^8+25 y^6+75 y^4+152 y^2+184+\dots) \q^2 \nn\\ &&
   + \dots \nn\\     
\hiVW^{dP_4}_{3,0\pm\pm \mp0}   &=& (y^4+5
   y^2+12+\dots) \q + (y^{10}+6 y^8+27 y^6+89 y^4+220 y^2+338+\dots)
   \q^2\nn \\ &&
  + \dots \nn\\  
   \hiVW^{dP_4}_{3,0\pm\pm \pm\pm}   &=& 1+(y^6+6 y^4+22
   y^2+42+\dots) \q \nn\\ &&
   + (y^{12}+6 y^{10}+28 y^8+99 y^6+287 y^4+629
   y^2+912+\dots) \q^2
   + \dots \nn\\       
  \hiVW^{dP_4}_{3,0\pm\pm \pm\mp}   &=& 1+(y^6+6 y^4+22
   y^2+43+\dots) \q  \nn\\ &&
   + (y^{12}+6 y^{10}+28 y^8+99 y^6+288 y^4+632
   y^2+920+\dots) \q^2
   + \dots \nn\\       
     \hiVW^{dP_4}_{3,0\pm\pm \mp\mp}   &=& (y^6+5 y^4+14
   y^2+17+\dots) \q  \nn\\ &&
   + (y^{12}+6 y^{10}+27 y^8+91 y^6+241 y^4+452
   y^2+542+\dots) \q^2
   + \dots
\eea

\subsubsection{$dP_5$}
\bea
%\q^{\frac{2}{3}} &&(y-1/y) h^{dP_5}_{2,00} 
  \hiVW^{dP_5}_{2,000000} &=& 
(y^5+6 y^3+21 y+\dots) \q^2 + 
(y^9+7 y^7+34 y^5+126
   y^3+291 y+\dots ) \q^3 + \dots
   \nn\\
%   \q^{\frac{5}{12}} &&(y-1/y) h^{dP_5}_{2,0C_1} =
  \hiVW^{dP_5}_{2,010000} &=& 
    (y^2+5+\dots) \q +
   (y^6+7 y^4+33
   y^2+79+\dots) \q^2  \nn\\&& + (y^{10}+7 y^8+36 y^6+144 y^4+430
   y^2+783+\dots) \q^3+ \dots
   \nn\\
 %  \q^{\frac{1}{6}} &&(y-1/y) h^{dP_5}_{2,0C_1+C_2} =
   \hiVW^{dP_5}_{2,011000} &=& 
    (y^3+5 y+\dots) \q +
   ( y^7+7 y^5+33 y^3+90 y+\dots) \q^2 \nn\\&&
    + ( y^{11}+7 y^9+36 y^7+144 y^5+449 y^3+941 y+\dots) \q^3 + \dots
   \nn\\
%   \q^{-\frac{1}{12}} &&(y-1/y) h^{dP_5}_{2,0C_1+C_2+C_3} =
   \hiVW^{dP_5}_{2,011100} &=& 
    1+ (y^4+7y^2+22+\dots)\q 
   + (y^8+7 y^6+36 y^4+127 y^2+260+\dots) \q^2\nn\\&&
    + (y^{12}+7 y^{10}+36 y^8+148 y^6+506
   y^4+1326 y^2+2218+\dots) \q^3+ \dots
   \nn\\
%      \q^{-\frac{1}{3}} &&(y-1/y) h^{dP_5}_{2,0C_1+C_2+C_3+C_4} = 
  \hiVW^{dP_5}_{2,011110} &=& 
      y+1/y + (y^5+7 y^3+25  y+) \q \nn\\&& 
      + (y^9+7 y^7+36 y^5+132 y^3+312
   y+\dots) \q^2 \nn\\&& + ( y^{13}+7 y^{11}+36 y^9+148 y^7+513
   y^5+1417 y^3+2738 y+\dots) \q^3 
       + \dots
   \nn\\
  %       \q^{-\frac{7}{12}} &&(y-1/y) h^{dP_5}_{2,0C_1+C_2+C_3+C_4+C_5} =
    \hiVW^{dP_5}_{2,011111} &=& 
           (y^2+5+\dots) + 
         (y^6+7 y^4+33
   y^2+78+\dots) \q \nn\\&& + (y^{10}+7 y^8+36 y^6+144 y^4+429
   y^2+779+\dots) \q^2  \nn\\&& + (y^{14}+7 y^{12}+36 y^{10}+148
   y^8+529 y^6+1598 y^4+3800 y^2+5938+\dots) + \dots
   \nn\\
\eea
\bea
\hiVW^{dP_5}_{3,000000} &=&  (y^{10}+6 y^8+27 y^6+92 y^4+192
   y^2+237+ \dots) \q^3+\dots \nn\\
\hiVW^{dP_5}_{3,0\pm0000}   &=&   (y^6+6 y^4+26 y^2+52+\dots) \q^2 + \dots \nn\\
\hiVW^{dP_5}_{3,0\pm\pm000}   &=&   (y^2+4+\dots) \q + (y^8+7 y^6+33 y^4+103 y^2+189+\dots) \q^2 + \dots\nn\\
\hiVW^{dP_5}_{3,0\pm\mp000}   &=& (y^8+6 y^6+26 y^4+63 y^2+78+\dots) \q^2 + \dots\nn\\
\hiVW^{dP_5}_{3,0\pm\pm \pm00}   &=& (y^4+4
   y^2+5+\dots) \q + (y^{10}+7 y^8+33 y^6+110 y^4+243 y^2+295+\dots) \q^2 \nn\\ &&
   + \dots \nn\\     
\hiVW^{dP_5}_{3,0\pm\pm \mp00}   &=& (y^4+6
   y^2+16+\dots) \q + (y^{10}+7 y^8+35 y^6+127 y^4+339 y^2+556+\dots) \q^2
   \nn\\ &&
   + \dots \nn\\
\hiVW^{dP_5}_{3,0\pm\pm \pm \pm0}   &=& 1+ (y^6+7 y^4+28
   y^2+59+\dots ) \q + \dots \nn\\ 
   \hiVW^{dP_5}_{3,0\pm\pm \pm \mp0}   &=& 1+ (y^6+7 y^4+28
   y^2+60+\dots ) \q + \dots
   \nn\\ 
   \hiVW^{dP_5}_{3,0\pm\pm \mp \mp0}   &=& (y^6+6 y^4+19
   y^2+23+\dots ) \q + \dots
   \nn\\  
   \hiVW^{dP_5}_{3,0\pm\pm \pm \pm\pm}   &=& (y^2+5+\dots) + (y^8+7 y^6+34 y^4+105
   y^2+196+\dots)  \q + \dots
   \nn\\ 
\hiVW^{dP_5}_{3,0\pm\pm \pm \pm\mp}   &=&    (y^2+1+\dots) + (y^8+7 y^6+30 y^4+77
   y^2+94+\dots) \q + \dots
   \nn\\
\hiVW^{dP_5}_{3,0\pm\pm \pm \mp\mp}   &=&   (y^2+4+\dots)+(y^8+7 y^6+33 y^4+102
   y^2+187+\dots) \q + \dots
\eea

\subsubsection{$dP_6$}
\bea
  \hiVW^{dP_6}_{2,0000000} &=& 
(y^5+7 y^3+28 y+\dots) \q^2 + 
(y^9+8 y^7+43 y^5+176 y^3+441 y+\dots) \q^3 + \dots
   \nn\\
  \hiVW^{dP_6}_{2,0100000} &=& 
(y^2+6+\dots)\q + (y^6+8 y^4+42
   y^2+109+\dots) \q^2 \nn\\ &&  + (y^{10}+8 y^8+45 y^6+197 y^4+635
   y^2+1227+\dots) \q^3 
    \dots
   \nn\\
  \hiVW^{dP_6}_{2,0110000} &=& 
(y^3+6 y+\dots) \q 
   +(y^7+8 y^5+42 y^3+125
   y +\dots) \q^2 \nn\\ &&  + (y^{11}+8 y^9+45 y^7+197 y^5+666 y^3+1488
   y+\dots)+ \dots
   \nn\\
  \hiVW^{dP_6}_{2,0111000} &=& 
1+ (y^4+8
   y^2+27+\dots) \q \nn\\ &&  + (y^8+8 y^6+45 y^4+172
   y^2+373+\dots) \q^2 \nn\\ && 
   + (y^{12}+8 y^{10}+45 y^8+201 y^6+740
   y^4+2068 y^2+3631+\dots) + \dots
  \nn \\
  \hiVW^{dP_6}_{2,0111100} &=&  (y+1/y) + 
      (y^5+8 y^3+31
   y+\dots) \q \nn\\ &&  + (y^9+8 y^7+45 y^5+180 y^3+453
   y+\dots) \q^2\nn\\ && 
   + (y^{13}+8 y^{11}+45 y^9+201 y^7+752
   y^5+2228 y^3+4537 y+\dots) \q^3
       + \dots
   \nn\\
           \hiVW^{dP_6}_{2,0111110} &=&
          (y^2+5+\dots)
         +( y^6+8 y^4+40
   y^2+100+\dots ) \q \nn\\&&  + (y^{10}+8 y^8+45 y^6+194 y^4+615
   y^2+1170+\dots ) \q^2 \nn\\&&  + (y^{14}+8 y^{12}+45 y^{10}+201
   y^8+771 y^6+2489 y^4+6251 y^2+10191+\dots)+\dots
   \nn\\
          \hiVW^{dP_6}_{2,0111111} &=&
       (y^3+6 y+\dots)
       + (y^7+8 y^5+42 y^3+123
   y+\dots) \q  \nn\\&&  +(y^{11}+8 y^9+45 y^7+197 y^5+663 y^3+1471
   y+\dots) \q^2 \nn\\&&  + (y^{15}+8 y^{13}+45 y^{11}+201 y^9+775
   y^7+2562 y^5+6863 y^3+12971 y+\dots) \q^3 \nn \\ &&
+ \dots
\eea

\bea
\hiVW^{dP_6}_{3,00000000} &=&   (y^{10}+7 y^8+35 y^6+133 y^4+300
   y^2+371+\dots) \q^3+\dots \nn\\
\hiVW^{dP_6}_{3,0\pm 00000}   &=&   (y^6+7 y^4+34 y^2+74+\dots) \q^2 + \dots \nn\\
\hiVW^{dP_6}_{3,0\pm\pm 0000}   &=&   (y^2+5+\dots) \q + (y^8+8 y^6+42 y^4+141 y^2+275+\dots) \q^2 + \dots\nn\\
\hiVW^{dP_6}_{3,0\pm\mp 0000}   &=& (y^8+7 y^6+34 y^4+90 y^2+111+\dots) \q^2 + \dots\nn\\
\hiVW^{dP_6}_{3,0\pm\pm \pm000}   &=& (y^4+5  y^2+6+\dots) \q + (y^{10}+8 y^8+42 y^6+152 y^4+358 y^2+435+\dots) \q^2 
   + \dots \nn\\     
\hiVW^{dP_6}_{3,0\pm\pm \mp000}   &=& (y^4+7
   y^2+20+\dots) \q + (y^{10}+8 y^8+44 y^6+172 y^4+486 y^2+835+\dots) \q^2
%   \nn\\ &&
   + \dots \nn\\
\hiVW^{dP_6}_{3,0\pm \pm \pm \pm00}   &=& 1 + ( y^6+8 y^4+34
   y^2+76+\dots) \q + \dots 
      \nn\\
\hiVW^{dP_6}_{3,0\pm \pm \pm  \mp00}   &=&1+ (y^6+8 y^4+34
   y^2+77+\dots) \q + \dots 
      \nn\\
 \hiVW^{dP_6}_{3,0\pm \pm \mp  \mp00}   &=&(y^6+7 y^4+24
   y^2+29+\dots) \q + \dots 
      \nn\\     
   \hiVW^{dP_6}_{3,0\pm \pm \pm \pm  \pm0}   &=& ( y^2+5+\dots) \q	 + ( y^8+8 y^6+41 y^4+134
   y^2+260+\dots) \q^2 + \dots \nn\\
    \hiVW^{dP_6}_{3,0\pm \pm \pm  \pm \mp0} &=& 
    (y^2+1+\dots) \q + (    y^8+8 y^6+37 y^4+101
   y^2+123+\dots ) \q^2 + \dots \nn\\
      \hiVW^{dP_6}_{3,0\pm  \pm \pm \pm  \pm \pm}   &=& 
  ( y^4+6  y^2+7+\dots) + \dots \nn\\
       \hiVW^{dP_6}_{3,0\pm \pm \pm \pm   \pm \mp}   &=& 
  ( y^4+6  y^2+17+\dots) + \dots \nn\\
       \hiVW^{dP_6}_{3,0\pm  \pm \pm \pm \mp \mp}   &=& 
  ( y^4+6  y^2+17+\dots) + \dots \nn\\
       \hiVW^{dP_6}_{3,0\pm \pm  \pm \mp \mp \mp}   &=& 
  ( y^4+5  y^2+6+\dots) + \dots
\eea

\subsubsection{$dP_7$}
\bea
  \hiVW^{dP_7}_{2,00000000} &=& (y^5+8 y^3+36 y+\dots) \q^2 + (y^9+9 y^7+53 y^5+237
   y^3+636 y+\dots) \q^3  + \dots
   \nn\\
  \hiVW^{dP_7}_{2,01000000} &=& (y^2+7+\dots) \q + ( y^6+9 y^4+52
   y^2+144+\dots) \q^2 \nn\\&& + ( y^{10}+9 y^8+55 y^6+261 y^4+897
   y^2+1819+\dots) \q^3 +   \dots
   \nn\\
\hiVW^{dP_7}_{2,01100000} &=& (y^3+7 y+\dots) \q + (y^7+9 y^5+52 y^3+166
   y+\dots) \q^2\nn\\ && + (y^{11}+9 y^9+55 y^7+261 y^5+945 y^3+2223
   y+\dots) + \dots
   \nn\\
  \hiVW^{dP_7}_{2,01110000} &=& 1 + (y^4+9
   y^2+32+\dots) \q + (y^8+9 y^6+55 y^4+224
   y^2+508+\dots) \q^2 \nn\\&&+ (y^{12}+9 y^{10}+55 y^8+265 y^6+1040
   y^4+3057 y^2+5582+\dots) \q^3
 + \dots
   \nn\\
  \hiVW^{dP_7}_{2,01111000} &=&  (y+1/y) + (y^5+9 y^3+37
   y+\dots) \q+(y^9+9 y^7+55 y^5+236 y^3+623
   y+\dots) \q^2 \nn\\ && 
   +(y^{13}+9 y^{11}+55 y^9+265 y^7+1060
   y^5+3322 y^3+7049 y+\dots) \q^3
  + \dots
   \nn\\
   \hiVW^{dP_7}_{2,01111100} &=&(y^2+5+\dots) + (y^6+9 y^4+47
   y^2+122+\dots)\q\nn\\ &&
   +(y^{10}+9 y^8+55 y^6+253 y^4+838
   y^2+1654+\dots) \q^2\nn\\ &&
   +(y^{14}+9 y^{12}+55 y^{10}+265
   y^8+1084 y^6+3694 y^4+9666 y^2+16306+\dots) \q^3+\dots
   \nn \\
    \hiVW^{dP_7}_{2,01111110} &=&(y^3+6 y+\dots) + (y^7+9 y^5+50 y^3+152
   y+\dots)\q \nn\\ &&
   + (y^{11}+9 y^9+55 y^7+258 y^5+915 y^3+2106
   y+\dots) \q^2\nn\\ &&
   +(y^{15}+9 y^{13}+55 y^{11}+265 y^9+1091
   y^7+3825 y^5+10739 y^3+21017 y+\dots) 
   + \dots
   \nn\\
    \hiVW^{dP_7}_{2,01111111} &=&(y^4+7 y^2+22+\dots) + \q ( y^8+9 y^6+52 y^4+199
   y^2+430+\dots) \q\nn\\ &&
    +(y^{12}+9 y^{10}+55 y^8+261 y^6+1001
   y^4+2855 y^2+5111+\dots) \q^2\nn\\ &&
   + (y^{16}+9 y^{14}+55
   y^{12}+265 y^{10}+1095 y^8+3950 y^6+12081 y^4+28850 y^2\nn\\ && + 45784+\dots) \q^3
   + \dots %\nn\\&&
\eea

\bea
\hiVW^{dP_7}_{3,00000000} &=&  (y^{10}+8 y^8+44 y^6+184 y^4+443
   y^2+548+\dots) \q^3 + \dots  \nn\\
\hiVW^{dP_7}_{3,0\pm 000000}   &=&   (y^6+8 y^4+43 y^2+100+\dots) \q^2 + \dots \nn\\
\hiVW^{dP_7}_{3,0\pm\pm 000000}   &=&   (y^2+6+\dots) \q + (y^8+9 y^6+52 y^4+185 y^2+378+\dots) \q^2 + \dots\nn\\
\hiVW^{dP_7}_{3,0\pm\mp 000000}   &=& (y^8+8 y^6+43 y^4+122 y^2+150+\dots) \q^2 + \dots\nn\\
\hiVW^{dP_7}_{3,0\pm\pm \pm 00000}   &=& (y^4+6 y^2+7+\dots) \q + (y^{10}+9 y^8+52 y^6+201  y^4+497 y^2+604+\dots) \q^2 
   + \dots \nn\\     
\hiVW^{dP_7}_{3,0\pm\pm \mp 00000}   &=& (y^4+8
   y^2+24+\dots) \q + (y^{10}+9 y^8+54 y^6+224 y^4+661
   y^2+1175+\dots) \q^2
%   \nn\\ &&
   + \dots  \nn\\
\hiVW^{dP_7}_{3,0\pm \pm \pm \pm 000}   &=& 1 + (y^6+9 y^4+40
   y^2+94+\dots) \q + \dots 
      \nn\\
\hiVW^{dP_7}_{3,0\pm \pm \pm  \mp 000}   &=&1+ (y^6+9 y^4+40
   y^2+94+\dots) \q + \dots 
      \nn\\
 \hiVW^{dP_7}_{3,0\pm \pm \mp  \mp 000}   &=&(y^8+9 y^6+53 y^4+159 y^2+195+\dots) \q + \dots 
      \nn\\      
   \hiVW^{dP_7}_{3,0\pm \pm \pm \pm  \pm 00}   &=& ( y^2+5+\dots) \q	 + (y^8+9 y^6+48 y^4+163
   y^2+324+\dots) \q^2 + \dots \nn\\
    \hiVW^{dP_7}_{3,0\pm \pm \pm  \pm \mp 00} &=& 
    (y^2+1+\dots) \q + (   y^8+9 y^6+44 y^4+125
   y^2+152+\dots ) \q^2 + \dots \nn\\
       \hiVW^{dP_7}_{3,0\pm \pm \pm  \mp \mp 00} &=& 
    (y^2+4+\dots) \q + (  y^8+9 y^6+47 y^4+158
   y^2+309+\dots ) \q^2 + \dots \nn
   \eea
   \bea
      \hiVW^{dP_7}_{3,0\pm  \pm \pm \pm  \pm \pm 0}   &=& 
  ( y^4+6   y^2+7+\dots) + (y^{10}+9 y^8+51 y^6+191 y^4+465
   y^2+564+\dots) +  \dots \nn\\
       \hiVW^{dP_7}_{3,0\pm \pm \pm \pm   \pm \mp 0}   &=& 
  ( y^4+6  y^2+17+\dots) +(y^{10}+9 y^8+51 y^6+201 y^4+565
   y^2+984+\dots) + \dots  \nn\\
       \hiVW^{dP_7}_{3,0\pm  \pm \pm \pm \mp \mp 0}   &=& 
  ( y^4+6  y^2+17+\dots) + (y^{10}+9 y^8+51 y^6+201 y^4+566
   y^2+987+\dots) + \dots \nn\\
       \hiVW^{dP_7}_{3,0\pm \pm  \pm \mp \mp \mp 0 }   &=& 
  ( y^4+5  y^2+6+\dots) +(y^{10}+9 y^8+50 y^6+186 y^4+447
   y^2+543+\dots) + \dots  \nn\\
       \hiVW^{dP_7}_{3,0\pm  \pm \pm \pm  \pm \pm\pm}   &=& 
  ( y^6+7 y^4+29 y^2+64+\dots) \nn\\ &&   + (y^{12}+9 y^{10}+53 y^8+229 y^6+756 y^4+1870
   y^2+2998+\dots) \q+  \dots \nn\\
         \hiVW^{dP_7}_{3,0\pm  \pm \pm \pm  \pm \pm \mp}   &=& 
  ( y^6+7 y^4+29
   y^2+64+\dots)   \nn\\ &&  +(y^{12}+9 y^{10}+53 y^8+229 y^6+756 y^4+1876
   y^2+3008+) \q +  \dots \nn\\
           \hiVW^{dP_7}_{3,0\pm  \pm \pm \pm  \pm \mp \mp}   &=& 
  ( y^6+7 y^4+24
   y^2+29+\dots)   \nn\\ &&  +(y^{12}+9 y^{10}+53 y^8+224 y^6+701 y^4+1505
   y^2+1818+\dots) \q +  \dots \nn\\
           \hiVW^{dP_7}_{3,0\pm  \pm \pm \pm  \mp \mp \mp}   &=& 
  ( y^6+7 y^4+28
   y^2+61+\dots)  \nn\\ && + (y^{12}+9 y^{10}+53 y^8+228 y^6+750 y^4+1850
   y^2+2960+) \q \nn \\ &&  + \dots 
\eea

\subsubsection{$dP_8$}
\bea
  \hiVW^{dP_8}_{2,0000 00000} &=& (y^5+9 y^3+45 y+\dots) \q^2 + (y^9+10 y^7+64 y^5+310
   y^3+882 y+\dots) \q^3 + \dots
   \nn\\
  \hiVW^{dP_8}_{2,0100 00000} &=& (y^2+8+\dots) \q + (y^6+10 y^4+63
   y^2+184+\dots) \q^2 \nn\\ && 
   + (y^{10}+10 y^8+66 y^6+337 y^4+1223
   y^2+2581+\dots) \q^3 + \dots
   \nn\\
\hiVW^{dP_8}_{2,0110 00000} &=& (y^3+8 y+\dots) \q + (y^7+10 y^5+63 y^3+213
   y+\dots) \q^2 \nn\\ &&  +(y^{11}+10 y^9+66 y^7+337 y^5+1294
   y^3+3175 y+\dots) \q^3 + \dots
   \nn\\
  \hiVW^{dP_8}_{2,0111 00000} &=& 1+ (y^4+10
   y^2+37+\dots) \q + (y^8+10 y^6+66 y^4+283
   y^2+665+\dots) \q^2  \nn\\&& 
   + ( y^{12}+10 y^{10}+66 y^8+341
   y^6+1415 y^4+4330 y^2+8164+\dots) \q^3  + \dots
   \nn\\
  \hiVW^{dP_8}_{2,01111 0000} &=&  (y+1/y) + (y^5+10 y^3+43
   y+\dots) \q  \nn\\&&  
   + (y^9+10 y^7+66 y^5+300 y^3+822
   y+\dots) \q^2  \nn\\&& 
   +(y^{13}+10 y^{11}+66 y^9+341 y^7+1447
   y^5+4745 y^3+10404 y+\dots) \q^3+ \dots
     \nn\\
     \hiVW^{dP_8}_{2,01111 1000} &=&(y^2+5+\dots) + (y^6+10 y^4+54
   y^2+144+\dots) \q  \nn\\ && 
   +(y^{10}+10 y^8+66 y^6+321 y^4+1098
   y^2+2231+\dots) \q^2  \nn\\ &&  
   +(y^{14}+10 y^{12}+66 y^{10}+341
   y^8+1479 y^6+5269 y^4+14221 y^2\nn\\ &&
   +24669+\dots) \q^3+\dots
   \nn\\
    \hiVW^{dP_8}_{2,01111 1100} &=&(y^3+6 y+\dots) + (y^7+10 y^5+58 y^3+181
   y+\dots) \q  \nn\\ && 
   + (y^{11}+10 y^9+66 y^7+329 y^5+1213
   y^3+2871 y+\dots) \q^2 \nn\\ && 
    + (y^{15}+10 y^{13}+66 y^{11}+341
   y^9+1491 y^7+5493 y^5+15985 y^3\nn\\ &&
   +32149 y+\dots) \q^3 + \dots\nn
 \eea
\bea
    \hiVW^{dP_8}_{2,01111 1110} &=&(y^4+7 y^2+22+\dots)+(y^8+10 y^6+61 y^4+236
   y^2+523+\dots) \q  \nn\\ && 
   + (y^{12}+10 y^{10}+66 y^8+334
   y^6+1332 y^4+3886 y^2+7137+\dots)  \q^2  \nn\\ && 
   + (y^{16}+10
   y^{14}+66 y^{12}+341 y^{10}+1498 y^8+5684 y^6+18025 y^4\nn\\ &&
   +44146 y^2+71854+\dots) \q^3
   + \dots
    \nn\\
    \hiVW^{dP_8}_{2,01111 1111} &=&(y^5+8 y^3+29 y+\dots) + (y^9+10 y^7+63 y^5+267
   y^3+687 y+\dots) \q \nn\\ && 
   +(y^{13}+10 y^{11}+66 y^9+337
   y^7+1396 y^5+4420 y^3+9412 y+\dots) \q^2 \nn\\ &&
   +(y^{17}+10
   y^{15}+66 y^{13}+341 y^{11}+1502 y^9+5781 y^7+19084 y^5 \nn\\ && +50531 y^3+94968 y+\dots) \q^3 
   + \dots
\eea

\bea
\hiVW^{dP_8}_{3,000000000} &=& (y^{10}+9 y^8+54 y^6+246 y^4+626
   y^2+774+\dots)  \q^3+\dots \nn\\
\hiVW^{dP_8}_{3,0\pm00000000 }   &=&   (y^6+9 y^4+53 y^2+130+\dots) \q^2 + \dots \nn\\
\hiVW^{dP_8}_{3,0\pm \pm 0000000}   &=&   (y^2+7+\dots) \q +
 (y^8+10 y^6+63 y^4+235 y^2+498+\dots) \q^2 + \dots\nn\\
\hiVW^{dP_8}_{3,0\pm \mp 0000000 }   &=& (y^8+8 y^6+43 y^4+122 y^2+150+\dots) \q^2 + \dots\nn\\ 
\hiVW^{dP_8}_{3,0\pm \pm  \pm 000000 }   &=& (y^4+7
   y^2+8+\dots) \q + (y^{10}+10 y^8+63 y^6+257 y^4+660
   y^2+802+\dots) \q^2 
   + \dots \nn\\     
\hiVW^{dP_8}_{3,0\pm \pm  \mp 000000}   &=& (y^4+9
   y^2+28+\dots) \q + (y^{10}+10 y^8+65 y^6+283 y^4+864
   y^2+1576+\dots) \q^2
   + \dots \nn\\
\hiVW^{dP_8}_{3,0\pm  \pm  \pm  \pm 0000}   &=& 1 + (y^6+10 y^4+46
   y^2+110+\dots) \q + \dots
      \nn\\
\hiVW^{dP_8}_{3,0\pm  \pm  \pm   \mp 0000}   &=&1 + (y^6+10 y^4+46
   y^2+111+\dots) \q + \dots
      \nn\\
 \hiVW^{dP_8}_{3,0\pm  \pm  \mp   \mp 0000}   &=&(y^6+9 y^4+34
   y^2+41+\dots) \q + \dots 
      \nn\\
  \hiVW^{dP_8}_{3,0\pm  \pm  \pm  \pm   \pm 000}   &=& ( y^2+5+\dots) \q	 + (y^8+10 y^6+55 y^4+192
   y^2+388+\dots) \q^2 + \dots \nn\\
    \hiVW^{dP_8}_{3,0\pm  \pm  \pm   \pm  \mp 000} &=& 
    (y^2+1+\dots) \q + ( y^8+10 y^6+55 y^4+192
   y^2+388+\dots ) \q^2 + \dots \nn\\
       \hiVW^{dP_8}_{3,0\pm  \pm  \pm   \mp  \mp 000 } &=& 
    (y^2+4+\dots) \q + ( y^8+10 y^6+54 y^4+186
   y^2+370+\dots ) \q^2 +  \dots \nn\\
 \hiVW^{dP_8}_{3,0\pm  \pm  \pm  \pm   \pm \pm 00}   &=& 
 (y^4+6   y^2+7+\dots) + (y^{10}+10 y^8+59 y^6+228 y^4+565
   y^2+686+\dots) \q + \dots \nn\\
    \hiVW^{dP_8}_{3,0\pm  \pm  \pm  \pm   \pm \mp 00}   &=& 
 (y^4+6  y^2+17+\dots) + (y^{10}+10 y^8+59 y^6+238 y^4+680
   y^2+1201+\dots) \q	 \nn\\ &&  + \dots \nn\\
    \hiVW^{dP_8}_{3,0\pm  \pm  \pm  \pm   \mp \mp 00}   &=& 
 (y^4+6  y^2+17+\dots) + (y^{10}+10 y^8+59 y^6+238 y^4+681
   y^2+1204+\dots) \q	  \nn\\ && + \dots\nn\\
      \hiVW^{dP_8}_{3,0\pm  \pm  \pm  \mp   \mp \mp 00 }   &=& 
 (y^4+5   y^2+6+\dots) + (y^{10}+10 y^8+58 y^6+222 y^4+543
   y^2+660+\dots) \q \nn\\ && + \dots
\nn\\
  \hiVW^{dP_8}_{3,0\pm  \pm  \pm  \pm   \pm \pm \pm 0} &=&(   y^6+7 y^4+29
   y^2+64+\dots)  \nn\\ &&  + (y^{12}+10 y^{10}+62 y^8+275 y^6+922 y^4+2311
   y^2+3747+\dots) \q  +\dots
     \nn \\
  \hiVW^{dP_8}_{3,0\pm  \pm  \pm  \pm   \pm \pm \mp 0} &=&(   y^6+7 y^4+29 y^2+64+\dots) 
   \nn\\ &&  + (y^{12}+10 y^{10}+62 y^8+275 y^6+922 y^4+2317
   y^2+3757+\dots)  +\dots
        \nn \\
  \hiVW^{dP_8}_{3,0\pm  \pm  \pm  \pm   \pm \mp \mp 0} &=&(  y^6+7 y^4+24
   y^2+29+\dots)  \nn\\ &&  + (y^{12}+10 y^{10}+62 y^8+270 y^6+861 y^4+1874
   y^2+2265+\dots) +\dots
        \nn \\
  \hiVW^{dP_8}_{3,0\pm  \pm  \pm  \pm   \mp \mp \mp 0} &=&(   y^6+7 y^4+28
   y^2+61+\dots)  \nn\\ &&  + (y^{12}+10 y^{10}+62 y^8+274 y^6+915 y^4+2284
   y^2+3697+\dots)  +\dots
 \nn
      \eea
     \bea
        \hiVW^{dP_8}_{3,0\pm  \pm  \pm  \pm   \pm \pm \pm \pm } &=&(  y^8+8 y^6+37 y^4+121
   y^2+227+\dots)  \nn\\ &&   + (y^{14}+10 y^{12}+64 y^{10}+301 y^8+1113 y^6+3268 y^4+7446
   y^2+11298+\dots) \q  \nn \\ && +\dots
     \nn \\
        \hiVW^{dP_8}_{3,0\pm  \pm  \pm  \pm   \pm \pm \pm \mp } &=&(  y^8+8 y^6+37 y^4+100
   y^2+122+\dots)   \nn\\ &&  + (y^{14}+10 y^{12}+64 y^{10}+301 y^8+1092 y^6+3058 y^4+6109
   y^2+7329+\dots) \q  \nn\\ && +\dots
     \nn \\
             \hiVW^{dP_8}_{3,0\pm  \pm  \pm  \pm   \pm \pm \mp \mp } &=&(  y^8+8 y^6+37 y^4+115
   y^2+217+\dots)  \nn\\ &&   + (y^{14}+10 y^{12}+64 y^{10}+301 y^8+1107 y^6+3234 y^4+7302
   y^2+11076+\dots) \q \nn\\ &&  +\dots
     \nn \\
        \hiVW^{dP_8}_{3,0\pm  \pm  \pm  \pm   \pm \mp \mp \mp } &=&( y^8+8 y^6+37 y^4+115
   y^2+217+\dots)   \nn\\ && + (y^{14}+10 y^{12}+64 y^{10}+301 y^8+1107 y^6+3235 y^4+7309
   y^2+11088+\dots) \q  \nn\\ && +\dots
       \nn \\
        \hiVW^{dP_8}_{3,0\pm  \pm  \pm  \pm   \mp \mp \mp \mp } &=&( y^8+8 y^6+36 y^4+96
   y^2+117+\dots) \nn\\ &&  + (y^{14}+10 y^{12}+64 y^{10}+300 y^8+1085 y^6+3024 y^4+6025
   y^2+7224+\dots) \q  \nn\\ && +\dots
\eea

\section{Vanishing of attractor indices for 3-block collections \label{sec_threeblockatt}}

Here, we show that the attractor indices $\Omstar(\gamma)$
 in  the attractor chamber 
\be
\label{stabdPa}
\varsigma_1^\star =  -c\, \cN_2 + b\, \cN_3\ ,\quad 
\varsigma_2^\star = -a\, \cN_3 + c\, \cN_1\ ,\quad 
\varsigma_3^\star = -b\, \cN_1 + a\, \cN_2\ ,
\ee
vanish for any positive dimension vector $\vec N$
such that $(\cN_1,\cN_2,\cN_3)\notin \{ (1,0,0), (0,1,0), (0,0,1) \}$
and $\vec N$ is not proportional to the dimension vector $(x,\dots;y,\dots; z,\dots)$ associated to pure D0-branes. 
When any of the $\cN_i$'s vanishes, the relevant quiver has no loop and the vanishing of  $\Omstar(\vec N)$ is well known.
For non-zero values of  the $\cN_i$'s, we show that the expected dimension in the attractor chamber is strictly negative and therefore the moduli space is empty.

\medskip

Similar to the case of $\IP^2$ in \S\ref{sec_P2moduli}, 
the expected dimension $\dstar_\IC$ in the attractor chamber 
can be written in one of the following ways, depending 
on the signs of the $\varsigma^\star_i$'s,
\begin{itemize}
\item when  $\varsigma^\star_3 \leq  0, \varsigma^\star_1 \geq  0$,  $\Phi_{31}^\alpha=0$ hence
\be
d_\IC= 1- \cQ(\vec N) + \frac{ 2 \cA }{ \cA + \cB + \cC } \cN_3   \varsigma^\star_3 -  \frac{ 2 \cC }{ \cA + \cB + \cC } \cN_1   \varsigma^\star_1
\ee  
\item when $\varsigma^\star_1 \leq  0, \varsigma^\star_2 \geq0$, $\Phi_{12}^\alpha=0$ hence
\be
d'_\IC= 1- \cQ(\vec N) + \frac{ 2 \cB }{ \cA + \cB + \cC } \cN_1   \varsigma^\star_1 -  \frac{ 2 \cA }{ \cA + \cB + \cC } \cN_2   \varsigma^\star_2 
\ee
\item when $\varsigma^\star_2 \leq 0,  \varsigma^\star_3 \geq 0$, $\Phi_{23}^\alpha=0$ hence
\be
d''_\IC= 1- \cQ(\vec N)  + \frac{ 2 \cC }{ \cA + \cB + \cC } \cN_2  \varsigma^\star_2 -  \frac{ 2 \cB }{ \cA + \cB + \cC } \cN_3   \varsigma^\star_3 
\ee
\end{itemize}
where $\cQ$ is the quadratic form
\be
\cQ(\vec N) =\sum_{i=1}^r N_i^2  -  \frac{\cA + \cB -\cC}{ \cA + \cB + \cC } c \,\cN_1 \cN_2   
-   \frac{ \ \cB + \cC-\cA}{ \cA + \cB + \cC } a\, \cN_2 \cN_3 -   \frac{\cC+ \cA-  \cB }{ \cA + \cB + \cC } 
b\, \cN_3 \cN_1
\ee
To show that the quadratic form $\cQ$ is positive,  we analyze the eigenvalue equation
\be
\begin{matrix} \uparrow \\ \alpha \\ \downarrow \\ \uparrow \\ \beta \\ \downarrow \\ \uparrow \\ \gamma \\ \downarrow  \end{matrix}
\begin{pmatrix}
 \begin{matrix} 1 & \cdots & 0\\ \vdots & \ddots & \vdots \\ 0 & \cdots & 1 \end{matrix} &  \begin{matrix}  &  & \\  & -\frac{c}{2} \frac{\cA + \cB -\cC}{ \cA + \cB + \cC } &  \\  &  &  \end{matrix} & 
 \begin{matrix}  &  & \\  & -\frac{b}{2} \frac{\cC+ \cA-  \cB }{ \cA + \cB + \cC }  &  \\  &  &  \end{matrix} \\
 \begin{matrix}  &  & \\  & -\frac{c}{2} \frac{\cA + \cB -\cC}{ \cA + \cB + \cC }  &  \\  &  &  \end{matrix} &  \begin{matrix} 1 & \cdots & 0\\ \vdots & \ddots & \vdots \\ 0 & \cdots & 1 \end{matrix} &  \begin{matrix}  &  & \\  & -\frac{a}{2} \frac{ \ \cB + \cC-\cA}{ \cA + \cB + \cC } &  \\  &  &  \end{matrix}\\
 \begin{matrix}  &  & \\  & - ]\frac{b}{2}\frac{\cC+ \cA-  \cB }{ \cA + \cB + \cC }  &  \\  &  &  \end{matrix} &  \begin{matrix}  &  & \\  & -\frac{a}{2} \frac{ \ \cB + \cC-\cA}{ \cA + \cB + \cC }&  \\  &  &  \end{matrix} &  \begin{matrix} 1 & \cdots & 0\\ \vdots & \ddots & \vdots \\ 0 & \cdots & 1 \end{matrix}
\end{pmatrix}
\cdot
\begin{pmatrix} e_1 \\ \vdots \\ e_\alpha \\ \tilde{e}_1 \\ \vdots \\ \tilde{e}_\beta \\ \bar{e}_1 \\ \vdots \\\bar{e}_\gamma
\end{pmatrix}
= \lambda \, \begin{pmatrix} e_1 \\ \vdots \\ e_\alpha \\ \tilde{e}_1 \\ \vdots \\ \tilde{e}_\beta \\ \bar{e}_1 \\ \vdots \\\bar{e}_\gamma
\end{pmatrix}
\ee
By subtracting rows, we see that the eigenvalue must be $\lambda=1$ unless the 
$e_i$'s, $\tilde e_i$'s, $\bar e_i$'s are equal within each block.
Denoting by $(e,\tilde{e}, \bar{e})$ the common values,  the system
 reduces to the three-dimensional eigenvalue problem  
\be \begin{pmatrix}
 1&-\frac{\sqrt{\cC}}{2} \frac{\cA+\cB-\cC}{\cA+\cB+\cC} 
 &  - \frac{\sqrt{\cB}}{2} \frac{\cC+\cA-\cB}{\cA+\cB+\cC}  \\
   -  \frac{\sqrt{\cC}}{2} \frac{\cA+\cB-\cC}{\cA+\cB+\cC}   & 1  
   &  -\frac{\sqrt{\cA}}{2}  \frac{\cB+\cC-\cA}{\cA+\cB+\cC} \\
 -\frac{\sqrt{\cB}}{2}   \frac{\cC+\cA-\cB}{\cA+\cB+\cC} 
 &- \frac{\sqrt{\cA}}{2} \frac{\cB+\cC-\cA}{\cA+\cB+\cC}
 & 1
\end{pmatrix}
\cdot 
 \begin{pmatrix} \sqrt{\alpha}\, e \\  
\sqrt{\beta}\, \tilde e \\ \sqrt{\gamma}\, \bar e \end{pmatrix}
= \lambda\,  \begin{pmatrix} \sqrt{\alpha}\, e \\  
\sqrt{\beta}\, \tilde e \\ \sqrt{\gamma}\, \bar e \end{pmatrix}
\ee
Under the condition \eqref{Markov2}, the determinant of the $3\times 3$ matrix vanishes 
while the determinants of the principal minors reduce to $(a,b,c)$ times $(2 \cA^2 + 2\cB^2+2\cC^2-\cA \cB\cC)$, which is positive for all cases of interest since $\frac{1}{\cA}+\frac{1}{\cB}+\frac{1}{\cC}\geq \frac14$. Hence, the matrix has two positive eigenvalues and one zero eigenvalue. 
More precisely, for  the 4 possible choices\footnote{up to permutations of $(\cA,\cB,\cC)$,
and corresponding permutations of $\frac{\cN_1}{\sqrt\alpha},\frac{\cN_2}{\sqrt\beta},\frac{\cN_3}{\sqrt\gamma}$.}
 of $(\cA,\cB,\cC)$, 
\begin{itemize}
\item $(\cA,\cB,\cC)=(9,9,9)$: the eigenvalues are $(\frac32,\frac32,0)$ with eigenvectors 
\be
(-1, 0, 1), (-1, 1, 0), (1, 1, 1)
\ee
leading to 
\be
\cQ(\vec N) = \cQ_0(\vec N) + \frac14\left(  \frac{\cN_1}{\sqrt{\alpha}} -\frac{2\cN_2}{\sqrt{\beta}}+\frac{\cN_3}{\sqrt{\gamma}}  \right)^2
+   \frac34\left( \frac{\cN_3}{\sqrt{\gamma}} -  \frac{\cN_1}{\sqrt{\alpha}}\right)^2
\ee
where 
\be
\cQ_0(\vec N) = 
\frac{1}{\alpha}\sum_{1\leq i<j\leq\alpha}( N_i-N_j)^2 +  
\frac{1}{\beta}\sum_{\alpha+1\leq i<j\leq\alpha+\beta}(N_i-N_j)^2+ 
\frac{1}{\gamma} \sum_{\alpha+\beta+1\leq i<j\leq r}(N_i-N_j)^2
\ee
\item $(\cA,\cB,\cC)=(8,8,16)$: the eigenvalues are $(2,1,0)$ with eigenvectors
\be
 \left(-1,-,\sqrt2 \right) ,  \left(-1,1,0\right) , \left(1,1,\sqrt{2}\right)
\ee
leading to 
\be
\cQ(\vec N) = \cQ_0(\vec N) + \frac12 \left(\frac{\cN_1}{\sqrt{\alpha}} + \frac{\cN_2}{\sqrt{\beta}} -  \frac{\cN_3 \sqrt2}{\sqrt{\gamma}}\right)^2
+  \frac12 \left( \frac{\cN_2}{\sqrt{\beta}} -  \frac{\cN_1}{\sqrt{\alpha}}\right)^2
\ee
\item $(\cA,\cB,\cC)=(6,12,18)$: the eigenvalues are $(2,1,0)$ with eigenvectors
\be
 \left(1,\sqrt2,-\sqrt3\right) ,  \left(-\sqrt{2},1,0\right) , 
   \left(1,\sqrt2,\sqrt3 \right) 
   \ee
leading to 
\be
\cQ(\vec N) = \cQ_0(\vec N) + \frac13 \left(\frac{\cN_1}{\sqrt{\alpha}} + \frac{\cN_2 \sqrt{2}}{\sqrt{\beta}} -  \frac{\cN_3\sqrt3}{\sqrt{\gamma}}\right)^2
+  \frac13 \left( \frac{\cN_2}{\sqrt{\beta}} -  \frac{\cN_1\sqrt2}{\sqrt{\alpha}}\right)^2
\ee
\item $(\cA,\cB,\cC)=(5,20,25)$: the eigenvalues are $(2,1,0)$ with eigenvectors
\be
 \left(1,2,-\sqrt{5}\right), (-2,1,0) ,  \left(1,2,\sqrt{5}\right)
\ee
leading to 
\be
\cQ(\vec N) = \cQ_0(\vec N) + \frac15 \left(\frac{\cN_1}{\sqrt{\alpha}} + \frac{2\cN_2}{\sqrt{\beta}} -  \frac{\cN_3\sqrt5}{\sqrt{\gamma}}\right)^2
+   \frac15\left( \frac{\cN_2}{\sqrt{\beta}} -  \frac{2\cN_1}{\sqrt{\alpha}}\right)^2
\ee
\end{itemize}
In all cases, $\cQ(\vec N)$ is positive, and degenerate along the direction   $\vec N\propto (x,\dots;y,\dots;z,\dots)$ corresponding to D0-branes. 
Thus, the expected dimension $d_\IC^\star$ is at most 1, and equal to 1
only when $\vec N\propto (x,\dots;y,\dots;z,\dots)$ corresponding to D0-branes, in which case 
$d_\IC^\star=1$. Using the  decompositions as sums of squares above, one can show that $d_\IC^\star=0$ only when $\vec N$ corresponds to a simple representation, and is strictly negative in other cases. 
We conclude that the attractor index $\Omstar(\vec N,y)$ vanishes unless  $\vec N$ corresponds to a simple representation or to a D0-brane.

\section{Toric weak Fano surfaces \label{sec_weakFano}}

In this section, we consider the 11 toric weak Fano surfaces which are not Fano. These surfaces, sometimes known as pseudo del Pezzo surfaces $PdP_k$, 
can all be constructed by a sequence of toric blow-ups from $\IP^2$ (see Figure \ref{figtoricblow}),
although the blown-up points may not be in generic position.
We use the toric system from \cite[Table 2]{hille2011exceptional} to determine the corresponding
quiver. Compared to the quiver for the del Pezzo surface $dP_k$ of the same degree, the quiver for $PdP_k$ often exhibits bidirectional arrows which are not visible from the Euler matrix, and we rely on earlier studies  \cite{Feng:2004uq,Hanany:2012hi} 
to identify them. Importantly, these
arrows are not lifted by quadratic terms in the superpotential, which is computable by brane tiling techniques.  Nonetheless,  since the blow-up formula is not sensitive to the genericity of the
blown-up points, and since the flow tree formula is blind  to the existence of bidirectional arrows, 
we expect that the VW invariants for $PdP_k$ are identical to those for $dP_k$, up to a change of polarisation and basis in $H_2(S,\IZ)$.

\subsection{$PdP_2$}
The surface $PdP_2$ arises a two-point blow-up of $\IP^2$, or a one-point blow-up 
of $\IF_2$ (see e.g. \cite[Fig. 4]{Morrison:2012js}).
The toric fan consists of 5 vectors, 
\be
v_i = 
\begin{pmatrix}
1 & -1 & -1 & -1 & 0 \\  
0 & 1 & 0 & -1 & -1 
\end{pmatrix}
\ee
The corresponding divisors $D_1,\dots D_5$ satisfy the linear relations 
\be
D_1 = D_2+D_3+D_4\ ,\quad D_2 = D_4 + D_5 
\ee
and have self-intersection  (corresponding to row $5b$ in \cite[Table 1]{hille2011exceptional})
\be
a_i = D_i^2 = ( 1, 0, -2, -1, -1)\ .
\ee
The products $c_1(S)\cdot D_i=a_1+2$ take the values $(3,2,0,1,1)$, so the canonical divisor is nef but not ample. Noting that the vectors $v_1,v_2,v_5$ generate the toric fan of $\IP^2$, 
we can identify 
\be
D_1=H, \ D_2=H-C_1, \ D_3=C_1-C_2, \ D_4 = C_2, \ D_5=H-C_1-C_2
\ee
where $H$ is the pull-back of  the hyperplane class on $\IP^2$ and 
 $C_i$ are the two exceptional divisors, such that $H^2=1, C_i\cdot C_j = -\delta_{ij}, H\cdot C_i=0$. 

\medskip

Following  \cite[Table 2, row 5b]{hille2011exceptional} (after a cyclic permutation to the right) let us consider the exceptional collection \eqref{toricol}
constructed from the toric system 
\be
\tilde D_i = ( H - C_2, H- C_1, C_1 , H-C_1-C_2, C_2)
\ee
The Chern vectors of the objects $E^i$ and dual objects $E_i$ are
\be
\begin{array}{ccl}
 \gamma^1 &=& \[1,\(0,0,0\),0\] \\
 \gamma^2 &=& \[1,\(1,0,-1\),0\] \\
 \gamma^3 &=& \[1,\(2,-1,-1\),1\] \\
 \gamma^4 &=& \[1,\(2,0,-1\),\frac{3}{2}\] \\
 \gamma^5 &=& \[1,\(3,-1,-2\),2\] \\
\end{array}
\ ,\quad 
\begin{array}{ccl}
 \gamma _1 &=& \[1,\(0,0,0\),0\] \\
 \gamma _2 &=& \[-1,\(1,0,-1\),0\] \\
 \gamma _3 &=& \[-1,\(0,-1,1\),1\] \\
 \gamma _4 &=& \[0,\(-1,1,1\),\frac{1}{2}\] \\
 \gamma _5 &=& \[1,\(0,0,-1\),-\frac{1}{2}\] \\
\end{array}
\ee 
The resulting Euler form 
\be
S=\left(
\begin{array}{ccccc}
 1 & 2 & 4 & 5 & 6 \\
 0 & 1 & 2 & 3 & 4 \\
 0 & 0 & 1 & 1 & 2 \\
 0 & 0 & 0 & 1 & 1 \\
 0 & 0 & 0 & 0 & 1 \\
\end{array}
\right)\ ,\quad
S^\vee=
\left(
\begin{array}{ccccc}
 1 & -2 & 0 & 1 & 1 \\
 0 & 1 & -2 & -1 & 1 \\
 0 & 0 & 1 & -1 & -1 \\
 0 & 0 & 0 & 1 & -1 \\
 0 & 0 & 0 & 0 & 1 \\
\end{array}
\right)
\ ,\quad
\kappa=
\left(
\begin{array}{ccccc}
 0 & 2 & 0 & -1 & -1 \\
 -2 & 0 & 2 & 1 & -1 \\
 0 & -2 & 0 & 1 & 1 \\
 1 & -1 & -1 & 0 & 1 \\
 1 & 1 & -1 & -1 & 0 \\
\end{array}
\right)
\ee
 turns out to be identical to the one for $dP_2$ model II  in \eqref{SdP2II}.
Comparing the adjacency matrix with the quivers in \cite[(3.4)]{Feng:2004uq}
and model 11 in \cite{Hanany:2012hi}, one finds agreement provided one includes an additional  bidirectional arrow between nodes 1 and 3,
consistent with the vanishing entry $S^\vee_{13}=0$. The latter arises due to a cancellation
between $\Ext^1(E_3^\vee,E_1^\vee)$ and $\Ext^2(E_3^\vee,E_1^\vee)$. Note that the superpotential, obtained from the brane tiling in
\cite[(13.1)]{Hanany:2012hi}, does not contain any quadratic terms which would lift 
the bidirectional arrow (this is presumably related to the fact that $PdP_2$ is a 
non-generic blow-up of $\IP^2$).  However,  for each of the internal perfect matchings, either
$\Phi_{31}$ or $\Phi_{13}=0$, so the dimension of quiver moduli space is unaffected, and the 
attractor indices {\bp should} vanish by the same argument as for  $dP_2$ model II.

\subsection{$PdP_{3a}$}
The cone over $PdP_{3a}$ is the orbifold $\IC^3/\IZ_6$ with orbifold action $(1,2,3)$.
The toric fan consists of 6 vectors, 
\be
v_i = 
\begin{pmatrix}
1 & 0& -1 & -1 & -1 & -1 \\  
0 &1 &  2 & 1 & 0 & -1 
\end{pmatrix}
\ee
The corresponding divisors $D_1,\dots D_6$ satisfy the linear relations 
\be
D_1 = D_3+D_4+D_5+D_6\ ,\quad D_6 = D_2+2D_3 + D_4 
\ee
and have self-intersection (corresponding to row $6d$ in \cite[Table 1]{hille2011exceptional})
\be
a_i= D_i^2 = ( 1,-2,-1,-2,-2,0 )
\ee
The products $a_i+2=c_1(S)\cdot D_i$ are $(3,0,1,0,0,2)$, so the canonical divisor is nef but not ample. Noting that the vectors $v_1,v_2,v_6$ span the toric fan of $\IP^2$, we can identify
\be
D_1=H, \ D_2=H-C_1-C_2-C_3, \ D_3= C_3, \ D_4=C_2-C_3, \ D_5 = C_1-C_2, \ D_6=H-C_1
\ee
where $H$ is the pull-back of  the hyperplane class on $\IP^2$ and 
 $C_i$ are the two exceptional divisors. 

\medskip

Following  \cite[Table 2, row 6d]{hille2011exceptional} (identifying  $P=H-C_3$, $Q=H$
where $P,Q$ are the pull-back of the toric divisors $C+F,C$ of $\IF_1$)  let us consider the exceptional collection \eqref{toricol}
constructed from the toric system 
\be
\label{toricsysPdP3a}
\tilde D_i = ( H- C_1-C_3, C_1 , H-C_1-C_2, C_2, H - C_2-C_3, C_3)
\ee
The Chern vectors of the objects $E^i$ and dual objects $E_i$ are now
\be
\begin{array}{ccl}
 \gamma^1 &=& \[1,\(0,0,0,0\),0\] \\
 \gamma^2 &=& \[1,\(1,-1,0,-1\),-\frac{1}{2}\] \\
 \gamma^3 &=& \[1,\(1,0,0,-1\),0\] \\
 \gamma^4 &=& \[1,\(2,-1,-1,-1\),\frac{1}{2}\] \\
 \gamma^5 &=& \[1,\(2,-1,0,-1\),1\] \\
 \gamma^6 &=& \[1,\(3,-1,-1,-2\),\frac{3}{2}\] \\
\end{array}
\ ,\quad
\begin{array}{ccl}
 \gamma _1 &=& \[1,\(0,0,0,0\),0\] \\
 \gamma _2 &=& \[0,\(1,-1,0,-1\),-\frac{1}{2}\] \\
 \gamma _3 &=& \[-1,\(0,1,0,0\),\frac{1}{2}\] \\
 \gamma _4 &=& \[-1,\(0,0,-1,1\),1\] \\
 \gamma _5 &=& \[0,\(-1,0,1,1\),\frac{1}{2}\] \\
 \gamma _6 &=& \[1,\(0,0,0,-1\),-\frac{1}{2}\] \\
\end{array}
\ee 
with slopes $0,1,2,3,4,5$ and $0,\infty,-1,0,\infty,-1$. 
The resulting Euler form 
\be
S=
\left(
\begin{array}{cccccc}
 1 & 1 & 2 & 3 & 4 & 5 \\
 0 & 1 & 1 & 2 & 3 & 4 \\
 0 & 0 & 1 & 1 & 2 & 3 \\
 0 & 0 & 0 & 1 & 1 & 2 \\
 0 & 0 & 0 & 0 & 1 & 1 \\
 0 & 0 & 0 & 0 & 0 & 1 \\
\end{array}
\right)\ ,
S^\vee=
\left(
\begin{array}{cccccc}
 1 & -1 & -1 & 0 & 1 & 1 \\
 0 & 1 & -1 & -1 & 0 & 1 \\
 0 & 0 & 1 & -1 & -1 & 0 \\
 0 & 0 & 0 & 1 & -1 & -1 \\
 0 & 0 & 0 & 0 & 1 & -1 \\
 0 & 0 & 0 & 0 & 0 & 1 \\
\end{array}
\right)\ ,
\kappa=
\left(
\begin{array}{cccccc}
 0 & 1 & 1 & 0 & -1 & -1 \\
 -1 & 0 & 1 & 1 & 0 & -1 \\
 -1 & -1 & 0 & 1 & 1 & 0 \\
 0 & -1 & -1 & 0 & 1 & 1 \\
 1 & 0 & -1 & -1 & 0 & 1 \\
 1 & 1 & 0 & -1 & -1 & 0 \\
\end{array}
\right)
\ee
turns out to be identical to the one  \eqref{EulerdP3HP} for $dP_3$ model I. 
The adjacency matrix agrees with the quiver for weighted projective space $\IP(1,2,3)$ 
in \cite[\S 4.2]{Herzog:2005sy}, or 
with model 7 in  \cite{Hanany:2012hi}, provided one includes 
three additional bidirectional arrows consistent with the vanishing 
of $S^\vee_{14}, S^\vee_{25}, S^\vee_{36}$. Again, the superpotential in \cite[(9.1)]{Hanany:2012hi} does not contain any quadratic terms which would lift 
the bidirectional arrows,  but  for each of the internal perfect matchings, one arrow in each of the bidirectional pairs vanishes,  so the dimension of quiver moduli space is unaffected, and the 
attractor indices {\bp should}vanish by the same argument as for  $dP_3$ model I.

\subsection{$PdP_{3b}$}
The toric fan consists of 6 vectors, 
\be
v_i = 
\begin{pmatrix}
1 & 0& -1 & -1 & -1 & 0 \\  
0 &1 &  1 & 0 & -1 & -1 
\end{pmatrix}
\ee
The corresponding divisors $D_1,\dots D_6$ satisfy the linear relations 
\be
D_1 = D_3+D_4+D_5\ ,\quad D_2+D_3=D_5+ D_6  
\ee
and have self-intersection (corresponding to row $6b$ in \cite[Table 1]{hille2011exceptional})
\be
D_i^2 = ( 0,-1,-1,-2,-1,-1 )
\ee
The products $c_1(S)\cdot D_i$ are $(2,1,1,0,1,1)$, so the canonical divisor is nef but not ample.
Noting that the vectors $v_1,v_3,v_6$ span the toric diagram of $\IP_2$, we may identify
\be
D_1=H-C_1, \ D_2=C_1, \ D_3= H-C_1-C_2\ D_4=C_2-C_3, \ D_5 = C_3, \ D_6=H-C_2-C_3
\ee
where $H$ is the pull-back of  the hyperplane class on $\IP^2$ and 
 $C_i$ are the two exceptional divisors. 
 
 \medskip
 
According to   \cite[Table 2, row 6b]{hille2011exceptional} a toric system
is given by
\be
\label{toricsysPdP3b}
\tilde D_i = ( H- C_1-C_3, C_1 , H-C_1-C_2, C_2, H - C_2-C_3, C_3)
\ee
This coincides with \eqref{toricsysPdP3a}, so the quiver is again identical to model I of $dP_3$  \eqref{EulerdP3HP},
possibly up to bidirectional arrows. Comparing with  model in model $9a$ in  \cite{Hanany:2012hi},
 with \cite[\S 3.3.1]{Feng:2004uq}, phase III  (see also \cite[\S 8]{Franco:2005rj}, model B), 
 we see that the relevant quiver has
one additional bidirectional arrow corresponding to the one of the vanishing entries $S^\vee_{14},S^\vee_{25}, S^\vee_{36}$ in the
adjacency matrix  \eqref{EulerdP3HP}.  The superpotential in \cite[(9.1)]{Hanany:2012hi} does not contain any quadratic terms which would lift 
the bidirectional arrows,  but  for each of the internal perfect matchings, one arrow in the bidirectional pair vanishes,  so the dimension of quiver moduli space is unaffected, and the 
attractor indices {\bp should} vanish by the same argument as for  $dP_3$ model I.

\subsection{$PdP_{3c}$}
This is also known as a $\IZ_2$ orbifold of the suspended pinched point singularity.
The toric fan consists of 6 vectors, 
\be
v_i = 
\begin{pmatrix}
1 & 0& -1 & -1 & -1 & 0 \\  
0 &1 &  2 & 1 & 0 & -1 
\end{pmatrix}
\ee
The corresponding divisors $D_1,\dots D_6$ satisfy the linear relations 
\be
D_1 = D_3+D_4+D_5\ ,\quad D_2+2D_3+D_4=D_6  
\ee
and have self-intersection (corresponding to row $6c$ in \cite[Table 1]{hille2011exceptional})
\be
D_i^2 = ( 0,-2,-1,-2,-1,0)
\ee
The products $c_1(S)\cdot D_i$ are $(2,0,1,0,1,2)$, 
so the canonical divisor is nef but not ample. 
Noting that the vectors $v_1,v_4,v_6$ span the toric fan of $\IP^2$, we may identify
\be
D_1=H-C_2,\ D_2=C_2-C_3, \ D_3=C_3, \ D_4=H-C_1-C_2-C_3, \ D_5=C_1, \ D_6=H-C_1\ .
\ee
According to   \cite[Table 2, row 6c]{hille2011exceptional}  a toric system
is  again given by \eqref{toricsysPdP3a}, so the quiver is again identical to model I of $dP_3$,  up to bidirectional arrows. Comparing with   \cite[\S 3.3.2]{Feng:2004uq} phase I and
model $8a$ in  \cite{Hanany:2012hi} (taking into
account the fact that the bidirectional arrows from between $1,2$ and $3,4$ should actually be oriented arrows $2\to 1, 3\to 4$),  we see that the relevant quiver has
one additional bidirectional arrow corresponding to one of the vanishing entries $S^\vee_{14},S^\vee_{25}, S^\vee_{36}$ in the
adjacency matrix  \eqref{EulerdP3HP}.  The superpotential in \cite[(8.1)]{Hanany:2012hi} does not contain any quadratic terms which would lift 
the bidirectional arrows,  but  for each of the internal perfect matchings, one arrow in the bidirectional pair vanishes,  so the dimension of quiver moduli space is unaffected, and the 
attractor indices {\bp should} vanish by the same argument as for  $dP_3$ model I.

\subsection{$PdP_{4a}$}
The toric fan consists of 7 vectors, 
\be
v_i = 
\begin{pmatrix}
1 & 0& -1 & -1 & -1 & 0 & 1\\  
0 &1 &  2 & 1 & 0 & -1 & -1
\end{pmatrix}
\ee
The corresponding divisors $D_1,\dots D_7$ satisfy the linear relations 
\be
D_1+D_7= D_3+D_4+D_5\ ,\quad D_2+2D_3+D_4=D_6  +D_7
\ee
and have self-intersection (corresponding to row $7a$ in \cite[Table 1]{hille2011exceptional})
\be
D_i^2 = (-1,-2,-1,-2,-1,-1,-1)
\ee
The products $c_1(S)\cdot D_i$ are $(1,0,1,0,1,1,1)$, so the canonical divisor is nef but not ample. 
Noting that the vectors $v_3,v_6,v_7$ span the toric fan of $\IP_2$, we may identify
\be
\begin{split} 
D_1=C_4, \ D_2=C_1-C_4, \ D_3=H-C_1-C_2, \ D_4=C_2-C_3, \\
 D_5=C_3, \ D_6=H-C_2-C_3,\ 
D_7=H-C_1-C_4\ .
\end{split}
\ee
According to   \cite[Table 2, row 7a]{hille2011exceptional} a toric system
is given by
\be
\label{toricsysPdP4a}
\tilde D_i = (H-C_1-C_2, C_2, C_1-C_2,H-C_1-C_3-C_4, C_4,C_3-C_4,H-C_3)
\ee
The Chern vectors for the projective and simple objects are then
\be
\begin{array}{ccl}
 \gamma^1 &=& \[1,\(0,0,0,0,0\),0\] \\
 \gamma^2 &=& \[1,\(1,-1,-1,0,0\),-\frac{1}{2}\] \\
 \gamma^3 &=& \[1,\(1,-1,0,0,0\),0\] \\
 \gamma^4 &=& \[1,\(1,0,-1,0,0\),0\] \\
 \gamma^5 &=& \[1,\(2,-1,-1,-1,-1\),0\] \\
 \gamma^6 &=& \[1,\(2,-1,-1,-1,0\),\frac{1}{2}\] \\
 \gamma^7 &=& \[1,\(2,-1,-1,0,-1\),\frac{1}{2}\] \\
\end{array}
\ ,\quad
\begin{array}{ccl}
 \gamma _1 &=& \[1,\(0,0,0,0,0\),0\] \\
 \gamma _2 &=& \[0,\(1,-1,-1,0,0\),-\frac{1}{2}\] \\
 \gamma _3 &=& \[-1,\(0,0,1,0,0\),\frac{1}{2}\] \\
 \gamma _4 &=& \[-1,\(0,1,0,0,0\),\frac{1}{2}\] \\
 \gamma _5 &=& \[-1,\(1,0,0,-1,-1\),\frac{1}{2}\] \\
 \gamma _6 &=& \[1,\(-1,0,0,0,1\),0\] \\
 \gamma _7 &=& \[1,\(-1,0,0,1,0\),0\] \\
\end{array}
\ee
The Euler matrix has a 4-block structure,
\be
S=\left(
\begin{array}{cccc}
 1 & 1 & 2 & 3 \\
 0 & 1 & 1  & 2 \\
 0 & 0 &  \mathbb{1}_3  & 1 \\
 0 & 0 &  0 &   \mathbb{1}_2   \\
\end{array}
\right),\ 
S^\vee=
\left(
\begin{array}{cccc}
1 & -1 & -1& 2 \\
 0 &1& -1  & 1 \\
 0 & 0 &  \mathbb{1}_3 & -1 \\
 0 & 0 &  0 &   \mathbb{1}_2   \\
\end{array}
\right),\ 
\kappa=
\left(
\begin{array}{cccc}
0 & 1 & 1 & -2 \\
 -1 &0 & 1  & -1 \\
 -1 & -1 &  \mathbb{0}_3  & 1 \\
 2 & 1 &  -1 &  \mathbb{0}_2  \\
\end{array}
\right)
\ee
corresponding to the same quiver as model 6, phase $c$ in 
 \cite{Hanany:2012hi}, where the superpotential
can be found. There are no bidirectional arrows in this case.

\subsection{$PdP_{4b}$}
The toric fan consists of 7 vectors, 
\be
v_i = 
\begin{pmatrix}
1 & 0& -1 & -1 & -1 & -1 & 0\\  
0 &1 &  2 & 1 & 0 & -1 & -1
\end{pmatrix}
\ee
The corresponding divisors $D_1,\dots D_7$ satisfy the linear relations 
\be
D_1= D_3+D_4+D_5+D_6\ ,\quad D_2+2D_3+D_4=D_6  +D_7
\ee
and have self-intersection (corresponding to row $7b$ in \cite[Table 1]{hille2011exceptional})
\be
D_i^2 = (0,-2,-1,-2,-2,-1,-1)
\ee
The products $c_1(S)\cdot D_i$ are $(2,0,1,0,0,1,1)$, so the canonical divisor is nef but not ample. 
Noting that the vectors $v_1,v_4,v_7$ span the toric fan of $\IP^2$, one may identify
\be
\begin{split} 
D_1=H-C_3, \ D_2=C_3-C_4, \ D_3=C_4, \ D_4=H-C_1-C_3-C_4, \\
 D_5=C_1-C_2, \ D_6=C_2,\ 
D_7=H-C_1-C_2\ .
\end{split}
\ee
 According to   \cite[Table 2, row 7b]{hille2011exceptional} a toric system
is given by
\be
\label{toricsysPdP4b}
\tilde D_i = (H-C_1-C_3, C_3, C_1-C_3,H-C_1-C_2-C_4, C_4,C_2-C_4,H-C_2)
\ee
This is identical to \eqref{toricsysPdP4a} up to an exchange of $C_2$ and $C_3$,
so we get the same quiver, possibly up to bidirectional arrows. In contrast, the 
quiver for model 5 in \cite{Hanany:2012hi} clusters into 4 blocks of size 2,2,2,1 (corresponding 
to the nodes labelled $(2,5), (3,6), (4,7), 1$) upon ignoring bi-directional arrows, 
and it is not clear how the two quivers are related.

\subsection{$PdP_{5a}$}
The cone over $PdP_{5a}$ is a $\IZ_2\times \IZ_2$ orbifold of the conifold, and can be viewed as 
a 5-point blow-up of $\IP^2$ in non-generic position, or as a 4-point blow-up of $\IP^1\times \IP^1$.
The toric fan consists of 8 vectors, 
\be
v_i = 
\begin{pmatrix}
1 & 1& 0 & -1 & -1 & -1 & 0 & 1 \\  
0 &1 &  1 & 1 & 0 & -1 & -1 & -1
\end{pmatrix}
\ee
The corresponding divisors $D_1,\dots D_8$ satisfy the linear relations 
\be
D_1+D_2+D_8=D_4+D_5+D_6 ,\quad D_2+D_3+D_4=D_6+D_7 +D_8
\ee
and have self-intersection (corresponding to row $8a$ in \cite[Table 1]{hille2011exceptional})
\be
D_i^2 = (-2,-1,-2,-1,-2,-1,-2,-1)
\ee
The products $c_1(S)\cdot D_i$ are $(0,1,0,1,0,1,0,1)$, 
so the canonical divisor is nef but not ample.
Since $v_1,v_3,v_5,v_7$ span the toric diagram of $\IP^1\times \IP^1$,
we may identify 
\be
\begin{split}
D_1=F-C_1-C_4,\ D_2=C_4, \ D_3=C-C_3-C_4, \ D_4 = C_3 \\
D_5 =F-C_2-C_3,\ D_6=C_2, \ D_7=C-C_1-C_2, \ D_8 = C_1
\end{split} 
\ee
where $F,C$ are the pull-back of the generators of $\IP^1\times \IP^1$,
with $F^2=C^2=0, F\cdot C=1$
and $C_1,\dots C_4$ are the exceptional divisors. 

According to   \cite[Table 2, row 8a]{hille2011exceptional} a toric system
is given by
\be
\label{toricsysPdP5a}
\tilde D_i = ( F-C_1-C_4,C_1,C-C_1-C_2,C_2,F-R_2-R_3,R_3,Q-R_3-R_4,R_4)
\ee
The Chern vectors are 
\be
\begin{array}{ccl}
 \gamma^1 &=& \[1,\(0,0,0,0,0,0\),0\] \\
 \gamma^2 &=& \[1,\(0,1,-1,0,0,-1\),-1\] \\
 \gamma^3 &=& \[1,\(0,1,0,0,0,-1\),-\frac{1}{2}\] \\
 \gamma^4 &=& \[1,\(1,1,-1,-1,0,-1\),-\frac{1}{2}\] \\
 \gamma^5 &=& \[1,\(1,1,-1,0,0,-1\),0\] \\
 \gamma^6 &=& \[1,\(1,2,-1,-1,-1,-1\),0\] \\
 \gamma^7 &=& \[1,\(1,2,-1,-1,0,-1\),\frac{1}{2}\] \\
 \gamma^8 &=& \[1,\(2,2,-1,-1,-1,-2\),\frac{1}{2}\] \\
\end{array}
\ ,\quad 
\begin{array}{ccl}
 \gamma _1 &=& \[1,\(0,0,0,0,0,0\),0\] \\
 \gamma _2 &=& \[1,\(0,1,-1,0,0,-1\),-1\] \\
 \gamma _3 &=& \[-1,\(0,0,1,0,0,0\),\frac{1}{2}\] \\
 \gamma _4 &=& \[-1,\(1,0,0,-1,0,0\),\frac{1}{2}\] \\
 \gamma _5 &=& \[-1,\(0,-1,0,1,0,1\),1\] \\
 \gamma _6 &=& \[-1,\(0,0,0,0,-1,1\),1\] \\
 \gamma _7 &=& \[1,\(-1,0,0,0,1,0\),-\frac{1}{2}\] \\
 \gamma _8 &=& \[1,\(0,0,0,0,0,-1\),-\frac{1}{2}\] \\
\end{array}
\ee
The Euler matrix has a four-block structure, 
\be
S=\left(
\begin{array}{cccc}
  \mathbb{1}_2  & 1 & 2 & 3 \\
 0 & \mathbb{1}_2 & 1  & 2 \\
 0 & 0 &  \mathbb{1}_2  & 1 \\
 0 & 0 &  0 &  \mathbb{1}_2  \\
\end{array}
\right),\ 
S^\vee=
\left(
\begin{array}{cccc}
  \mathbb{1}_2  & -1 & 0 & 1 \\
 0 & \mathbb{1}_2 & -1  & 0 \\
 0 & 0 &  \mathbb{1}_2  & -1 \\
 0 & 0 &  0 &  \mathbb{1}_2  \\
\end{array}
\right),\ 
\kappa=
\left(
\begin{array}{cccc}
  \mathbb{0}_2  & 1 & 0 & -1 \\
 -1 & \mathbb{0}_2 & 1  & 0 \\
 0 & -1 &  \mathbb{0}_2  & 1 \\
 1 & 0 &  -1 &  \mathbb{0}_2  \\
\end{array}
\right)
\ee
This agrees with the quiver for model $4a$ in \cite{Hanany:2012hi}, which has no bidirectional arrows.

\subsection{$PdP_{5b}$}
The cone over $PdP_{5b}$ is the orbifold $L_{131}/\IZ_2$ with orbifold action $(0,1,1,1)$ .
The toric fan consists of 8 vectors, 
\be
v_i = 
\begin{pmatrix}
1 & 0& -1 & -1 & -1 & -1 & 0 & 1 \\  
0 &1 &  2 & 1 & 0 & -1 & -1 & -1
\end{pmatrix}
\ee
The corresponding divisors $D_1,\dots D_8$ satisfy the linear relations 
\be
D_1+D_8=D_3+D_4+D_5+D_6+D_7\ ,\quad D_2+2D_3+D_4=D_6+D_7 +D_8
\ee
and have self-intersection (corresponding to row $8b$ in \cite[Table 1]{hille2011exceptional})
\be
D_i^2 = (-1,-2,-1,-2,-2,-1,-2,-1)
\ee
The products $c_1(S)\cdot D_i$ are $(1,0,1,0,0,1,0,1)$, so the canonical divisor is nef but not ample. 
Since $v_1,v_4,v_7$ span the toric diagram of $\IP^2$,
we may identify 
\be
\begin{split}
D_1=H-C_2-C_3-C_4,\ D_2=C_3, \ D_3=C_2-C_3, \ D_4 = H-C_1-C_2 \\
D_5 =C_1-C_5,\ D_6=C_5, \ D_7=H-C_1-C_4-C_5, \ D_8 = C_4
\end{split} 
\ee
According to \cite[Table 2, row 8b]{hille2011exceptional} a toric system
is given by
\be
\label{toricsysPdP5b}
\tilde D_i = ( H-C_1-C_2-C_4, C_4,C_2-C_4,C_1-C_2,H-C_1-C_3,C_3-C_5,C_5,H-C_3-C_5)
\ee
The Chern vectors are 
\be
\begin{array}{ccl}
 \gamma^1 &=& \[1,\(0,0,0,0,0,0\),0\] \\
 \gamma^2 &=& \[1,\(1,-1,-1,0,-1,0\),-1\] \\
 \gamma^3 &=& \[1,\(1,-1,-1,0,0,0\),-\frac{1}{2}\] \\
 \gamma^4 &=& \[1,\(1,-1,0,0,-1,0\),-\frac{1}{2}\] \\
 \gamma^5 &=& \[1,\(1,0,-1,0,-1,0\),-\frac{1}{2}\] \\
 \gamma^6 &=& \[1,\(2,-1,-1,-1,-1,0\),0\] \\
 \gamma^7 &=& \[1,\(2,-1,-1,0,-1,-1\),0\] \\
 \gamma^8 &=& \[1,\(2,-1,-1,0,-1,0\),\frac{1}{2}\] \\
\end{array}
\ ,\quad
\begin{array}{ccl}
 \gamma _1 &=& \[1,\(0,0,0,0,0,0\),0\] \\
 \gamma _2 &=& \[1,\(1,-1,-1,0,-1,0\),-1\] \\
 \gamma _3 &=& \[-1,\(0,0,0,0,1,0\),\frac{1}{2}\] \\
 \gamma _4 &=& \[-1,\(0,0,1,0,0,0\),\frac{1}{2}\] \\
 \gamma _5 &=& \[-1,\(0,1,0,0,0,0\),\frac{1}{2}\] \\
 \gamma _6 &=& \[0,\(0,0,0,-1,0,0\),\frac{1}{2}\] \\
 \gamma _7 &=& \[0,\(0,0,0,0,0,-1\),\frac{1}{2}\] \\
 \gamma _8 &=& \[1,\(-1,0,0,1,0,1\),-\frac{1}{2}\] \\
\end{array}
\ee
The Euler matrix has a four-block structure, 
\be
S=\left(
\begin{array}{cccc}
  \mathbb{1}_2  & 1 & 2 & 3 \\
 0 & \mathbb{1}_3 & 1  & 2 \\
 0 & 0 &  \mathbb{1}_2  & 1 \\
 0 & 0 &  0 &  1  \\
\end{array}
\right),\ 
S^\vee=
\left(
\begin{array}{cccc}
  \mathbb{1}_2  & -1 & 1& 1 \\
 0 & \mathbb{1}_3 & -1  & 0 \\
 0 & 0 &  \mathbb{1}_2  & -1 \\
 0 & 0 &  0 & 1  \\
\end{array}
\right),\ 
\kappa=
\left(
\begin{array}{cccc}
  \mathbb{0}_2  & 1 & -1 & -1 \\
 -1 & \mathbb{0}_3 & 1  & 0 \\
 1 & -1 &  \mathbb{0}_2  & 1 \\
 1 & 0 &  -1 &  \mathbb{0}_1  \\
\end{array}
\right)
\ee
This agrees with the quiver for model 3b in \cite{Hanany:2012hi}, which reduces to a 4-block collection upon ignoring the  bidirectional arrow (with the nodes labelled 2,5,8  in \cite{Hanany:2012hi} clustered into a single block of size 3). Again, the superpotential in \cite[(5.11)]{Hanany:2012hi} does not contain any quadratic terms which would lift 
the bidirectional arrow,  but  for each of the internal perfect matchings, one arrow in  the bidirectional pair vanishes. We have not demonstrated that the attractor indices vanish,
but this should be doable by the same techniques as for the other models in this paper.

\subsection{$PdP_{5c}$}
The cone over $PdP_{5c}$ is the orbifold $\IC^3/\IZ_4\times \IZ_2$ with orbifold action $(1,0,3)\times(0,1,1)$, and can be viewed as 
a 5-point blow-up of $\IP^2$ in non-generic position, or as a 4-point blow-up of $\IP^1\times \IP^1$.
The toric fan consists of 8 vectors, 
\be
v_i = 
\begin{pmatrix}
1 & 0& -1 & -1 & -1 & -1 & -1 & 0 \\  
-1 &1 &  2 & 1 & 0 & -1 & -2 & -1
\end{pmatrix}
\ee
The corresponding divisors $D_1,\dots D_8$ satisfy the linear relations 
\be
D_1=D_3+D_4+D_5+D_6+D_7\ ,\quad D_2+3D_3+2D_4+D_5=D_1+D_7 +D_8
\ee
and have self-intersection (corresponding to row $8c$ in \cite[Table 1]{hille2011exceptional})
\be
D_i^2 = (0,-2,-1,-2,-2,-2,-1,-2)
\ee
The products $c_1(S)\cdot D_i$ are $(2,0,1,0,0,0,1,0)$, so the canonical divisor is nef but not ample.
Since $v_1,v_2,v_5,v_8$ span the toric diagram of $\IP^1\times \IP^1$,
we may identify 
\be
\begin{split}
D_1=C, \ D_2=F-C_3-C_4, \ D_3=C_4, \ D_4 = C_3-C_4 \\
D_5 =C-C_1-C_2-C_3,\ D_6=C_2, \ D_7=C_1-C_2, \ D_8 = F-C_1
\end{split} 
\ee
where $F,C$ are the pull-back of the generators of $\IP^1\times \IP^1$,
with $F^2=C^2=0, F\cdot C=1$
and $C_1,\dots C_4$ are the exceptional divisors.

According to   \cite[Table 2, row 8c]{hille2011exceptional} a toric system
is given by
\be
\label{toricsysPdP5a}
\tilde D_i = ( F-C_1-C_4,C_4,C_1-C_4,C+F-C_1-C_3,C_3-C_2,C_2,F-R_2-R_3,C-F)
\ee
The Chern vectors are 
\be
\begin{array}{ccl}
 \gamma^1 &=& \[1,\(0,0,0,0,0,0\),0\] \\
 \gamma^2 &=& \[1,\(0,1,-1,0,0,-1\),-1\] \\
 \gamma^3 &=& \[1,\(0,1,-1,0,0,0\),-\frac{1}{2}\] \\
 \gamma^4 &=& \[1,\(0,1,0,0,0,-1\),-\frac{1}{2}\] \\
 \gamma^5 &=& \[1,\(1,2,-1,0,-1,-1\),\frac{1}{2}\] \\
 \gamma^6 &=& \[1,\(1,2,-1,-1,0,-1\),\frac{1}{2}\] \\
 \gamma^7 &=& \[1,\(1,2,-1,0,0,-1\),1\] \\
 \gamma^8 &=& \[1,\(1,3,-1,-1,-1,-1\),1\] \\
\end{array}
\ ,\quad
\begin{array}{ccl}
 \gamma _1 &=& \[1,\(0,0,0,0,0,0\),0\] \\
 \gamma _2 &=& \[1,\(0,1,-1,0,0,-1\),-1\] \\
 \gamma _3 &=& \[-1,\(0,0,0,0,0,1\),\frac{1}{2}\] \\
 \gamma _4 &=& \[-1,\(0,0,1,0,0,0\),\frac{1}{2}\] \\
 \gamma _5 &=& \[-1,\(1,-1,0,0,-1,0\),\frac{3}{2}\] \\
 \gamma _6 &=& \[-1,\(1,-1,0,-1,0,0\),\frac{3}{2}\] \\
 \gamma _7 &=& \[1,\(-1,0,0,1,1,0\),-1\] \\
 \gamma _8 &=& \[1,\(-1,1,0,0,0,0\),-1\] \\
\end{array}
\ee
leading to the 4-block form
\be
S=\left(
\begin{array}{cccc}
  \mathbb{1}_2  & 1 & 3 & 4 \\
 0 & \mathbb{1}_2 & 2  & 3 \\
 0 & 0 &  \mathbb{1}_2  & 1 \\
 0 & 0 &  0 &  \mathbb{1}_2  \\
\end{array}
\right),\ 
S^\vee=
\left(
\begin{array}{cccc}
  \mathbb{1}_2  & -1 & 1 & 0 \\
 0 & \mathbb{1}_2 & -2  & 1 \\
 0 & 0 &  \mathbb{1}_2  & -1 \\
 0 & 0 &  0 &  \mathbb{1}_2  \\
\end{array}
\right),\ 
\kappa=
\left(
\begin{array}{cccc}
  \mathbb{0}_2  & 1 &- 1 & 0 \\
 1 & \mathbb{0}_2 & 2  & -1 \\
 -1 & -2 &  \mathbb{0}_2  & 1 \\
 0 & 1 &  -1 &  \mathbb{0}_2  \\
\end{array}
\right)
\ee
This agrees with the quiver for model 2 in \cite{Hanany:2012hi}, which clusters into 4 blocks of size 2 (the nodes labelled (1,6), (2,5), (3,8), (4,7) in \cite{Hanany:2012hi})  upon ignoring bidirectional arrows.
Again, the superpotential in \cite[(4.1)]{Hanany:2012hi} does not contain any quadratic terms which would lift 
the bidirectional arrow,  but  for each of the internal perfect matchings, one arrow in  the bidirectional pair vanishes. We have not demonstrated that the attractor indices vanish,
but this should be doable by the same techniques as for the other models in this paper. 

\subsection{$PdP_{6}$}
The cone over $PdP_{6}$ is the orbifold $\IC^3/\IZ_3\times \IZ_3$ with orbifold action $(1,0,2)\times(0,1,2)$ .
The toric fan consists of 9 vectors, 
\be
v_i = 
\begin{pmatrix}
2 & 1& 0 & -1 & -1 & -1 & -1 & 0 & 1 \\  
-1 &0 & 1 & 2 & 1 & 0 & -1 & -1 & -1
\end{pmatrix}
\ee
The corresponding divisors $D_1,\dots D_9$ satisfy the linear relations 
\be
2D_1+D_2+D_9=D_4+D_5+D_6+D_7\ ,\quad D_3+2D_4+D_5=D_1+D_7 +D_8+D_9
\ee
and have self-intersection (corresponding to row $9$ in \cite[Table 1]{hille2011exceptional})
\be
D_i^2 = (-1,-2,-2,-1,-2,-2,-1,-2,-2)
\ee
The products $c_1(S)\cdot D_i$ are $(1,0,0,1,0,0,1,0,0)$, 
so the canonical divisor is nef but not ample. 
Since $v_3,v_6,v_9$ span the toric diagram of $\IP^2$,
we may identify 
\be
\begin{split}
D_1=C_6,\ D_2=C_5-C_6, \ D_3=H-C_3-C_4-C_5, \ D_4 = C_4 \\
D_5 =C_3-C_4,\ D_6=H-C_2-C_3, \ D_7=C_2, \ D_8 = C_1-C_2
\end{split} 
\ee
According to \cite[Table 2, row 9]{hille2011exceptional} a toric system
is given (after a cyclic rotation to the right)  by 
\be
\label{toricsysPdP6}
\tilde D_i = ( C_5-C_2, H-C_1-C_4-C_5, C_4,C_1-C_4,H-C_1-C_3-C_6,C_3-C_6,H-C_2-C_3-C_5,C_2)
\ee
The Chern vectors for the primitive and simple objects are given by
\be
\begin{array}{ccl}
 \gamma^1 &=& \[1,\(0,0,0,0,0,0,0\),0\] \\
 \gamma^2 &=& \[1,\(0,0,-1,0,0,1,0\),-1\] \\
 \gamma^3 &=& \[1,\(1,-1,-1,0,-1,0,0\),-1\] \\
 \gamma^4 &=& \[1,\(1,-1,-1,0,0,0,0\),-\frac{1}{2}\] \\
 \gamma^5 &=& \[1,\(1,0,-1,0,-1,0,0\),-\frac{1}{2}\] \\
 \gamma^6 &=& \[1,\(2,-1,-1,-1,-1,0,-1\),-\frac{1}{2}\] \\
 \gamma^7 &=& \[1,\(2,-1,-1,-1,-1,0,0\),0\] \\
 \gamma^8 &=& \[1,\(2,-1,-1,0,-1,0,-1\),0\] \\
 \gamma^9 &=& \[1,\(3,-1,-2,-1,-1,-1,-1\),0\] \\
 \end{array}
\ ,\quad
\begin{array}{ccl}
 \gamma _1 &=& \[1,\(0,0,0,0,0,0,0\),0\] \\
 \gamma _2 &=& \[1,\(0,0,-1,0,0,1,0\),-1\] \\
 \gamma _3 &=& \[1,\(1,-1,-1,0,-1,0,0\),-1\] \\
 \gamma _4 &=& \[-2,\(0,0,1,0,1,-1,0\),\frac{3}{2}\] \\
 \gamma _5 &=& \[-2,\(0,1,1,0,0,-1,0\),\frac{3}{2}\] \\
 \gamma _6 &=& \[-2,\(1,0,1,-1,0,-1,-1\),\frac{3}{2}\] \\
 \gamma _7 &=& \[1,\(-1,0,0,0,0,1,1\),-\frac{1}{2}\] \\
 \gamma _8 &=& \[1,\(-1,0,0,1,0,1,0\),-\frac{1}{2}\] \\
 \gamma _9 &=& \[1,\(0,0,-1,0,0,0,0\),-\frac{1}{2}\] \\
\end{array}
\ee
leading to the 3-block  form identical to the one  for $dP_6$ in \eqref{SdP61},
\be
\label{SdP61}
S=\left(
\begin{array}{ccc}
  \mathbb{1}_3 & 1 & 2 \\
 0 &   \mathbb{1}_3  & 1 \\
 0 & 0 &   \mathbb{1}_3  \\
\end{array}
\right),\ 
S^\vee=\left(
\begin{array}{ccc}
   \mathbb{1}_3  & -1 & 1 \\
 0 &   \mathbb{1}_3  & -1 \\
 0 & 0 &   \mathbb{1}_3  \\
\end{array}
\right), 
\kappa=\left(
\begin{array}{ccc}
   \mathbb{0}_3  & 1 & -1 \\
 -1 &   \mathbb{0}_3  & 1 \\
 1 & -1 &   \mathbb{0}_3  \\
\end{array}
\right)
\ee
This coincides with the quiver for model 1 in \cite{Hanany:2012hi}, which has no bidirectional arrows.
The vanishing of attractor indices follows by the same arguments as for $dP_6$.

%\bibliography{combined}
%\bibliographystyle{utphys}

\providecommand{\href}[2]{#2}\begingroup\raggedright\endgroup

\end{document}